\documentclass[twocolumn,tighten,times,twocolappendix,numberedappendix]{aastex631}

\usepackage{amsmath}
\usepackage{float}
\usepackage{color}
\usepackage{url}
\usepackage{etoolbox}
\usepackage{placeins}

\usepackage{graphicx}

\usepackage[flushleft]{threeparttable}

\hypersetup{breaklinks=true,colorlinks=true,linkcolor=blue,citecolor=blue,filecolor=blue,urlcolor=blue}
\makeatletter
\patchcmd{\NAT@citex}
  {\@citea\NAT@hyper@{%
     \NAT@nmfmt{\NAT@nm}%
     \hyper@natlinkbreak{\NAT@aysep\NAT@spacechar}{\@citeb\@extra@b@citeb}%
     \NAT@date}}
  {\@citea\NAT@nmfmt{\NAT@nm}%
   \NAT@aysep\NAT@spacechar\NAT@hyper@{\NAT@date}}{}{}

\patchcmd{\NAT@citex}
  {\@citea\NAT@hyper@{%
     \NAT@nmfmt{\NAT@nm}%
     \hyper@natlinkbreak{\NAT@spacechar\NAT@@open\if*#1*\else#1\NAT@spacechar\fi}%
       {\@citeb\@extra@b@citeb}%
     \NAT@date}}
  {\@citea\NAT@nmfmt{\NAT@nm}%
   \NAT@spacechar\NAT@@open\if*#1*\else#1\NAT@spacechar\fi\NAT@hyper@{\NAT@date}}
  {}{}
  
\usepackage{array}
\newcolumntype{C}[1]{>{\centering\let\newline\\\arraybackslash\hspace{0pt}}m{#1}}

\usepackage{xspace}

\def\aj{AJ}
\def\araa{ARA\&A}
\def\apj{ApJ}
\def\apjl{ApJ}
\def\apjs{ApJS}
\def\ao{Appl.~Opt.}
\def\apss{Ap\&SS}
\def\aap{A\&A}

\def\aaps{A\&AS}

\def\mnras{MNRAS}

\def\pasa{Publ.~Astron.~Soc.~Australia}
\def\pasp{PASP}

\def\ssr{Space~Sci.~Rev.}

\def\nat{Nature}

\def\gca{Geochim.~Cosmochim.~Acta}

\def\physscr{Phys.~Scr}

\def\nar{New~Astro.~Rev.}

\def\adndt{Atom.~Data~Nucl.~Data~Tabl.}
\def\cajpj{Can.~J.~Phys.}

\newcommand{\ionic}[2]{#1$\,${\scshape{#2}}\xspace}
\newcommand{\ionf}[2]{#1$\,${\scshape{#2}}}

\def\arcdeg{\hbox{$^\circ$}}
\def\arcmin{\hbox{$^\prime$}}

\newcommand{\foiii}{[O\,{\sc iii}]\xspace}

\newcommand{\nii}{N\,{\sc ii}\xspace}
\newcommand{\niii}{N\,{\sc iii}\xspace}

\newcommand{\oii}{O\,{\sc ii}\xspace}

\newcommand{\cii}{C\,{\sc ii}\xspace}

\newcommand{\heii}{He\,{\sc ii}\xspace}

\definecolor{burgundy}{rgb}{0.5, 0.0, 0.13}

\usepackage{xspace}

\newcommand{\orcidicon}{\includegraphics[width=0.26cm]{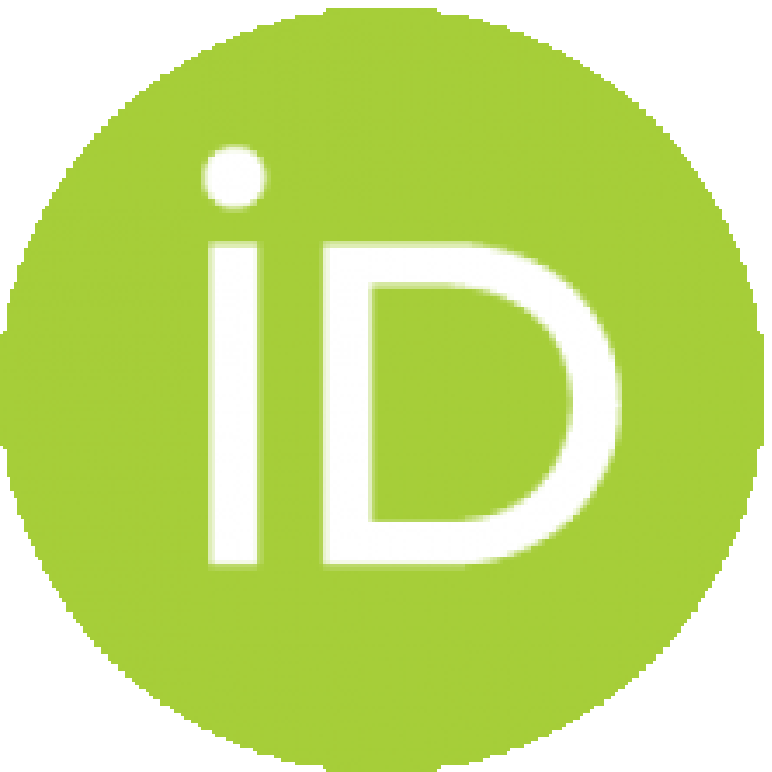}}

\newcommand{\orcidauthor}[1]{\href{https://orcid.org/#1}{\orcidicon}}

\shorttitle{Physical and chemical properties of WR planetary nebulae}
\shortauthors{Danehkar}

\makeatletter
\patchcmd{\frontmatter@RRAP@format}{(}{}{}{}
\patchcmd{\frontmatter@RRAP@format}{)}{}{}{}
\renewcommand\Dated@name{}
\makeatother

\begin{document}

\title{Physical and chemical properties of Wolf-Rayet planetary nebulae}

\correspondingauthor{A.~Danehkar}
\email{danehkar@umich.edu}

\author[0000-0003-4552-5997]{A.~Danehkar}
\affiliation{Department of Astronomy, University of Michigan, 1085 S. University Ave, Ann Arbor, MI 48109, USA}

\date[ ]{\footnotesize\textit{Received 2021 June 16; accepted 2021 August 25}}

\begin{abstract}
Wolf-Rayet ([WR]) and weak emission-line (\textit{wels}) central stars of planetary nebulae (PNe) have hydrogen-deficient atmospheres, whose origins are not well understood. In the present study, we have conducted plasma diagnostics and abundance analyses of 18 Galactic PNe surrounding [WR] and \textit{wels} nuclei, using collisionally excited lines (CELs) and optical recombination lines (ORLs) measured with the Wide Field Spectrograph on the ANU 2.3-m telescope at the Siding Spring Observatory complemented with optical archival data. Our plasma diagnostics imply that the electron densities and temperatures derived from CELs are correlated with the intrinsic nebular H$\beta$ surface brightness and excitation class, respectively. Self-consistent plasma diagnostics of heavy element ORLs of N${}^{2+}$ and  O${}^{2+}$ suggest that a small fraction of cool ($\lesssim$\,7000\,K), dense ($\sim$\,$10^4$--$10^5$\,cm$^{-3}$) materials may be present in some objects, though with large uncertainties. Our abundance analyses indicate that the abundance discrepancy factors (ADF\,$\equiv$\,ORLs/CELs) of O${}^{2+}$ are correlated with the dichotomies between forbidden-line and He\,{\sc i} temperatures. Our results likely point to the presence of a tiny fraction of cool, oxygen-rich dense clumps within the diffuse warm ionized nebulae.  Moreover, our elemental abundances derived from CELs are mostly consistent with AGB models in the range of initial masses from 1.5 to 5M$_{\odot}$. Further studies are necessary to understand better the origins of abundance discrepancies in PNe around [WR] and \textit{wels} stars.
\end{abstract}


\keywords{\href{https://astrothesaurus.org/uat/1249}{Planetary nebulae (1249)};
\href{https://astrothesaurus.org/uat/847}{Interstellar medium (847)};
\href{https://astrothesaurus.org/uat/224}{Chemical abundances (224)};
\href{https://astrothesaurus.org/uat/1806}{Wolf-Rayet stars(1806)}
\vspace{4pt}
\newline
\textit{Supporting material:} figure sets, machine-readable tables
}


\section{Introduction}
\label{wc:sec:introduction}

Planetary nebulae (PNe) are important astrophysical objects because they can be used as tracers of the composition of the interstellar medium (ISM) in galaxies, as well as to probe the uncertain physics of asymptotic giant branch (AGB) stars. Observations of PNe are employed to determine the elemental abundances of the ISM present in our own and other galaxies \citep[e.g.][]{Aller1983,Kingsburgh1994,Stasinska1998,Garcia-Rojas2012}. Mixing processes during the progenitor's life (e.g., first and second dredge up prior to the AGB, and third dredge up and hot bottom burning during the AGB) will change the envelope composition of He, C, N and possibly O and Ne \citep{Pequignot2000,Karakas2003,Karakas2009}. Other elements such as S, Ar, and Cl are left untouched by the evolution and nucleosynthesis in low- to intermediate-mass (1--8\,M$_{\odot}$) stars. For this reason, PN elemental abundances not only reflect the composition of the ISM at the time, when the progenitor was born, but also can be used to constrain the nucleosynthesis and mixing in AGB stars \citep[e.g.][]{Straniero1997,Werner2006,Karakas2009,Stasinska2013}.

Historically, strong and easy to measure collisionally excited lines (CELs) provided reliable chemical tracers, and they have been extensively used to derive the abundances of heavy elements such as N, O, Ne, Ar and S relative to H \citep[see e.g.][]{Kingsburgh1994,Kwitter2001,Tsamis2003a,Henry2004,Liu2004b}. CEL emissivities depend exponentially on
the electron temperature, so temperature variations introduce uncertainties into their results \citep[e.g.][]{Garnett1992,Stasinska2005}. On the other hand, optical recombination lines (ORLs) have a much weaker dependence on the electron temperature and density, thus in principle resulting in more reliable abundance analyses. ORLs from heavy element ions are extremely weak relative to H$\beta$, but are observable in deep spectra of nearby PNe. Numerous studies indicated that the abundances derived from ORLs are systematically higher than those obtained from CELs in PNe \citep{Liu2000,Liu2001,Luo2001,Wesson2003,Tsamis2004,Wesson2004,Wesson2005,Tsamis2008,Garcia-Rojas2009}. The causes of the CEL/ORL abundance discrepancies are not fully understood and remain the open problem in nebular astrophysics. This abundance discrepancy problem was already found in gaseous nebulae about eighty years ago \citep{Wyse1942,Aller1945}.

\begin{table*}
\caption{Journal of the ANU observations, including the stellar characteristics.
\label{wc:tab:obs:journal}
}
\centering
\footnotesize
\begin{tabular}{llcllcll}
\hline\hline\noalign{\smallskip}
PN & PN\,G & R.A. Dec./J2000 & CSPN & $T_{\rm eff}$ & Aperture   & Exp. & Obs. Date\\ 
   &  {(A92)}    &         & {(A03)}    & (kK)  & (arcsec$^{2}$)  & (sec) & \\ 
\noalign{\smallskip}
\hline 
\noalign{\smallskip}
PB\,6 	&278.8$+$04.9 	& 10$^{\rm h}$13$^{\rm m}$15\fs9 $-$50\degr19\arcmin59\farcs1 & [WO\,1]   & 103(K91) & $14 \times 14$ & 1200 		& 2010 Apr 20 \\
\noalign{\smallskip}
M\,3-30 &017.9$-$04.8	& 18$^{\rm h}$41$^{\rm m}$14\fs9 $-$15\degr33\arcmin43\farcs6 & [WO\,1]   & 49(A03) & $14 \times 15$ & 1200 		& 2010 Apr 21 \\
\noalign{\smallskip}
Hb\,4 (shell)	&003.1$+$02.9	& 17$^{\rm h}$41$^{\rm m}$52\fs7 $-$24\degr42\arcmin08\farcs0  & [WO\,3]   & 85(A03) & $10 \times 10$ & 300,\,1200& 2010 Apr 21 \\
\noalign{\smallskip}
Hb\,4 (N-knot)	&003.1$+$02.9	& 17$^{\rm h}$41$^{\rm m}$52\fs7 $-$24\degr42\arcmin08\farcs0  & [WO\,3]   & 85(A03) & $6 \times 5$ & 300,\,1200	& 2010 Apr 21 \\
\noalign{\smallskip}
Hb\,4 (S-knot)	&003.1$+$02.9	& 17$^{\rm h}$41$^{\rm m}$52\fs7 $-$24\degr42\arcmin08\farcs0  & [WO\,3]   & 85(A03)  & $6 \times 4$ & 300,\,1200 & 2010 Apr 21 \\
\noalign{\smallskip}
IC\,1297&358.3$-$21.6	& 19$^{\rm h}$17$^{\rm m}$23\fs5 $-$39\degr36\arcmin46\farcs4 & [WO\,3]   & 91(A03) & $12 \times 13$ & 60,\,1200	& 2010 Apr 21 \\
\noalign{\smallskip}
Th\,2-A	&306.4$-$00.6   & 13$^{\rm h}$22$^{\rm m}$33\fs8 $-$63\degr21\arcmin01\farcs3 & [WO3]{\scriptsize pec}(W08)& 157(P89) & $19 \times 16$ & 1200 		& 2010 Apr 20 \\
\noalign{\smallskip}
Pe\,1-1	&285.4$+$01.5 	& 10$^{\rm h}$38$^{\rm m}$27\fs6 $-$56\degr47\arcmin06\farcs5 & [WO\,4]   & 85(A02) & $9 \times 7$ & 60,\,1200 	& 2010 Apr 21 \\
\noalign{\smallskip}
M\,1-32 &011.9$+$04.2	& 17$^{\rm h}$56$^{\rm m}$20\fs1 $-$16\degr29\arcmin04\farcs6 & [WO\,4]{\scriptsize pec}& 56\,$^{\mathrm{a}}$ & $15 \times 12$ & 1200 		& 2010 Apr 20 \\
\noalign{\smallskip}
M\,3-15 &006.8$+$04.1	& 17$^{\rm h}$45$^{\rm m}$31\fs7 $-$20\degr58\arcmin01\farcs6  & [WC\,4]   & 55(A03) & $6 \times 8 $ & 60,\,1200 	& 2010 Apr 20 \\
\noalign{\smallskip}
M\,1-25 &004.9$+$04.9	& 17$^{\rm h}$38$^{\rm m}$30\fs3 $-$22\degr08\arcmin38\farcs8  & [WC\,5-6] & 56(A03) & $6 \times 8 $ & 60,\,1200 	& 2010 Apr 20 \\
\noalign{\smallskip}
Hen\,2-142 &327.1$-$02.2 & 15$^{\rm h}$59$^{\rm m}$57\fs6 $-$55\degr55\arcmin32\farcs9 & [WC\,9]   & 35(A03) & $7 \times 7$ & 60,\,1200 & 2010 Apr 20 \\
\noalign{\smallskip}
Hen\,3-1333 &332.9$-$09.9 & 17$^{\rm h}$09$^{\rm m}$00\fs9 $-$56\degr54\arcmin48\farcs1 & [WC\,10] & 30(D98) & $7 \times 7$ & 1200 	& 2010 Apr 20 \\
\noalign{\smallskip}
Hen\,2-113 &321.0$+$03.9 & 14$^{\rm h}$59$^{\rm m}$53\fs5 $-$54\degr18\arcmin07\farcs5 & [WC\,10]  & 30(D98) & $9 \times 9$   & 60,\,1200 & 2010 Apr 20 \\
\noalign{\smallskip}
K\,2-16 &352.9$+$11.4 & 16$^{\rm h}$44$^{\rm m}$49\fs1 $-$28\degr04\arcmin04\farcs7 & [WC\,11]   & 19(A03) &  $19 \times 21$ & 1200 	& 2010 Apr 20 \\
\noalign{\smallskip}
NGC\,6578 &010.8$-$01.8	& 18$^{\rm h}$16$^{\rm m}$16\fs5 $-$20\degr27\arcmin02\farcs7 & \textit{wels}(T93)  & 63(S89) &  $10 \times 10$ & 60,\,1200 & 2010 Apr 22 \\
\noalign{\smallskip}
M\,2-42   &008.2$-$04.8	& 18$^{\rm h}$22$^{\rm m}$32\fs0 $-$24\degr09\arcmin28\farcs4 & \textit{wels}(D11)  & 75(P89) &  $6 \times 7$  & 1200 	& 2010 Apr 22 \\
\noalign{\smallskip}
NGC\,6567 &011.7$-$00.6	& 18$^{\rm h}$13$^{\rm m}$45\fs2 $-$19\degr04\arcmin34\farcs2 & \textit{wels}(T93)  & 47(G97) &  $10 \times 10$ & 60,\,1200 & 2010 Apr 22 \\
\noalign{\smallskip}
NGC\,6629 &009.4$-$05.0	& 18$^{\rm h}$25$^{\rm m}$42\fs5 $-$23\degr12\arcmin10\farcs2 & \textit{wels}(T93)  & 35(G97) &  $10 \times 10$ & 60,\,1200 & 2010 Apr 22 \\
\noalign{\smallskip}
Sa\,3-107 &358.0$-$04.6	& 17$^{\rm h}$59$^{\rm m}$55\fs0 $-$32\degr59\arcmin11\farcs8 & \textit{wels}(D11)  & 46\,$^{\mathrm{a}}$ &  $7 \times 7$ & 1200 	& 2010 Apr 22 \\
\noalign{\smallskip}    
\hline
\end{tabular}
\begin{list}{}{}
\item[$^{\mathrm{a}}$]Calculated from the nebular excitation class (EC) and the EC--$T_{\rm eff}$ correlation given by \citet{Dopita1990,Dopita1991}. 
\end{list}
\begin{tablenotes}
\item[1]\textbf{Note.} References are as follows: 
A92 -- \citet{Acker1992}; 
A02 -- \citet{Acker2002}; 
A03 -- \citet{Acker2003}; 
D98 -- \citet{DeMarco1998a}; 
D11 -- \citet{Depew2011}; 
G97 -- \citet{Gorny1997}; 
K91 -- \citet{Kaler1991};
P89 -- \citet{Preite-Martinez1989};
S89 -- \citet{Shaw1989}; 
T93 -- \citet{Tylenda1993}; 
W08 -- \citet{Weidmann2008}. 
\end{tablenotes}
\end{table*}

\begin{figure*}
\begin{center}
\includegraphics[width=0.165\textwidth, trim = 0 0 0 0, clip, angle=0]{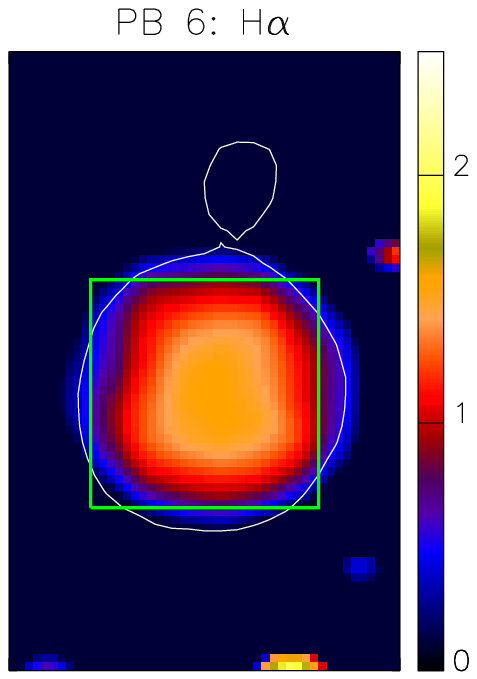}%
\includegraphics[width=0.165\textwidth, trim = 0 0 0 0, clip, angle=0]{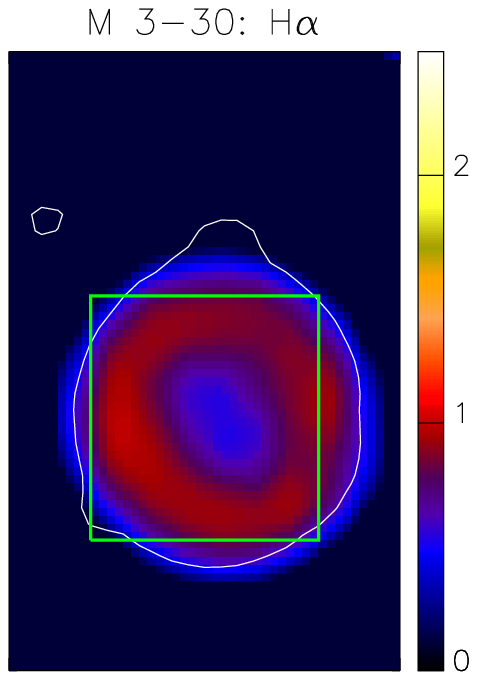}%
\includegraphics[width=0.165\textwidth, trim = 0 0 0 0, clip, angle=0]{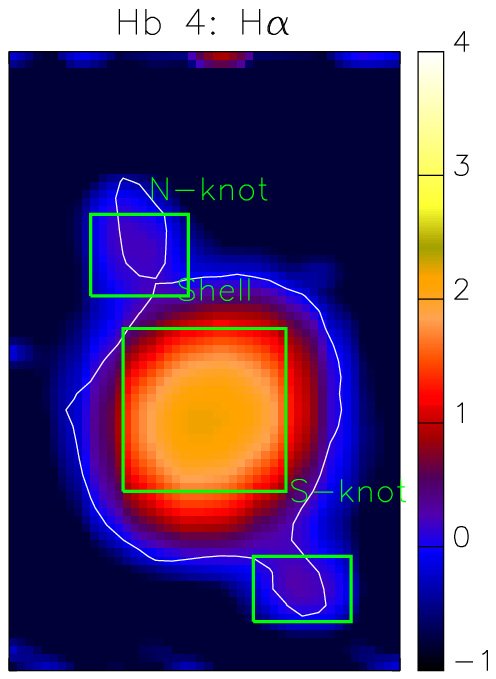}%
\includegraphics[width=0.165\textwidth, trim = 0 0 0 0, clip, angle=0]{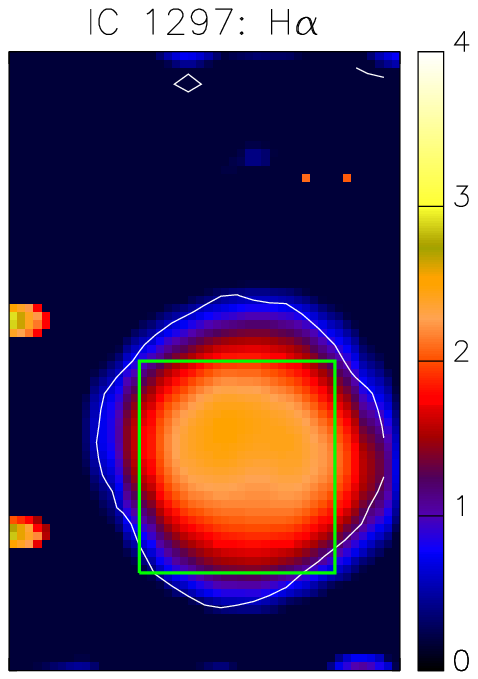}%
\includegraphics[width=0.165\textwidth, trim = 0 0 0 0, clip, angle=0]{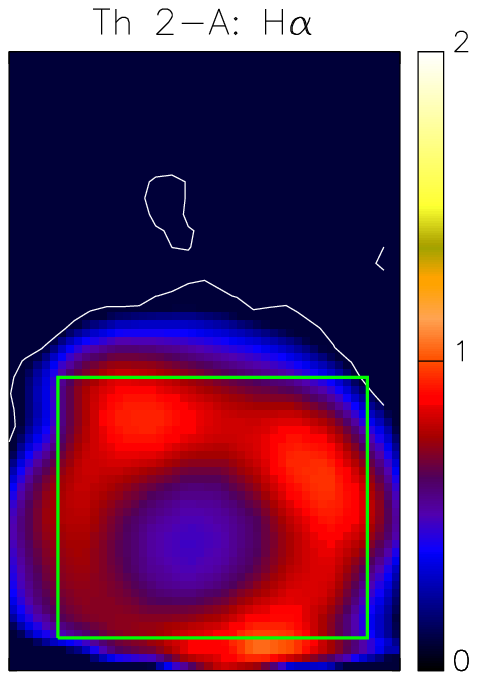}%
\includegraphics[width=0.165\textwidth, trim = 0 0 0 0, clip, angle=0]{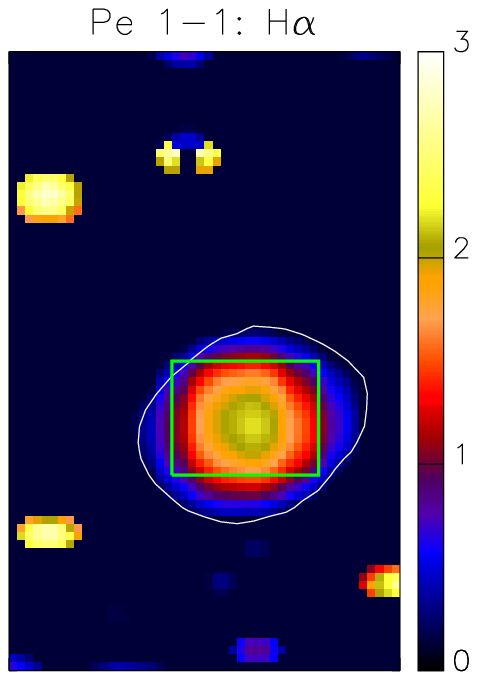}\\
\includegraphics[width=0.165\textwidth, trim = 0 0 0 0, clip, angle=0]{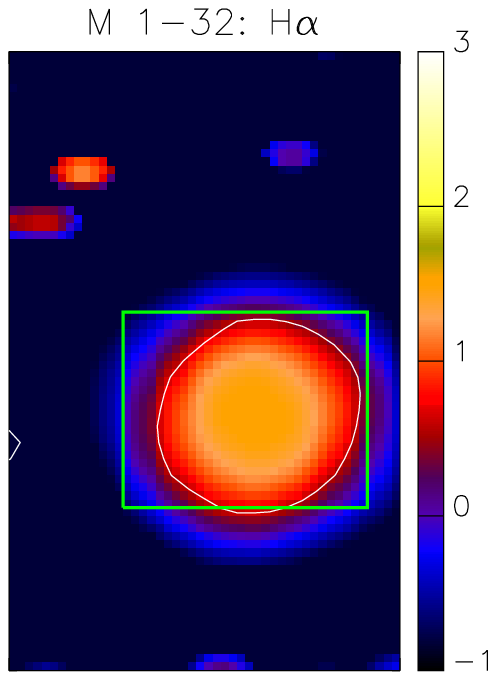}%
\includegraphics[width=0.165\textwidth, trim = 0 0 0 0, clip, angle=0]{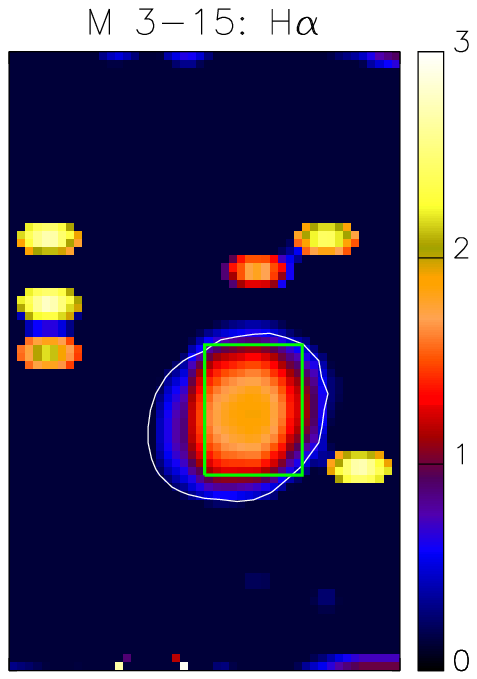}%
\includegraphics[width=0.165\textwidth, trim = 0 0 0 0, clip, angle=0]{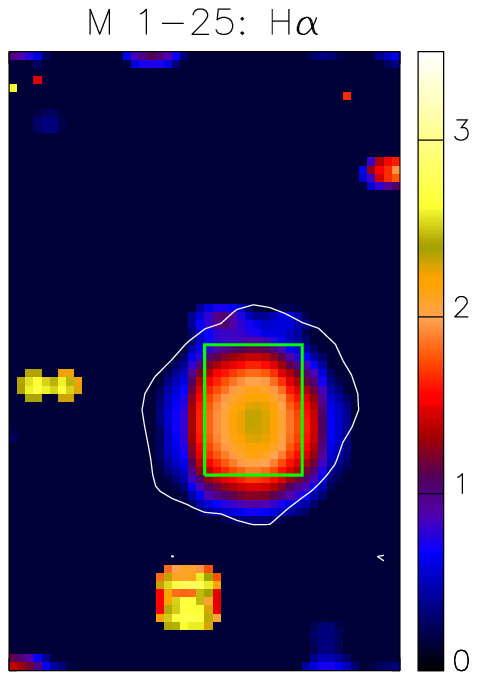}%
\includegraphics[width=0.165\textwidth, trim = 0 0 0 0, clip, angle=0]{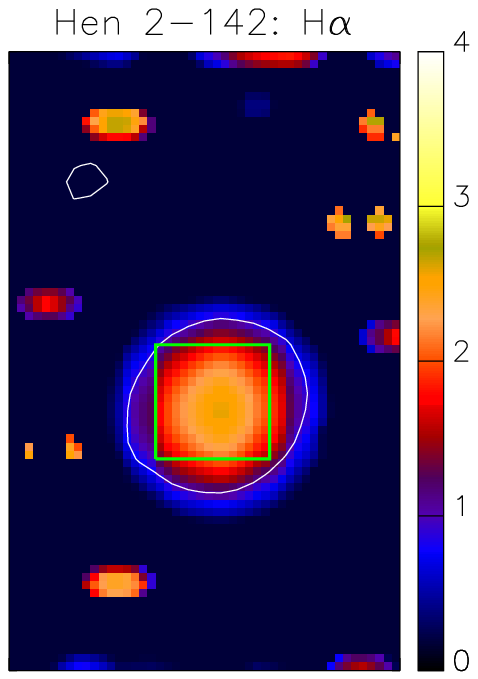}%
\includegraphics[width=0.165\textwidth, trim = 0 0 0 0, clip, angle=0]{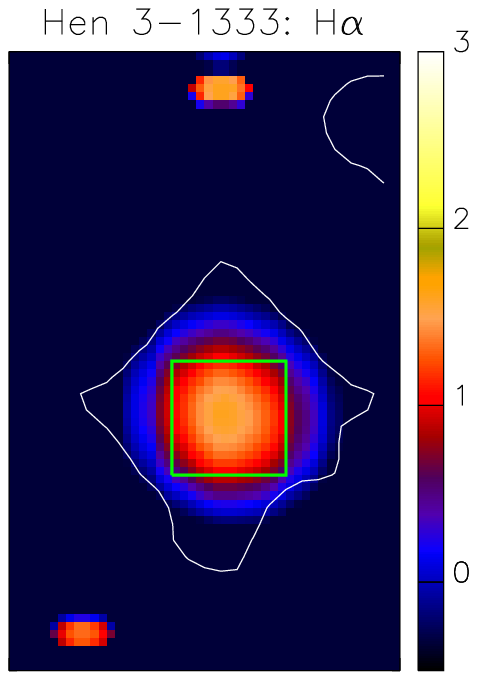}%
\includegraphics[width=0.165\textwidth, trim = 0 0 0 0, clip, angle=0]{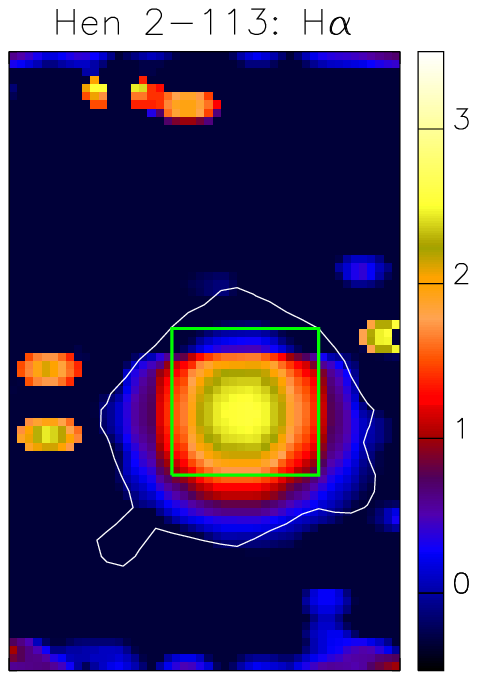}\\
\includegraphics[width=0.165\textwidth, trim = 0 0 0 0, clip, angle=0]{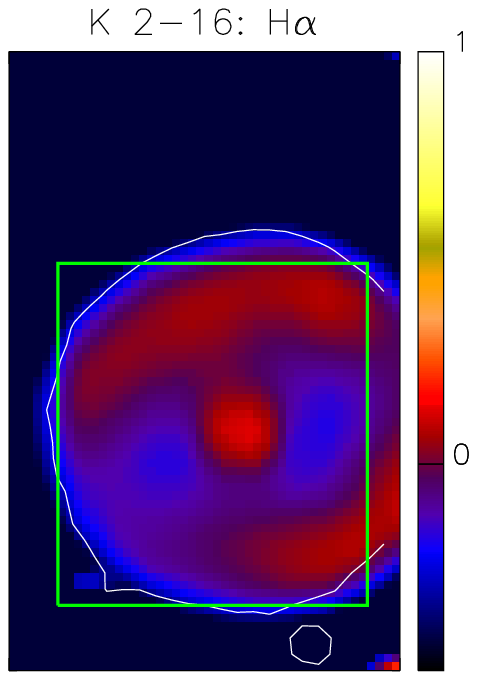}%
\includegraphics[width=0.165\textwidth, trim = 0 0 0 0, clip, angle=0]{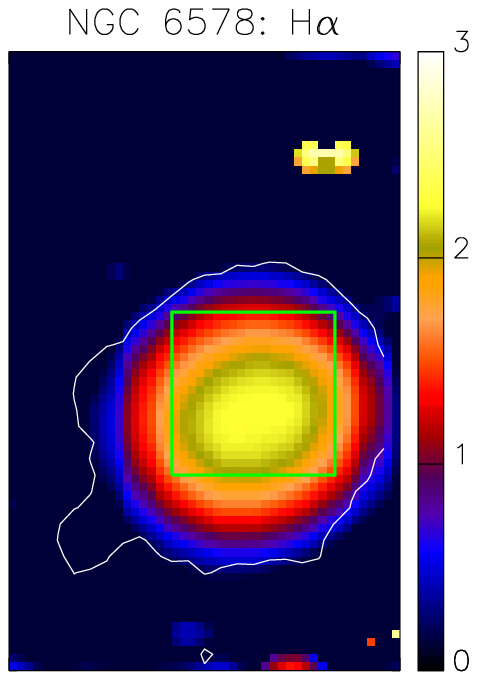}%
\includegraphics[width=0.165\textwidth, trim = 0 0 0 0, clip, angle=0]{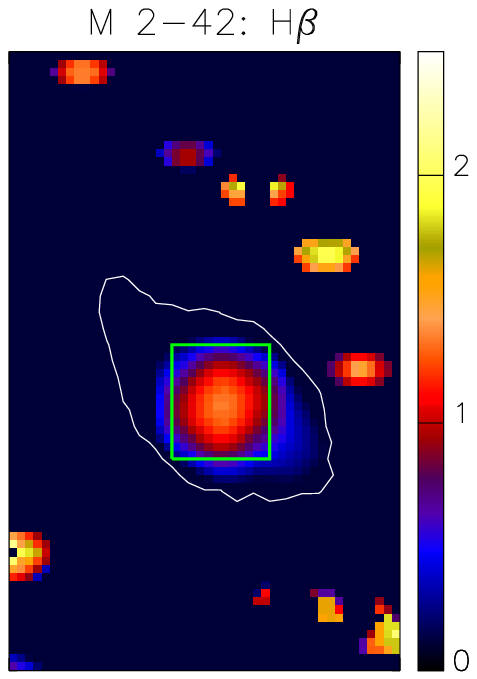}%
\includegraphics[width=0.165\textwidth, trim = 0 0 0 0, clip, angle=0]{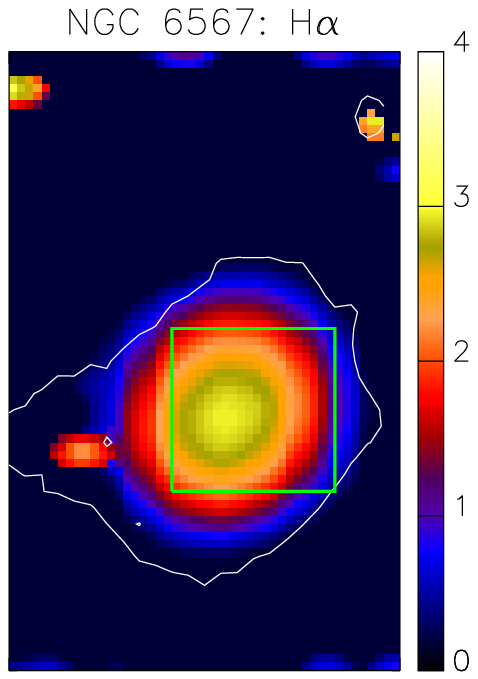}%
\includegraphics[width=0.165\textwidth, trim = 0 0 0 0, clip, angle=0]{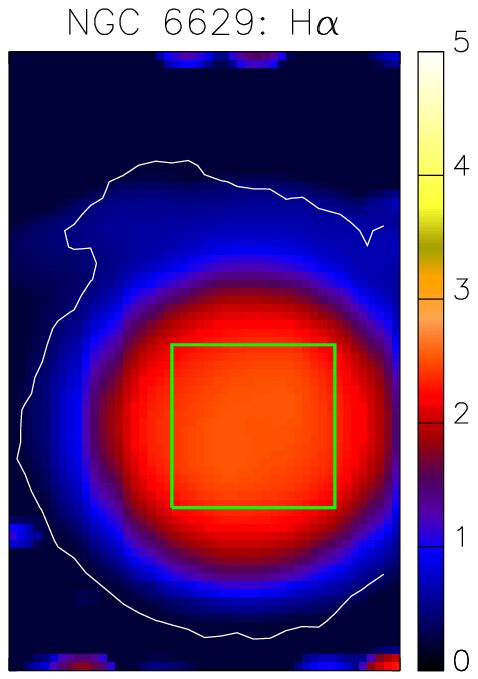}%
\includegraphics[width=0.165\textwidth, trim = 0 0 0 0, clip, angle=0]{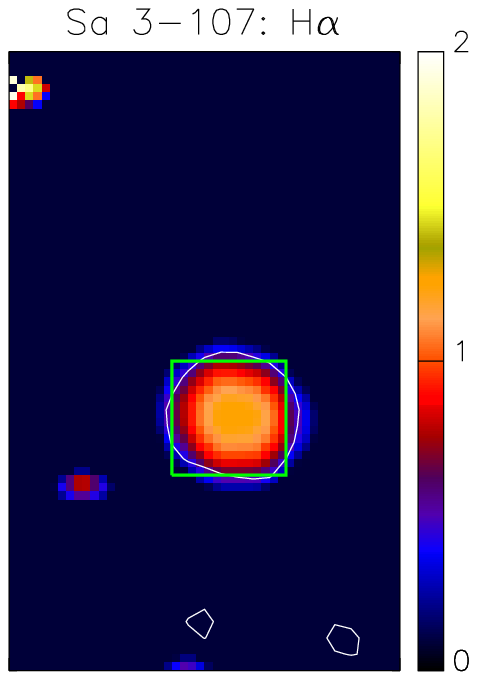}
\caption{From left to right, and top to bottom, flux maps of the H$\alpha$ emission in the WiFeS FOV ($24\times38$\,arcsec$^2$) for PB\,6, M\,3-30, Hb\,4, IC\,1297, Th\,2-A, Pe\,1-1, M\,1-32 M\,3-15, M\,1-25, Hen\,2-142, Hen\,3-1333, Hen\,2-113, K\,2-16, NGC\,6578, M\,2-42 (H$\beta$ emission), NGC\,6567, NGC\,6629, and Sa\,3-107 on logarithmic scales of $10^{-15}$ erg\,cm$^{-2}$\,s$^{-1}$\,spaxel$^{-1}$. The green rectangles show apertures with the sizes listed in Table\,\ref{wc:tab:obs:journal} used to extract the integrated spectra.  
The white contours show $\sim 10$ percent of the nebular average surface brightness in the H$\alpha$ emission 
from the SuperCOSMOS H$\alpha$ Sky Survey \citep[SHS;][]{Parker2005}, or in the $r$-band from the SuperCOSMOS Sky Surveys  \citep[SSS;][]{Hambly2001}. 
North is up and east is toward the left-hand side.
\label{wc:ifu:fov}%
}%
\end{center}
\end{figure*}

The dichotomy between electron temperatures measured from CELs and those from ORLs is another long-standing problem in the study of PNe, which may be closely linked to the abundance discrepancy problem. 
Nearly five decades ago, \citet{Peimbert1967,Peimbert1971} found that the difference between [O\,{\sc iii}] and H\,{\sc i} Balmer jump (BJ) temperatures, $T_{\rm e}$([O\,{\sc iii}]$)>T_{\rm e}$(BJ), in H\,{\sc ii} regions and PNe, and suggested the presence of temperature fluctuations. These temperature fluctuations could lead to overestimating the electron temperature deduced from CELs,  resulting in  underestimated ionic abundances \citep{Peimbert1967}. However, the detailed analysis of NGC 6543 by \citet{Wesson2004} for example showed that the temperature fluctuations are too small to explain the abundance discrepancy problem. Moreover, \citet{Wesson2005} found that the derived temperatures mostly follow the relation $T_{\rm e}$([O\,{\sc iii}]$)>T_{\rm e}$(BJ$)>T_{\rm e}$(He\,{\sc i}$)>T_{\rm e}($O\,{\sc ii}). This relation was predicted by the two-phase models \citep{Liu2003,Liu2004b}, containing some cold ($T_{\rm e}\sim10^3$\,K) hydrogen-deficient small-scale structures, highly enriched in helium and heavy elements, embedded in the diffuse warm ($T_{\rm e}\sim10^4$\,K) nebular gas of normal abundances. The study of Abell~30 by \citet{Wesson2003} pointed to the presence of cold ionized material in its hydrogen-deficient knots.

\begin{table*}
\caption{Comparison of the observed fluxes of the diagnostic emission lines He\,{\sc ii}, [O\,{\sc iii}], and [N\,{\sc ii}] measured by the WiFeS observations with those from the adopted archival data, on a scale relative to H$\beta$, where H$\beta=100$.
\label{wc:tab:archives}
}
\footnotesize
\centering
\begin{tabular}{llccccccccllll}
\hline\hline
\noalign{\smallskip}
PN	&PN G	&\multicolumn{2}{c}{He\,{\sc ii}\,$\lambda$4686}&&\multicolumn{2}{c}{[O\,{\sc iii}]\,$\lambda$5007}	&& \multicolumn{2}{c}{[N\,{\sc ii}]\,$\lambda$6584} & Lit. & \multicolumn{3}{c}{Lit. Observing Log} \\
\cline{3-4}  \cline{6-7} \cline{9-10} \cline{12-14}
\noalign{\smallskip}
	&	&D21	&Lit.	&&D21	&Lit.	&&D21	&Lit.	&Ref. & Slit (arcsec$^{2}$) & Exp. (sec) & Obs. Date \\
\noalign{\smallskip}
\hline 
\noalign{\smallskip} 
PB 6         	&278.8$+$04.9	&130.2	&147.0	&&1243.4	&1010.0	&&341.3	&306.0	&K91 & 4$\times$370, $45^{\circ}$	& 1200	& 1986 Mar 16\\
M 3-30       	&017.9$-$04.8	&77.3	&78.5	&&598.3	&604.5	&&71.0	&53.9	&P01 & 4$\times$13 & 2$\times$900 &1996 Jun 17\\
M 3-30       	&017.9$-$04.8	&77.3	&76.3	&&598.3	&428.0	&&71.0	&76.7	&G07 & 2.1$\times$270 &2400,300&1994 Mar/Jul\\
             	&           	&    	&   	&&   	&   	&&   	&   	&    &                   &        & 1995 Jul\\
Hb 4 (shell) 	&003.1$+$02.9	&11.9	&10.3	&&1381.8	&1481.4	&&419.9	&620.0	&P01 & 4$\times$13 &2$\times$900,3$\times$600&1996 Jul 14\\
Hb 4 (shell) 	&003.1$+$02.9	&11.9	&13.8	&&1381.8	&1321.0	&&419.9	&345.0	&G07 & 2.1$\times$270&2400,300&1994 Mar/Jul\\
             	&           	&    	&   	&&   	&   	&&   	&   	&    &                   &        & 1995 Jul\\
IC 1297      	&358.3$-$21.6	&37.0	&34.4	&&1342.3	&1367.0	&&53.9	&56.3	&M02 & 5$\times$320, $45^{\circ}$&210 (R/B) & 1997 Mar/Apr\\
Th 2-A       	&306.4$-$00.6	&49.5	&46.8	&&1672.9	&1719.0	&&296.9	&260.0	&M02 & 5$\times$320, $45^{\circ}$ &1800(B),1200(R) &1997 Mar/Apr\\
Pe 1-1       	&285.4$+$01.5	&0.0	&0.0	&&1186.2	&1140.1	&&494.1	&375.1	&G12 & 1$\times$5 &60,900,2$\times$1200 & 2010 Jun 5\\
M 1-32       	&011.9$+$04.2	&0.0	&0.4	&&533.5	&475.6	&&1455.5	&1230.6	&G12 & 1$\times$5 &60,3$\times$1500 & 2010 Jun 6\\ 
M 3-15       	&006.8$+$04.1	&0.0	&0.0	&&1188.9	&1122.5	&&274.3	&249.1	&P01& 4$\times$13&3$\times$900 & 1996 Jun 17\\
M 1-25       	&004.9$+$04.9	&0.0	&0.0	&&548.0	&530.0	&&714.4	&625.0	&G07 & 2.1$\times$270 & 2400,300 &1994 Mar/Jul \\
             	&           	&    	&   	&&   	&   	&&   	&   	&    &                   &        & 1995 Jul\\
Hen 2-142    	&327.1$-$02.2	&0.0	&0.0	&&6.0	&6.3	&&896.2	&1195.0	&G07 & 2.1$\times$270 & 2400,300 &1994 Mar/Jul
\\
             	&           	&    	&   	&&   	&   	&&   	&   	&    &                   &        & 1995 Jul\\
Hen 3-1333   	&332.9$-$09.9	&0.0	&0.0	&&0.0	&0.0	&& 207.6$^{\mathrm{a}}$	& 131.3$^{\mathrm{a}}$	&D97 & 1.5$\times$330 & 3$\times$600,2$\times$1200 &1993 Mar 14/15\\
Hen 2-113    	&321.0$+$03.9	&0.0	&0.0	&&0.0	&0.0	&& 121.2$^{\mathrm{a}}$	& 117.6$^{\mathrm{a}}$	&D97 & 1.5$\times$330 & 500,4$\times$600,2$\times$1200 & 1993 Mar 14/15 \\
K 2-16       	&352.9$+$11.4	&0.0	&0.0	&&139.0	&120.9	&&446.1	&277.4	&P01 & 4$\times$13 & 3$\times$900 & 1996 Jun 14\\
             	&           	&    	&   	&&   	&   	&&   	&   	&    &                   &  2$\times$900  & 1997 Aug 3\\
NGC 6578     	&010.8$-$01.8	&0.5	&1.4	&&864.3	&944.0	&&40.5	&50.7	&K03 & 5$\times$320, $45^{\circ}$& 1180(B),300(R) & 1997 Mar\\
M 2-42       	&008.2$-$04.8	&0.3	&0.3	&&707.2	&879.2	&&80.1	&90.8	&W07 & 2.1$\times$270 &60,300 &1996 Jul\\
NGC 6567     	&011.7$-$00.6	&0.6	&1.4	&&1011.9	&1016.0	&&26.1	&26.2	&K03 & 5$\times$320, $45^{\circ}$& 390(B),330 (R) & 1997 Mar\\
NGC 6629     	&009.4$-$05.0	&0.0	&0.0	&&712.8	&723.0	&&17.6	&21.1	&M02 & 5$\times$320, $45^{\circ}$ & 420(B),360(R)& 1997 Mar/Apr\\
\noalign{\smallskip}  
\hline
\end{tabular}
\begin{list}{}{}
\item[$^{\mathrm{a}}$]For Hen 3-1333 and Hen 2-113, [N\,{\sc ii}]\,$\lambda$6548 emission fluxes are listed.
\end{list}
\begin{tablenotes}
\item[1]\textbf{Note.} References for line fluxes from the literature are as follows: 
D21, This work; 
D97, \citet{DeMarco1997};
G07, \citet{Girard2007}; 
G12, \citet{Garcia-Rojas2012}; 
K91, \citet{Kaler1991};
K03, \citet{Kwitter2003}; 
M02, \citet{Milingo2002a}; 
P01, \citet{Pena2001} and \citet{Pena1998};  
W07, \citet{Wang2007}.
Archival spectra associated with nebular regions were extracted from long-slits passing through the central stars, oriented with the position angles (given in degrees `$^{\circ}$' on the right-hand side of the slit dimensions) if specified, and some of them were taken using different exposures for blue (B) and red (R) arms.
\end{tablenotes}
\end{table*}

Although most central stars of PNe (CSPNe) have `hydrogen-rich' surface abundances, a considerable fraction ($\lesssim25$\%) of them show `hydrogen-deficient' fast expanding atmospheres characterized by a large mass-loss rate \citep{Tylenda1993,Leuenhagen1996,Leuenhagen1998,Acker2003}. Their surface abundances exhibit helium, carbon, oxygen and neon, products of the helium-burning phase and a post-helium flash \citep{Werner2006}. Most of these CSPNe were classified as the carbon sequence of Wolf-Rayet (or [WR]) stars, resembling those of massive Wolf--Rayet (WR) stars \citep[][]{vanderHucht1981,vanderHucht2001}, where the square bracket distinguishes them from massive counterparts. About half of them show very high effective temperature, ranging from 80\,000~K to 150\,000~K, and are identified as the early-type ([WCE]), including spectral class [WO1]--[WC5] \citep{Koesterke1997,Pena1998}. Others having surface temperature between 20-80~kK are called the late-type ([WCL]), containing spectral class [WC6--11] \citep{Leuenhagen1996,Leuenhagen1998}. A few central stars of PNe show narrower and weaker emission lines (C\,{\sc iv} $\lambda$5805 and C\,{\sc iii} $\lambda$5695), which are not identical to those of [WR] classes. They were named weak emission-line stars (\textit{wels}) by \citet{Tylenda1993}. It has been suggested that [WR] stars are produced by a born-again scenario, i.e., a (very-) late thermal pulse \citep[see e.g.][]{Blocker2001,Herwig2001,Koesterke2001,Werner2006}. Therefore, one might expect a link between the hydrogen-deficient inclusions within the nebulae and their central stars. 

In this paper, we perform detailed plasma diagnostics and abundance analyses using CELs and ORLs for 18 Galactic PNe around hydrogen-rich [WR] stars and \textit{wels}, which might provide clues about the origin of their hydrogen-rich central stars. 
In Section\,\ref{wc:sec:observation}, we describe briefly our observations. 
In Section\,\ref{wc:sec:tempdens}, we present physical conditions derived from CELs and ORLs. In Section\,\ref{wc:sec:abundances}, we present ionic and elemental abundances, followed by a discussion of the ORL/CEL abundance discrepancy and CEL--ORL temperature dichotomy in Section\,\ref{wc:sec:adfs}. 
In Section\,\ref{wc:sec:agbmodels}, we discuss the implication of our observations for AGB stellar models. 
Finally, our conclusions and discussions are presented in Section\,\ref{wc:sec:conclusion}. 

\begin{table*}
\caption{Basic data for the nebulae, including the absolute total H$\beta$ flux, the 1.4 and 5~GHz radio flux densities, the optical and radio angular-dimensions, and the interstellar extinctions $c({\rm H}\beta)$ from the Balmer flux ratio H$\alpha$/H$\beta$, the radio-H$\beta$ method, and the literature.
\label{wc:tab:obs:data}
}
\centering
\footnotesize
\begin{tabular}{lllcccclll}
\hline\hline
\noalign{\smallskip}
{\scriptsize Name} & PN\,G &{\scriptsize $\log F({\rm H}\beta)$(C92)} & {\scriptsize $F$(1.4\,GHz)}  & {\scriptsize $F$(5\,GHz)} & \multicolumn{2}{c}{\scriptsize Angular diameter (arcsec)}  & \multicolumn{3}{c}{\scriptsize $c({\rm H}\beta)$}  \\ 
\cline{8-10}
\noalign{\smallskip}
     &   & {\scriptsize (${\rm erg}\,{\rm cm}^{-2}\,{\rm s^{-1}}$)} & {\scriptsize (C98)(mJy)} & {\scriptsize (S10)(mJy)} & {\scriptsize (optical)} & {\scriptsize (radio)} & {\scriptsize (Balmer)}  & {\scriptsize (radio)} & {\scriptsize (literature) } \\
\noalign{\smallskip}
\hline 
\noalign{\smallskip} 
PB\,6 	&278.8$+$04.9  & $-11.87$ & --            & 30.0  & $11.0$(A92) & --  & $0.543     _{-0.031    }^{+0.036     }$ & $0.738$ & 0.52(A03)\\
\noalign{\smallskip}
M\,3-30 &017.9$-$04.8  & $-12.29$ & 8.6           & 7.3   & $19.2 \times 18.5$(T03) & 22.0(A92) & $1.005     _{-0.013    }^{+0.013     }$ & $0.602$ & 1.30(A03) \\ 
\noalign{\smallskip}
Hb\,4 (shell)	   &003.1$+$02.9  & $-11.96$      & 158.0 & 166.0 & $11.4 \times 7.4$(T03) & 7.5(A92) & $1.851     _{-0.009    }^{+0.008     }$ & $1.689$ & 1.99(A03)  \\ 
\noalign{\smallskip}
Hb\,4 (N-knot)	   &003.1$+$02.9  & --            &  --   &  --   &  --  & --  & $2.089     _{-0.033    }^{+0.029     }$ &--        & --  \\ 
\noalign{\smallskip}
Hb\,4 (S-knot)     &003.1$+$02.9  & --            &  --   &  --   &  --  & --  & $1.908     _{-0.057    }^{+0.045     }$ &--        & --  \\ 
\noalign{\smallskip}
IC\,1297           &358.3$-$21.6  & $-10.95$      & 59.9  & 69.0 & $10.9 \times 9.9$(T03) & -- & $0.221     _{-0.013    }^{+0.015     }$ & $0.284$ & 0.19(A03) \\
\noalign{\smallskip}
Th\,2-A	           &306.4$-$00.6  & $-12.80$(A92) & --    & 60.0 & $27.7 \times 25.2$(T03) & -- & $1.079     _{-0.021    }^{+0.022     }$ & $2.059$ & 0.93(M02) \\
\noalign{\smallskip}
Pe\,1-1	           &285.4$+$01.5  & $-12.26$ & -- & 125.3 & $3.0$(A92) & -- & $1.943     _{-0.007    }^{+0.005     }$ & $1.886$ & 1.87(A03) \\
\noalign{\smallskip}
M\,1-32            &011.9$+$04.2  & $-12.20$(A92) & 70.5  & 64.0 & $9.4 \times 8.3$(T03) & 9.0(A92) & $1.347     _{-0.028    }^{+0.024     }$ & $1.542$ & 1.59(A03) \\
\noalign{\smallskip}
M\,3-15            &006.8$+$04.1  & $-12.45$(A92) & 48.4  & 65.0 & $4.2$(A92) & 5.0(A92) & $2.251     _{-0.011    }^{+0.010     }$ & $1.828$ & 2.10(A03) \\
\noalign{\smallskip}
M\,1-25            &004.9$+$04.9  & $-11.90$      & 40.3  & 55.0 & $4.6$(A92) & 3.2(A92) & $1.596     _{-0.004    }^{+0.005     }$ & $1.197$ & 1.46(A03)\\
\noalign{\smallskip}
Hen\,2-142         &327.1$-$02.2  & $-11.85$      & --    & 65.0 & $4.4 \times 3.5$(T03) & -- & $1.554     _{-0.087    }^{+0.091     }$ & $1.234$ & 1.73(A03) \\
\noalign{\smallskip}
Hen\,3-1333        &332.9$-$09.9  & $-12.15$      & --    & 26.0(P82) & $3.6 \times 3.4$(T03) & -- & $1.064     _{-0.021    }^{+0.020     }$ &--        & 1.00(A03) \\
\noalign{\smallskip}
Hen\,2-113         &321.0$+$03.9  & $-11.82$      & --    & 115.0(P82) & $3.7 \times 4.8$ (D15)  & -- & $1.335     _{-0.025    }^{+0.028     }$ &--        & 1.48(A03) \\
\noalign{\smallskip}
K\,2-16            &352.9$+$11.4  & $-12.62$(F13)\,$^{\mathrm{a}}$ & 2.5 & -- &  $26.6 \times 24.3$(T03) & -- & $0.499     _{-0.040    }^{+0.043     }$ &--        & 0.97(A03) \\
\noalign{\smallskip}
NGC\,6578          &010.8$-$01.8  & $-11.57$      & 162.4 & 166.0 &  $12.1 \times 11.8$(T03) & -- & $1.510     _{-0.009    }^{+0.010     }$ & $1.341$ & 1.39(K03) \\
\noalign{\smallskip}
M\,2-42            &008.2$-$04.8  & $-12.12$      & 9.8   & 14.0 &  $4.0 \times 4.0$(S08)  & -- & $0.979     _{-0.028    }^{+0.031     }$ & $0.806$ & 1.03(A91) \\
\noalign{\smallskip}
NGC\,6567          &011.7$-$00.6  & $-10.95$      & 163.3 & 161.0 &  $8.1 \times 6.4$(T03) & -- & $0.770     _{-0.006    }^{+0.008     }$ & $0.672$ & 0.70(K03) \\
\noalign{\smallskip}
NGC\,6629          &009.4$-$05.0  & $-10.93$      & 264.0 & 265.8 &  $16.6 \times 15.5$(T03) & -- & $0.975     _{-0.009    }^{+0.008     }$ & $0.905$ & 0.90(P11) \\
\noalign{\smallskip}
Sa\,3-107          &358.0$-$04.6  & $-12.95$(F13)\,$^{\mathrm{a}}$ & 5.2(C99) & -- &  $8.0 \times 8.0$\,$^{\mathrm{b}}$  & -- & $1.616     _{-0.010    }^{+0.009     }$ &--        & -- \\
\noalign{\smallskip}
\hline
\end{tabular}
\begin{list}{}{}
\item[$^{\mathrm{a}}$]Calculated from the observed intensity of H$\alpha$ using the logarithmic extinction formula. 
\item[$^{\mathrm{b}}$]Determined at $\sim10$ percent of mean surface brightness isophote of H$\alpha$ images obtained from \citet{Parker2005}. 
\end{list}
\begin{tablenotes}
\item[1]\textbf{Note.} References are as follows: 
A91 -- \citet{Acker1991b};
A92 -- \citet{Acker1992}; 
A03 -- \citet{Acker2003}; 
C92 -- \citet{Cahn1992}; 
C98 -- \citet{Condon1998};
C99 -- \citet{Condon1999}; 
D15 -- \citet{Danehkar2015a}; 
F13 -- \citet{Frew2013a}; 
K03 -- \citet{Kwitter2003};
M02 -- \citet{Milingo2002a}; 
P11 -- \citet{Pottasch2011}; 
P82 -- \citet{Purton1982}; 
S08 -- \citet{Stanghellini2008}; 
S10 -- \citet{Stanghellini2010};
T03 -- \citet{Tylenda2003}. 
\end{tablenotes}
\end{table*}


\section{Observations}
\label{wc:sec:observation}

The optical integral field unit (IFU) spectra of PNe for this work were obtained at Siding Spring Observatory, Australia, using the Wide Field Spectrograph \citep[WiFeS;][]{Dopita2007,Dopita2010} mounted on the 2.3-m Australian National University (ANU) telescope in April 2010 under program number 1100147 (PI: Q.\,A.~Parker). WiFeS is an image-slicing IFU developed and built for the ANU 2.3-m telescope, feeding a double-beam spectrograph. WiFeS samples 0.5 arcsec along each of twenty five $38$\,arcsec \,$\times$ \,$1$\,arcsec slitlets that provides a field-of-view (FOV) of $25$\,arcsec\,$\times$\,$38$\,arcsec  and a spatial resolution element of  $1.0$\,arcsec\,$\times$\,$0.5$\,arcsec. The spectrograph uses volume phase holographic gratings to provide a spectral resolution of $R\sim3000$ and $R\sim7000$. WiFeS has recently been used for a number of southern Galactic PNe \citep[e.g.][]{Ali2015,Ali2016,Basurah2016,Dopita2017}.

Our observations were carried out with the spectral resolution of $R\sim7000$, covering  $\lambda\lambda$4415--5589\,{\AA} in the blue channel and $\lambda\lambda$5222--7070\,{\AA} in the red channel. 
Exposure times ranged from  60--1200\,sec depending on the nebular H$\beta$ surface brightness. Spectroscopic standard stars were observed for the flux calibration purposes, notably EG\,274 and LTT 3864. We also acquired series of bias, dome flat-field frames, twilight sky flats, arc lamp exposures, and wire frames for data reduction, flat-fielding, wavelength calibration and spatial calibration. 

Table~\ref{wc:tab:obs:journal} presents a journal of the ANU observations that includes spectral classes of the central stars (Column 4), the stellar effective temperature (Column 5), the observing aperture (Column 6, see Figure~\ref{wc:ifu:fov}), the exposure time (Column 7) used for each PN, and the observing date (Column 8). The rectangle aperture used to extract the integrated spectrum of each object is shown on the WiFeS FOV in Figure~\ref{wc:ifu:fov}. 
The same aperture and FOV were employed for different gratings and exposure times in each target.
The spectral classes of the [WR]-type stars are based on the classification schemes by \citet{Crowther1998} and \citet{Acker2003}, and those of the \textit{wels} according to \citet{Tylenda1993} and \citet{Depew2011}. Although the CSPN M\,2-42 was identified as \textit{wels} \citep{Depew2011}, \niii\ lines and \heii\ identified in its stellar spectrum might associate it with [WN\,8] stars \citep{Danehkar2016}. The CSPN M\,1-32 was classified under the \textit{peculiar} [WO\,4]${}_{\rm pec}$ subclass according to the width of C\,{\sc iv}-5801/12 doublet \citep{Acker2003}. Similarly, the Th\,2-A contain a [WO\,3]${}_{\rm pec}$ star \citep{Weidmann2008}, with collimated bipolar outflows discovered recently \citep{Danehkar2015}. Bipolar collimated outflows were also present in the other PNe of our sample: M\,1-32, M\,3-15 \citep{Akras2012}, M\,2-42 \citep{Danehkar2016}, and Hb\,4 \citep{Derlopa2019,Danehkar2021}.

The spectra were reduced using the \textsc{iraf} pipeline \textsf{wifes} \citep[version 2.14; see][]{Dopita2010} . The reduction involves flat-fielding, wavelength calibration, spatial calibration, sky subtraction and flux calibration \citep[described in detail by][]{Danehkar2013a,Danehkar2014a,Danehkar2014b}.  

To extract flux intensities and uncertainties based on the root-mean-squared (RMS) deviation, we applied 
the IDL library MGFIT 
to the spectrum of each object, which matches multiple Gaussian functions to a list of emission lines using a random walk method optimized based on a genetic algorithm \citep[][]{Wesson2016}. 
We also verified the automatically identified lines and manually removed any misidentified lines from the final list. The RMS deviation of the continuum near each line is quantified in order to estimate uncertainties of each fitted line according to the signal-dependent noise models \citep{Landman1982,Lenz1992}. 

We explored the literature to include those optical lines outside the wavelength coverage of the ANU/WiFeS observations that are necessary for our comprehensive plasma diagnostics and abundance analyses. 
As ionization structures of PNe are typically inhomogeneous  \citep[see e.g.][]{Danehkar2018,Akras2016,Akras2020},
different observations of a nebula can be combined if they are roughly 
associated with the same region having the same excitation conditions. 
Thus, we adopted those archival data, which have similar ionization properties based on 
observed fluxes of the diagnostic emission lines He\,{\sc ii}\,$\lambda$4686, [O\,{\sc iii}]\,$\lambda$5007, and [N\,{\sc ii}]\,$\lambda$6584 (or $\lambda$6548). 
These observations mostly employed the slits placed over the entire nebulae passing through the central stars, which are similar to the apertures covering the nebulae used to extract our WiFeS spectra (see Figure~\ref{wc:ifu:fov}).
Table~\ref{wc:tab:archives} compares the observed fluxes of the diagnostic lines He\,{\sc ii}, [O\,{\sc iii}], and [N\,{\sc ii}] measured by the WiFeS with those from 
the selected archival data. It can be seen that there are generally reasonable agreements between the WiFeS and archival data included in our study. Table~\ref{wc:tab:archives} also presents the observing logs of the archival data,
including the observing long-slit dimensions (Column 10; also the position angle if it is specified), exposure time (Column 11), and observing date (Column 12).
 
Basic data for the nebulae are presented in Table~\ref{wc:tab:obs:data}, including the absolute total flux of H$\beta$ (Column 3), the radio flux densities at 1.4 and 5~GHz (Columns 4 and 5) and the nebular angular-dimensions measured in the optical and in radio observations (Columns 6 and 7), and the interstellar extinctions  (Columns 8--10; described in \S\,\ref{wc:sec:line_extinction}). The nebular angular-dimensions are mostly based on \citet{Tylenda2003}, one of them is from $\sim10$ percent of mean H$\alpha$ surface brightness \citep{Parker2005}. 

\setcounter{table}{3}
\begin{table*}
\caption{Observed and dereddened line fluxes on a scale relative to H$\beta$, where H$\beta=100$. Observed fluxes are denoted by $F(\lambda)$ and dereddened fluxes by $I(\lambda)$. 
The symbol `*' in the observed and dereddened fluxes indicates that the listed line is blended with the above listed line.
\label{wc:tab:nebula:observations}
}
\centering
\footnotesize
\begin{tabular}{lllccccccccc}
\hline\hline
\noalign{\smallskip}
 $\lambda_{\rm lab}$  & Ion  & $\lambda_{\rm obs}$ & $F(\lambda)$ & $\varepsilon_{F(\lambda)}$(\%) & $I(\lambda)$ & $\varepsilon_{I(\lambda)}$(\%) & Mult & Lower term & Upper term & g1 & g2 \\
\noalign{\smallskip}
\hline
\noalign{\smallskip}
\multicolumn{12}{c}{PB 6          (PNG278.8$+$04.9) }\\
\noalign{\smallskip}
\hline
\noalign{\smallskip}
   3726.03 & [O\,{\sc ii}]$^{\mathrm{a}}$   &    3726.03 &     50.700 & $\pm  0.0$ &     69.898 & $\pm  2.6$ & F1       & 2p3 4S*          & 2p3 2D*          & 4        & 4        \\
\noalign{\smallskip}
   3728.82 & [O\,{\sc ii}]$^{\mathrm{a}}$   &    3728.82 &          * &       &          * &       & F1       & 2p3 4S*          & 2p3 2D*          & 4        & 6        \\
\noalign{\smallskip}
   3868.75 & [Ne\,{\sc iii}]$^{\mathrm{a}}$ &    3868.75 &     84.000 & $\pm  0.0$ &    112.065 & $\pm  2.3$ & F1       & 2p4 3P           & 2p4 1D           & 5        & 5        \\
\noalign{\smallskip}
   3889.05 & H\,{\sc 8}$^{\mathrm{a}}$      &    3889.05 &     11.500 & $\pm  0.0$ &     15.266 & $_{ -1.9}^{+  2.3}$ & H8       & 2p+ 2P*          & 8d+ 2D           & 8        & *        \\
\noalign{\smallskip}
   3967.46 & [Ne\,{\sc iii}]$^{\mathrm{a}}$ &    3967.46 &     30.800 & $\pm  0.0$ &     40.079 & $_{ -1.8}^{+  2.1}$ & F1       & 2p4 3P           & 2p4 1D           & 3        & 5        \\
\noalign{\smallskip}
   4101.74 & H\,{\sc 6}$^{\mathrm{a}}$      &    4101.74 &     22.500 & $\pm  0.0$ &     28.240 & $\pm  1.8$ & H6       & 2p+ 2P*          & 6d+ 2D           & 8        & 72       \\
\noalign{\smallskip}
   4340.47 & H\,{\sc 5}$^{\mathrm{a}}$      &    4340.47 &     44.200 & $\pm  0.0$ &     51.795 & $\pm  1.3$ & H5       & 2p+ 2P*          & 5d+ 2D           & 8        & 50       \\
\noalign{\smallskip}
   4363.21 & [O\,{\sc iii}]$^{\mathrm{a}}$  &    4363.21 &     17.100 & $\pm  0.0$ &     19.904 & $_{ -1.0}^{+  1.3}$ & F2       & 2p2 1D           & 2p2 1S           & 5        & 1        \\
\noalign{\smallskip}
   4452.37 & O\,{\sc ii}                    &    4453.76 &      0.053 & $\pm 16.7$ &      0.060 & $_{-22.8}^{+ 20.9}$ & V5       & 3s 2P            & 3p 2D*           & 4        & 4        \\
\noalign{\smallskip}
   4471.50 & He\,{\sc i}                    &    4472.51 &      2.061 & $\pm  1.5$ &      2.322 & $\pm  2.2$ & V14      & 2p 3P*           & 4d 3D            & 9        & 15       \\
\noalign{\smallskip}
   4491.23 & O\,{\sc ii}                    &    4493.19 &      0.019 & $\pm 23.1$ &      0.021 & $_{-31.2}^{+ 28.7}$ & V86a     & 3d 2P            & 4f D3*           & 4        & 6        \\
\noalign{\smallskip}
   4510.91 & N\,{\sc iii}                   &    4511.90 &      0.207 & $\pm  4.5$ &      0.230 & $\pm  5.8$ & V3       & 3s' 4P*          & 3p' 4D           & 2        & 4        \\
\noalign{\smallskip}
   4514.86 & N\,{\sc iii}                   &    4516.17 &      0.066 & $\pm 19.8$ &      0.073 & $_{-26.7}^{+ 24.4}$ & V3       & 3s' 4P*          & 3p' 4D           & 6        & 8        \\
\noalign{\smallskip}
   4541.59 & He\,{\sc ii}                   &    4542.59 &      4.340 & $\pm  0.6$ &      4.787 & $_{ -1.0}^{+  1.1}$ & 4.9      & 4f+ 2F*          & 9g+ 2G           & 32       & *        \\
\noalign{\smallskip}
   4562.60 & Mg\,{\sc i}]                   &    4563.59 &      0.112 & $\pm  7.1$ &      0.123 & $_{ -9.3}^{+  8.8}$ &          & 3s2 1S           & 3s3p 3P*         & 1        & 5        \\
\noalign{\smallskip}
   4571.10 & Mg\,{\sc i}]                   &    4572.04 &      0.174 & $\pm  5.4$ &      0.190 & $\pm  6.7$ &          & 3s2 1S           & 3s3p 3P*         & 1        & 3        \\
\noalign{\smallskip}
   4607.03 & [Fe\,{\sc iii}]                &    4607.70 &      0.233 & $\pm  5.3$ &      0.252 & $_{ -7.1}^{+  6.5}$ & F3       & 3d6 5D           & 3d6 3F2          & 9        & 7        \\
\noalign{\medskip}
\multicolumn{12}{c}{                          \ldots }\\
\noalign{\medskip}
\noalign{\medskip}
\hline
\end{tabular}
\begin{list}{}{}
\item[$^{\mathrm{a}}$]Fluxes adopted from the literature: 
PB 6 (K91), 
M 3-30 (P01,G07),
Hb 4 shell (P01,G07), 
Hb 4 N-knot (H97), 
Hb 4 S-knot (H97), 
IC 1297 (M02), 
Th 2-A (M02), 
Pe 1-1 (G12), 
M 1-32 (G12), 
M 3-15 (P01), 
M 1-25 (G07), 
Hen 2-142 (G07), 
Hen 3-1333 (D97), 
Hen 2-113 (D97), 
K 2-16 (P01), 
NGC 6578 (K03), 
M 2-42 (W07), 
NGC 6567 (K03), 
NGC 6629 (M02).  
References for line fluxes from the literature are as follows: 
D97, \citet{DeMarco1997};
G07, \citet{Girard2007}; 
G12, \citet{Garcia-Rojas2012}; 
H97, \citet{Hajian1997}; 
K91, \citet{Kaler1991};
K03, \citet{Kwitter2003}; 
M02, \citet{Milingo2002a}; 
P01, \citet{Pena2001} and \citet{Pena1998};  
W07, \citet{Wang2007}.
\item[\textbf{Note:}]Table \ref{wc:tab:nebula:observations} is published in its entirety in the machine-readable format. A portion is shown here for guidance regarding its form and content.
\end{list}
\end{table*}


\subsection{Flux Measurement and Interstellar Extinction}
\label{wc:sec:line_extinction}

Table~\ref{wc:tab:nebula:observations} presents a full list of observed lines and their measured fluxes (the full table is available in the machine-readable format). 
The laboratory wavelength, the emission line identification, and the observed wavelength are given in the first 3 columns, followed by
the observed fluxes with the RMS errors (in percentage), and the fluxes after correction for interstellar extinction with the associated errors (in percentage) at the 90\% confidence levels in the next columns.
The multiplet number, the lower and upper terms of the transition, and the statistical weights of the lower and upper levels are presented in the ending columns. All fluxes are given relative to H$\beta$, on a scale where ${\rm H}\beta=100$. 
The line fluxes adopted from the archival data are also listed in Table~\ref{wc:tab:nebula:observations}.

The logarithmic extinction $c({\rm H}\beta)$ at H$\beta$ was obtained from the observed Balmer emission line H$\alpha$/H$\beta$ flux ratio and its theoretical line ratio for case B recombination \citep[][based on the physical conditions of low-excited CELs derived in \S\,\ref{wc:sec:tempdens}]{Storey1995}. For M\,2-42, the H$\alpha$ emission was saturated in some spaxels over the main shell in the WiFeS observation, so we adopted the H$\alpha$ emission flux measured by \citet{Wang2007}. 
Each flux intensity was then dereddened using the formula,
$I(\lambda)=10^{c({\rm H}\beta)[1+f(\lambda)]}\,F(\lambda)$, where $F(\lambda)$ and $I(\lambda)$ are the observed and intrinsic line flux, respectively, and $f(\lambda)$ is the standard Galactic extinction law of $R_{\mathrm{V}}=3.1$ \citep{Seaton1979a,Howarth1983} normalized such that $f({\rm H}\beta)=0$.

The radio-H$\beta$ extinction was also determined from the observed radio free--free continuum radiation at 5 GHz and the measured H$\beta$ flux using the formula given by \citet{Milne1975}:
\begin{align}
c({\rm H}\beta)  = & \log\left( \frac{3.28\times10^{-9}\, t^{-0.4} [S_{{\rm 5 GHz}} / F({\rm H}\beta)] }
{\ln(9900 t^{3/2})[1+(1-x'')y+3.7x''y]}\right) 
 \label{wc2:eq_reddening_radio}%
\end{align}%
where $S_{{\rm 5 GHz}}$ in Jy is the observed 5 GHz flux density (Column 5 in Table~\ref{wc:tab:obs:data}), $F({\rm H}\beta)$ the observed H$\beta$ flux in erg\,cm$^{-2}$\,s$^{-1}$ (Column 4 in Table~\ref{wc:tab:obs:data}), $t\equiv T_{\rm e}($X$^{+})/10^4$ is the electron temperature of singly-ionized forbidden lines in $10^4$\,K (derived in \S\,\ref{wc:sec:tempdens:cel}), $y=N({\rm He})/N({\rm H})$ the number abundance of helium, and $x''=N({\rm He^{++}})/N({\rm He})$ the fraction of doubly ionized helium atoms (derived in \S\,\ref{wc:sec:abundances:orl}). We assume that hydrogen is fully ionized. 

Table~\ref{wc:tab:obs:data} compares $c({\rm H}\beta)$ derived from the Balmer flux ratio H$\alpha$/H$\beta$ (Column 8) with those from the radio-H$\beta$ method (Column 9) and the literature (Column 10).  We see that there are generally good agreements between them. However, there are some discrepancies in a small number, which could be due to the uncertainties in the measured values of $F({\rm H}\beta)$, derived electron temperatures, helium ionic abundances and  observed 5-GHz continuum fluxes.


\section{Plasma Diagnostics}
\label{wc:sec:tempdens}

\subsection{CEL Plasma Diagnostics}
\label{wc:sec:tempdens:cel}

Nebular electron temperatures $T_{\rm e}$ and densities $N_{\rm e}$ were obtained from the intrinsic intensities of CELs by solving level populations for an $n$-level ($\geqslant5$) atomic model using the 
IDL library proEQUIB \citep{Danehkar2018b}.
To propagate uncertainties in fluxes, we utilized an IDL implementation of the affine-invariant Markov chain Monte Carlo (MCMC) ensemble sampler 
proposed by \citet{Goodman2010}. 
For the MCMC computations, we adopted a confidence level of 90\% and a uniform distribution on the range covered by the errors of each observed line flux to propagate flux uncertainties into the extinction, dereddened line fluxes, physical conditions and chemical abundances. 

The atomic data references are listed in Table~\ref{wc:tab:atomicdata}. For the abundance analyses, we carefully adopted the atomic data sets from the CHIANTI database version 7.0 \citep{Landi2012} and 9.0 \citep{Dere2019}, assembled into the AtomNeb library \citep{Danehkar2019}, 
including the energy levels ($E_{j}$), collision strengths ($\Omega_{ij}$), and  transition probabilities ($A_{ij}$) for CELs, and effective recombination coefficients ($\alpha_{\rm eff}$) and branching ratios ($Br$) for ORLs. 
The diagnostic procedure was done in an iterative way to provide self-consistent results for $N_{\rm e}$([S\,{\sc ii}]) and $T_{\rm e}$([N\,{\sc ii}]), i.e., a representative initial $T_{\rm e}$([N\,{\sc ii}]) was assumed to calculate $N_{\rm e}$; then $T_{\rm e}$ was derived in conjunction with the derived $N_{\rm e}$([S\,{\sc ii}]), and the procedure iterated to provide self-consistent results. We used the temperature $T_{\rm e}$([N\,{\sc ii}]) for deriving $N_{\rm e}$([O\,{\sc ii}]), 
whereas the temperature $T_{\rm e}$([O\,{\sc iii}]) from high-excitation CELs was adopted for $N_{\rm e}$([Ar\,{\sc iv}]) and $N_{\rm e}$([Cl\,{\sc iii}]) where adequate lines were available. The density from high-excitation CELs, either [Ar\,{\sc iv}] or [Cl\,{\sc iii}], was also used to calculate $T_{\rm e}$([O\,{\sc iii}]).

The derived electron temperatures and densities for our sample of PNe are presented in Table~\ref{wc:tab:diagnostic:cels}. 
The diagnostic type ($T_{\rm e}$ or $N_{\rm e}$), ion, diagnostic lines are given in Columns 1--3, respectively. The derived values of $T_{\rm e}$ are $N_{\rm e}$ with the corresponding uncertainties are given for each PN in the last column. Figure~\ref{wc:fig:dd:cels} shows the $N_{\rm e}$--$T_{\rm e}$ diagnostic diagram of PB\,6
(the corresponding diagrams for all the other objects can be found in the online version of the journal).

\begin{table*}
\caption{References for atomic data.
\label{wc:tab:atomicdata}
}
\footnotesize
\centering
\begin{tabular}{lll}
\hline\hline
\noalign{\smallskip}
Ion 	& Transition probabilities & Collision strengths \\
\noalign{\smallskip}
\hline 
\noalign{\smallskip} 
N${}^{0}$   & \citet{Tachiev2002} & \citet{Tayal2006}  \\ 
N${}^{+}$   & \citet{Tachiev2001} & \citet{Tayal2011} \\ 
\noalign{\smallskip} 
O${}^{0}$   & \citet{FroeseFischer2004} & \citet{Zatsarinny2003}, \citet{Bell1998} \\ 
O${}^{+}$   & \citet{Zeippen1982} & \citet{Kisielius2009}  \\ 
O${}^{2+}$  & \citet{Storey2000}, \citet{Tachiev2001} &  \citet{Lennon1994}, \citet{Bhatia1993} \\ 
\noalign{\smallskip} 
Ne${}^{2+}$ & \citet{Daw2000}, \citet{Landi2005}  &  \citet{McLaughlin2000}\\ 
Ne${}^{3+}$ & \citet{Merkelis1999}   &   \citet{Ramsbottom1998} \\ 
\noalign{\smallskip} 
S${}^{+}$   & Nahar (unpublished, 2001) & \citet{Ramsbottom1996}  \\ 
S${}^{2+}$  & \citet{FroeseFischer2006}, \citet{Tayal1997} & \citet{Hudson2012} \\
\noalign{\smallskip} 
Cl${}^{2+}$  & \citet{Mendoza1982} &  \cite{Ramsbottom2001} \\ 
\noalign{\smallskip} 
Ar${}^{2+}$ & \citet{Biemont1986} & \citet{Galavis1995} \\ 
Ar${}^{3+}$ & \citet{Dere2019} & \citet{Ramsbottom1997}, \citet{Ramsbottom1997a} \\ 
Ar${}^{4+}$ & \citet{Biemont1983} & \citet{Galavis1995} \\ 
\noalign{\smallskip}  
Fe${}^{2+}$ & \citet{Ercolano2008} & \citet{Ercolano2008} \\ 
\noalign{\smallskip}
\hline
\noalign{\smallskip}
Ion 	& Effective recombination coefficients & Case \\
\noalign{\smallskip}
\hline 
\noalign{\smallskip} 
H${}^{+}$   & \citet{Storey1995} &  B \\ 
\noalign{\smallskip} 
He${}^{+}$   & \citet{Porter2013} &  B \\ 
\noalign{\smallskip} 
He${}^{2+}$  & \citet{Storey1995} & B \\ 
\noalign{\smallskip} 
C${}^{2+}$  & \citet{Davey2000} & A,\,B \\ 
C${}^{3+}$  & \citet{Pequignot1991} & A \\ 
\noalign{\smallskip} 
N${}^{2+}$  & \citet{Fang2011,Fang2013a} & B \\ 
N${}^{3+}$  & \citet{Pequignot1991} & A \\ 
\noalign{\smallskip} 
O${}^{2+}$  & \citet{Storey2017} & B  \\ 
\noalign{\smallskip}  
\hline
\end{tabular}
\end{table*}

\subsubsection{Electron Densities}
\label{wc:sec:tempdens:cel:dens}

The electron densities deduced from various CEL diagnostic ratios, [S\,{\sc ii}], [O\,{\sc ii}], [Ar\,{\sc iv}] and [Cl\,{\sc iii}], are presented in Table~\ref{wc:tab:diagnostic:cels}. The ionization potential of S$^{+}$ and O$^{+}$, 10.4 and 13.6\,eV, are below those of Cl$^{2+}$ and Ar$^{3+}$, 23.8 and 40.7\,eV, respectively, so these density-diagnostic lines are emitted from dissimilar ionization zones. 
For PB\,6, the electron density derived from [Ar\,{\sc iv}] is lower than those from [Cl\,{\sc iii}], while these lines arise from high excitation regions. 
For the shell of Hb\,4, the density derived from [O\,{\sc ii}] doublet is lower by a factor of 2 than that from [S\,{\sc ii}] and [Cl\,{\sc iii}]. IC\,1297 does not seem to have a large density variation, since densities from different ions are roughly close. 
The discrepancy between densities from low- and high-excited CELs in some objects could be related to inhomogeneous condensations. 
We notice that the [Ar\,{\sc iv}] $\lambda\lambda$4711,\,4740 doublet lines have the highest critical densities among all the density-diagnostic lines.\footnote{Critical densities $N_{\rm cr}$ with $T_{\rm e} =10\,000$\,K and the atomic data from Table~\ref{wc:tab:atomicdata}:
[S\,{\sc ii}] $\lambda\lambda$6717,\,6731, $N_{\rm cr}=1730$, 5010\,cm$^{-3}$; 
[O\,{\sc ii}] $\lambda\lambda$3726,\,3729, $N_{\rm cr}=4090$, 1210\,cm$^{-3}$; 
[Ar\,{\sc iv}] $\lambda\lambda$4711,\,4740, $N_{\rm cr}=12\,510$, 96\,920\,cm$^{-3}$; 
[Cl\,{\sc iii}] $\lambda\lambda$5518,\,5538, $N_{\rm cr}=5200$, 24\,950\,cm$^{-3}$, respectively.
} With the relatively low densities prevailing in PB\,6, M\,3-30, and IC\,1297, the $\lambda$4711/$\lambda$4740 flux ratios are less sensitive to density, so small errors in the measurement of those lines give rise to very high uncertainties in the derived densities. The [S\,{\sc ii}] and [O\,{\sc ii}] doublets yield roughly the similar density in Pe\,1-1. However, for M\,3-30, Hb\,4, M\,1-32 and M\,3-15, the densities derived from the [S\,{\sc ii}] and [O\,{\sc ii}] are very different, while they are emitted from similar ionization zone. This might be due to the poor quality of the [O\,{\sc ii}] $\lambda\lambda$ 3726,3729 emission lines measured from the blue end of the spectrum \citep[see e.g.][]{Rodriguez2020}. Alternatively, it may be explained by the atomic data \citep[see][]{Kisielius2009}. For PB\,6, the density derived from the [Cl\,{\sc iii}] diagnostic line ratios is slightly higher than that from the [S\,{\sc ii}] doublet. The [Cl\,{\sc iii}] diagnostic lines with higher critical densities could preferentially be emitted from a higher density medium. This behavior suggests the presence of density inhomogeneities in this object.

Fig.\,\ref{wc:plasma:diagnostics} (top panel) shows the electron density $N_{\rm e}$ from singly-ionized CELs ([S\,{\sc ii}];  apart from [O\,{\sc ii}] in K\,2-16) plotted against the intrinsic nebular H$\beta$ surface brightness. The dashed line represents a linear fit to the logarithmic values for the 18 PNe, which has a Pearson correlation coefficient of $r=0.77$ and a null-hypothesis testing $p$-value of $0.0002$:
\begin{equation}
\log N_{\rm e} =  (4.582 \pm 0.219)  + (0.508 \pm  0.105)\, \log S({\rm H}\beta)
\label{wc:eq_ne_shb}
\end{equation}
where the dereddened nebular H$\beta$ surface brightness is defined as the integrated H$\beta$ flux divided by the nebular area, $S({\rm H}\beta)=I({\rm H}\beta)/(\pi r^2)$, in unit of erg\,cm$^{-2}$\,s$^{-1}$\,sr$^{-1}$, the intrinsic H$\beta$ flux $I({\rm H}\beta)=10^{c({\rm H}\beta)}\,F({\rm H}\beta)$, and $r$ is the nebular optical angular radius (see Table~\ref{wc:tab:obs:journal}). 
It is in agreement with the theoretical relation approximated by \citet{Odell1962}, $S({\rm H}\beta) \propto \varepsilon\,r\,N_{\rm e}^{2} \propto \varepsilon^{2/3} M^{1/3} N_{\rm e}^{5/3}$, where $M$ is the total mass of the nebula and $\varepsilon$ is the filling factor. 
The nebular H$\beta$ surface brightness was found to decline with radii 
as the nebula expands and the density drops 
\citep{Stanghellini2002,Stanghellini2003,Stanghellini2008,Frew2016}, 
so it can represent an evolutionary indicator of the nebula.

\setcounter{table}{5}
\begin{table}
\caption{Plasma diagnostics based on CELs.
\label{wc:tab:diagnostic:cels}}
\centering
\begin{tabular}{llcc}
\hline\hline
\noalign{\smallskip}
Type& Ion  & Diagnostic & Value \\
\noalign{\smallskip}
\hline
\noalign{\smallskip}
\multicolumn{4}{c}{PB 6          (PNG278.8$+$04.9) }\\
\noalign{\smallskip}
\hline
\noalign{\smallskip}
$T_{\rm e}$                &[N~{\sc ii}]     &6548.10+6583.50     /5754.60              & $     11300_{      -260}^{+       200}$ \\
\noalign{\smallskip}
$T_{\rm e}$\,$_{\rm rc}$   &[N~{\sc ii}]     &6548.10+6583.50     /5754.60              & $     10270_{      -220}^{+       220}$ \\
\noalign{\smallskip}
$T_{\rm e}$                &[O~{\sc iii}]    &4958.91+5006.84     /4363.21              & $     14220_{       -90}^{+       110}$ \\
\noalign{\smallskip}
$T_{\rm e}$\,$_{\rm rc}$   &[O~{\sc iii}]    &4958.91+5006.84     /4363.21              & $     14040_{      -130}^{+        90}$ \\
\noalign{\smallskip}
$N_{\rm e}$                &[S~{\sc ii}]     &6730.82             /6716.44              & $      1800_{      -320}^{+       400}$ \\
\noalign{\smallskip}
$N_{\rm e}$                &[Ar~{\sc iv}]    &4740.17             /4711.37              & $      1370_{      -100}^{+        70}$ \\
\noalign{\smallskip}
$N_{\rm e}$                &[Cl~{\sc iii}]   &5537.60             /5517.66              & $      2190_{      -270}^{+       390}$ \\
\noalign{\medskip}
\multicolumn{4}{c}{                          \ldots }\\
\noalign{\medskip}
\noalign{\medskip}
\hline
\end{tabular}
\begin{list}{}{}
\item[\textbf{Note:}]Table \ref{wc:tab:diagnostic:cels} is published in its entirety in the machine-readable format. A portion is shown here for guidance regarding its form and content. 
The label ``rc'' indicates that the auroral lines, $[$N\,{\sc ii}$]$ $\lambda5755$ and $[$O\,{\sc iii}$]$ $\lambda4363$, are corrected for recombination contribution.
\end{list}
\end{table}

\setcounter{figure}{1}
\begin{figure*}
\begin{center}
\includegraphics[width=0.7\textwidth, trim = 0 10 0 10, clip, angle=0]{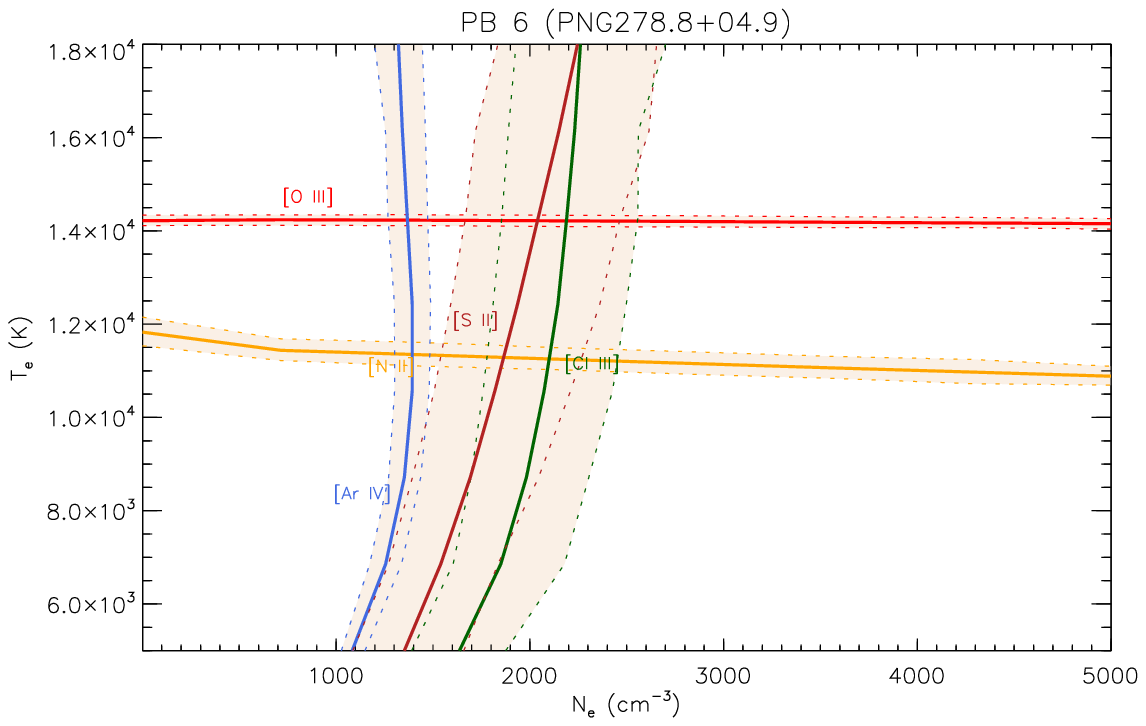}
\caption{$N_{\rm e}$--$T_{\rm e}$ diagnostic diagram of PB\,6 based on CELs. 
The upper and lower limits at the 90\% confidence level are plotted by the dotted lines. 
The complete figure set (19 images) is available in the online journal. 
\label{wc:fig:dd:cels}}
\end{center}

\figsetstart
\figsetnum{2}
\figsettitle{$N_{\rm e}$--$T_{\rm e}$ diagnostic diagram based on CELs. The upper and lower limits at the 90\% confidence level are plotted by the dotted lines. }

\figsetgrpstart
\figsetgrpnum{2.1}
\figsetgrptitle{PB\,6.}
\figsetplot{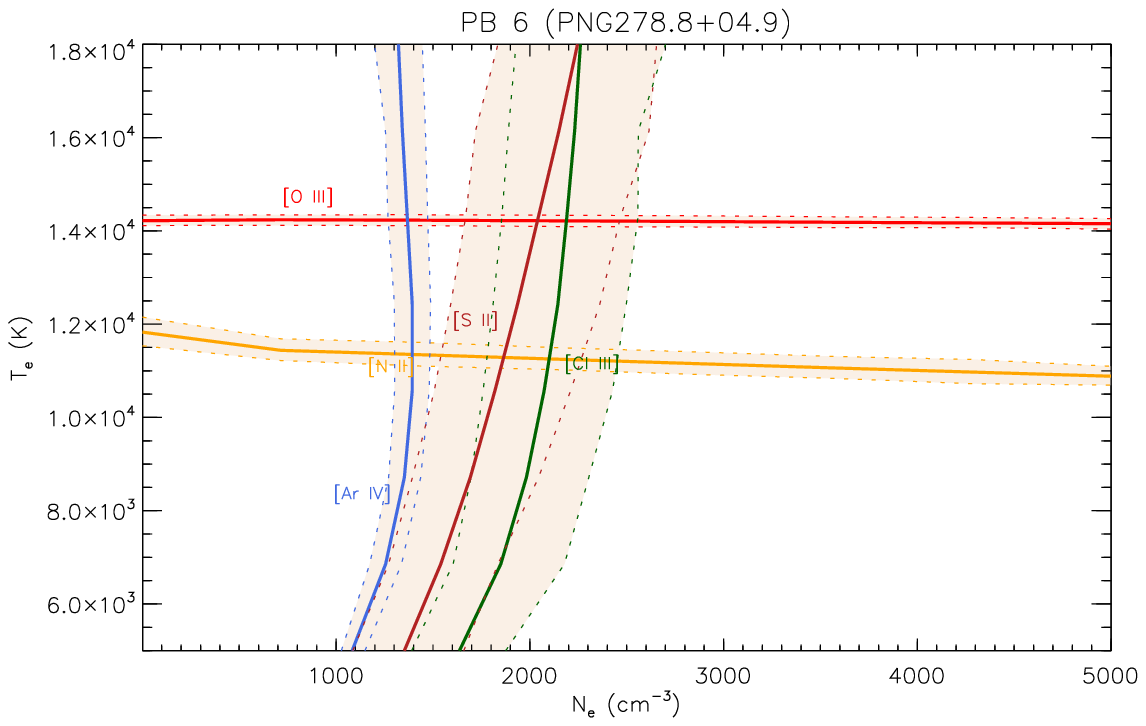}
\figsetgrpnote{$N_{\rm e}$--$T_{\rm e}$ diagnostic diagram of PB\,6 based on CELs. The upper and lower limits at the 90\% confidence level are plotted by the dotted lines.}
\figsetgrpend

\figsetgrpstart
\figsetgrpnum{2.2}
\figsetgrptitle{M\,3-30.}
\figsetplot{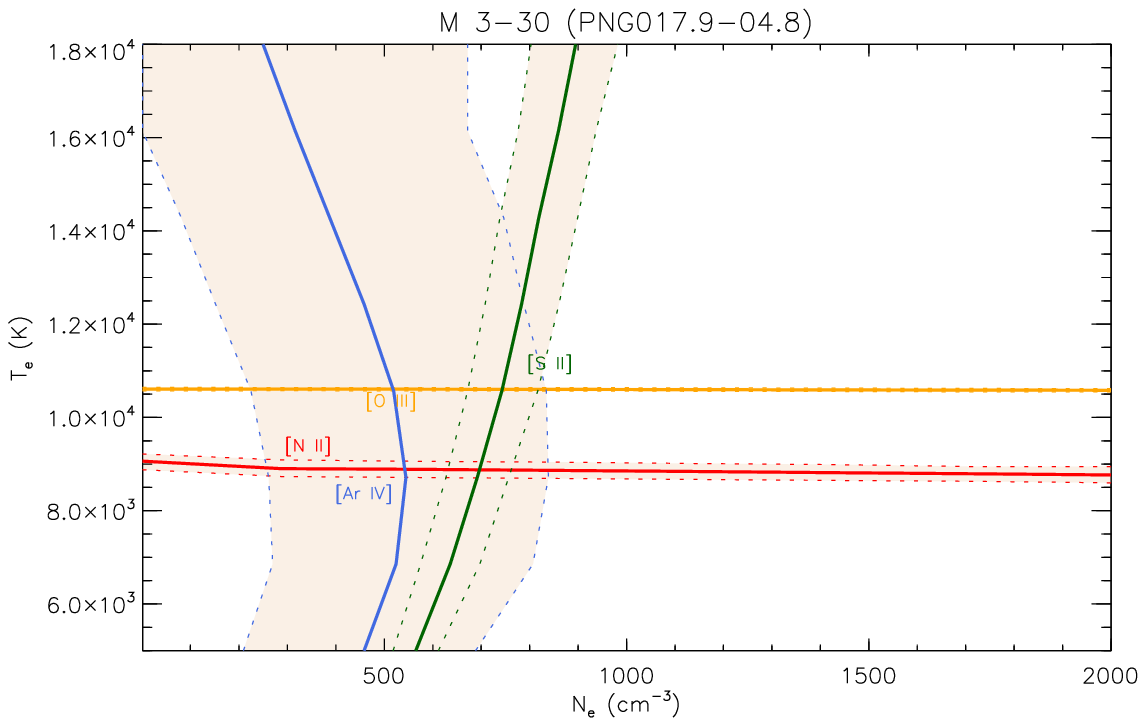}
\figsetgrpnote{$N_{\rm e}$--$T_{\rm e}$ diagnostic diagram of M\,3-30 based on CELs. The upper and lower limits at the 90\% confidence level are plotted by the dotted lines.}
\figsetgrpend

\figsetgrpstart
\figsetgrpnum{2.3}
\figsetgrptitle{Hb\,4 (shell).}
\figsetplot{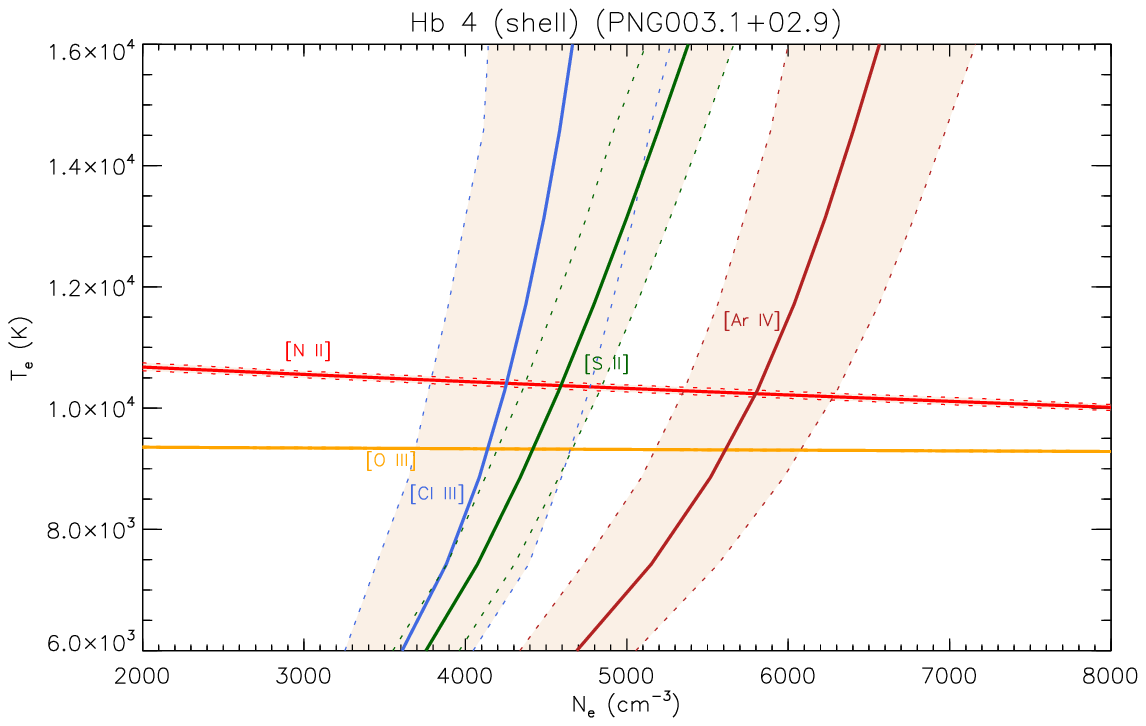}
\figsetgrpnote{$N_{\rm e}$--$T_{\rm e}$ diagnostic diagram of Hb\,4 (shell) based on CELs. The upper and lower limits at the 90\% confidence level are plotted by the dotted lines.}
\figsetgrpend

\figsetgrpstart
\figsetgrpnum{2.4}
\figsetgrptitle{Hb\,4 (N-knot).}
\figsetplot{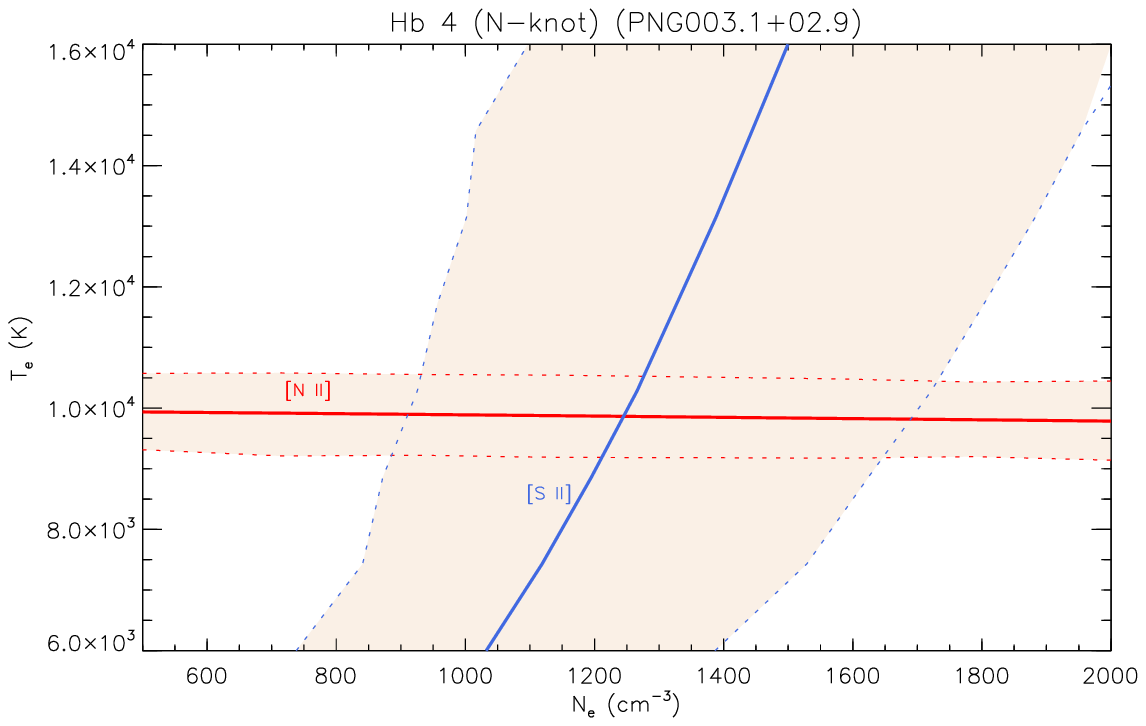}
\figsetgrpnote{$N_{\rm e}$--$T_{\rm e}$ diagnostic diagram of Hb\,4 (N-knot) based on CELs. The upper and lower limits at the 90\% confidence level are plotted by the dotted lines.}
\figsetgrpend

\figsetgrpstart
\figsetgrpnum{2.5}
\figsetgrptitle{IC\,1297.}
\figsetplot{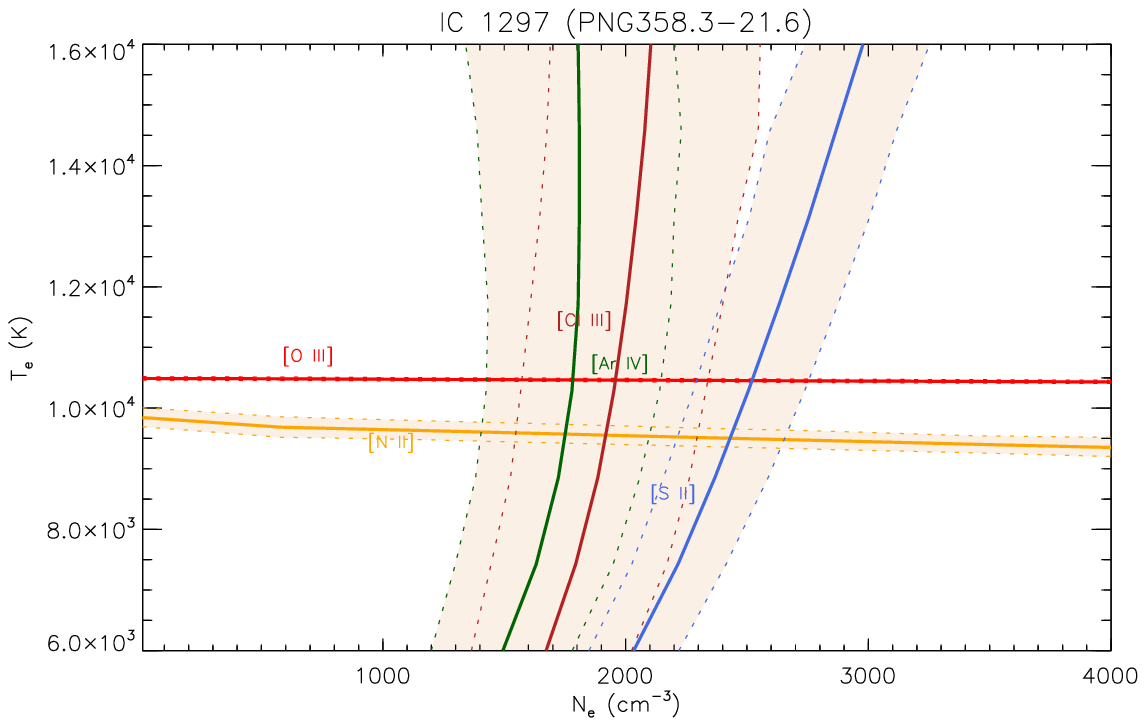}
\figsetgrpnote{$N_{\rm e}$--$T_{\rm e}$ diagnostic diagram of IC\,1297 based on CELs. The upper and lower limits at the 90\% confidence level are plotted by the dotted lines.}
\figsetgrpend

\figsetgrpstart
\figsetgrpnum{2.6}
\figsetgrptitle{Th\,2-A.}
\figsetplot{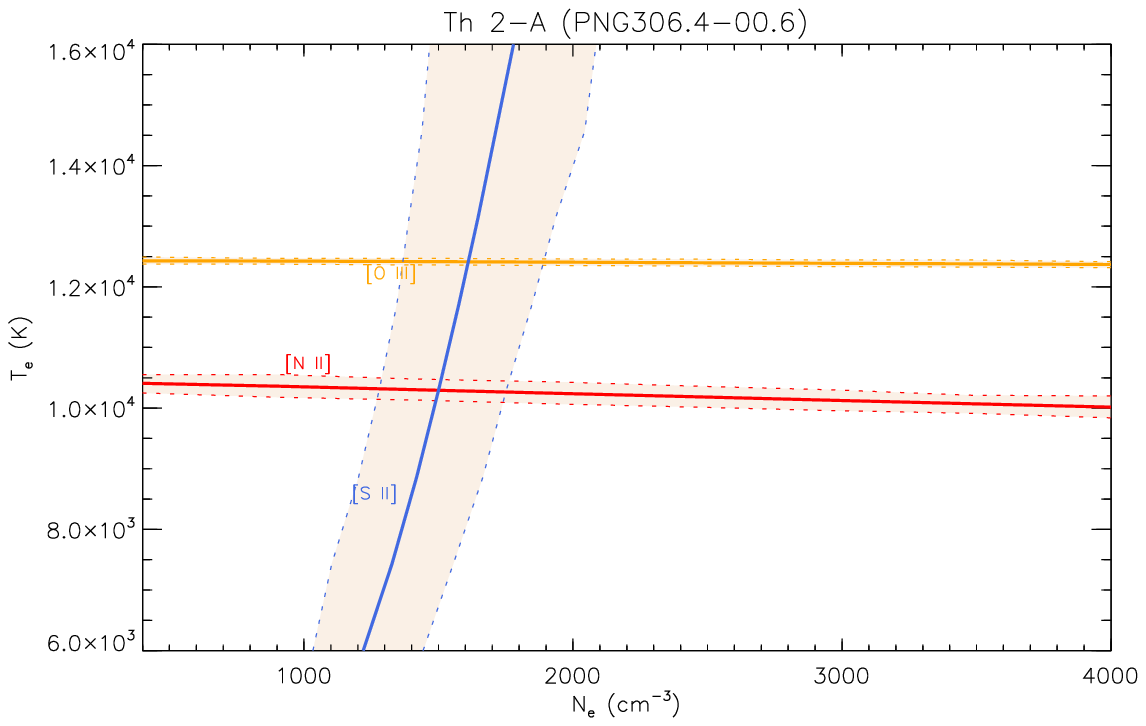}
\figsetgrpnote{$N_{\rm e}$--$T_{\rm e}$ diagnostic diagram of Th\,2-A based on CELs. The upper and lower limits at the 90\% confidence level are plotted by the dotted lines.}
\figsetgrpend

\figsetgrpstart
\figsetgrpnum{2.7}
\figsetgrptitle{Pe\,1-1.}
\figsetplot{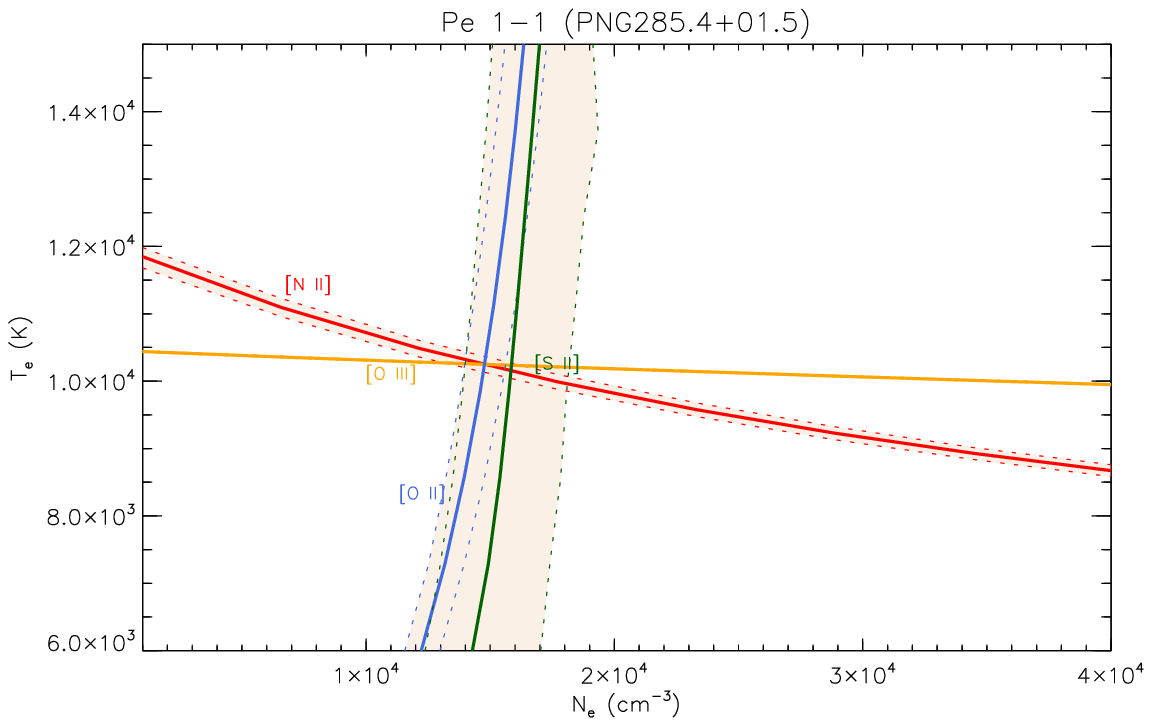}
\figsetgrpnote{$N_{\rm e}$--$T_{\rm e}$ diagnostic diagram of Pe\,1-1 based on CELs. The upper and lower limits at the 90\% confidence level are plotted by the dotted lines.}
\figsetgrpend

\figsetgrpstart
\figsetgrpnum{2.8}
\figsetgrptitle{M\,1-32.}
\figsetplot{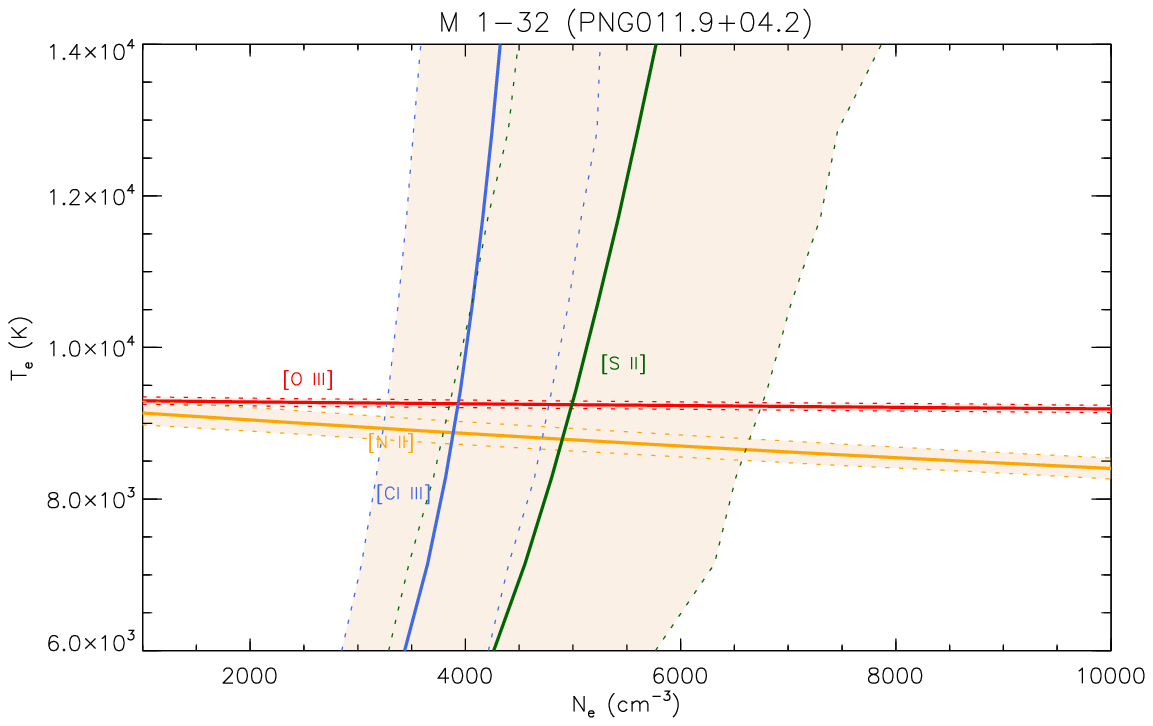}
\figsetgrpnote{$N_{\rm e}$--$T_{\rm e}$ diagnostic diagram of M\,1-32 based on CELs. The upper and lower limits at the 90\% confidence level are plotted by the dotted lines.}
\figsetgrpend

\figsetgrpstart
\figsetgrpnum{2.9}
\figsetgrptitle{M\,3-15.}
\figsetplot{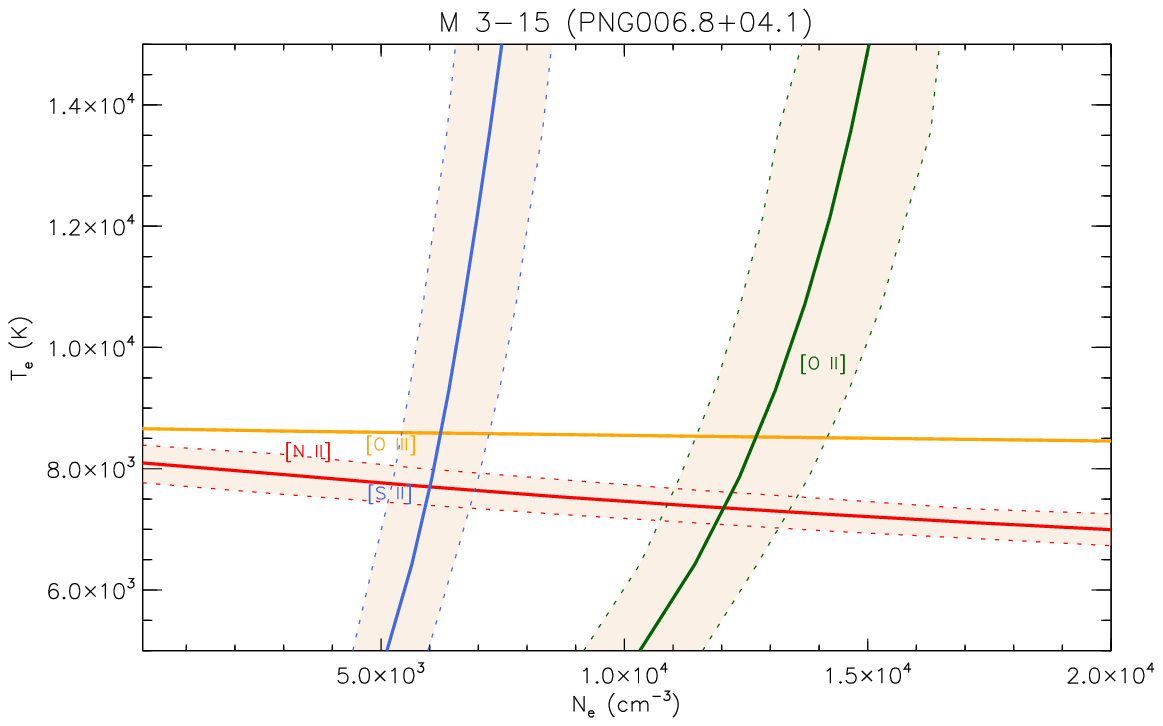}
\figsetgrpnote{$N_{\rm e}$--$T_{\rm e}$ diagnostic diagram of M\,3-15 based on CELs. The upper and lower limits at the 90\% confidence level are plotted by the dotted lines.}
\figsetgrpend

\figsetgrpstart
\figsetgrpnum{2.10}
\figsetgrptitle{M\,1-25.}
\figsetplot{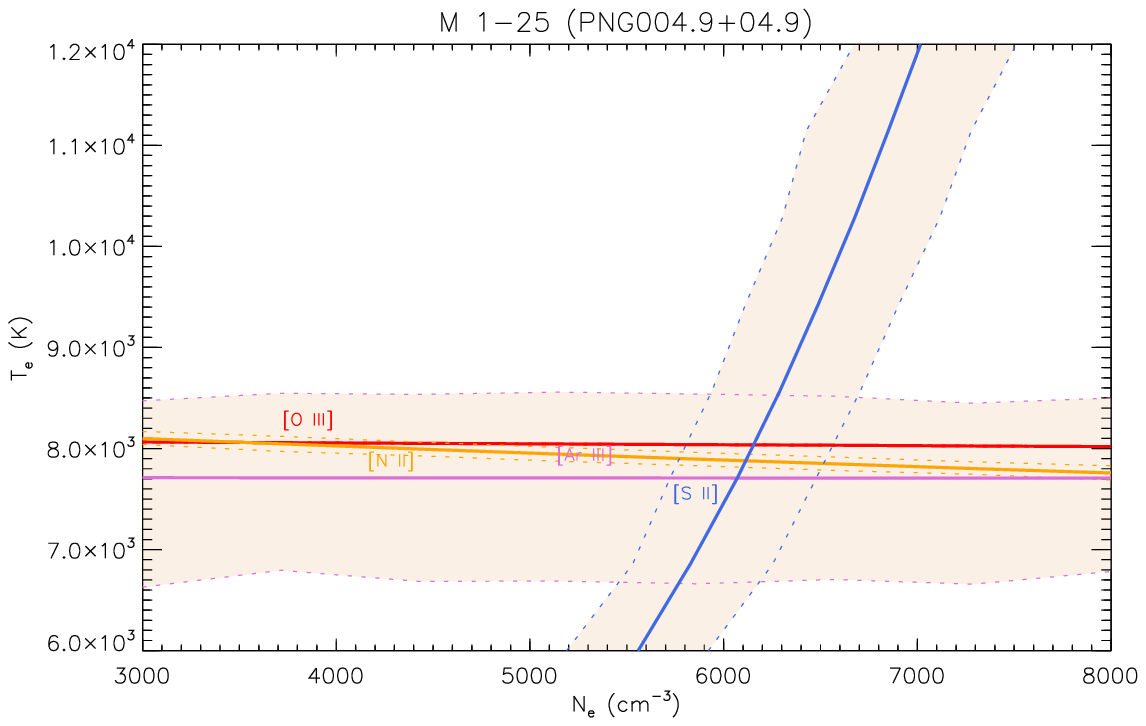}
\figsetgrpnote{$N_{\rm e}$--$T_{\rm e}$ diagnostic diagram of M\,1-25 based on CELs. The upper and lower limits at the 90\% confidence level are plotted by the dotted lines.}
\figsetgrpend

\figsetgrpstart
\figsetgrpnum{2.11}
\figsetgrptitle{Hen\,2-142.}
\figsetplot{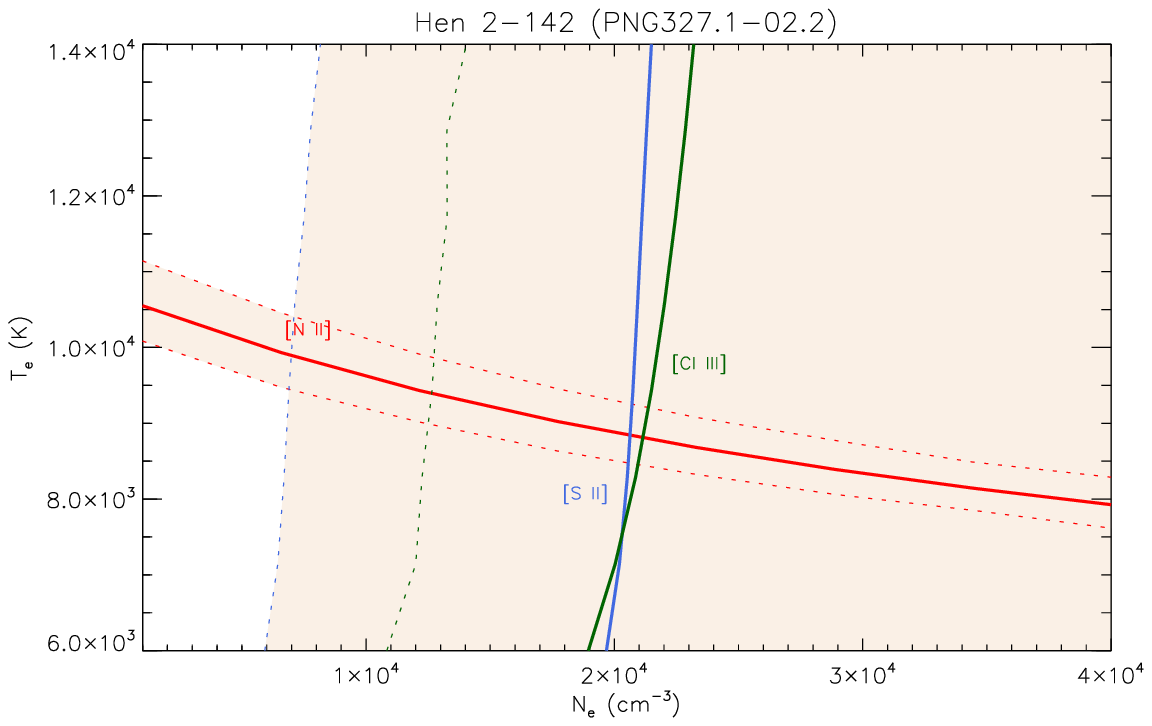}
\figsetgrpnote{$N_{\rm e}$--$T_{\rm e}$ diagnostic diagram of Hen\,2-142 based on CELs. The upper and lower limits at the 90\% confidence level are plotted by the dotted lines.}
\figsetgrpend

\figsetgrpstart
\figsetgrpnum{2.12}
\figsetgrptitle{Hen\,3-1333.}
\figsetplot{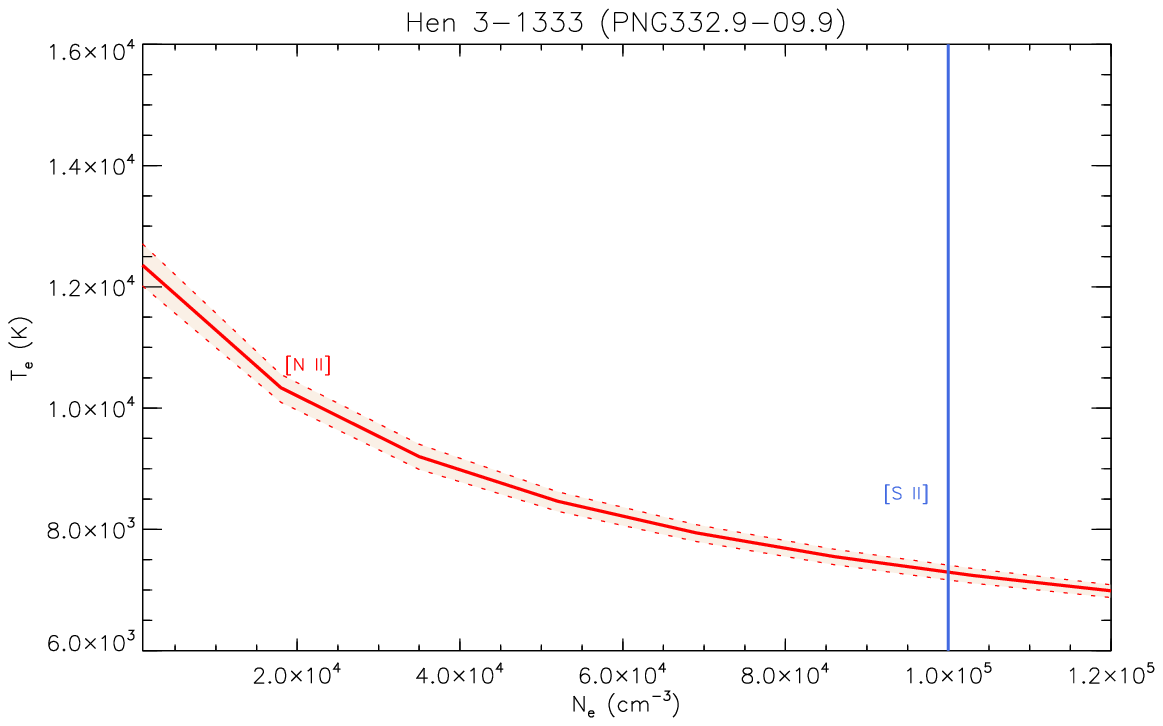}
\figsetgrpnote{$N_{\rm e}$--$T_{\rm e}$ diagnostic diagram of Hen\,3-1333 based on CELs. The upper and lower limits at the 90\% confidence level are plotted by the dotted lines.}
\figsetgrpend

\figsetgrpstart
\figsetgrpnum{2.13}
\figsetgrptitle{Hen\,2-113.}
\figsetplot{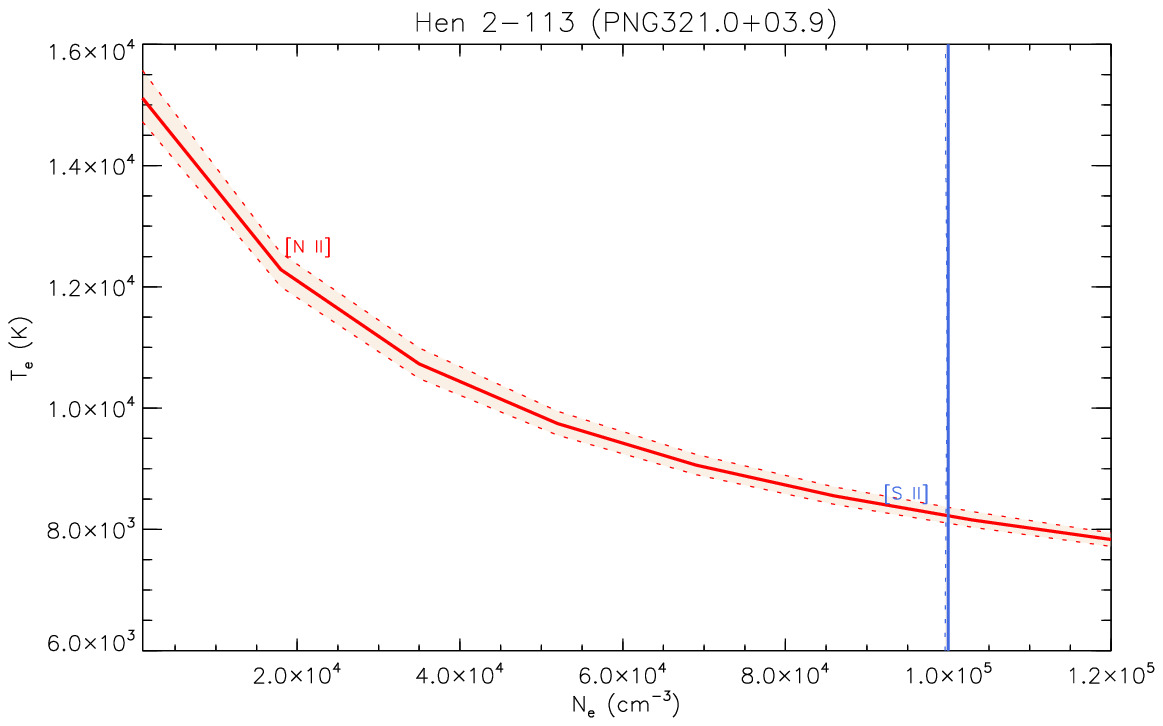}
\figsetgrpnote{$N_{\rm e}$--$T_{\rm e}$ diagnostic diagram of Hen\,2-113 based on CELs. The upper and lower limits at the 90\% confidence level are plotted by the dotted lines.}
\figsetgrpend

\figsetgrpstart
\figsetgrpnum{2.14}
\figsetgrptitle{K\,2-16.}
\figsetplot{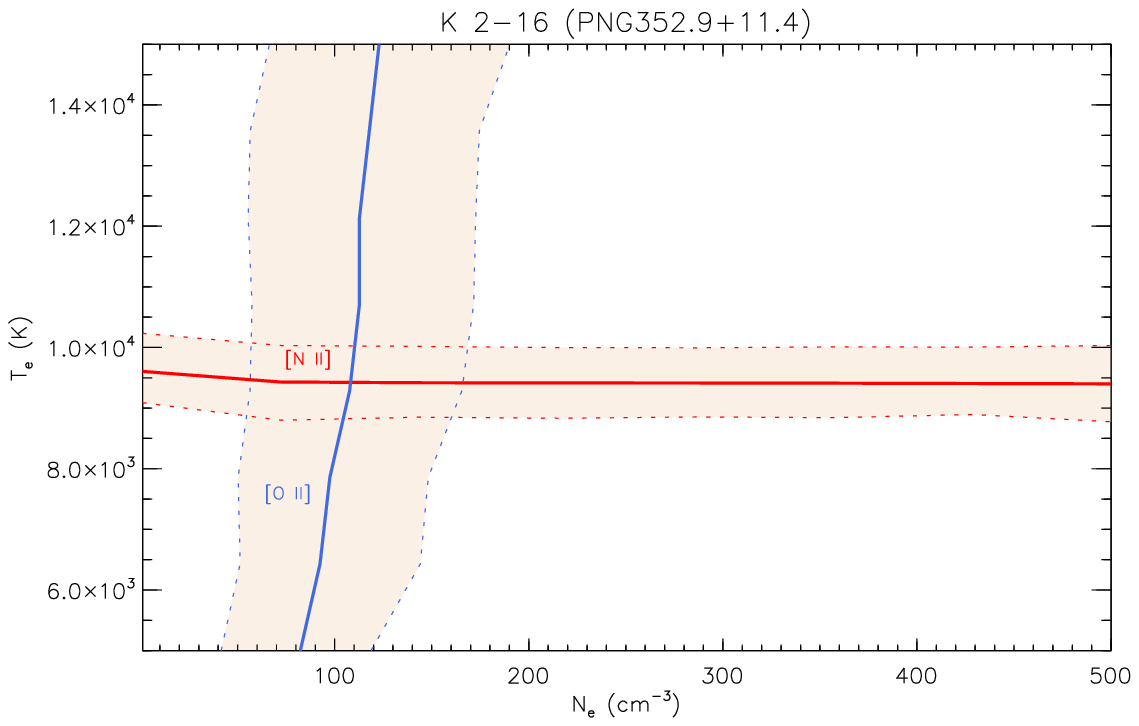}
\figsetgrpnote{$N_{\rm e}$--$T_{\rm e}$ diagnostic diagram of K\,2-16 based on CELs. The upper and lower limits at the 90\% confidence level are plotted by the dotted lines.}
\figsetgrpend

\figsetgrpstart
\figsetgrpnum{2.15}
\figsetgrptitle{NGC\,6578.}
\figsetplot{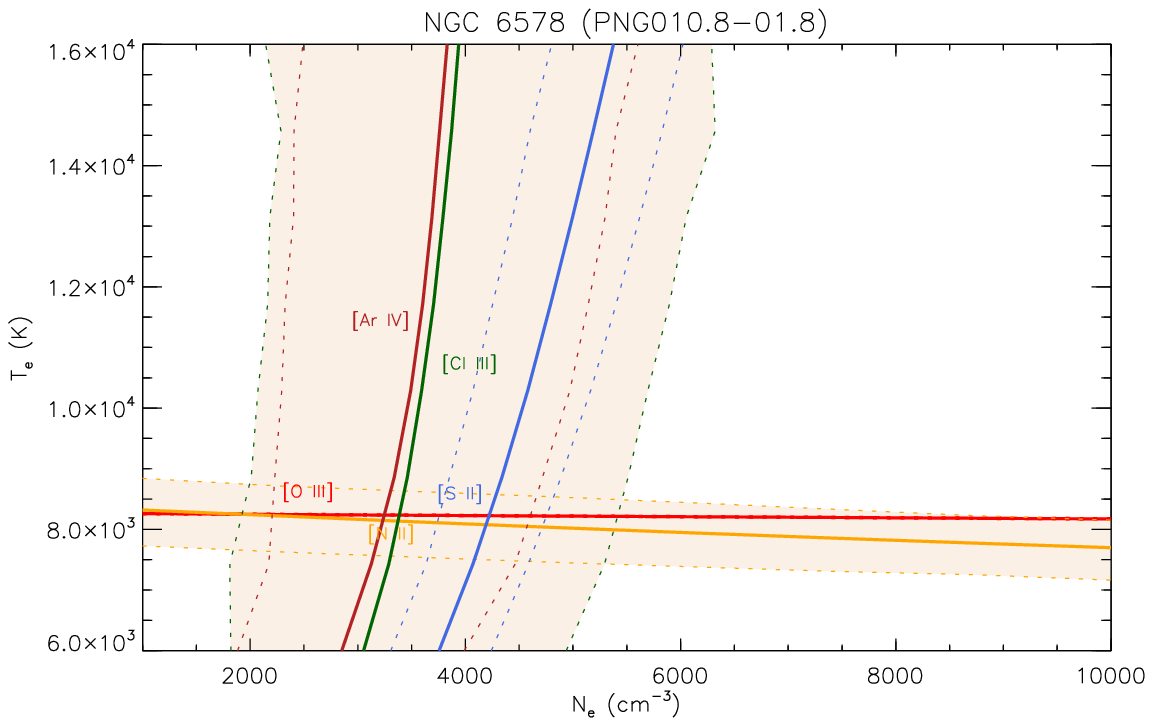}
\figsetgrpnote{$N_{\rm e}$--$T_{\rm e}$ diagnostic diagram of for NGC\,6578 based on CELs. The upper and lower limits at the 90\% confidence level are plotted by the dotted lines.}
\figsetgrpend

\figsetgrpstart
\figsetgrpnum{2.16}
\figsetgrptitle{M\,2-42.}
\figsetplot{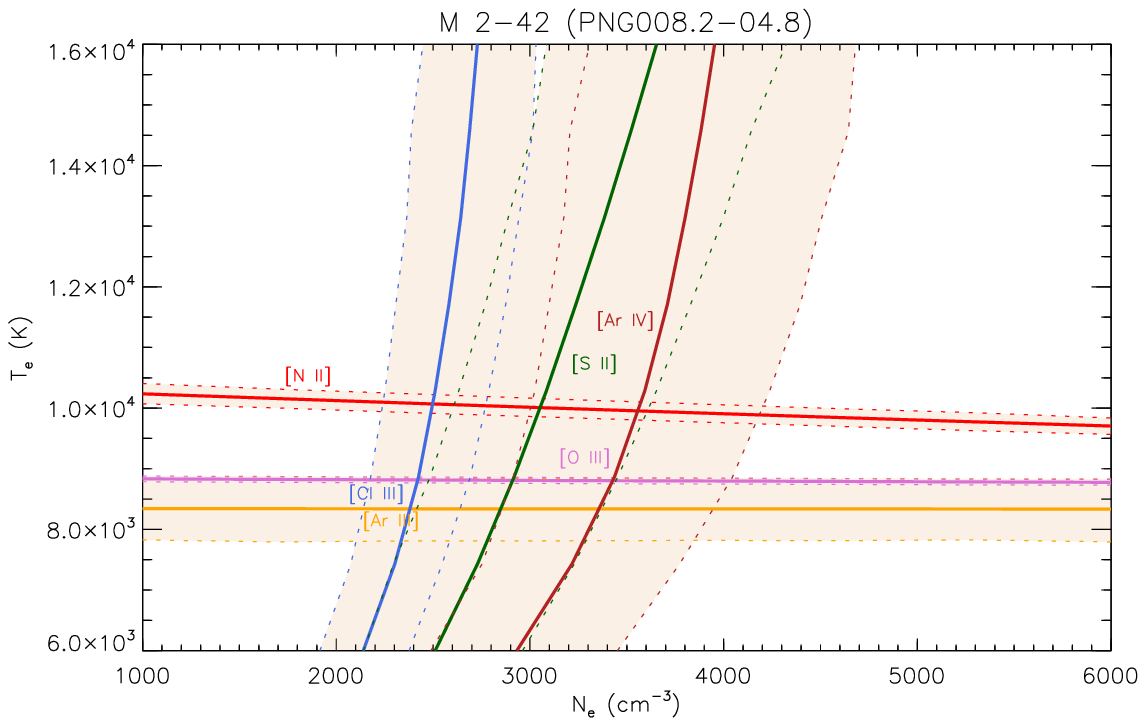}
\figsetgrpnote{$N_{\rm e}$--$T_{\rm e}$ diagnostic diagram of  M\,2-42 based on CELs. The upper and lower limits at the 90\% confidence level are plotted by the dotted lines.}
\figsetgrpend

\figsetgrpstart
\figsetgrpnum{2.17}
\figsetgrptitle{NGC\,6567.}
\figsetplot{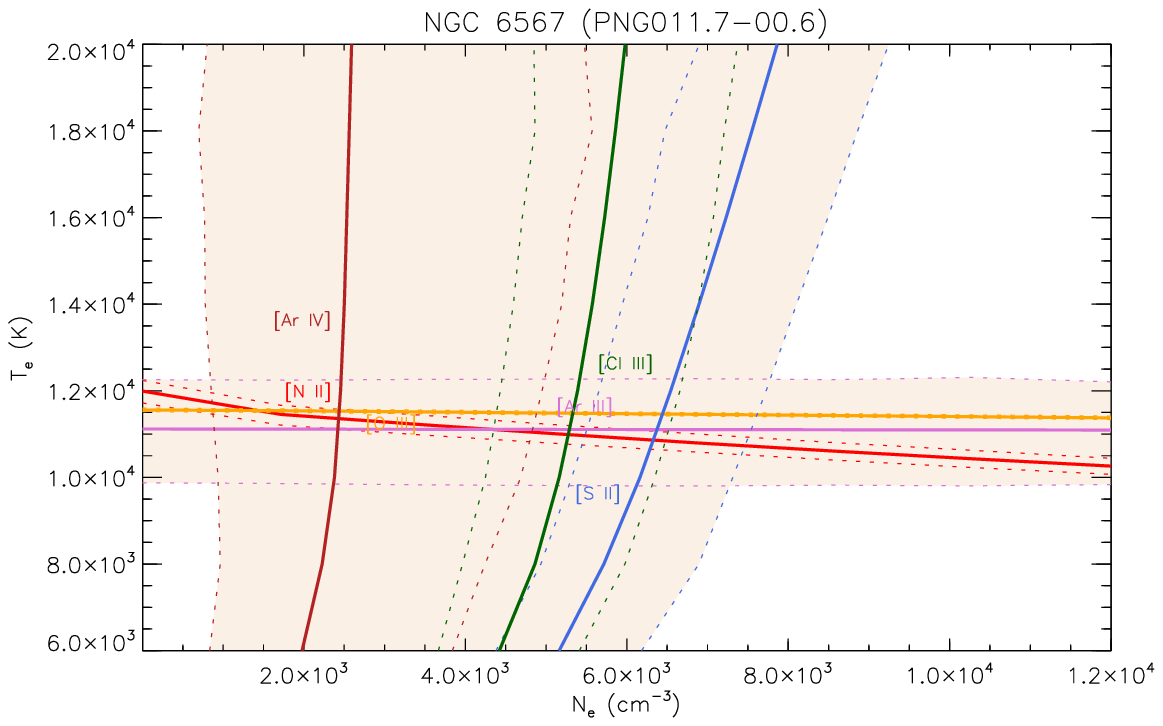}
\figsetgrpnote{$N_{\rm e}$--$T_{\rm e}$ diagnostic diagram of NGC\,6567 based on CELs. The upper and lower limits at the 90\% confidence level are plotted by the dotted lines.}
\figsetgrpend

\figsetgrpstart
\figsetgrpnum{2.18}
\figsetgrptitle{NGC\,6629.}
\figsetplot{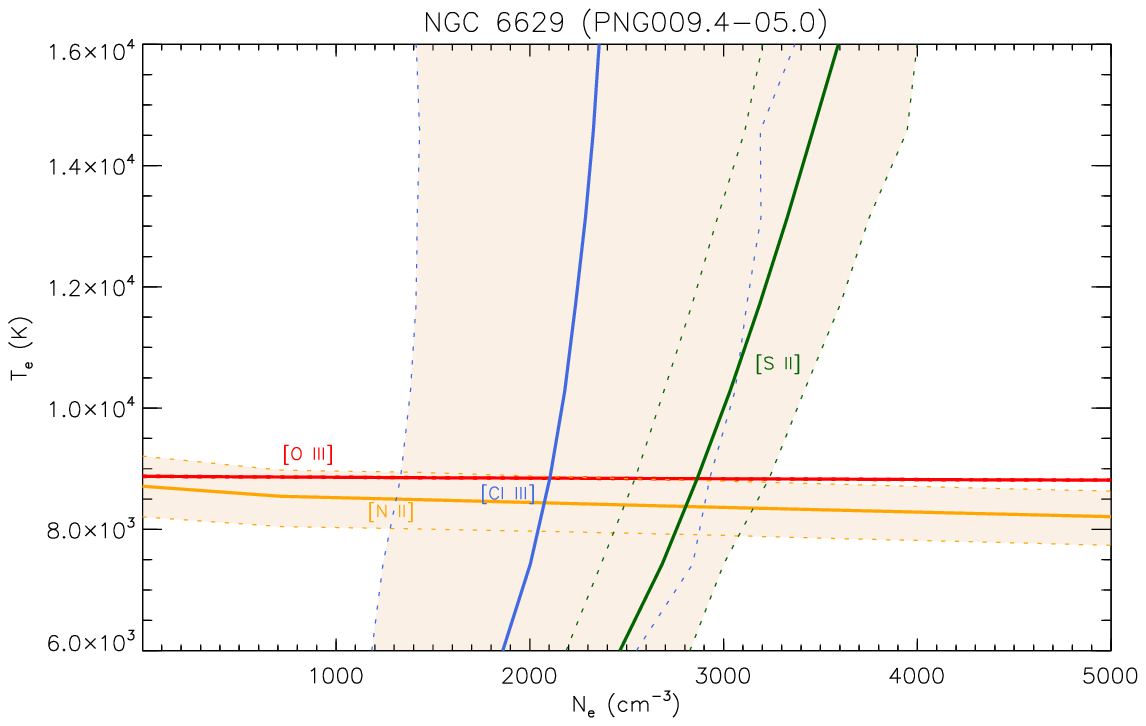}
\figsetgrpnote{$N_{\rm e}$--$T_{\rm e}$ diagnostic diagram of NGC\,6629 based on CELs. The upper and lower limits at the 90\% confidence level are plotted by the dotted lines.}
\figsetgrpend

\figsetgrpstart
\figsetgrpnum{2.19}
\figsetgrptitle{Sa\,3-107.}
\figsetplot{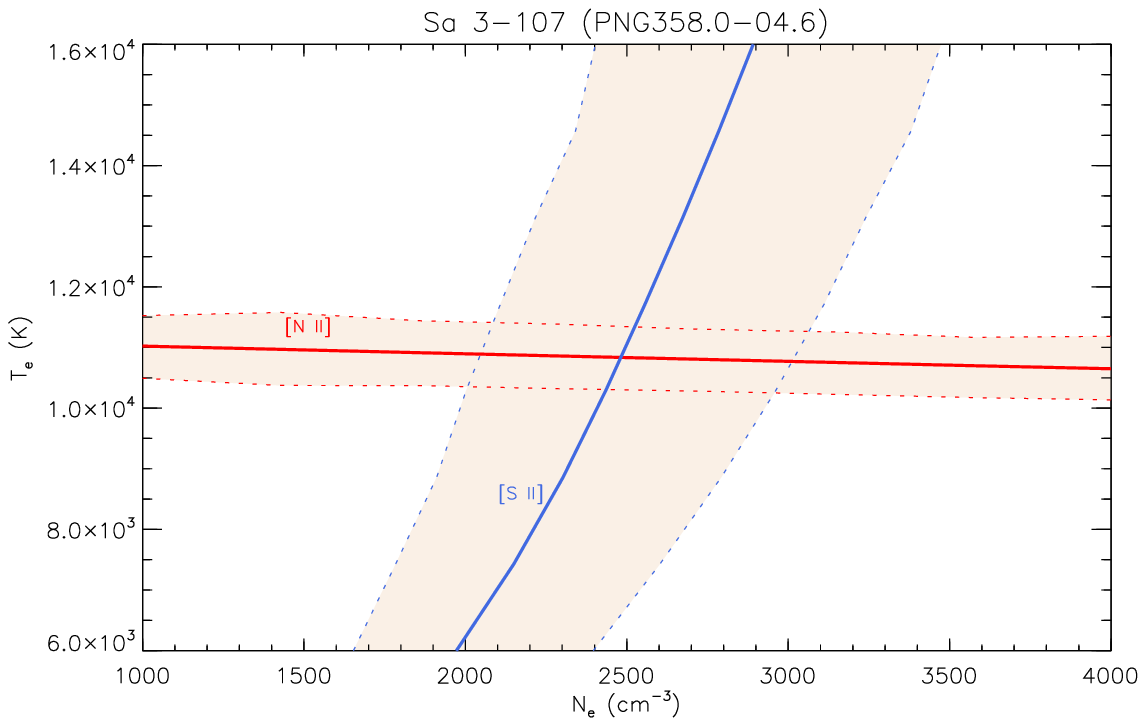}
\figsetgrpnote{$N_{\rm e}$--$T_{\rm e}$ diagnostic diagram of Sa\,3-107 based on CELs. The upper and lower limits at the 90\% confidence level are plotted by the dotted lines.}
\figsetgrpend

\figsetend

\end{figure*}


\subsubsection{Electron Temperatures }
\label{wc:sec:tempdens:cel:temp}

The electron temperatures deduced from the ratios of the nebular lines to auroral lines are presented in Table~\ref{wc:tab:diagnostic:cels}. We also derived $T_{\rm e}$([N\,{\sc ii}]) and $T_{\rm e}$([O\,{\sc iii}]) after removing recombination contributions from the auroral lines where adequate recombination lines were available (labeled by ``rc'').

Recombination excitation can make a significant contribution to the [N\,{\sc ii}] $\lambda$5755 auroral line, the [O\,{\sc ii}] $\lambda\lambda$3726,3729 nebular lines, and the [O\,{\sc ii}] $\lambda\lambda$7320,7330 auroral lines; but only a small contribution to the [O\,{\sc iii}] $\lambda$4363 auroral line. Most nitrogen is in the form of N$^{2+}$, so the recombination contribution to the [N\,{\sc ii}] auroral line is significant. As most oxygen is in the form of O$^{2+}$, the recombination contribution from O$^{3+}$ to the [O\,{\sc iii}] auroral line is insignificant. The recombination contribution can lead to apparently high temperatures deduced from the [N\,{\sc ii}] ($\lambda$6548+$\lambda$6584)/$\lambda$5755 ratio. 
Furthermore, this could also lead to over-estimated temperatures in nebulae containing inhomogeneous condensations, whose density is higher than the critical densities of the nebular lines \citep{Viegas1994}. Owing to the relatively low critical densities of the [N\,{\sc ii}] nebular lines,\footnote{Critical densities $N_{\rm cr}$ with $T_{\rm e} =10\,000$\,K and the atomic data from Table~\ref{wc:tab:atomicdata}:
[N\,{\sc ii}] $\lambda\lambda$5755,\,6584, 
$N_{\rm cr}=1.646\times 10^{7}$, $8.88\times 10^{4}$\,cm$^{-3}$; 
[O\,{\sc iii}] $\lambda\lambda$4363,\,5007, $N_{\rm cr}=2.554\times 10^{7}$, $6.874\times 10^{5}$\,cm$^{-3}$, respectively.} the presence of density inhomogeneities may lead to apparently high electron temperatures. Because of the fairly low critical densities of the [O\,{\sc ii}] $\lambda\lambda$3726,3729 nebular lines, their recombination emissivities depend also on electron density, in addition to electron temperature and abundance, which make these lines difficult to interpret when the nebula contains inhomogeneous condensations and/or chemical inhomogeneities.

The recombination contribution to the [N\,{\sc ii}] $\lambda$5755 auroral line can be estimated using the formula given by \citet{Liu2000}: 
\begin{equation}
\frac{I_{\rm R}(\lambda5755)}{I({\rm H}\beta)}=3.19 \,\, {t^{0.30}} \, \bigg( \frac{\rm N^{2+}}{\rm H^{+}} \bigg)_{\rm ORLs},
\label{wc:temp:nii:correction}
\end{equation}
where $t\equiv T_{\rm e}$(ORL$)/10^4$ is the electron temperature adopted for the ORL abundance analysis in $10^4$\,K (\S\,\ref{wc:sec:abundances:orl}), and ${\rm N^{2+}}/{\rm H^{+}}$ is derived from the N\,{\sc ii} ORLs (described in \S\,\ref{wc:sec:abundances:orl}). 

\setcounter{figure}{2}
\begin{figure}
\begin{center}
\includegraphics[width=0.49\textwidth, trim = 0 5 15 5, clip, angle=0]{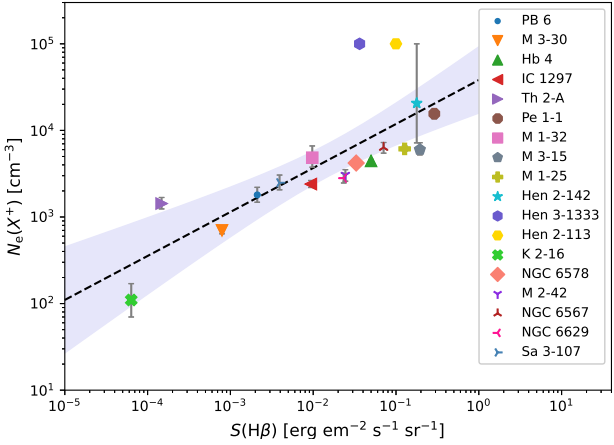}\\
\includegraphics[width=0.49\textwidth, trim = 0 5 15 5, clip, angle=0]{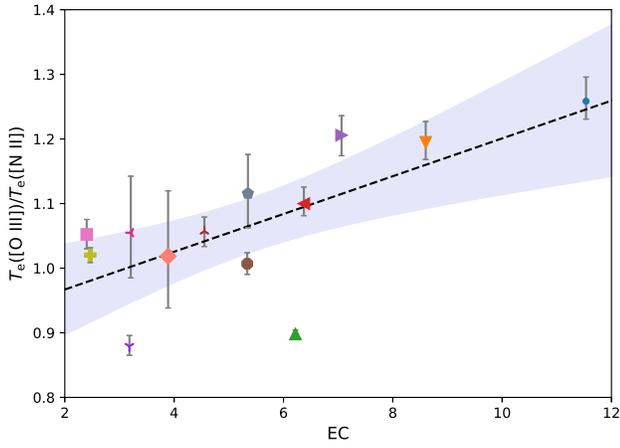}\\
\includegraphics[width=0.49\textwidth, trim = 0 5 15 5, clip, angle=0]{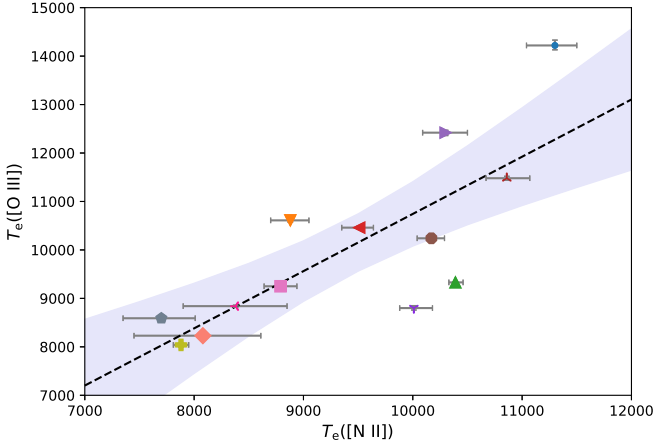}%
\caption{\textit{Top Panel:} The electron density $N_{\rm e}$(X$^{+}$)(cm$^{-3}$) from the singly-ionized forbidden lines ([S\,{\sc ii}], except for [O\,{\sc ii}] in K\,2-16) plotted against the intrinsic nebular H$\beta$ surface brightness $S$(H$\beta$)(erg\,cm$^{-2}$\,s$^{-1}$\,sr$^{-1}$). The dashed line is a linear fit to $\log N_{\rm e}$(X$^{+}$) as a function of $\log S$(H$\beta$) for the 18 PNe, discussed in the text. 
\textit{Middle Panel:} Variation of $T_{\rm e}(\mbox{[O\,{\sc iii}]})$/$T_{\rm e}(\mbox{[N\,{\sc ii}]})$ with excitation class (EC). The dashed line is a linear fit to the 13 points in the figure, discussed in the text.  
\textit{Bottom Panel:} The electron temperature $T_{\rm e}(\mbox{[O\,{\sc iii}]})$ plotted against $T_{\rm e}(\mbox{[N\,{\sc ii}]})$, along with a linear fit shown by a dashed line.
The gray shaded area in each panel corresponds to the 90\% confidence interval of the linear fit. 
\label{wc:plasma:diagnostics}%
}%
\end{center}
\end{figure}

We estimate the recombination contribution to the [O\,{\sc iii}] $\lambda$4363 auroral line using the following formula by \citet{Liu2000}: 
\begin{equation}
\frac{I_{\rm R}(\lambda4363)}{I({\rm H}\beta)}=12.4 \,\, {t^{0.79}} \, \bigg( \frac{\rm O^{3+}}{\rm H^{+}} \bigg)_{\rm ORLs},
\label{wc:temp:oiii:correction}
\end{equation}
where the ${\rm O^{3+}}/{\rm H^{+}}$ ratio is computed using ${\rm O^{3+}}/{\rm H^{+}}=\lbrack ({\rm He}/{\rm H^{+}})^{2/3} -1 \rbrack \times ( {\rm {\rm O^{+}}/{\rm H^{+}} + O^{2+}}/{\rm H^{+}} )$. The ${\rm O^{2+}}/{\rm H^{+}}$ ratio is obtained from the O\,{\sc ii} ORLs (described in \S\,\ref{wc:sec:abundances:orl}), while the ${\rm O^{+}}/{\rm H^{+}}$ ratio is excluded.

For example in PB~6, which has $T_{\rm e}$([N\,{\sc ii}])$=11300$\,K, the observed flux of N\,{\sc ii} lines yields N$^{2+}$/H$^{+}=3.455 \times 10^{-3}$ for $T_{\rm e}$(He\,{\sc i})$=5580$\,K  and $N_{\rm e}= 1800$\,cm$^{-3}$. Inserting them into equation (\ref{wc:temp:nii:correction}), we estimate $I_{\rm R}(\lambda5754)= 0.925$ (where H$\beta=100$), or 20 per cent of the observed intensity of the $\lambda$5755 line. After subtracting the recombination contribution from the observed intensity, the [N\,{\sc ii}] line ratio yields $T_{\rm e}$([N\,{\sc ii}])$=10270$\,K. 
We summarize our findings for all objects in Table~\ref{wc:tab:diagnostic:cels}.

Fig.\,\ref{wc:plasma:diagnostics} (middle panel) plots $T_{\rm e}(\mbox{[O\,{\sc iii}]})$/$T_{\rm e}(\mbox{[N\,{\sc ii}]})$ versus the excitation class (EC), where a trend with increasing the EC is seen. 
We employ the EC defined by \citet{Dopita1990} as ${\rm EC} = 0.45 \times F($\foiii$\lambda 5007)/F({\rm H}\beta)$ if $F($\heii$\lambda 4685)/F({\rm H}\beta)\leqslant0.12$, otherwise ${\rm EC} = 5.54 \times [F($\heii$\lambda 4685)/F({\rm H}\beta) + 0.78]$.
A linear fit (dashed line) to the 13 PNe in the figure yields 
\begin{equation}
\frac{T_{\rm e}(\mbox{[O\,{\sc iii}]})}{T_{\rm e}(\mbox{[N\,{\sc ii}]})} = (0.908 \pm 0.055)+(0.0292\pm 0.0092)\,{\rm EC},
\label{wc:eq_te_ec1}
\end{equation}
with a Pearson $r$-value of $0.69$ and a null-hypothesis $p$-value of 0.009. 
However, if we instead use the electron temperatures 
corrected for recombination contribution where available, 
we obtain a weaker correlation ${T_{\rm e}(\mbox{[O\,{\sc iii}]})}/{T_{\rm e}(\mbox{[N\,{\sc ii}]})} = (0.949 \pm 0.071)+(0.0277\pm 0.0119)\,{\rm EC}$ with $r=0.57$ ($p=0.041$). We caution that 
the recombination-corrected [N\,{\sc ii}] temperature is not always reliable due to the
${\rm N^{2+}}/{\rm H^{+}}$ overestimated from the N\,{\sc ii} ORLs contaminated by fluorescence \citep[][]{Escalante2005,Escalante2012}.

Eq.~(\ref{wc:eq_te_ec1}) is similar to the correlations between $T_{\rm e}(\mbox{[O\,{\sc iii}]})$/$T_{\rm e}(\mbox{[N\,{\sc ii}]})$ and $I(\lambda4686)$ obtained by \citet{Kingsburgh1994} and \citet{Wang2007}. 
This thermal trend with the EC can be explained by radiation fields from the central stars.
For the same 13 PNe in Fig.\,\ref{wc:plasma:diagnostics} (bottom panel), $T_{\rm e}(\mbox{[O\,{\sc iii}]})$ is plotted against $T_{\rm e}(\mbox{[N\,{\sc ii}]})$ having the following linear correlation (dashed line):
\begin{equation}
T_{\rm e}(\mbox{[O\,{\sc iii}]}) = -(1064 \pm 2703) + (1.181 \pm 0.285)\,T_{\rm e}(\mbox{[N\,{\sc ii}]})
\label{wc:eq_te_n2_te_n3}
\end{equation}
with $r=0.78$ and $p=0.002$. If we instead employ the recombination-corrected temperatures  
where applicable, we derive a weaker correlation $T_{\rm e}(\mbox{[O\,{\sc iii}]}) = -(40 \pm 3433) + (1.125 \pm 0.381)\,T_{\rm e}(\mbox{[N\,{\sc ii}]})$ with $r=0.67$ ($p=0.013$).

Eq.~(\ref{wc:eq_te_n2_te_n3}) confirms that the electron 
temperatures derived from the [N\,{\sc ii}] and [O\,{\sc iii}] lines could be associated with the excitation class.
However, the recombination-corrected [N\,{\sc ii}] temperatures may not correctly describe the ionization structure due to the N\,{\sc ii} lines created by resonance fluorescence in low-excited PNe.

\setcounter{figure}{3}
\begin{figure*}
\begin{center}
\includegraphics[width=0.325\linewidth]{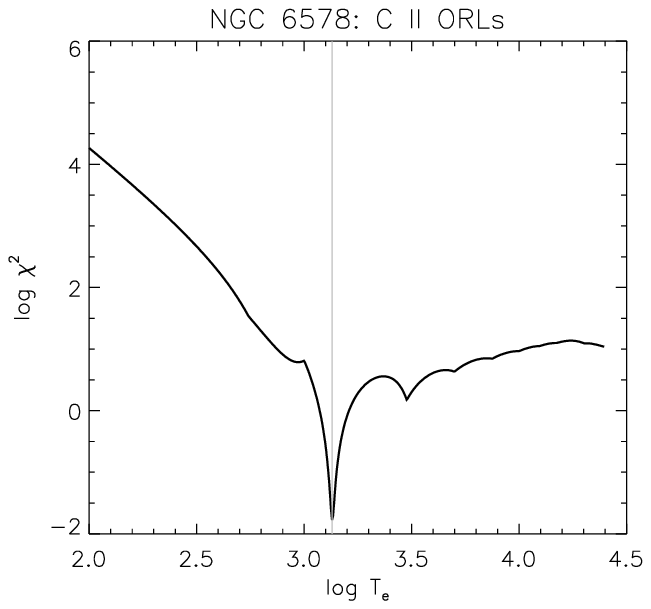}%
\includegraphics[width=0.325\linewidth]{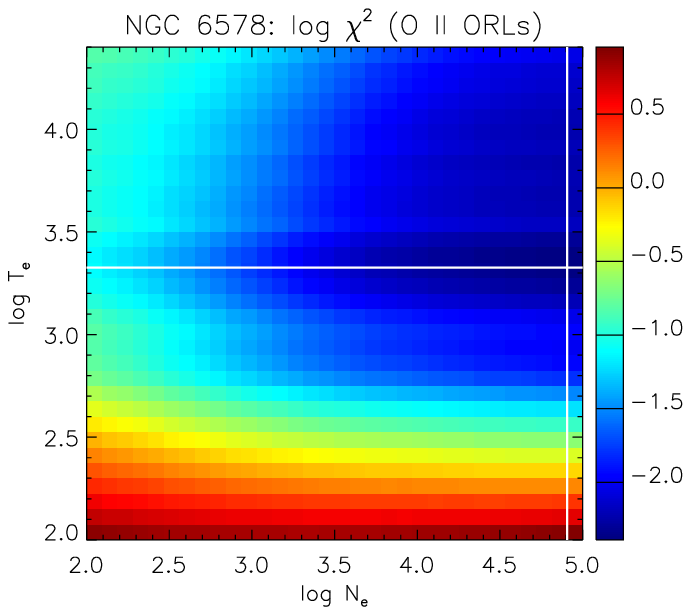}
\includegraphics[width=0.325\linewidth]{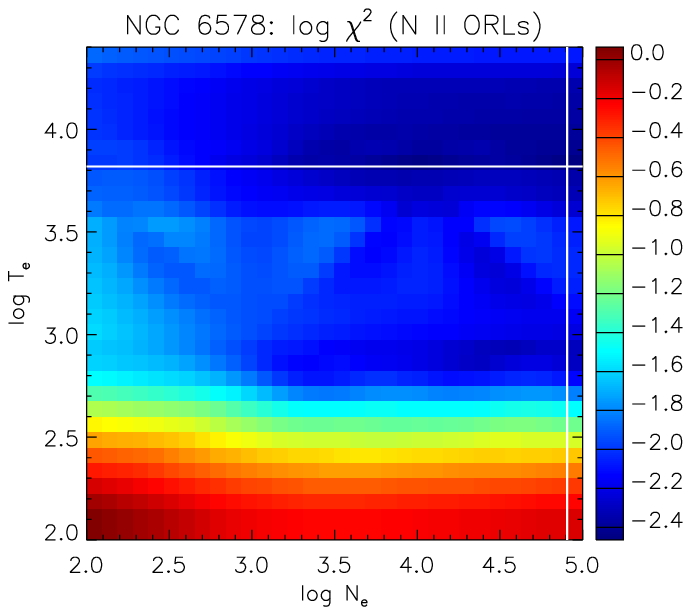}%
\caption{$T_{\rm e}$ diagnostic diagrams of NGC\,6578 based on C\,{\sc ii} ORLs, $N_{\rm e}$--$T_{\rm e}$ diagnostic diagrams based on O\,{\sc ii} and N\,{\sc ii} ORLs. The best-fitting physical conditions at $\chi_{\rm min}^2$ are shown by the solid lines.
The complete figure set (21 images) is available in the online journal.
\label{wc:fig:dd:orls}}
\end{center}

\figsetstart
\figsetnum{4}
\figsettitle{$T_{\rm e}$ diagnostic diagrams based on C\,{\sc ii} ORLs, and $N_{\rm e}$--$T_{\rm e}$ diagnostic diagrams based on O\,{\sc ii} and N\,{\sc ii} ORLs. The best-fitting physical conditions at $\chi_{\rm min}^2$ are shown by the solid lines. }

\figsetgrpstart
\figsetgrpnum{4.1}
\figsetgrptitle{PB\,6.}
\figsetplot{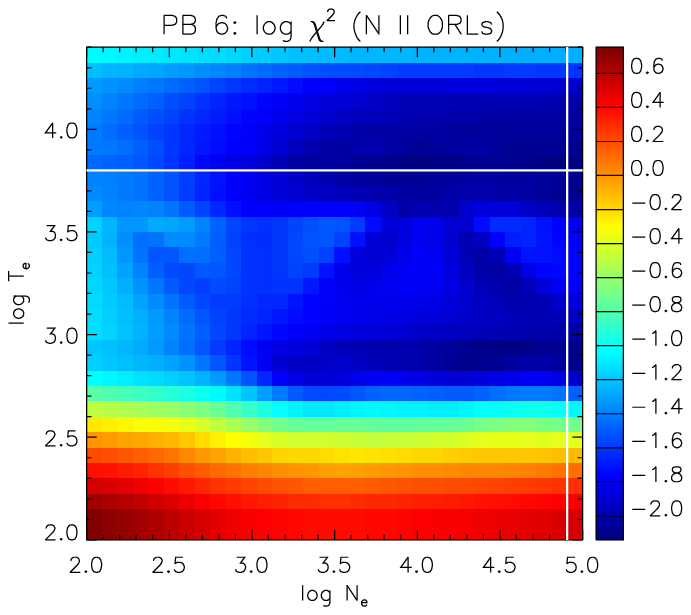}
\figsetgrpnote{$N_{\rm e}$--$T_{\rm e}$ diagnostic diagram of PB\,6 based on N\,{\sc ii} ORLs. }
\figsetgrpend

\figsetgrpstart
\figsetgrpnum{4.2}
\figsetgrptitle{M\,3-30.}
\figsetplot{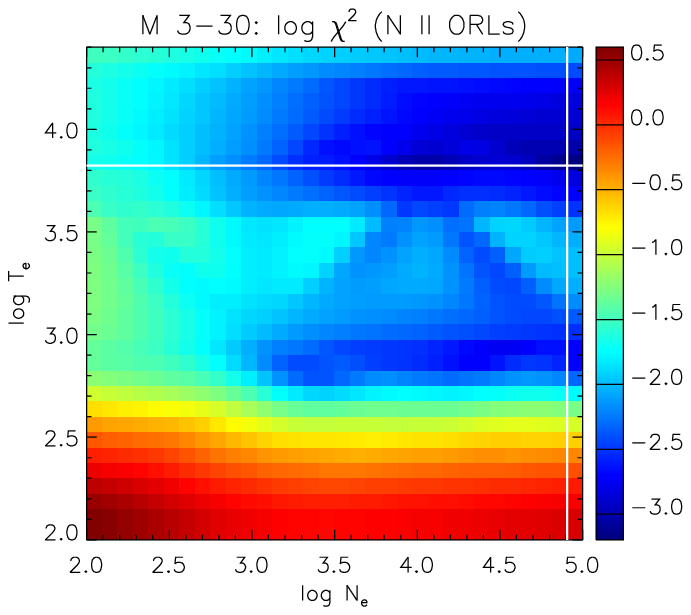}
\figsetgrpnote{$N_{\rm e}$--$T_{\rm e}$ diagnostic diagram of M\,3-30 based on N\,{\sc ii} ORLs. }
\figsetgrpend

\figsetgrpstart
\figsetgrpnum{4.3}
\figsetgrptitle{Hb\,4 (shell).}
\figsetplot{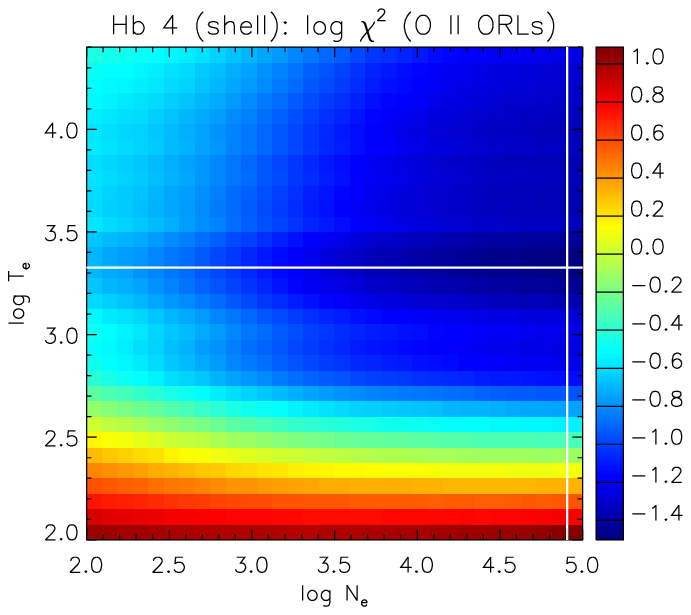}
\figsetgrpnote{$N_{\rm e}$--$T_{\rm e}$ diagnostic diagram of Hb\,4 (shell) based on O\,{\sc ii}. }
\figsetgrpend

\figsetgrpstart
\figsetgrpnum{4.4}
\figsetgrptitle{Hb\,4 (shell).}
\figsetplot{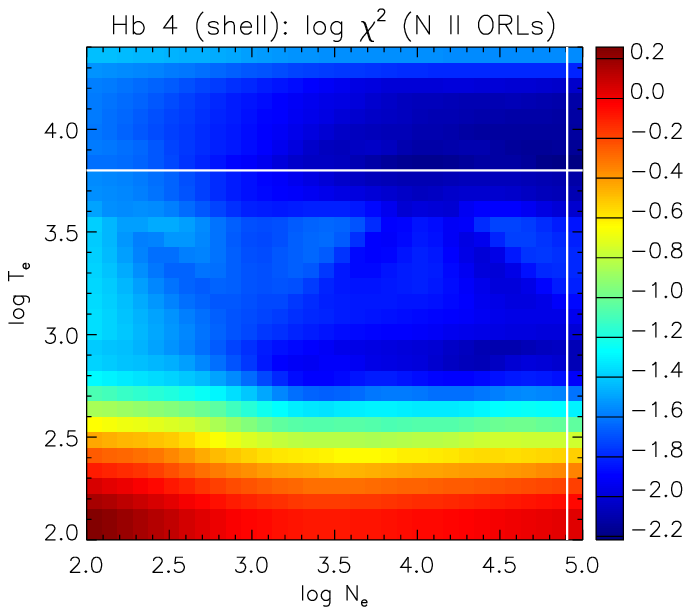}
\figsetgrpnote{$N_{\rm e}$--$T_{\rm e}$ diagnostic diagram of NGC\,6578 based on N\,{\sc ii} ORLs. }
\figsetgrpend

\figsetgrpstart
\figsetgrpnum{4.5}
\figsetgrptitle{IC\,1297.}
\figsetplot{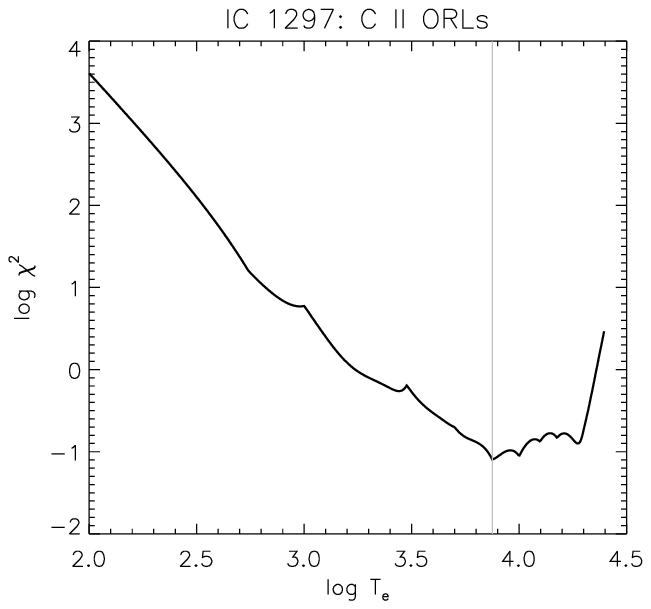}
\figsetgrpnote{$T_{\rm e}$ diagnostic diagram of IC\,1297 based on C\,{\sc ii} ORLs. }
\figsetgrpend

\figsetgrpstart
\figsetgrpnum{4.6}
\figsetgrptitle{IC\,1297.}
\figsetplot{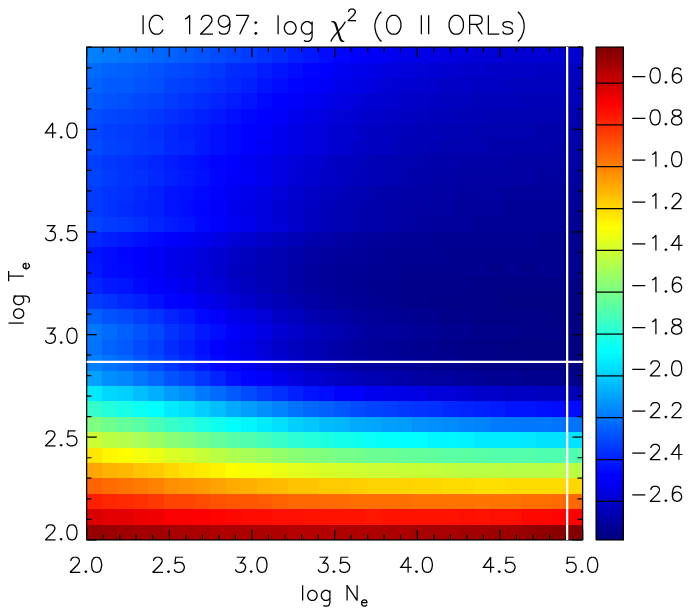}
\figsetgrpnote{$N_{\rm e}$--$T_{\rm e}$ diagnostic diagram of IC\,1297 based on O\,{\sc ii}. }
\figsetgrpend

\figsetgrpstart
\figsetgrpnum{4.7}
\figsetgrptitle{IC\,1297.}
\figsetplot{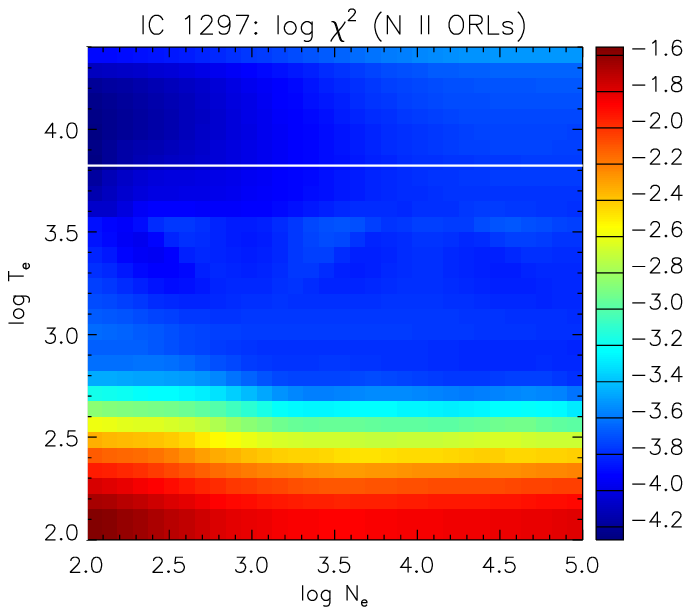}
\figsetgrpnote{$N_{\rm e}$--$T_{\rm e}$ diagnostic diagram of IC\,1297 based on N\,{\sc ii} ORLs. }
\figsetgrpend

\figsetgrpstart
\figsetgrpnum{4.8}
\figsetgrptitle{Th\,2-A.}
\figsetplot{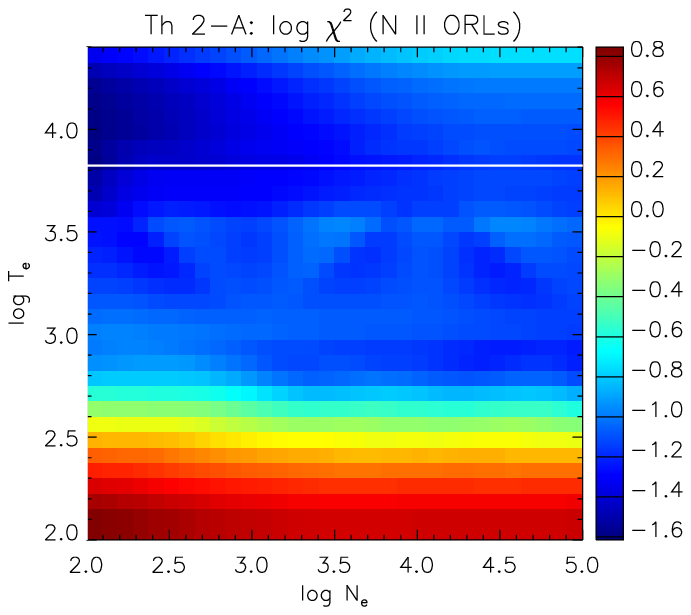}
\figsetgrpnote{$N_{\rm e}$--$T_{\rm e}$ diagnostic diagram of Th\,2-A based on N\,{\sc ii} ORLs. }
\figsetgrpend

\figsetgrpstart
\figsetgrpnum{4.9}
\figsetgrptitle{Pe\,1-1.}
\figsetplot{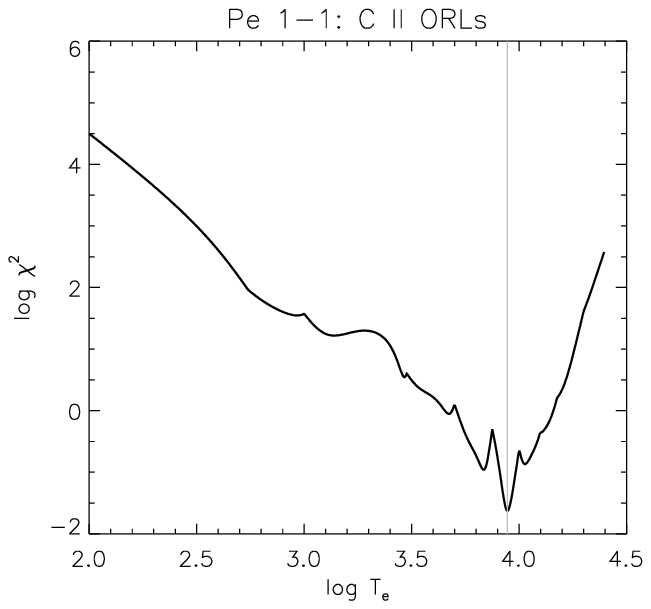}
\figsetgrpnote{$T_{\rm e}$ diagnostic diagram of Pe\,1-1 based on C\,{\sc ii} ORLs. }
\figsetgrpend

\figsetgrpstart
\figsetgrpnum{4.10}
\figsetgrptitle{M\,1-25.}
\figsetplot{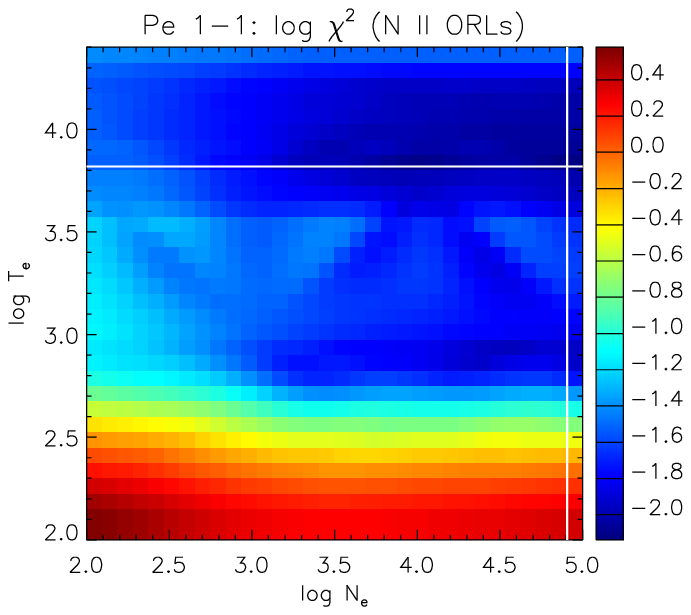}
\figsetgrpnote{$N_{\rm e}$--$T_{\rm e}$ diagnostic diagram of Pe\,1-1 based on N\,{\sc ii} ORLs. }
\figsetgrpend

\figsetgrpstart
\figsetgrpnum{4.11}
\figsetgrptitle{M\,1-32.}
\figsetplot{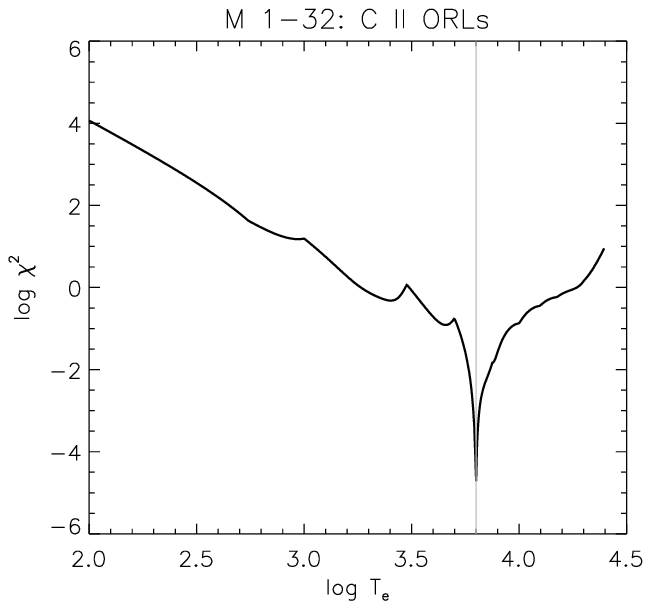}
\figsetgrpnote{$T_{\rm e}$ diagnostic diagram of M\,1-32 based on C\,{\sc ii} ORLs. }
\figsetgrpend

\figsetgrpstart
\figsetgrpnum{4.12}
\figsetgrptitle{M\,1-32.}
\figsetplot{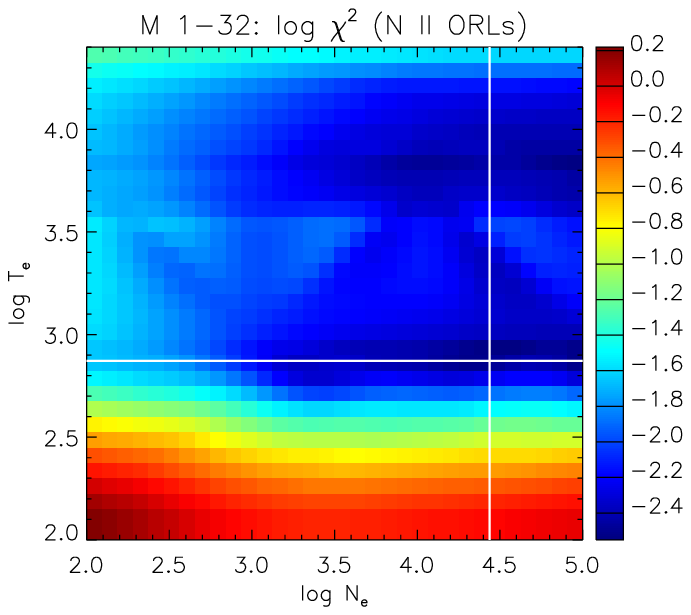}
\figsetgrpnote{$N_{\rm e}$--$T_{\rm e}$ diagnostic diagram of M\,1-32 based on N\,{\sc ii} ORLs. }
\figsetgrpend

\figsetgrpstart
\figsetgrpnum{4.13}
\figsetgrptitle{M\,3-15.}
\figsetplot{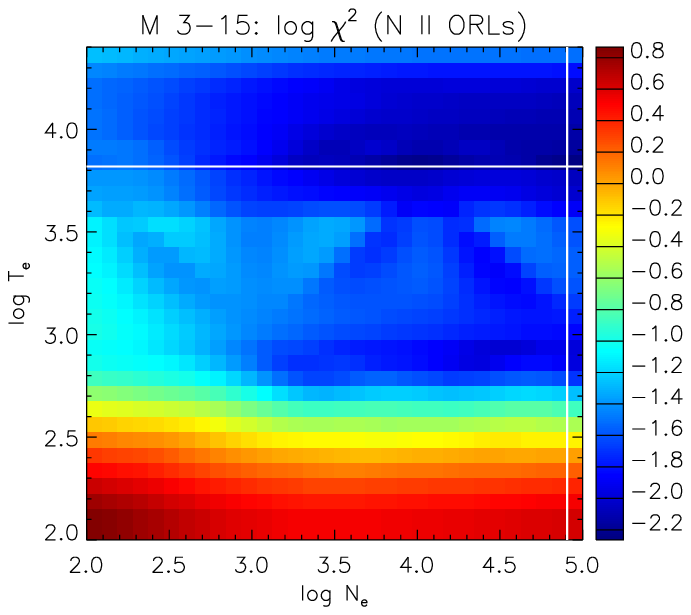}
\figsetgrpnote{$N_{\rm e}$--$T_{\rm e}$ diagnostic diagram of M\,3-15 based on N\,{\sc ii} ORLs. }
\figsetgrpend

\figsetgrpstart
\figsetgrpnum{4.14}
\figsetgrptitle{NGC\,6578.}
\figsetplot{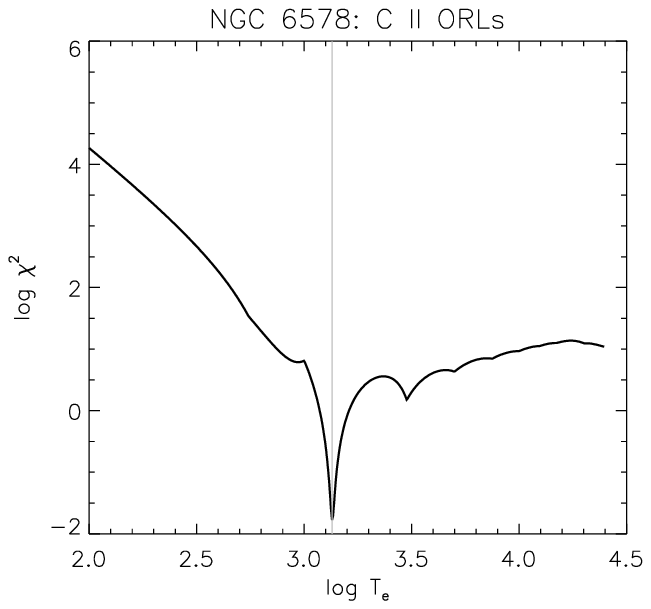}
\figsetgrpnote{$T_{\rm e}$ diagnostic diagram of NGC\,6578 based on C\,{\sc ii} ORLs.}
\figsetgrpend

\figsetgrpstart
\figsetgrpnum{4.15}
\figsetgrptitle{NGC\,6578.}
\figsetplot{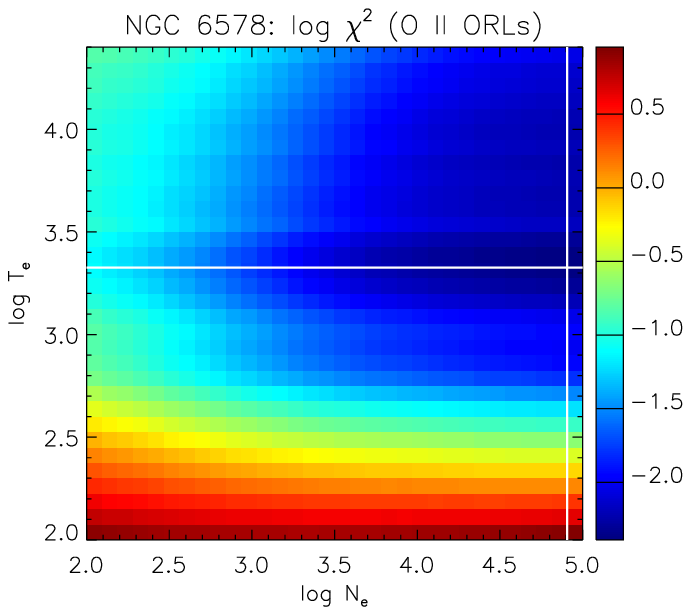}
\figsetgrpnote{$N_{\rm e}$--$T_{\rm e}$ diagnostic diagram of NGC\,6578 based on O\,{\sc ii} ORLs. }
\figsetgrpend

\figsetgrpstart
\figsetgrpnum{4.16}
\figsetgrptitle{NGC\,6578.}
\figsetplot{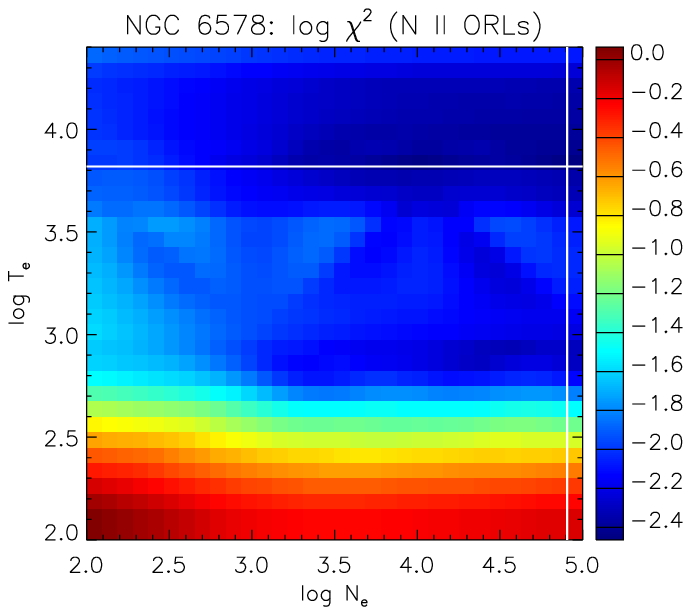}
\figsetgrpnote{$N_{\rm e}$--$T_{\rm e}$ diagnostic diagram of NGC\,6578 based on N\,{\sc ii} ORLs. }
\figsetgrpend

\figsetgrpstart
\figsetgrpnum{4.17}
\figsetgrptitle{M\,2-42.}
\figsetplot{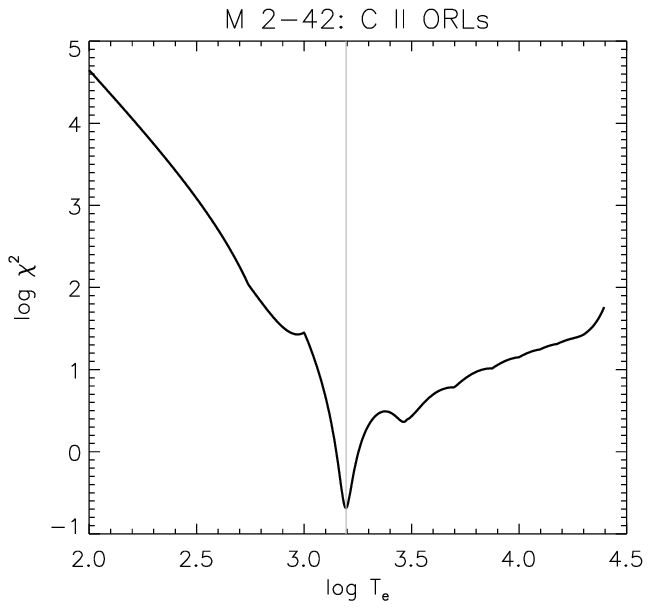}
\figsetgrpnote{$T_{\rm e}$ diagnostic diagram of M\,2-42 based on C\,{\sc ii} ORLs. }
\figsetgrpend

\figsetgrpstart
\figsetgrpnum{4.18}
\figsetgrptitle{M\,2-42.}
\figsetplot{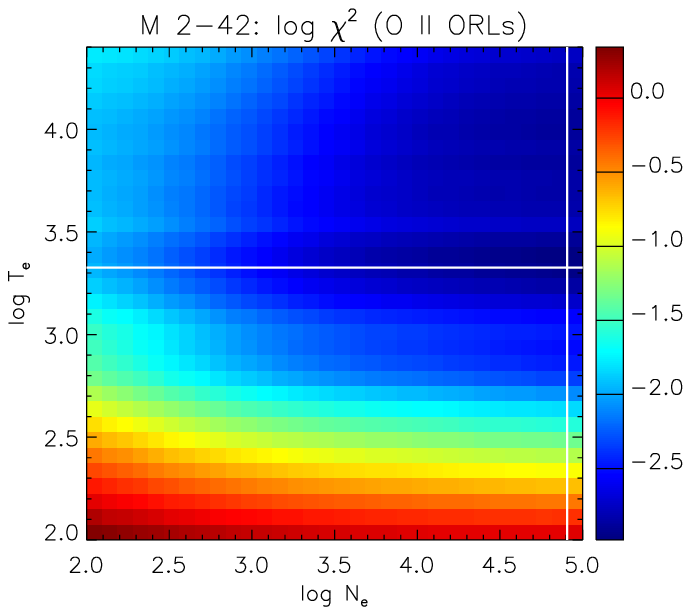}
\figsetgrpnote{$N_{\rm e}$--$T_{\rm e}$ diagnostic diagram of M\,2-42 based on O\,{\sc ii} ORLs. }
\figsetgrpend

\figsetgrpstart
\figsetgrpnum{4.19}
\figsetgrptitle{M\,2-42.}
\figsetplot{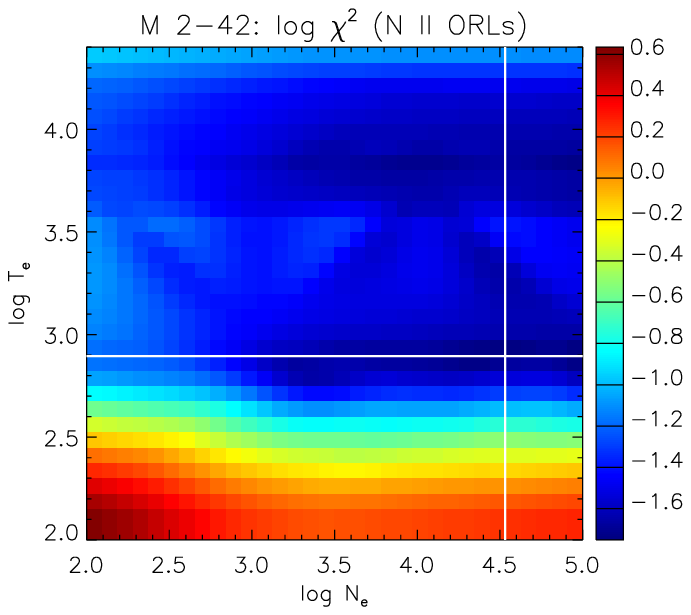}
\figsetgrpnote{$N_{\rm e}$--$T_{\rm e}$ diagnostic diagram of M\,2-42 based on N\,{\sc ii} ORLs. }
\figsetgrpend

\figsetgrpstart
\figsetgrpnum{4.20}
\figsetgrptitle{NGC\,6567.}
\figsetplot{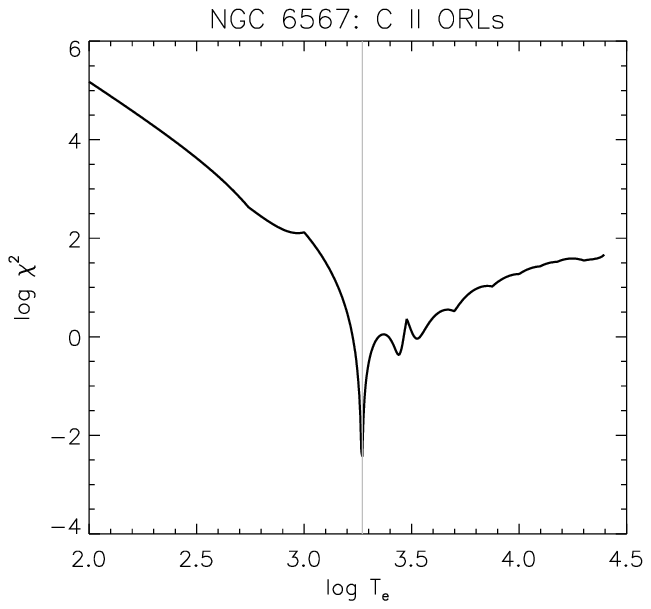}
\figsetgrpnote{$T_{\rm e}$ diagnostic diagram of NGC\,6567 based on C\,{\sc ii} ORLs. }
\figsetgrpend

\figsetgrpstart
\figsetgrpnum{4.21}
\figsetgrptitle{Sa\,3-107.}
\figsetplot{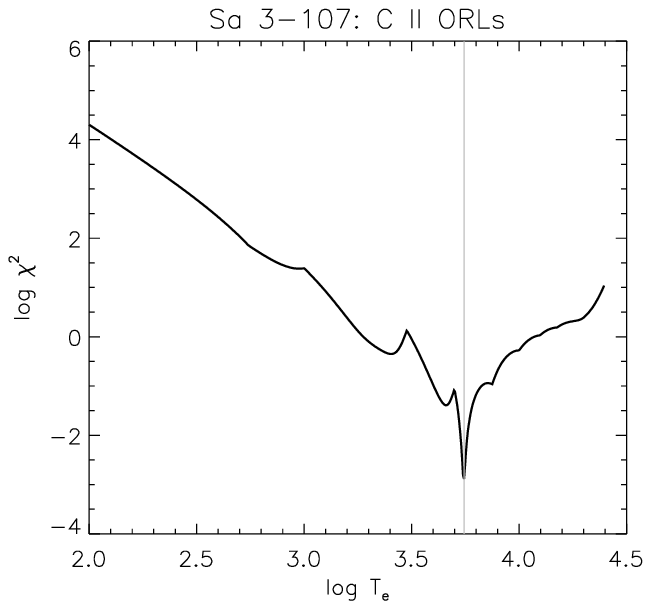}
\figsetgrpnote{$T_{\rm e}$ diagnostic diagram of Sa\,3-107 based on C\,{\sc ii} ORLs. }
\figsetgrpend

\figsetend
\end{figure*}

\subsection{ORL Plasma Diagnostics}
\label{wc:sec:tempdens:orl}

\subsubsection{He\,{\sc i} Electron Temperatures}
\label{wc:sec:temp:orl:hei}

Table~\ref{wc:tab:diagnostic:hei} presents helium temperatures derived from the flux ratios He\,{\sc i} $\lambda\lambda$5876/4472, $\lambda\lambda$6678/4472, $\lambda\lambda$7281/4472, $\lambda\lambda$7281/5876 and $\lambda\lambda$7281/6678. 
To obtain the electron temperature from the helium line ratios, we used the analytic formula given for the emissivities of He~I lines by \citet{Benjamin1999} and fitting parameters calculated by \citet{Zhang2005a}. The method by \citet{Benjamin1999} is valid for temperatures from 5000--20,000~K. However, \citet{Zhang2005a} combined the He \,{\sc i} recombination model of \citet{Smits1996} and the collisional excitation rates for the 2s$^3$S and 2s$^1$S meta-stable levels by \citet{Sawey1993}, and provided a new electron temperature diagnostic based on the method developed by \citet{Benjamin1999}, but also applicable in the case of $T_{\rm e}< 5000$\,K. 
Similarly, we employ the MCMC-based approach to propagate the flux errors of the He\,{\sc i} lines into our calculations of the helium temperatures. 
It can be seen that our derived He\,{\sc i} temperatures are lower than the forbidden-line temperatures in 8 PNe (see Table~\ref{wc:plasma:summary}). However, the helium temperatures is roughly in the range of the temperatures derived from CELs in Pe\,1-1, M\,3-15, M\,1-25, and NGC\,6629, and higher than the CEL temperatures in M\,2-42.


\subsubsection{Physical Conditions from Heavy Element ORLs}
\label{wc:sec:temp:orl:cii}

Using the fact that emissivities of heavy element ORLs have a relatively weak, power-law dependence on the electron temperature, the relative intensities of ORLs can be used to determine electron temperatures \citep[see e.g.][]{McNabb2013,Storey2013}. 
Although the plasma diagnostics based on the flux intensity ratio of two different ORLs are the most common way, least squares minimization, relying on a number of lines, can be used as an alternative method. This method is based on minimizing the difference between the normalized intrinsic line flux intensities and the normalized values calculated by the theoretical model.
In the least squares minimization method, a single fitting parameter is used, i.e. the electron temperature. We can also use this method to determine the electron density from ORLs. The electron temperatures derived from ORLs were adopted to calculate the ORL ionic abundances in some PNe where applicable (see Table~\ref{wc:tab:adopted:TeNe:2}).

To determine the physical conditions from heavy element ORLs, we employed a self-consistent 
method based on a least-squares minimization (see Fig.~\ref{wc:fig:dd:orls}), which is similar to the ORL plasma diagnostics used in \citet[][]{McNabb2013} and \citet[][]{Storey2013}. For the ORL plasma diagnostics and abundance analyses, we utilized the effective recombination coefficients $\alpha_{\rm eff}$ of the ions 
C${}^{2+}$, N${}^{2+}$  (case B), and O${}^{2+}$  (case B) listed in Table~\ref{wc:tab:atomicdata}. 
Note that N${}^{2+}$ and O${}^{2+}$ are the only ions whose density-dependent recombination coefficients have recently been calculated in the density range from $10^2$ to $10^5$\,cm${}^{-3}$, as well as in the temperature range from 125 to 20,000\,K for the N${}^{2+}$ ion, and from 100 to 25,000\,K for the O${}^{2+}$ ion.

\setcounter{table}{6}
\begin{table}
\caption{Plasma diagnostics based on \ionic{He}{i} lines.
\label{wc:tab:diagnostic:hei}}
\centering
\begin{tabular}{llcc}
\hline\hline
\noalign{\smallskip}
Type& Ion  & Diagnostic & Value \\
\noalign{\smallskip}
\hline
\noalign{\smallskip}
\multicolumn{4}{c}{PB 6          (PNG278.8$+$04.9) }\\
\noalign{\smallskip}
\hline
\noalign{\smallskip}
$T_{\rm e}$                &He~{\sc i}       &5875.66             /4471.50              & $      4160_{      -620}^{+       540}$ \\
\noalign{\smallskip}
$T_{\rm e}$                &He~{\sc i}       &6678.16             /4471.50              & $      6990_{     -1190}^{+      1690}$ \\
\noalign{\smallskip}
$T_{\rm e}$                &He~{\sc i}       &Mean                                     & $      5580_{     -1020}^{+      1110}$ \\
\noalign{\medskip}
\multicolumn{4}{c}{                          \ldots }\\
\noalign{\medskip}
\noalign{\medskip}
\hline
\end{tabular}
\begin{list}{}{}
\item[\textbf{Note:}]Table \ref{wc:tab:diagnostic:hei} is published in its entirety in the machine-readable format. A portion is shown here for guidance regarding its form and content. 
\end{list}
\end{table}

\setcounter{table}{7}
\begin{table}
\caption{Plasma diagnostics based on heavy element ORLs.
\label{wc:tab:diagnostic:orls}}
\centering
\begin{tabular}{llcc}
\hline\hline
\noalign{\smallskip}
Type & Ion  & Diagnostic & Value \\
\noalign{\smallskip}
\hline
\noalign{\smallskip}
\multicolumn{4}{c}{PB 6          (PNG278.8$+$04.9) }\\
\noalign{\smallskip}
\hline
\noalign{\smallskip}
$T_{\rm e}$                &N~{\sc ii}       &{ 4613.87,4788.13,5676.02,5931.78}                                                     & $      6300_{     -3100}^{+      9700}$ \\
\noalign{\smallskip}
$N_{\rm e}$                &N~{\sc ii}       &{ 4613.87,4788.13,5676.02,5931.78}                                                       & $     80600_{    -79900}^{+         0}$ \\
\noalign{\medskip}
\multicolumn{4}{c}{                          \ldots }\\
\noalign{\medskip}
\noalign{\medskip}
\hline
\end{tabular}
\begin{list}{}{}
\item[\textbf{Note:}]Table \ref{wc:tab:diagnostic:orls} is published in its entirety in the machine-readable format. A portion is shown here for guidance regarding its form and content. 
\end{list}
\end{table}

\setcounter{table}{8}
\begin{table*}
\caption{Summary of electron temperatures ($T_{\rm e}$) and densities ($N_{\rm e}$) obtained from CELs and ORLs.
\label{wc:plasma:summary}
}
\centering
\scriptsize
\begin{tabular}{llccccccccccc}
\hline\hline
\noalign{\smallskip}
Name        & PN\,G & $T_{\rm e}$/$T_{\rm e\,rc}$[\ionf{N}{ii}]& $T_{\rm e}$/$T_{\rm e\,rc}$[\ionf{O}{iii}]& $N_{\rm e}$[\ionf{S}{ii}]& 
$N_{\rm e}$[\ionf{Cl}{iii}]& $N_{\rm e}$[\ionf{Ar}{iv}]& $T_{\rm e}$ \ionf{He}{i}& $T_{\rm e}$ \ionf{N}{ii} & $T_{\rm e}$ \ionf{O}{ii} &$T_{\rm e}$ \ionf{C}{ii} &$N_{\rm e}$ \ionf{N}{ii}&$N_{\rm e}$ \ionf{O}{ii} \\
             &  & (CEL)& (CEL) & (CEL) &  (CEL) &  (CEL) &  (ORL) &  (ORL) &     (ORL) & (ORL) &     (ORL) & (ORL)\\
\noalign{\smallskip}
\hline
\noalign{\smallskip}
PB 6        & 278.8$+$04.9 & 11300 / 10270& 14220 / 14040&  1800&  2190&  1370&  5580&  6300&      &      & $10^{4.9}$&      \\
M 3-30      & 017.9$-$04.8 &  8880        & 10610 /  9170&   700&      &   540&  2640&  6700&      &      & $10^{4.9}$&      \\
Hb 4        & 003.1$+$02.9 & 10390 /  9410&  9330 /  9180&  4440&  4130&  5590&  2790&  6300&  2100&      & $10^{4.9}$& $10^{4.9}$\\
IC 1297     & 358.3$-$21.6 &  9510 /  9160& 10460 / 10400&  2400&  1960&  1780&  6560&  6700&   700&  7500&   100& $10^{4.9}$\\
Th 2-A      & 306.4$-$00.6 & 10300 /  9060& 12420 / 11740&  1430&      &      &      &  6700&      &      &   100&      \\
Pe 1-1      & 285.4$+$01.5 & 10170 /  9190& 10240        & 15580&      &      &  9390&  6600&      &  8800& $10^{4.9}$&      \\
M 1-32      & 011.9$+$04.2 &  8790 /  8480&  9250        &  4840&  3930&     &  1320&   700&      &  6300& 27400&      \\
M 3-15      & 006.8$+$04.1 &  7700        &  8590        &  6010&     &  5540&  8600&  6600&      &      & $10^{4.9}$&     \\
M 1-25      & 004.9$+$04.9 &  7880 /  6950&  8040        &  6120&  4090&      &  7560&      &      &      &      &      \\
Hen 2-142   & 327.1$-$02.2 &  8810       &             & 20630& 21160&      &      &      &      &      &      &      \\
Hen 3-1333  & 332.9$-$09.9 &  7290       &             &$10^5$&      &      &      &      &      &      &      &      \\
Hen 2-113   & 321.0$+$03.9 &  8220       &             &$10^5$&      &      &      &      &      &      &      &      \\
K 2-16      & 352.9$+$11.4 &  9420       &             & 110\,$^{\mathrm{a}}$     &      &      &      &      &      &      &      &      \\
NGC 6578    & 010.8$-$01.8 &  8080       &  8230 /  8220&  4200&  3390&  3250&  4710&  6600&  2100&  1300& $10^{4.9}$& $10^{4.9}$\\
M 2-42      & 008.2$-$04.8 & 10010       &  8800   &  3050&  2420&  3430& 11750&   800&  2100&  1600& 34000& $10^{4.9}$\\
NGC 6567    & 011.7$-$00.6 & 10860       & 11480 / 11470&  6340&  5340&  2440&  2590&      &      &  1900&      &      \\
NGC 6629    & 009.4$-$05.0 &  8380       &  8840       &  2810&  2100&      &  7540&     &      &      &       &      \\
Sa 3-107    & 358.0$-$04.6 & 10840       &             &  2480&  1400&      &  2160&      &      &  5500&      &      \\
\noalign{\smallskip}
\hline 
\footnotesize
\end{tabular}
\begin{list}{}{}
\item[$^{\mathrm{a}}$] Electron density derived from the [\ionf{O}{ii}] lines. 
\end{list}
\begin{tablenotes}
\item[1]\textbf{Note.} The label ``rc'' indicates that the auroral lines $[$N\,{\sc ii}$]$ and $[$O\,{\sc iii}$]$  are corrected for recombination contribution. 
Uncertainties are presented in Tables~\ref{wc:tab:diagnostic:cels}--\ref{wc:tab:diagnostic:orls}.
\end{tablenotes}
\end{table*}

The effective recombination coefficients $\alpha_{\rm eff}$ are used
to calculate the theoretical flux intensity of each emission line at the wavelength $\lambda$ in the full ranges of the physical conditions $N_{\rm e}$ and $T_{\rm e}$ for the N${}^{2+}$ and O${}^{2+}$ ions (using $N_{\rm e}$-dependent $\alpha_{\rm eff}$), and in the full ranges of $T_{\rm e}$ for C${}^{2+}$:
\begin{equation}
I_{\rm theo}\left(\lambda_j\right) = 
\frac{\alpha_{\rm eff}\left(\lambda_j\right)}
{\alpha_{\rm eff}\left(\rm{H}\beta\right)}
\frac{4861.33}{\lambda_j({\rm {\AA}})}\frac{N({\rm X}^{i+})}{N({\rm H}^{+})}\times{100},
\label{wc:diagn:orl1}
\end{equation}
where ${N({\rm X}^{i+})}/{N({\rm H}^{+})}$ is the ionic abundance of ion ${\rm X}^{i+}$ derived from ORLs, whose initial value in the first iteration is calculated using the temperature and density from the CELs. 

The physical conditions ($N_{\rm e}$ and $T_{\rm e}$ for N${}^{2+}$ and O${}^{2+}$, and $T_{\rm e}$ for C${}^{2+}$) are identified at the minimum value of the following weighted least-squares for ion ${\rm X}^{i+}$:
\begin{equation}
\chi^2 = \frac{1}{w_{\rm sum}}
\sum_{j=1}^{N}
w_{j}\big(I_{\rm obs}\left(\lambda_j\right) - I_{\rm theo}\left(\lambda_j\right)\big)^2,
\label{wc:diagn:orl2}
\end{equation}
where ${\chi^2}$ is the sum of the weighted least-squares over $N$ ORLs used for ion ${\rm X}^{i+}$, $I_{\rm theo}$ is the theoretical flux intensity of the emission line $\lambda_j$ predicted by Eq.~(\ref{wc:diagn:orl1}), $I_{\rm obs}$ is the dereddened flux intensity of the emission line $\lambda_j$ measured from the observation, the weight is given by $w_{j}=1/\sigma^2_{I_{\rm obs}(\lambda_j)}$, $\sigma_{I_{\rm obs}(\lambda_j)}$ is the absolute error of the dereddened flux intensity $I_{\rm obs}(\lambda_j)$ of the observed line $\lambda_j$, and ${w_{\rm sum}}=\sum_{j=1}^{N}w_{j}$ is the total of weights. 

The physical conditions ($T_{\rm e}$ and if possible $N_{\rm e}$) derived from the weighted least-squares minimization method are used to calculate the ionic abundance ${N({\rm X}^{i+})}/{N({\rm H}^{+})}$. The new ionic abundance is then substituted into Eq.~(\ref{wc:diagn:orl1}) to calculate the theoretical intensities of the emission lines in the full ranges of the physical conditions. Again, new physical conditions are determined from the minimization value of the weighted least-squares, $\chi_{\rm min}^2$, expressed by Eq.~(\ref{wc:diagn:orl2}). This iterative procedure was performed in a self-consistent manner until there are no variation in the physical conditions, which constrains the temperature (and density for O${}^{2+}$ and N${}^{2+}$). 

Figure~\ref{wc:fig:dd:orls} shows the least-square distributions ($\log \chi^2$) calculated for \oii\ and \nii\ ORLs in the $T_{\rm e}$--$N_{\rm e}$ space with the minimum value of least-squares $\chi_{\rm min}^2$ at the crossing point of the two solid lines, and the least-square diagram ($\log \chi^2$) for \cii\ ORLs in the $T_{\rm e}$ range with $\chi_{\rm min}^2$ located at the solid line. Following \citet{Storey2013}, the uncertainties are determined at the lower and upper limits $\chi_{\rm min}^2 + 1$ of the least-squares normalized by $\chi_{\rm min}^2$. 

Table~\ref{wc:tab:diagnostic:orls} presents the electron temperatures ($T_{\rm e}$) and densities ($N_{\rm e}$) derived from heavy element ORLs for those PNe where adequate recombination lines were available. 
It can be seen that the ORLs are emitted from ionized regions having temperatures lower than the regions from which CELs originate in several PNe. 
Seven PNe have detection of the multiplet V17.04 ($\lambda$6461.95; 6g--4f) and V16.04 ($\lambda$6151.43; 6f--4d). The $\lambda$6462 line is the strongest C\,{\sc ii} recombination line which was detected in these PNe. 
We used the Case A effective recombination for $\lambda$6462, and the Case B for $\lambda$6151 from the atomic data of \citet{Davey2000} covering 500 to 20\,000\,K. 
The Case A effective recombination for $\lambda$6151 differs from its Case B value by only 2 per cent, indicating that this transition is case insensitive. The contribution from the blended N\,{\sc ii} $\lambda$6150.75 (4p--3d) line is typically negligible.

Table~\ref{wc:tab:diagnostic:orls} lists the physical condition derived from N\,{\sc ii} ORLs for 10 PNe. The effective recombination coefficients of N\,{\sc ii} calculated by \citet{Fang2011,Fang2013a} are in the range from 125 to 20,000\,K,  which allow us to identify any cold ionized regions. The well-detected $\lambda$5931.78 line in 5 PNe (PB 6, Hb 4, IC 1297, Th 2-A, and Pe 1-1) 
are likely to provide the most reliable diagnostic line as this line can mostly be produced by recombination from the 3p$^{3}$P--3d$^{3}$D$^{\circ}$ level of N$^{2+}$ and the V28 multiplet in high-excited nebulae. Using all lines from the same multiplet might reduce the effect of any errors caused by atomic data, which were not available in all the objects.  
The lines $\lambda$5666.63 and $\lambda$5679.56 can be produced by recombination from the 3p$^{3}$D--3s$^{3}$P$^{\circ}$ level of N$^{2+}$ and the V3 multiplet, but
these lines can be largely contaminated by fluorescence \citep{Escalante2005,Escalante2012}, especially
in low-excited PNe.
The line $\lambda$5679.56 in M\,3-30, M\,1-32, M\,3-15, NGC\,6578, and NGC\,6567 with the $T_{\rm eff} \sim 50$--60\,kK could be produced by fluorescence, so they were not employed in our analysis of these objects.
Similarly, the line $\lambda$5666.63 detected in M\,1-32, M\,1-25, and NGC\,6567 was excluded. 
For M\,2-42, we did not use the strong lines $\lambda$5666.63 and $\lambda$5679.56, which could be largely excited by resonance fluorescence. 
The lines $\lambda$5679.56, 5666.63, 5931.78 and 5710.77, which are possibly produced by fluorescence, were also excluded in Sa\,3-107. 
However, we kept these lines in other objects with $T_{\rm eff} \gtrsim 75$\,kK where they could be excited by recombination process.
The $\lambda$5676.02 line from the V3 multiplet (3p$^{3}$D--3s$^{3}$P$^{\circ}$) is the strongest N\,{\sc ii} recombination line with low uncertainty detected in Hb\,4, whereas 
$\lambda$4442.02 (V55a) has a higher uncertainty, and $\lambda$4621.39 (V92) and $\lambda$5931.78 (V28) are relatively very weak. For Th\,2-A with $T_{\rm eff} \sim 157$\,kK, the $\lambda$5679.56 line is produced by recombination from the 3p$^{3}$D--3s$^{3}$P$^{\circ}$ level of N$^{2+}$ and the V3 multiplet, and $\lambda$5931.78 line from the V28 multiplet. The $\lambda\lambda$5676.02 ORL from the V3 multiplet is stronger than those from $\lambda$5931.78 from the V28 multiplet in the spectra of Hb\,4. 

\setcounter{table}{9}
\begin{table*}
\caption{Electron density and temperatures adopted for our CEL and ORL abundance analyses.
\label{wc:tab:adopted:TeNe:2}}
\centering
\scriptsize
\begin{tabular}{llllllll}
\hline\hline
\noalign{\smallskip}
Nebula         & PN\,G & $T_{\rm e}$ & $N_{\rm e}$ & $T_{\rm e}$ & $N_{\rm e}$ & $T_{\rm e}$ & $N_{\rm e}$ \\
               &  & CEL (Low)~  & CE (Low)~   & CEL (High)   & CEL (High)    & ORL ~~~~~~~  & ORL ~~~~~~~ \\
\noalign{\smallskip}
\hline  
\noalign{\smallskip}	
PB 6           & 278.8$+$04.9 &  [\ionf{N}{ii}]$_{\rm rc}$  &  [\ionf{S}{ii}]  &  [\ionf{O}{iii}]$_{\rm rc}$  &  [\ionf{Cl}{iii}]  &  \ionf{He}{i}  &  [\ionf{S}{ii}]  \\
M\,3-30        & 017.9$-$04.8 &  [\ionf{N}{ii}]  &  [\ionf{S}{ii}]  &  [\ionf{O}{iii}]$_{\rm rc}$  &  [\ionf{Ar}{iv}]  &  \ionf{N}{ii}  &  [\ionf{S}{ii}]  \\
Hb\,4 (shell)  & 003.1$+$02.9 &  [\ionf{N}{ii}]$_{\rm rc}$  &  [\ionf{S}{ii}]  &  [\ionf{O}{iii}]$_{\rm rc}$  &  [\ionf{Cl}{iii}]  &   \ionf{N}{ii}  &  [\ionf{S}{ii}]  \\
Hb\,4 (N-knot) & 003.1$+$02.9 &  [\ionf{N}{ii}]  &  [\ionf{S}{ii}]  &  [\ionf{N}{ii}]  &  [\ionf{S}{ii}]  &  [\ionf{N}{ii}]  &  [\ionf{S}{ii}]  \\
Hb\,4 (S-knot) & 003.1$+$02.9 &  [\ionf{N}{ii}]  &  [\ionf{S}{ii}]  &  [\ionf{N}{ii}]  &  [\ionf{S}{ii}]  &  [\ionf{N}{ii}]  &  [\ionf{S}{ii}]  \\
IC\,1297       & 358.3$-$21.6 &  [\ionf{N}{ii}]$_{\rm rc}$  &  [\ionf{S}{ii}]  &  [\ionf{O}{iii}]$_{\rm rc}$  &  [\ionf{Cl}{iii}]  &  \ionf{C}{ii}   &  [\ionf{S}{ii}]   \\
Th\,2-A        & 306.4$-$00.6 &  [\ionf{N}{ii}]$_{\rm rc}$  &  [\ionf{S}{ii}]  &  [\ionf{O}{iii}]$_{\rm rc}$  &  [\ionf{S}{ii}]  &  \ionf{N}{ii}  &  [\ionf{S}{ii}]  \\
Pe\,1-1        & 285.4$+$01.5 &  [\ionf{N}{ii}]$_{\rm rc}$  &  [\ionf{S}{ii}]  &  [\ionf{O}{iii}]  &  [\ionf{S}{ii}]  &  \ionf{He}{i}  &  [\ionf{S}{ii}] \\
M\,1-32        & 011.9$+$04.2 &  [\ionf{N}{ii}]$_{\rm rc}$  &  [\ionf{S}{ii}]  &  [\ionf{O}{iii}]  &  [\ionf{S}{ii}]  &  \ionf{C}{ii}  &  [\ionf{S}{ii}]  \\
M\,3-15        & 006.8$+$04.1 &  [\ionf{N}{ii}]  &  [\ionf{S}{ii}]  &  [\ionf{O}{iii}]  &  [\ionf{S}{ii}]  &  \ionf{He}{i}  &  [\ionf{S}{ii}]  \\
M\,1-25        & 004.9$+$04.9 &  [\ionf{N}{ii}]  &  [\ionf{S}{ii}]  &  [\ionf{O}{iii}]  &  [\ionf{S}{ii}]  &  \ionf{He}{i}   &  [\ionf{S}{ii}]  \\
Hen\,2-142     & 327.1$-$02.2 &  [\ionf{N}{ii}]  &  [\ionf{Cl}{iii}]  &  [\ionf{N}{ii}]  &  [\ionf{Cl}{iii}]  &  [\ionf{N}{ii}]  &  [\ionf{Cl}{iii}]  \\
Hen\,3-1333    & 332.9$-$09.9 &  [\ionf{N}{ii}]  &  [\ionf{S}{ii}]  &  [\ionf{N}{ii}]  &  [\ionf{S}{ii}]  &  [\ionf{N}{ii}]  &  [\ionf{S}{ii}]  \\
Hen\,2-113     & 321.0$+$03.9 &  [\ionf{N}{ii}]  &  [\ionf{S}{ii}]  &  [\ionf{N}{ii}]  &  [\ionf{S}{ii}]  &  [\ionf{N}{ii}]  &  [\ionf{S}{ii}]  \\
K\,2-16        & 352.9$+$11.4 &  [\ionf{N}{ii}]  &  [\ionf{O}{ii}]  &  [\ionf{N}{ii}]  &  [\ionf{O}{ii}]  &  [\ionf{N}{ii}]  &  [\ionf{O}{ii}]  \\
NGC\,6578      & 010.8$-$01.8 &  [\ionf{N}{ii}]  &  [\ionf{S}{ii}]  &  [\ionf{O}{iii}]$_{\rm rc}$  &  [\ionf{S}{ii}]  &  \ionf{He}{i}  &  [\ionf{S}{ii}]  \\
M\,2-42        & 008.2$-$04.8 &  [\ionf{N}{ii}]  &  [\ionf{S}{ii}]  &  [\ionf{O}{iii}]  &  [\ionf{Ar}{iv}]  &  [\ionf{N}{ii}]  &  [\ionf{S}{ii}]  \\
NGC\,6567      & 011.7$-$00.6 &  [\ionf{N}{ii}]  &  [\ionf{S}{ii}]  &  [\ionf{O}{iii}]  &  [\ionf{Cl}{iii}]  &  [\ionf{N}{ii}]  &  [\ionf{S}{ii}]  \\
NGC\,6629      & 009.4$-$05.0 &  [\ionf{N}{ii}]  &  [\ionf{S}{ii}]  &  [\ionf{O}{iii}]  &  [\ionf{S}{ii}]  &  \ionf{He}{i}  &  [\ionf{S}{ii}]  \\
Sa\,3-107      & 358.0$-$04.6 &  [\ionf{N}{ii}]  &  [\ionf{S}{ii}]  &  [\ionf{N}{ii}]  &  [\ionf{S}{ii}]  &  \ionf{C}{ii}  &  [\ionf{S}{ii}]  \\
\noalign{\smallskip} 	
\hline
\end{tabular}
\begin{list}{}{}
\item[\textbf{Note:}]The label ``rc'' indicates that the auroral lines are corrected for recombination contribution.
The flux uncertainties are directly propagated into our abundance analysis without considering the uncertainties in the physical conditions. 
The electron temperature $T_{\rm e}$ derived for the N-knot is assumed for the S-knot in Hb\,4.
\end{list}
\end{table*}

Table~\ref{wc:tab:diagnostic:orls} also lists the physical conditions obtained from O\,{\sc ii} ORLs for 4 PNe. 
The recombination lines from the V1 multiplet (3p$^{4}$D$^{\circ}$--3s$^{4}$P), 
here $\lambda$4641.81, $\lambda$4661.63 and $\lambda$4676.24, are likely to provide the most reliable temperature diagnostics \citep[see e.g.][]{Wesson2005,McNabb2013}.
Using several ORLs from the same multiplet might reduce any effects of deviation from local thermodynamic equilibrium (LTE) at low densities. \citet{Tsamis2003} found that the relative intensities  of O\,{\sc ii} V1 multiplet components deviate from LTE predictions for those PNe having electron densities lower than than 1000 cm$^{-3}$ (e.g. NGC 3132 with $N_{\rm e}=600$\,cm$^{-3}$). This effect may be reduced by using all the lines from the V1 multiplet \citep[e.g.][]{Wesson2005}. 

\setcounter{table}{10}
\begin{table}
\caption{Ionic abundances derived from CELs.
\label{wc:tab:abundances:cels}
}
\footnotesize
\centering
\begin{tabular}{llcc}
\hline\hline
\noalign{\smallskip}
 Ion  & Line & Weight & Abund. \\
\noalign{\smallskip}
\hline
\noalign{\smallskip}
\multicolumn{4}{c}{PB 6          (PNG278.8$+$04.9) }\\
\noalign{\smallskip}
\hline
\noalign{\smallskip}
N$^{+}$    &[N~{\sc ii}]     $\lambda$6548.10  &    1 & $ 4.225_{-0.164}^{+ 0.142} \times 10^{  -5}$ \\
\noalign{\smallskip}
N$^{+}$    &[N~{\sc ii}]     $\lambda$6583.50  &    3 & $ 4.426_{-0.188}^{+ 0.185} \times 10^{  -5}$ \\
\noalign{\smallskip}
O$^{0}$    &[O~{\sc i}]      $\lambda$6300.34  &    3 & $ 6.456_{-0.240}^{+ 0.203} \times 10^{  -6}$ \\
\noalign{\smallskip}
O$^{0}$    &[O~{\sc i}]      $\lambda$6363.78  &    1 & $ 6.947_{-0.273}^{+ 0.255} \times 10^{  -6}$ \\
\noalign{\smallskip}
O$^{+}$    &[O~{\sc ii}]     $\lambda$3726.03  &    1 & $ 4.928_{-0.142}^{+ 0.164} \times 10^{  -5}$ \\
\noalign{\smallskip}
O$^{2+}$   &[O~{\sc iii}]    $\lambda$4958.91  &    1 & $ 1.378_{-0.010}^{+ 0.008} \times 10^{  -4}$ \\
\noalign{\smallskip}
O$^{2+}$   &[O~{\sc iii}]    $\lambda$5006.84  &    3 & $ 1.477_{-0.021}^{+ 0.018} \times 10^{  -4}$ \\
\noalign{\smallskip}
Ne$^{2+}$  &[Ne~{\sc iii}]   $\lambda$3868.75  &    3 & $ 3.645_{-0.090}^{+ 0.105} \times 10^{  -5}$ \\
\noalign{\smallskip}
Ne$^{2+}$  &[Ne~{\sc iii}]   $\lambda$3967.46  &    1 & $ 4.327_{-0.092}^{+ 0.110} \times 10^{  -5}$ \\
\noalign{\smallskip}
Ne$^{3+}$  &[Ne~{\sc iv}]    $\lambda$4724.15  &    1 & $ 2.825_{-0.148}^{+ 0.134} \times 10^{  -4}$ \\
\noalign{\smallskip}
S$^{+}$    &[S~{\sc ii}]     $\lambda$6716.44  &    1 & $ 3.867_{-0.162}^{+ 0.156} \times 10^{  -7}$ \\
\noalign{\smallskip}
S$^{+}$    &[S~{\sc ii}]     $\lambda$6730.82  &    1 & $ 3.869_{-0.158}^{+ 0.139} \times 10^{  -7}$ \\
\noalign{\smallskip}
S$^{2+}$   &[S~{\sc iii}]    $\lambda$6312.10  &    1 & $ 2.229_{-0.082}^{+ 0.072} \times 10^{  -6}$ \\
\noalign{\smallskip}
Cl$^{2+}$  &[Cl~{\sc iii}]   $\lambda$5517.66  &    1 & $ 2.935_{-0.120}^{+ 0.105} \times 10^{  -8}$ \\
\noalign{\smallskip}
Cl$^{2+}$  &[Cl~{\sc iii}]   $\lambda$5537.60  &    1 & $ 2.935_{-0.092}^{+ 0.077} \times 10^{  -8}$ \\
\noalign{\smallskip}
Ar$^{2+}$  &[Ar~{\sc iii}]   $\lambda$7135.80  &    1 & $ 5.410_{-0.258}^{+ 0.232} \times 10^{  -7}$ \\
\noalign{\smallskip}
Ar$^{3+}$  &[Ar~{\sc iv}]    $\lambda$4711.37  &    1 & $ 7.817_{-0.058}^{+ 0.061} \times 10^{  -7}$ \\
\noalign{\smallskip}
Ar$^{3+}$  &[Ar~{\sc iv}]    $\lambda$4740.17  &    1 & $ 7.205_{-0.045}^{+ 0.051} \times 10^{  -7}$ \\
\noalign{\smallskip}
Ar$^{4+}$  &[Ar~{\sc v}]     $\lambda$4625.53  &    1 & $ 4.082_{-1.224}^{+ 1.177} \times 10^{  -7}$ \\
\noalign{\smallskip}
Ar$^{4+}$  &[Ar~{\sc v}]     $\lambda$7005.67  &   72 & $ 3.711_{-0.175}^{+ 0.165} \times 10^{  -7}$ \\
\noalign{\smallskip}
Fe$^{2+}$  &[Fe~{\sc iii}]   $\lambda$4607.03  &    1 & $ 5.833_{-0.483}^{+ 0.457} \times 10^{  -7}$ \\
\noalign{\smallskip}
Fe$^{2+}$  &[Fe~{\sc iii}]   $\lambda$4881.11  &    2 & $ 5.353_{-1.484}^{+ 1.437} \times 10^{  -8}$ \\
\noalign{\smallskip}
Fe$^{2+}$  &[Fe~{\sc iii}]   $\lambda$5270.40  &    2 & $ 6.421_{-1.044}^{+ 1.082} \times 10^{  -8}$ \\
\noalign{\medskip}
\multicolumn{4}{c}{                          \ldots }\\
\noalign{\medskip}
\noalign{\medskip}
\hline
\end{tabular}
\begin{list}{}{}
\item[\textbf{Note:}]Table \ref{wc:tab:abundances:cels} is published in its entirety in the machine-readable format. A portion is shown here for guidance regarding its form and content. 
\end{list}
\end{table}

\setcounter{table}{11}
\begin{table}
\caption{Ionic abundances derived from ORLs.
\label{wc:tab:abundances:orls}
}
\centering
\footnotesize
\begin{tabular}{llcc}
\hline\hline
\noalign{\smallskip}
 Ion  & Line & Weight & Abund. \\
\noalign{\smallskip}
\hline
\noalign{\smallskip}
\multicolumn{4}{c}{PB 6          (PNG278.8$+$04.9) }\\
\noalign{\smallskip}
\hline
\noalign{\smallskip}
He$^{+}$   &He~{\sc i}       $\lambda$4471.50  &    1 & $ 4.508_{-0.078}^{+ 0.065} \times 10^{  -2}$ \\
\noalign{\smallskip}
He$^{+}$   &He~{\sc i}       $\lambda$5875.66  &    3 & $ 4.670_{-0.064}^{+ 0.057} \times 10^{  -2}$ \\
\noalign{\smallskip}
He$^{+}$   &He~{\sc i}       $\lambda$6678.16  &    1 & $ 4.338_{-0.123}^{+ 0.096} \times 10^{  -2}$ \\
\noalign{\smallskip}
He$^{2+}$  &He~{\sc ii}      $\lambda$4685.68  &    1 & $ 1.028_{-0.004}^{+ 0.003} \times 10^{  -1}$ \\
\noalign{\smallskip}
C$^{2+}$   &C~{\sc ii}       $\lambda$6461.95  &    1 & $ 3.546_{-0.197}^{+ 0.221} \times 10^{  -4}$ \\
\noalign{\smallskip}
N$^{2+}$   &N~{\sc ii}       $\lambda$4613.87  &    1 & $ 7.217_{-1.962}^{+ 1.653} \times 10^{  -4}$ \\
\noalign{\smallskip}
N$^{2+}$   &N~{\sc ii}       $\lambda$4788.13  &    2 & $ 5.221_{-0.283}^{+ 0.242} \times 10^{  -3}$ \\
\noalign{\smallskip}
N$^{2+}$   &N~{\sc ii}       $\lambda$5676.02  &    3 & $ 4.043_{-0.115}^{+ 0.094} \times 10^{  -3}$ \\
\noalign{\smallskip}
N$^{2+}$   &N~{\sc ii}       $\lambda$5931.78  &    2 & $ 2.171_{-0.115}^{+ 0.094} \times 10^{  -3}$ \\
\noalign{\smallskip}
N$^{3+}$   &N~{\sc iii}      $\lambda$4640.64  &    1 & $ 3.310_{-0.200}^{+ 0.192} \times 10^{  -4}$ \\
\noalign{\smallskip}
O$^{2+}$   &O~{\sc ii}       $\lambda$4491.23  &    1 & $ 6.238_{-1.566}^{+ 1.245} \times 10^{  -4}$ \\
\noalign{\medskip}
\multicolumn{4}{c}{                          \ldots }\\
\noalign{\medskip}
\noalign{\medskip}
\hline
\end{tabular}
\begin{list}{}{}
\item[\textbf{Note:}]Table \ref{wc:tab:abundances:orls} is published in its entirety in the machine-readable format. A portion is shown here for guidance regarding its form and content.
\end{list}
\end{table}


\section{Abundance Analyses}
\label{wc:sec:abundances}

\subsection{Ionic Abundances from CELs}
\label{wc:sec:abundances:cel}

We determined abundances for ionic species of N, O, Ne, S, Cl, Ar and Fe from CELs. 
To derive ionic abundances, we solve the statistical equilibrium equations for each ion using the IDL library proEQUIB, giving level population and line emissivities for specified $T_{\rm e}$ and $N_{\rm e}$.  
In Table~\ref{wc:plasma:summary}, we summarize physical conditions obtained from CELs and ORLs.
Table~\ref{wc:tab:adopted:TeNe:2} also lists the $T_{\rm e}$ and $N_{\rm e}$ adopted for our CEL abundance analysis. 
The electron temperatures derived from [N\,{\sc ii}] and [O\,{\sc iii}] in \S\,\ref{wc:sec:tempdens:cel:temp} were employed for 
low- and high-excited lines, respectively, where they are available and reliable.
We were careful to use the recombination-corrected temperatures due to 
potentially overestimated ${\rm N^{2+}}$ as discussed in \S\,\ref{wc:sec:tempdens:cel:temp}.
Similarly, the electron densities obtained in \S\,\ref{wc:sec:tempdens:cel:dens} were utilized for 
low and high-excited CELs where they are available and applicable.
The CELs used to determine ionic abundances are listed in Table \ref{wc:tab:abundances:cels} (the full table
is available in the machine-readable format).
The uncertainties in ionic abundances are determined using the MCMC ensemble sampler from the flux errors without accounting for the uncertainties estimated for $T_{\rm e}$ and $N_{\rm e}$. As the CEL emissivity calculation has an exponential dependence on $T_{\rm e}$,
including the uncertainties of the physical conditions leads to large uncertainties in abundances. 
The accurate determination of abundances depends on the adopted physical conditions. Once the level population are solved, the ionic abundances X${}^{i+}$/H${}^{+}$ are calculated from the intrinsic intensities of CELs as follows:
\begin{equation}
\frac{N({\rm X}^{i+})}{N({\rm H}^{+})}=\frac{I(\lambda_{ij})}{I({\rm H}\beta)} 
\frac{\lambda_{ij}({\rm {\AA}})}{4861.33} \frac{\alpha_{\rm eff}({\rm H}\beta)}{A_{ij}} 
\frac{N_{\rm e}}{n_i},
\label{wc:eq_cel1}
\end{equation}
where $I(\lambda_{ij})$ is the dereddened flux of the emission line $\lambda_{ij}$ emitted by ion ${\rm X}^{i+}$ following the transition from the upper level $i$ to the lower level $j$,  
$I({\rm H}\beta)$ the dereddened flux of H$\beta$, $\alpha_{\rm eff}({\rm H}\beta)$ the effective recombination coefficient of H$\beta$, $n_i(T_{\rm e},N_{\rm e},A_{ij},\Omega_{ij})$ the fractional population of the upper level $i$, 
$A_{ij}$ the Einstein spontaneous transition probability, and  $\Omega_{ij}$ the collision strength of the transition.

The weighted-average ionic abundances of $N$ lines (see Table~\ref{wc:tab:abundances:total} and \S\,\ref{wc:sec:abundances:total}) are calculated as follows:
\begin{equation}
\left( \frac{N({\rm X}^{i+})}{N({\rm H}^{+})} \right)_{\rm mean} = \frac{1}{W_{\rm tot}}
\sum_{j=1}^{N}
\frac{N({\rm X}^{i+})}{N({\rm H}^{+})},
\end{equation}
where ${W_{\rm tot}}=\sum_{j=1}^{N}W_{j}$ is the sum of the weights, and $W_{j}$ are the weights given in Table \ref{wc:tab:abundances:cels} calculated based on the predicted intrinsic fluxes at the given physical conditions and normalized by the minimum value of $W_{j}$.

\begin{figure}
\begin{center}
\includegraphics[width=0.5\textwidth, trim = 0 5 15 30, clip, angle=0]{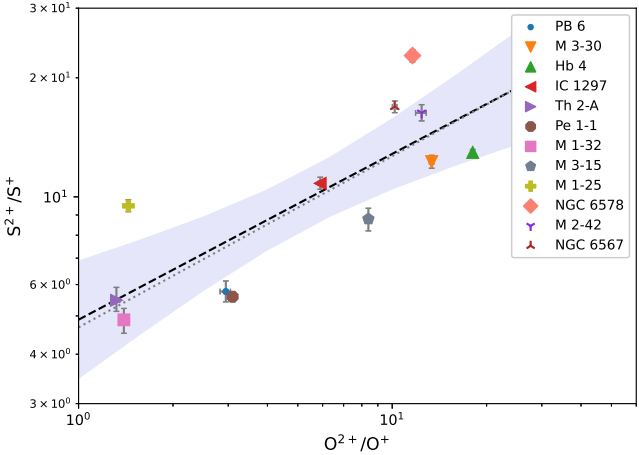}%
\caption{S$^{2+}$/S$^{+}$ versus O$^{2+}$/O$^{+}$. The dashed line is a least-squares fit that yields 
${\rm S}^{2+}/{\rm S}^{+} = 4.892 \, ({\rm O}^{2+}/{\rm O}^{+})^{0.419}$ with  
the 90\% confidence level shown by the lightly shaded area. 
The dotted gray line shows ${\rm S}^{2+}/{\rm S}^{+} = 4.677 \, ({\rm O}^{2+}/{\rm O}^{+})^{0.433}$ from \citet{Kingsburgh1994}. 
\label{wc:abudances:oppop:sppsp}%
}%
\end{center}
\end{figure}

For the CEL abundance analysis of each
object, we adopted the density and temperature based on the results from our CEL plasma diagnostics in Section~\ref{wc:sec:tempdens}, as listed in Table~\ref{wc:plasma:summary}. Following \citet{Kingsburgh1994}, we mostly adopted $T_{\rm e}$([N\,{\sc ii}]) for singly ionized 
species and $T_{\rm e}$([O\,{\sc iii}]) for ions of higher excitation ions in the CEL abundance calculations where applicable (see Table~\ref{wc:tab:adopted:TeNe:2}). 

The forbidden lines of [O\,{\sc iii}] $\lambda\lambda$4959,5007 were used to derive O$^{2+}$/H$^{+}$ ionic ratios. For O$^{+}$/H$^{+}$ ionic abundances, we adopted the observed fluxes of the [O\,{\sc ii}] $\lambda$3727 doublet and [O\,{\sc ii}] $\lambda\lambda$7320,7330 lines from the literature. 
We should note that the O$^{+}$/H$^{+}$ abundance ratios derived from the [O\,{\sc ii}] $\lambda$3727 doublet may be more reliable than those derived from the [O\,{\sc ii}] $\lambda\lambda$7320,7330 lines, which may contain recombination contributions  and/or being biased towards higher density regions ($N_{\rm cr}=3.3$\,-\,$4.9 \times 10^{6}$\,cm$^{-3}$).
Moreover, \citet{Rodriguez2020} found that the O$^{+}$/H$^{+}$ ionic abundances derived from [O\,{\sc ii}] $\lambda$3727 and $\lambda$7325 are largely different due to observational uncertainties.
The N${}^{+}$/H${}^{+}$ abundance ratio was derived from the $[$N\,{\sc ii}$]$ $\lambda$6548 and $\lambda$6584 lines.  We did not use the weak auroral line $[$N\,{\sc ii}$]$ $\lambda$5755 due to its high uncertainty and the recombination contribution from  N$^{2+}$. 
For most PNe, we determined the S$^{+}$/H$^{+}$ abundance ratio from the [S\,{\sc ii}] $\lambda\lambda$6716,6731 
doublets, and S$^{2+}$/H$^{+}$ from the [S\,{\sc iii}] $\lambda$6312. 
We should note that the [S\,{\sc ii}] $\lambda\lambda$4068,4076 lines are usually weak and affected by recombination processes and density effects.
For most PNe, we were able to determine the Ar${}^{2+}$/H${}^{+}$ abundance ratio from the $[$Ar\,{\sc iii}$]$ $\lambda$7136 and $\lambda$7751 lines and the Ar${}^{3+}$/H${}^{+}$ abundance ratio from the $[$Ar\,{\sc iv}$]$ $\lambda\lambda$4711,4740 doublet. 
For some PNe with [WO] central stars, we also derived the Ar${}^{4+}$/H${}^{+}$ abundance ratio from the $[$Ar\,{\sc v}$]$ $\lambda$4625 and $\lambda$7005 lines.

For Sa\,3-107, where O$^{2+}$ but not O$^{+}$ is observed, the O$^{+}$/H$^{+}$ ionic ratio was estimated by using a correlation between log(O$^{2+}$/O$^{+}$) and log(S$^{2+}$/S$^{+}$). 
A least-squares fit to the 12 PNe (around stars with $T_{\rm eff} > 35$\,kK) plotted in Fig.\,\ref{wc:abudances:oppop:sppsp} yields 
${\rm S}^{2+}/{\rm S}^{+} = 4.892^{+1.009}_{-0.837} \, ({\rm O}^{2+}/{\rm O}^{+})^{0.419 \pm 0.099}$ with 
$r=0.8$ ($p=0.0017$) that can be used to estimated O$^{+}$ from S$^{2+}$/S$^{+}$ when only O$^{2+}$ is available. Previously, \citet{Kingsburgh1994} obtained ${\rm S}^{2+}/{\rm S}^{+} = 4.677 \, ({\rm O}^{2+}/{\rm O}^{+})^{0.433}$ using a least-squares fit to the 22 PNe, which was then used to estimate S$^{2+}$ when only S$^{+}$ was observed, and vice versa. 

\setcounter{table}{12}
\begin{table*}
\caption{Mean ionic and total elemental abundances relative to hydrogen derived from ORLs and CELs. \label{wc:tab:abundances:total}}
\centering
\scriptsize
\begin{tabular}{lclc}
\hline\hline
\noalign{\smallskip}
 Ion~~~~~~~~~~~  & Typ. & Ref. & Abund. \\
\noalign{\smallskip}
\hline
\noalign{\smallskip}
\multicolumn{4}{c}{PB 6          (PNG278.8$+$04.9) }\\
\noalign{\smallskip}
\hline
\noalign{\smallskip}
He$^{+}$/H       & ORL     &          & $       4.571_{      -0.057}^{+       0.051} \times 10^{  -2}$ \\
\noalign{\smallskip}
He$^{2+}$/H      & ORL     &          & $       1.028_{      -0.004}^{+       0.003} \times 10^{  -1}$ \\
\noalign{\smallskip}
He/H             & ORL     &          & $       1.486_{      -0.008}^{+       0.008} \times 10^{  -1}$ \\
\noalign{\smallskip}
\noalign{\medskip}
C$^{2+}$/H       & ORL     &          & $       3.546_{      -0.197}^{+       0.221} \times 10^{  -4}$ \\
\noalign{\smallskip}
{\it icf}(C)     & ORL     & WL07     & $       2.938_{      -0.081}^{+       0.103}$ \\
\noalign{\smallskip}
C/H              & ORL     & WL07     & $       1.042_{      -0.071}^{+       0.103} \times 10^{  -3}$ \\
\noalign{\smallskip}
\noalign{\medskip}
{\it icf}(C)     & ORL     & DMS14    & $       3.728_{      -1.978}^{+       1.793}$ \\
\noalign{\smallskip}
C/H              & ORL     & DMS14    & $       1.322_{      -0.899}^{+       0.735} \times 10^{  -3}$ \\
\noalign{\smallskip}
\noalign{\medskip}
N$^{2+}$/H       & ORL     &          & $       3.455_{      -0.113}^{+       0.098} \times 10^{  -3}$ \\
\noalign{\smallskip}
N$^{3+}$/H       & ORL     &          & $       3.310_{      -0.200}^{+       0.192} \times 10^{  -4}$ \\
\noalign{\smallskip}
{\it icf}(N)     & ORL     & WL07     & $       1.131_{      -0.007}^{+       0.007}$ \\
\noalign{\smallskip}
N/H              & ORL     & WL07     & $       4.280_{      -0.166}^{+       0.176} \times 10^{  -3}$ \\
\noalign{\smallskip}
\noalign{\medskip}
O$^{2+}$/H       & ORL     &          & $       6.238_{      -1.566}^{+       1.245} \times 10^{  -4}$ \\
\noalign{\smallskip}
{\it icf}(O)     & ORL     & WL07     & $       2.938_{      -0.038}^{+       0.051}$ \\
\noalign{\smallskip}
O/H              & ORL     & WL07     & $       1.833_{      -0.503}^{+       0.496} \times 10^{  -3}$ \\
\noalign{\smallskip}
\noalign{\medskip}
N$^{+}$/H        & CEL     &          & $       4.376_{      -0.162}^{+       0.172} \times 10^{  -5}$ \\
\noalign{\smallskip}
{\it icf}(N)     & CEL     & KB94     & $       8.661_{      -0.337}^{+       0.408}$ \\
\noalign{\smallskip}
N/H              & CEL     & KB94     & $       3.790_{      -0.215}^{+       0.332} \times 10^{  -4}$ \\
\noalign{\smallskip}
\noalign{\medskip}
{\it icf}(N)     & CEL     & DMS14    & $       7.525_{      -4.423}^{+       5.819}$ \\
\noalign{\smallskip}
N/H              & CEL     & DMS14    & $       3.293_{      -2.089}^{+       3.016} \times 10^{  -4}$ \\
\noalign{\smallskip}
\noalign{\medskip}
O$^{0}$/H        & CEL     &          & $       6.579_{      -0.229}^{+       0.219} \times 10^{  -6}$ \\
\noalign{\smallskip}
O$^{+}$/H        & CEL     &          & $       4.928_{      -0.142}^{+       0.164} \times 10^{  -5}$ \\
\noalign{\smallskip}
O$^{2+}$/H       & CEL     &          & $       1.453_{      -0.018}^{+       0.017} \times 10^{  -4}$ \\
\noalign{\smallskip}
{\it icf}(O)     & CEL     & KB94     & $       2.194_{      -0.020}^{+       0.023}$ \\
\noalign{\smallskip}
O/H              & CEL     & KB94     & $       4.268_{      -0.085}^{+       0.107} \times 10^{  -4}$ \\
\noalign{\smallskip}
\noalign{\medskip}
{\it icf}(O)     & CEL     & DMS14    & $       2.402_{      -0.673}^{+       0.673}$ \\
\noalign{\smallskip}
O/H              & CEL     & DMS14    & $       4.672_{      -1.452}^{+       1.562} \times 10^{  -4}$ \\
\noalign{\smallskip}
\noalign{\medskip}
Ne$^{2+}$/H      & CEL     &          & $       3.816_{      -0.083}^{+       0.111} \times 10^{  -5}$ \\
\noalign{\smallskip}
Ne$^{3+}$/H      & CEL     &          & $       2.825_{      -0.148}^{+       0.134} \times 10^{  -4}$ \\
\noalign{\smallskip}
{\it icf}(Ne)    & CEL     & KB94     & $       2.938_{      -0.067}^{+       0.103}$ \\
\noalign{\smallskip}
\noalign{\medskip}
\noalign{\smallskip}
\hline
\noalign{\smallskip}
\end{tabular}
\begin{tabular}{lclc}
\hline\hline
\noalign{\smallskip}
 Ion  & ~~Typ.~~ & Ref. & Abund. \\
\noalign{\smallskip}
\hline
\noalign{\smallskip}
Ne/H             & CEL     & KB94     & $       1.121_{      -0.041}^{+       0.064} \times 10^{  -4}$ \\
\noalign{\smallskip}
\noalign{\medskip}
{\it icf}(Ne)    & CEL     & DMS14    & $       3.457_{      -1.765}^{+       1.789}$ \\
\noalign{\smallskip}
Ne/H             & CEL     & DMS14    & $       1.319_{      -0.804}^{+       0.822} \times 10^{  -4}$ \\
\noalign{\smallskip}
\noalign{\medskip}
S$^{+}$/H        & CEL     &          & $       3.868_{      -0.134}^{+       0.136} \times 10^{  -7}$ \\
\noalign{\smallskip}
S$^{2+}$/H       & CEL     &          & $       2.229_{      -0.082}^{+       0.072} \times 10^{  -6}$ \\
\noalign{\smallskip}
{\it icf}(S)     & CEL     & KB94     & $       1.481_{      -0.019}^{+       0.022}$ \\
\noalign{\smallskip}
S/H              & CEL     & KB94     & $       3.873_{      -0.175}^{+       0.182} \times 10^{  -6}$ \\
\noalign{\smallskip}
\noalign{\medskip}
{\it icf}(S)     & CEL     & DMS14    & $       2.626_{      -1.441}^{+       1.365}$ \\
\noalign{\smallskip}
S/H              & CEL     & DMS14    & $       6.869_{      -4.240}^{+       4.523} \times 10^{  -6}$ \\
\noalign{\smallskip}
\noalign{\medskip}
Cl$^{2+}$/H      & CEL     &          & $       2.935_{      -0.091}^{+       0.083} \times 10^{  -8}$ \\
\noalign{\smallskip}
{\it icf}(Cl)    & CEL     & L00      & $       1.738_{      -0.104}^{+       0.138}$ \\
\noalign{\smallskip}
Cl/H             & CEL     & L00      & $       5.101_{      -0.413}^{+       0.479} \times 10^{  -8}$ \\
\noalign{\smallskip}
\noalign{\medskip}
{\it icf}(Cl)    & CEL     & DMS14    & $       3.931_{      -2.433}^{+       2.162}$ \\
\noalign{\smallskip}
Cl/H             & CEL     & DMS14    & $       1.154_{      -0.840}^{+       0.780} \times 10^{  -7}$ \\
\noalign{\smallskip}
\noalign{\medskip}
Ar$^{2+}$/H      & CEL     &          & $       5.410_{      -0.258}^{+       0.232} \times 10^{  -7}$ \\
\noalign{\smallskip}
Ar$^{3+}$/H      & CEL     &          & $       7.511_{      -0.042}^{+       0.050} \times 10^{  -7}$ \\
\noalign{\smallskip}
Ar$^{4+}$/H      & CEL     &          & $       3.716_{      -0.201}^{+       0.199} \times 10^{  -7}$ \\
\noalign{\smallskip}
{\it icf}(Ar)    & CEL     & KB94     & $       1.131_{      -0.016}^{+       0.013}$ \\
\noalign{\smallskip}
Ar/H             & CEL     & KB94     & $       1.881_{      -0.060}^{+       0.060} \times 10^{  -6}$ \\
\noalign{\smallskip}
\noalign{\medskip}
{\it icf}(Ar)    & CEL     & DMS14    & $       2.870_{      -2.870}^{+       2.944}$ \\
\noalign{\smallskip}
Ar/H             & CEL     & DMS14    & $       1.552_{      -1.552}^{+       1.966} \times 10^{  -6}$ \\
\noalign{\smallskip}
\noalign{\medskip}
Fe$^{2+}$/H      & CEL     &          & $       1.638_{      -0.138}^{+       0.141} \times 10^{  -7}$ \\
\noalign{\smallskip}
{\it icf}(Fe)    & CEL     & ITL94    & $      10.826_{      -0.497}^{+       0.566}$ \\
\noalign{\smallskip}
Fe/H             & CEL     & ITL94    & $       1.773_{      -0.182}^{+       0.212} \times 10^{  -6}$ \\
\noalign{\smallskip}
\noalign{\medskip}
{\it adf}(N$^{2+}$)       & ORL/CEL &          & $      26.782_{      -2.402}^{+       2.406}$ \\
\noalign{\smallskip}
{\it adf}(N)              & ORL/CEL & KB94     & $      11.292_{      -1.345}^{+       1.123}$ \\
\noalign{\smallskip}
{\it adf}(N)              & ORL/CEL & DMS14    & $      12.997_{     -12.997}^{+      48.697}$ \\
\noalign{\smallskip}
\noalign{\medskip}
{\it adf}(O$^{2+}$)       & ORL/CEL &          & $       4.295_{      -1.245}^{+       1.097}$ \\
\noalign{\smallskip}
{\it adf}(O)              & ORL/CEL & KB94     & $       4.295_{      -1.470}^{+       1.452}$ \\
\noalign{\smallskip}
{\it adf}(O)              & ORL/CEL & DMS14    & $       3.923_{      -2.141}^{+       3.117}$ \\
\noalign{\smallskip}
\multicolumn{4}{c}{                          \ldots }\\
\noalign{\medskip}
\hline
\end{tabular}
\begin{list}{}{}
\item[\textbf{Note:}]Table \ref{wc:tab:abundances:total} is published in its entirety in the machine-readable format. A portion is shown here for guidance regarding its form and content.
References for \textit{icf} formulas are as follows: 
DMS14 -- \citet{Delgado-Inglada2014}; 
ITL94 -- \citet{Izotov1994}; 
KB94 -- \citet{Kingsburgh1994}; 
L00 -- \citet{Liu2000}; 
WL07 -- \citet{Wang2007}. 
For Sa\,3-107, the CEL O${}^{+}$ ionic abundance estimated by using the correlation 
${\rm S}^{2+}/{\rm S}^{+} = 4.892 \, ({\rm O}^{2+}/{\rm O}^{+})^{0.419}$ derived in Section~\ref{wc:sec:abundances:cel} (see Fig.\,\ref{wc:abudances:oppop:sppsp}).
\end{list}
\end{table*}


\subsection{Ionic Abundances from ORLs}
\label{wc:sec:abundances:orl}

We determined abundances for ionic species of He, C, N and O from ORLs for our sample where adequate observed lines were available. In our calculation, we adopted the electron temperature and density listed in Table~\ref{wc:plasma:summary}
(selected $T_{\rm e}$ and $N_{\rm e}$ listed in Table~\ref{wc:tab:adopted:TeNe:2}). 
The temperatures derived from either CELs or ORLs were delicately chosen for 
our ORL abundance analysis where they are reliable and applicable.
As the densities derived from heavy element ORLs are highly uncertain, 
we carefully adopted the densities obtained from CELs. 
The atomic data sets used for the effective recombination coefficients of ORLs are listed in Table~\ref{wc:tab:atomicdata}. Using these effective recombination coefficients, we determine ionic abundances from the measured intensities of ORLs as follows:
\begin{equation}
\frac{N({\rm X}^{i+})}{N({\rm H}^{+})}=\frac{I(\lambda)}{I({\rm H}\beta)} 
\frac{\lambda({\rm {\AA}})}{4861.33} \frac{\alpha_{\rm eff}({\rm H}\beta)}{\alpha_{\rm eff}(\lambda)},
\label{wc:eq_orl1}%
\end{equation}
where $I(\lambda)$ is the intrinsic line flux of the emission line $\lambda$ emitted by ion ${\rm X}^{i+}$,  
$I({\rm H}\beta)$ is the intrinsic line flux of H$\beta$, $\alpha_{\rm eff}({\rm H}\beta)$ the effective recombination coefficient of H$\beta$, and $\alpha_{\rm eff}(\lambda)$ the effective recombination coefficient for the emission line $\lambda$.

The ionic  abundances derived from the ORLs are given in Table~\ref{wc:tab:abundances:orls}. To obtain the weighted-average He$^{+}$/H$^{+}$ ionic abundance, the ionic abundances derived from the He\,{\sc i} ORLs were averaged with weights according to the predicted intrinsic intensity ratios of the available lines. 
The He$^{2+}$/H$^{+}$ ionic abundance was derived from the He\,{\sc ii} $\lambda$4686 line. The C$^{2+}$/H$^{+}$ ratios were derived from some high-excitation C\,{\sc ii} ORLs 
(4f--6g $\lambda$6461.95 and 4d--6f $\lambda$6151.43). 

\setcounter{table}{13}
\begin{table*}
\caption{Summary of elemental abundances obtained from ORLs and CELs, on a logarithmic scale where H = 12.
\label{wc:abund:summary}
}
\centering
\scriptsize
\begin{tabular}{lllccccccccccc}
\hline\hline
\noalign{\smallskip}
Name        & PN\,G &\multicolumn{1}{c}{X}&  He/H&   C/H&N/H&O/H&   N/H&   O/H&  Ne/H&   S/H&  Cl/H&  Ar/H&  Fe/H \\
            &  &\multicolumn{1}{c}{\textit{icf}(X)}&  (ORL)& (ORL)&(ORL)&(ORL)&  (CEL)&  (CEL)&  (CEL)&   (CEL)&  (CEL)&  (CEL)&  (CEL)  \\
\noalign{\smallskip}
\hline
\noalign{\smallskip}
PB\,6       & 278.8$+$04.9 &KB94& 11.17&  9.02&  9.63&  9.26&  8.58&  8.63&  8.05&  6.59&  4.71&  6.27&  6.25\\
            &  &DMS14& 11.17&  9.12&      &      &  8.52&  8.67&  8.12&  6.84&  5.06&  6.19&      \\
\noalign{\smallskip}
M\,3-30     & 017.9$-$04.8 &KB94& 11.18&  9.42&  9.65&  9.79&  8.30&  8.61&  8.17&  7.03&  5.10&  6.34&  7.76\\
            &  &DMS14& 11.18&  9.61&      &      &  8.20&  8.61&  8.20&  7.13&  5.97&  6.52&      \\
\noalign{\smallskip}
Hb\,4       & 003.1$+$02.9 &KB94& 11.06&  8.92&  9.23&  9.79&  8.73&  8.80&  7.97&  7.28&  5.30&  6.53&  6.73\\
            &  &DMS14& 11.06&  8.86&      &      &  8.73&  8.80&  8.00&  7.28&  6.45&  6.61&      \\
\noalign{\smallskip}
IC\,1297    & 358.3$-$21.6 &KB94& 11.05&  9.07&  8.92&  9.09&  8.01&  8.76&  8.27&  6.89&  4.93&  6.25&  6.17\\
            &  &DMS14& 11.05&  9.15&      &      &  7.94&  8.75&  8.30&  6.93&  5.65&  6.28&      \\
\noalign{\smallskip}
Th\,2-A     & 306.4$-$00.6 &KB94& 11.05& 10.01&  9.52& 10.27&  8.05&  8.88&  8.49&  6.56&  5.07&  6.41&  6.70\\
            &  &DMS14& 11.05&  9.95&      &      &  7.99&  8.88&  8.49&  6.62&  5.35&  6.32&      \\
\noalign{\smallskip}
Pe\,1-1     & 285.4$+$01.5 &KB94& 11.01&  9.45&  9.39&  9.43&  8.16&  8.64&  8.05&  6.71&  4.51&  7.43&  6.75\\
            &  &DMS14& 11.01&  9.43&      &      &  8.64&  8.64&  8.31&  6.67&  4.97&  6.29&      \\
\noalign{\smallskip}
M\,1-32     & 011.9$+$04.2 &KB94& 11.11&  9.51&  9.59& 10.02&  8.64&  8.57&  7.49&  7.12&  5.15&  6.77&  7.38\\
            &  &DMS14& 11.11&  9.47&      &      &  9.01&  8.57&  7.81&  7.07&  5.42&  6.55&      \\
\noalign{\smallskip}
M\,3-15     & 006.8$+$04.1 &KB94& 11.04&  9.81&  9.45&      &  8.35&  8.83&  7.93&  6.99&      &  5.58&  6.73\\
            &  &DMS14& 11.04&  9.34&      &      &  8.92&  8.83&  8.12&  6.98&      &      &      \\
\noalign{\smallskip}
M\,1-25     & 004.9$+$04.9 &KB94& 11.07&  9.33&  9.58&  9.85&  8.36&  8.81&  7.56&  7.15&  4.85&  6.81&  6.59\\
            &  &DMS14& 11.07&  9.10&      &      &  8.74&  8.81&  7.87&  7.11&  5.17&  6.58&      \\
\noalign{\smallskip}
Hen\,2-142  & 327.1$-$02.2 &KB94& 10.34&      &      &      &  8.02&  8.65&      &  6.55&  5.09&      &      \\
            &  &DMS14& 10.34&      &      &      &  8.02&  8.65&      &  6.53&  4.97&      &      \\
\noalign{\smallskip}
Hen\,3-1333 & 332.9$-$09.9 &KB94&      &      &      &      &  8.54&  9.50&      &  7.67&      &      &      \\
            &  &DMS14&      &      &      &      &  8.54&  9.50&      &  7.64&      &      &      \\
\noalign{\smallskip}
Hen\,2-113  & 321.0$+$03.9 &KB94&      &      &      &      &  8.07&  8.62&      &  6.84&      &      &      \\
            &  &DMS14&      &      &      &      &  8.07&  8.62&      &  6.82&      &      &      \\
\noalign{\smallskip}
K\,2-16     & 352.9$+$11.4 &KB94&      &      &      &      &  7.98&  8.30&      &  7.08&      &      &      \\
            &  &DMS14&      &      &      &      &  8.16&  8.30&      &  6.46&      &      &      \\
\noalign{\smallskip}
NGC\,6578   & 010.8$-$01.8 &KB94& 11.07&  9.36&  8.94&  9.57&  7.81&  8.78&  8.38&  6.94&  5.06&  6.50&  7.50\\
            &  &DMS14& 11.07&  9.20&      &      &  8.03&  8.78&  9.63&  6.93&  5.93&  6.55&      \\
\noalign{\smallskip}
M\,2-42     & 008.2$-$04.8 &KB94& 10.99&  9.60&  9.45&  9.15&  8.03&  8.59&  7.97&  6.98&  5.18&  6.23&  6.79\\
            &  &DMS14& 10.99&  8.88&      &      &  8.26&  8.59&  9.39&  6.98&  6.05&  6.26&      \\
\noalign{\smallskip}
NGC\,6567   & 011.7$-$00.6 &KB94& 10.96&  9.50&  9.25&  9.81&  7.46&  8.36&  7.69&  6.27&  4.48&  5.71&  6.13\\
            &  &DMS14& 10.96&  9.22&      &      &  7.64&  8.36&  8.28&  6.27&  5.30&  5.73&      \\
\noalign{\smallskip}
NGC\,6629   & 009.4$-$05.0 &KB94& 11.00&  8.69&      &  8.99&  7.24&  8.63&  8.11&  6.42&  4.85&  6.29&  6.50\\
            &  &DMS14& 11.00&  8.81&      &      &  7.77&  8.63&  8.34&  6.39&  5.48&  6.29&      \\
\noalign{\smallskip}
Sa\,3-107   & 358.0$-$04.6 &KB94& 11.08&  9.17&  8.99&      &  7.52&  7.94&      &  6.01&  4.48&  5.76&  6.83\\
            &  &DMS14& 11.08&  9.13&      &      &  7.67&  7.94&      &  6.00&  5.43&  5.87&      \\
\noalign{\smallskip}
\hline 
\end{tabular}
\begin{list}{}{}
\item[\textbf{Note.}]Total abundances in the first row for each object calculated using the ionization correction factors (\textit{icf}) from \citet[][KB94]{Kingsburgh1994}, except for Cl \citep{Liu2000}, Fe \citep{Izotov1994}, ORLs \citep{Wang2007}, and O ORLs \citep{Wesson2005}, described in Appendix~\ref{wc:sec:icf:2}, and in the second row calculated with the \textit{icf} formulas from \citet[][DMS14]{Delgado-Inglada2014}.
\end{list}
\end{table*}

We have detected a number of N\,{\sc ii} multiplets in most PNe. They were used to calculate ORL N$^{2+}$/H$^{+}$ ionic ratios, as presented in Table~\ref{wc:tab:abundances:orls}. 
The multiplet V3 lines are more reliable as they are less sensitive to optical depth effects, and they have been detected in several objects. Other multiplets are extremely case-sensitive, and also quite weak, with large flux uncertainties, so they are less reliable. We detected the extremely case-senstive multiplet V28 in many PNe, which sometimes has a departure from the case B approximation for the triplets, so its calculated N$^{2+}$/H$^{+}$ ionic ratio is usually unreliable. 
However, the N\,{\sc ii} lines can be largely contaminated by fluorescence in low-excited PNe \citep{Escalante2012}, so
their ionic ratios could be unreliable in those with cool central stars ($T_{\rm eff} \lesssim 60$\,kK). 
In the case of the detected $\lambda$4640.64 (V2) N\,{\sc iii} recombination line, N$^{3+}$ abundance is also available in 6 PNe. 
However, the $\lambda$4640.64 N\,{\sc iii} recombination line is usually affected by continuum fluorescence \citep{Ferland1992}. Therefore, the ionic ratio derived from this line is unreliable. 

Table~\ref{wc:tab:abundances:orls} lists the ORL O$^{2+}$/H$^{+}$ ionic ratios calculated from the O\,{\sc ii} lines of mostly multiplet V1 and some multiplet V28. The abundances from the quartet-quartet transition of multiplet V1 are less case-sensitive, and it has only 4 per cent difference between case A and B. However, multiplet V28 is extremely case-sensitive, and the case B effective recombination coefficient is 20 times the case A value. The ORL O$^{2+}$/H$^{+}$ ionic ratio derived from the case-sensitive multiplet V28 is higher than those from  multiplet V1, indicating a departure from Case B conditions. The faint O\,{\sc ii} lines of multiplet V28 with high flux uncertainties make the derived ionic abundances quite uncertain.


\subsection{Total Elemental Abundances} 
\label{wc:sec:abundances:total}

Table~\ref{wc:tab:abundances:total} presents the mean ionic abundances, the ionization correction factors (\textit{icf}), and the total elemental abundances of the PNe in our sample, with their corresponding uncertainties (the full table is available in the machine readable format).
Total elemental abundances were calculated from the mean ionic abundances using the conventional \textit{icf} schemes of \citet{Kingsburgh1994}, except for \textit{icf}(CEL~Cl) from \citet{Liu2000}, \textit{icf}(CEL~Fe) from \citet{Izotov1994}, \textit{icf}(ORLs) from \citet{Wang2007}, and \textit{icf}(ORL~O) from \citet{Wesson2005} (see Appendix~\ref{wc:sec:icf:2}).
For comparison, we also derived total elemental abundances using the \textit{icf} formulas given by \citet{Delgado-Inglada2014}. 
In Table~\ref{wc:abund:summary}, we summarize the elemental abundances by number in logarithmic units relative to hydrogen where $\log N($H$) = 12$. We see that the total abundances derived from the methods by \citet{Delgado-Inglada2014} are mostly in agreement with those made with the conventional \textit{icf} schemes. 

\begin{figure*}
\begin{center}
\includegraphics[width=0.5\textwidth, trim = 0 0 15 30, clip, angle=0]{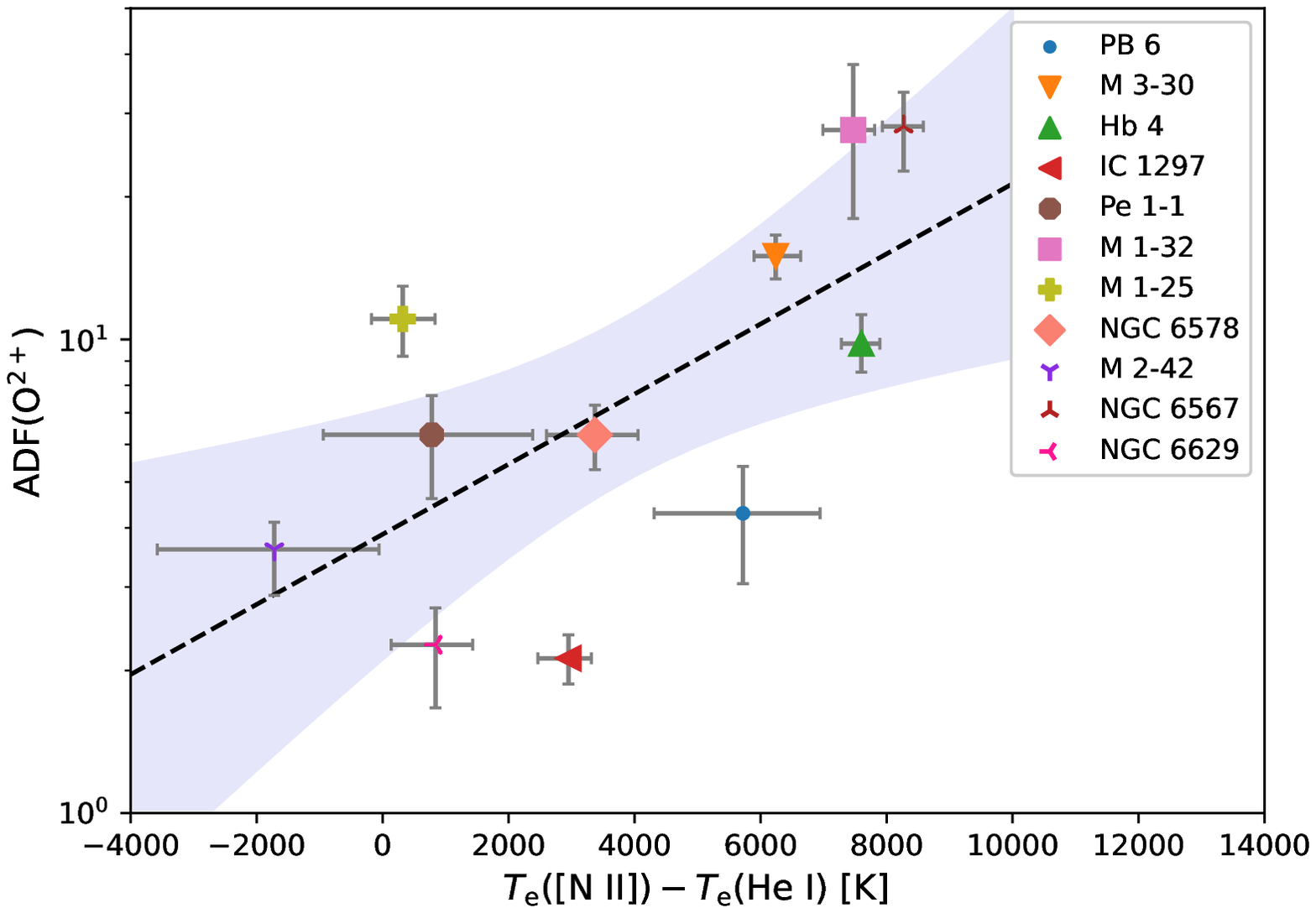}%
\includegraphics[width=0.5\textwidth, trim = 0 0 15 30, clip, angle=0]{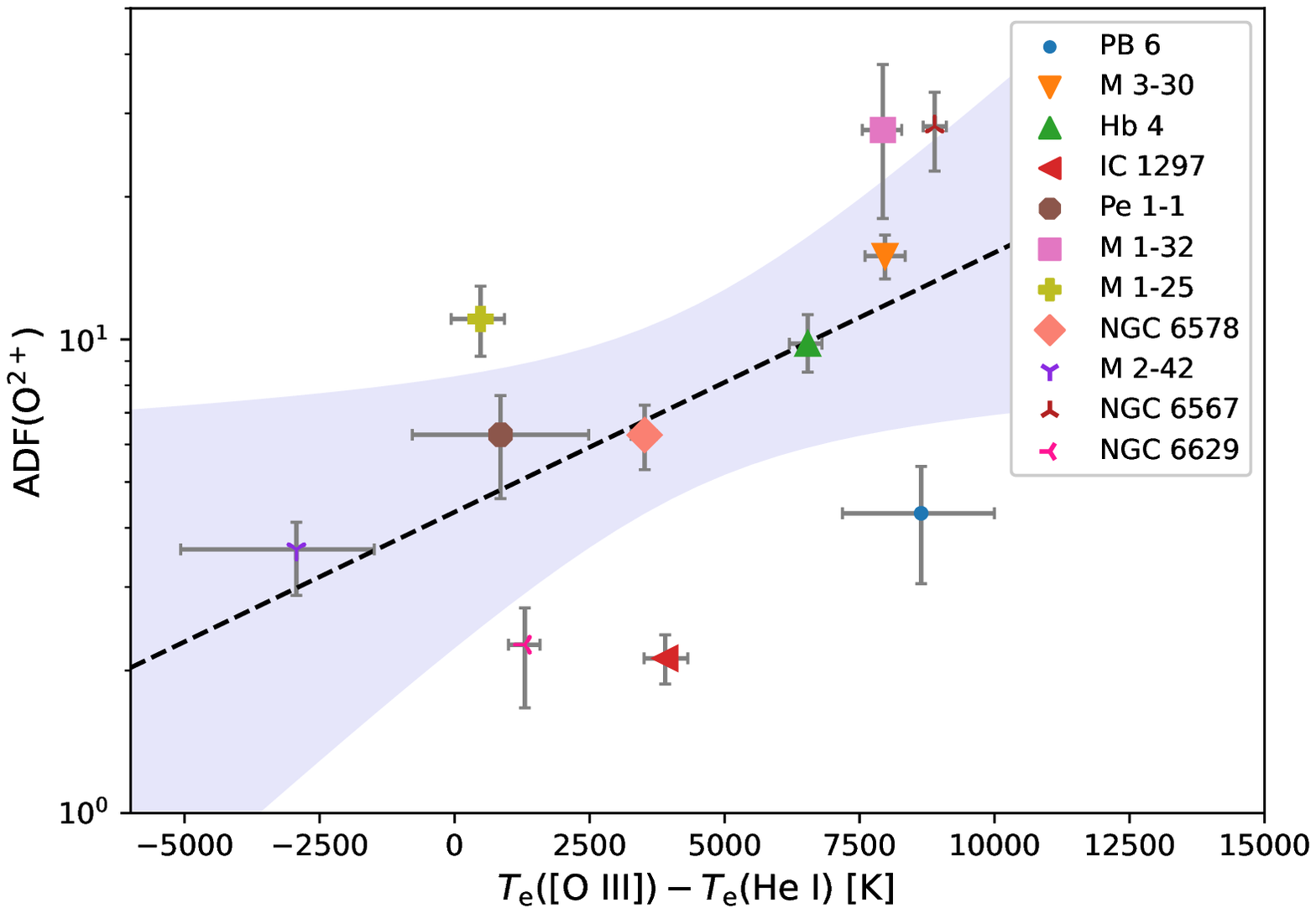}\\
\includegraphics[width=0.5\textwidth, trim = 0 0 15 30, clip, angle=0]{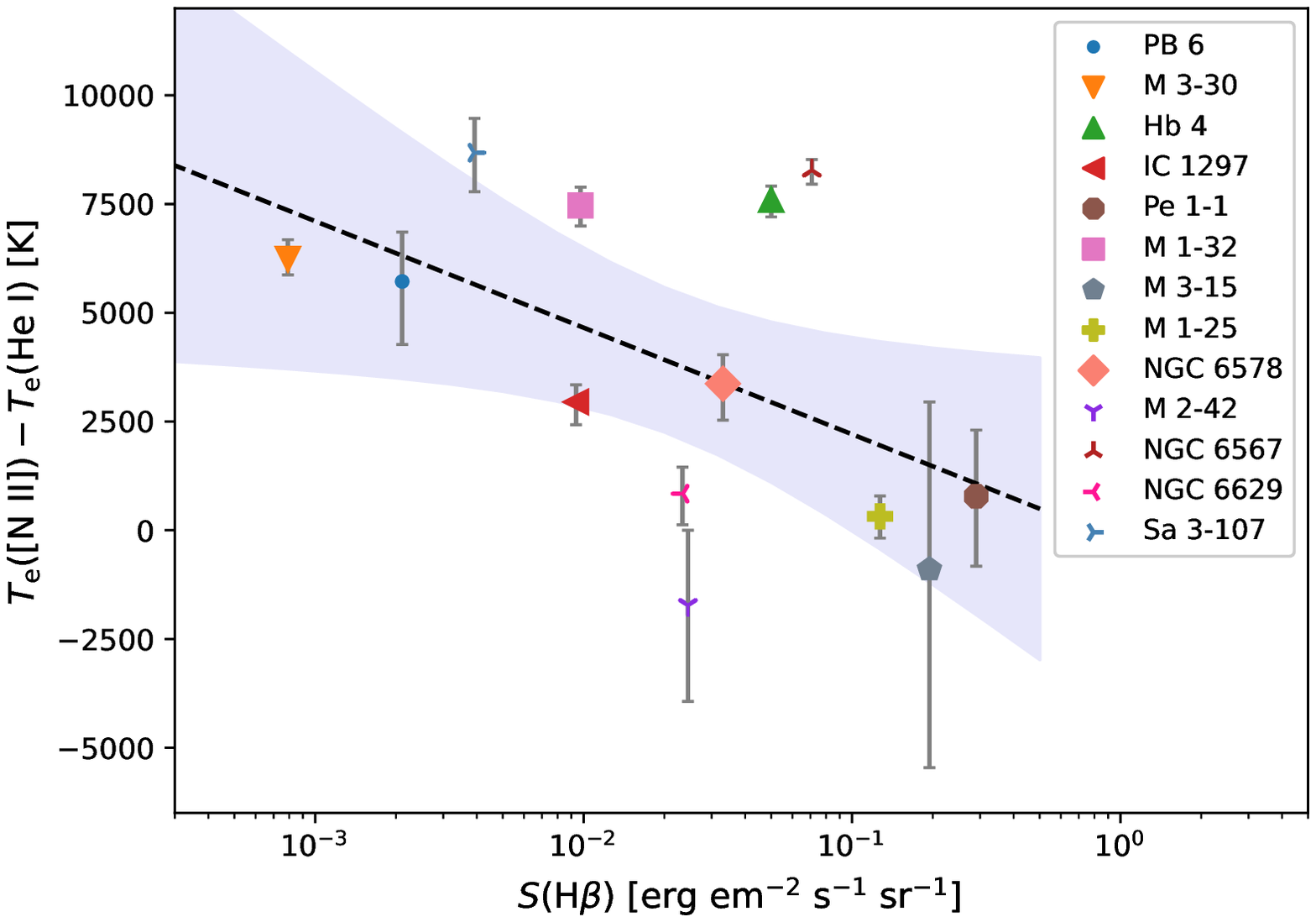}%
\includegraphics[width=0.5\textwidth, trim = 0 0 15 30, clip, angle=0]{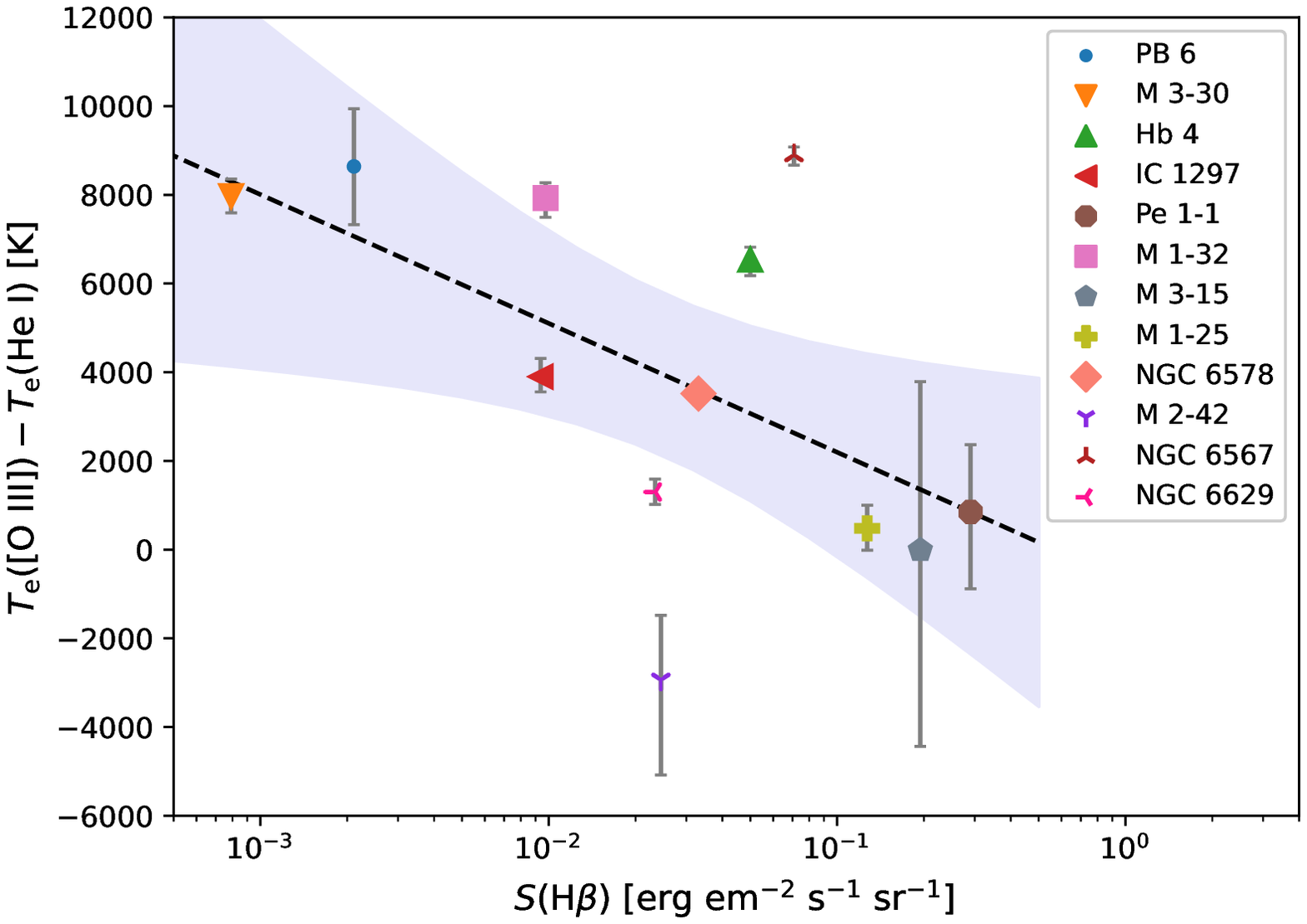}\\
\caption{\textit{Top Panels:} The ADF for O$^{2+}$ plotted against the dichotomy between the [N\,{\sc ii}] and He\,{\sc i} temperatures (left), and between the [O\,{\sc iii}] and He\,{\sc i} temperatures (right). 
The dashed lines show linear fits to $\log$~ADF(O$^{2+}$) as a function of the CEL--He\,{\sc i} temperature dichotomy, discussed in the text. 
\textit{Bottom Panels:} The dichotomy between the [N\,{\sc ii}] and He\,{\sc i} temperatures (left), and between the [O\,{\sc iii}] and He\,{\sc i} temperatures (right) plotted against the intrinsic nebular H$\beta$ surface brightness $ S$(H$\beta$)(erg\,cm$^{-2}$\,s$^{-1}$\,sr$^{-1}$). 
The dashed lines show linear fits to the temperature dichotomy as a function of $\log S$(H$\beta$), discussed in the text. 
The lightly shaded area in each panel shows the 90\% confidence level of the linear fit. 
\label{wc:tedef_adf}%
}%
\end{center}
\end{figure*}

\begin{figure}
\begin{center}
\includegraphics[width=0.5\textwidth, trim = 0 0 15 30, clip, angle=0]{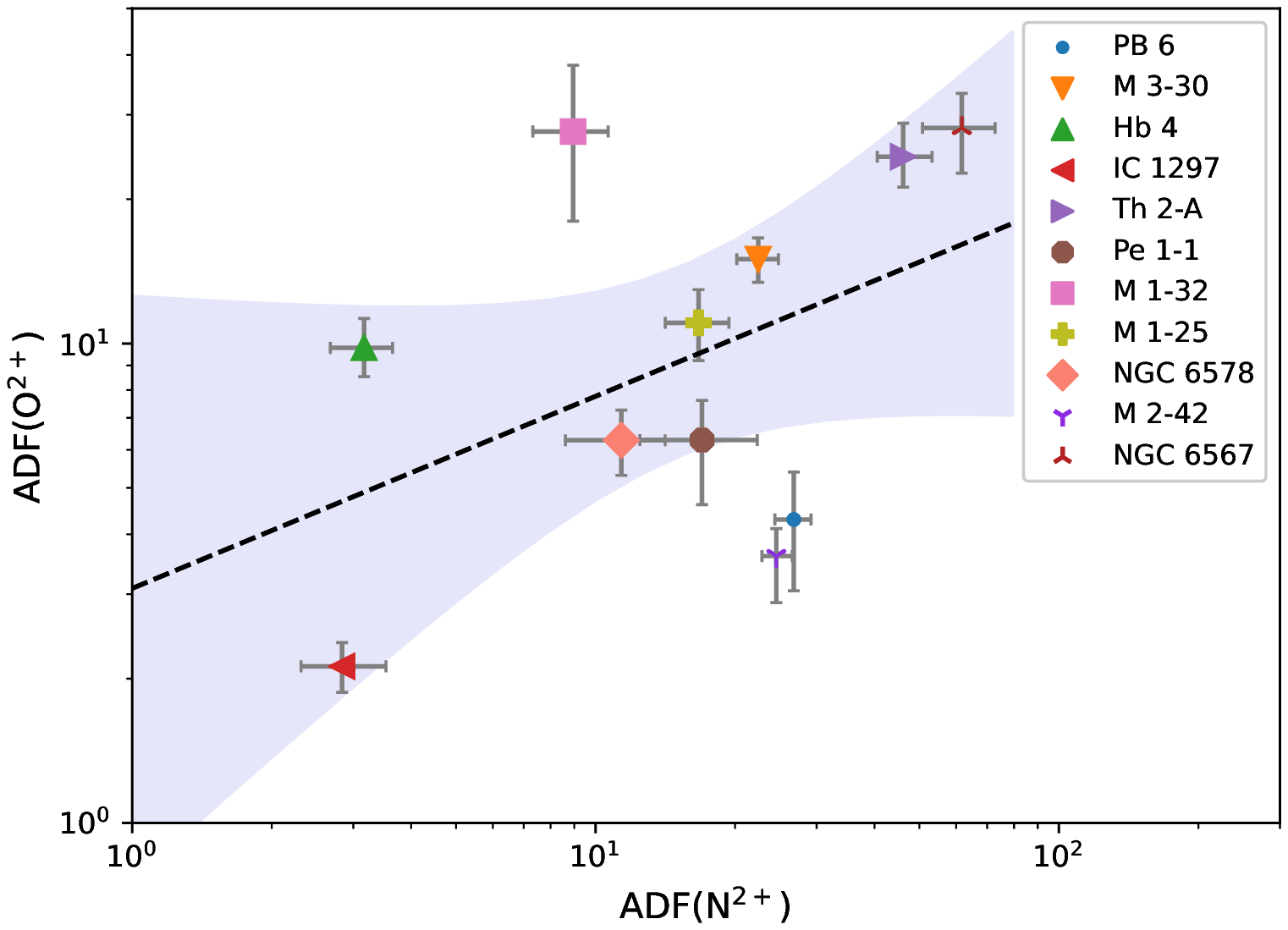}%
\caption{ADF(O$^{2+}$) versus ADF(N$^{2+}$). 
The dashed line is a least-squares fit, discussed in the text, with the 90\% confidence level shown by the gray area.
\label{wc:adf_n2:adf_o2}%
}%
\end{center}
\end{figure}

\section{ORL/CEL Discrepancy and Temperature Dichotomy} 
\label{wc:sec:adfs}

In this section, we explore possible correlations among various parameters. The ORL/CEL abundance discrepancies have been found to be correlated with various nebular physical quantities such as surface brightness, diameter, metallicity, density and excitation class \citep{Liu2001,Tsamis2004,Liu2004b,Zhang2004,Wesson2005,Wang2007,Tsamis2008,Garcia-Rojas2013}. 
We define the abundance discrepancy factors as
${\rm ADF}({\rm X}^{2+}) \equiv ({\rm X}^{2+}/{\rm H}^{+})_{\rm ORL} / ({\rm X}^{2+}/{\rm H}^{+})_{\rm CEL}$, 
and the temperature dichotomies as differences between the forbidden-line and recombination-line temperatures $T_{\rm e}($CEL$) - T_{\rm e}($ORL$)$. 
We estimate the CEL N$^{2+}$ ionic abundances from N$^{+}$ by assuming ${\rm N}^{2+}/{\rm N}^{+} = {\rm O}^{2+}/{\rm O}^{+}$ based on their ionization potentials.

The dichotomy $T_{\rm e}(\mbox{CEL}) - T_{\rm e}(\mbox{He\,{\sc i}})$ and ${\rm ADF}({\rm O}^{2+})$ for the objects plotted in Fig.\,\ref{wc:tedef_adf} (top panels) are fitted by,
\begin{align}
\log {\rm ADF}({\rm O}^{2+})  = & (0.589 \pm 0.144)+ (7.401 \pm 2.860) \times 10^{-5} \nonumber\\
 & \times \big[T_{\rm e}(\mbox{[N\,{\sc ii}]}) - T_{\rm e}(\mbox{He\,{\sc i}})\big], \label{wc:eq:tedef_adf1} \\
\log {\rm ADF}({\rm O}^{2+})  = & (0.636 \pm 0.155)+ (5.477 \pm 2.694) \times 10^{-5} \nonumber\\
 & \times \big[T_{\rm e}(\mbox{[O\,{\sc iii}]}) - T_{\rm e}(\mbox{He\,{\sc i}})\big], \label{wc:eq:tedef_adf2}
\end{align}
with $r=0.65$ ($p=0.03$) and $r=0.56$ ($p=0.07$), respectively. 

Previously, \citet{Liu2001} found that ADF(O$^{2+}$) is directly correlated with the difference between the [O\,{\sc iii}] forbidden-line and H\,{\sc i} Balmer jump electron temperatures. \citet{Tsamis2004} also derived a similar correlation for a sample of 16 PNe.  As seen in Fig.\,\ref{wc:tedef_adf} (top-left), the ADFs of O$^{2+}$ are correlated with the dichotomy between the nebular to auroral forbidden-line [N\,{\sc ii}] temperature and the He\,{\sc i} temperature. We see that higher values of ADFs are associated with higher CEL-ORL temperature dichotomies. 
These correlations likely corroborate the previous findings that there could be an intimate connection between the nebular thermal structure and the ORL/CEL abundance discrepancy. 
However, we did not find any robust associations between the ADFs of N$^{2+}$ and $T_{\rm e}(\mbox{CEL}) - T_{\rm e}(\mbox{He\,{\sc i}})$.

Previously, an anti-correlation between $T_{\rm e}(\mbox{[O\,{\sc iii}]}) - T_{\rm e}(\mbox{BJ})$ and $\log S({\rm H}\beta)$ of the 25 PNe was also found by \cite{Tsamis2004}. This suggests that the temperature dichotomies could be increased by decreasing nebular surface brightness. In Fig.~\ref{wc:tedef_adf} (bottom panels), we also plot the CEL--He\,{\sc i} temperature dichotomy as a function of the intrinsic nebular H$\beta$ surface brightness $S$(H$\beta$), and the following anti-correlations exist:
\begin{align}
\big[T_{\rm e}(\mbox{[N\,{\sc ii}]}) - T_{\rm e}(\mbox{He\,{\sc i}})\big]  = & (-243 \mp 2270)  - (2450 \pm 1251) \nonumber\\
 & \times \log S({\rm H}\beta),\label{wc:eq:shb_tedef_adf4} \\
\big[T_{\rm e}(\mbox{[O\,{\sc iii}]}) - T_{\rm e}(\mbox{He\,{\sc i}})\big]  = &  (-710 \mp 2410) - (2905 \pm 1373) \nonumber\\
 & \times \log S({\rm H}\beta), 
\label{wc:eq:shb_tedef_adf3}
\end{align}
with $r=-0.51$ ($p=0.08$) and $r=-0.56$ ($p=0.06$), respectively. 
The nebular surface brightness is an indicator of the nebular evolution, since it decreases due to the expansion of the nebula as the density drops. 
Although the ADFs of some PNe were also found to be anti-correlated with their surface brightness \citep{Liu2004b,Tsamis2004,Wang2007}, we did not obtain any statistically significant correlations between ADFs and surface brightness.

In Fig.\,\ref{wc:adf_n2:adf_o2}, the ADF(O$^{2+}$) and ADF(N$^{2+}$) for 11 PNe are fitted by
\begin{align}
{\rm ADF}({\rm O}^{2+})  = & 3.086^{+3.551}_{-1.651} \, [{\rm ADF}({\rm N}^{2+})]^{0.401 \pm 0.266},
\label{wc:eq:adf_n2:adf_o2}
\end{align}
with a Pearson $r$-value of $0.449$ and a null-hypothesis $p$-value of $0.166$. 

The correlations (\ref{wc:eq:tedef_adf1})--(\ref{wc:eq:tedef_adf2}) likely confirm previous suggestions that a fraction of cool, oxygen-rich dense clumps may be present in diffuse warm ionized gases. 
The correlation (\ref{wc:eq:adf_n2:adf_o2}) also depicts a weak dependence of ADF(N${}^{2+}$) on ADF(O${}^{2+}$), while N\,{\sc ii} ORLs could be contaminated by resonance fluorescence in some objects. Moreover, the anti-correlations (\ref{wc:eq:shb_tedef_adf4})--(\ref{wc:eq:shb_tedef_adf3}) possibly suggest that the temperature
dichotomy is weakly related to the nebular evolution, though we did not find any remarkable correlations between the ADF and $S$(H$\beta$). 

\begin{figure}
\begin{center}
\includegraphics[width=0.47\textwidth, trim = 0 0 15 5, clip, angle=0]{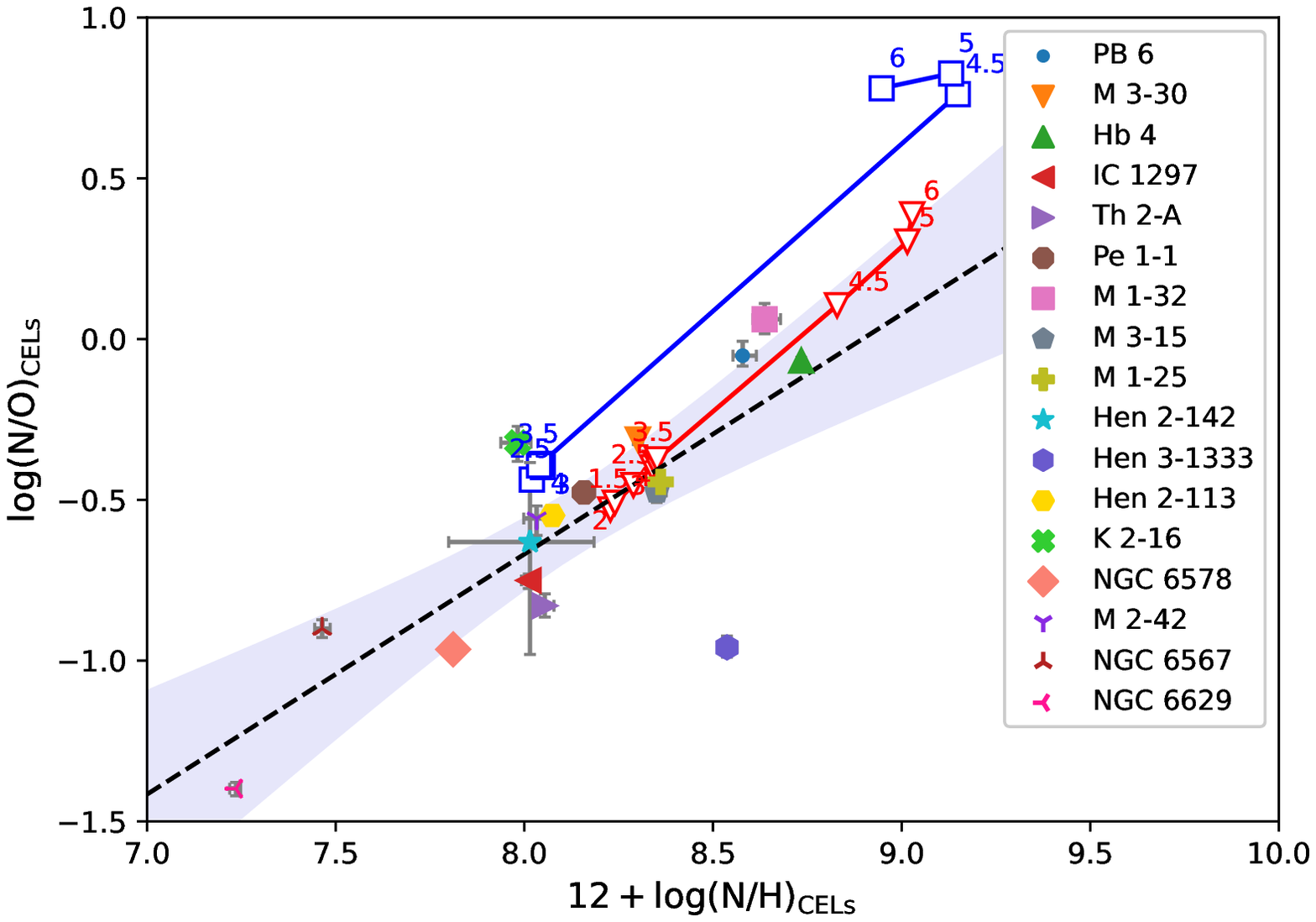}\\
\includegraphics[width=0.47\textwidth, trim = 0 0 15 5, clip, angle=0]{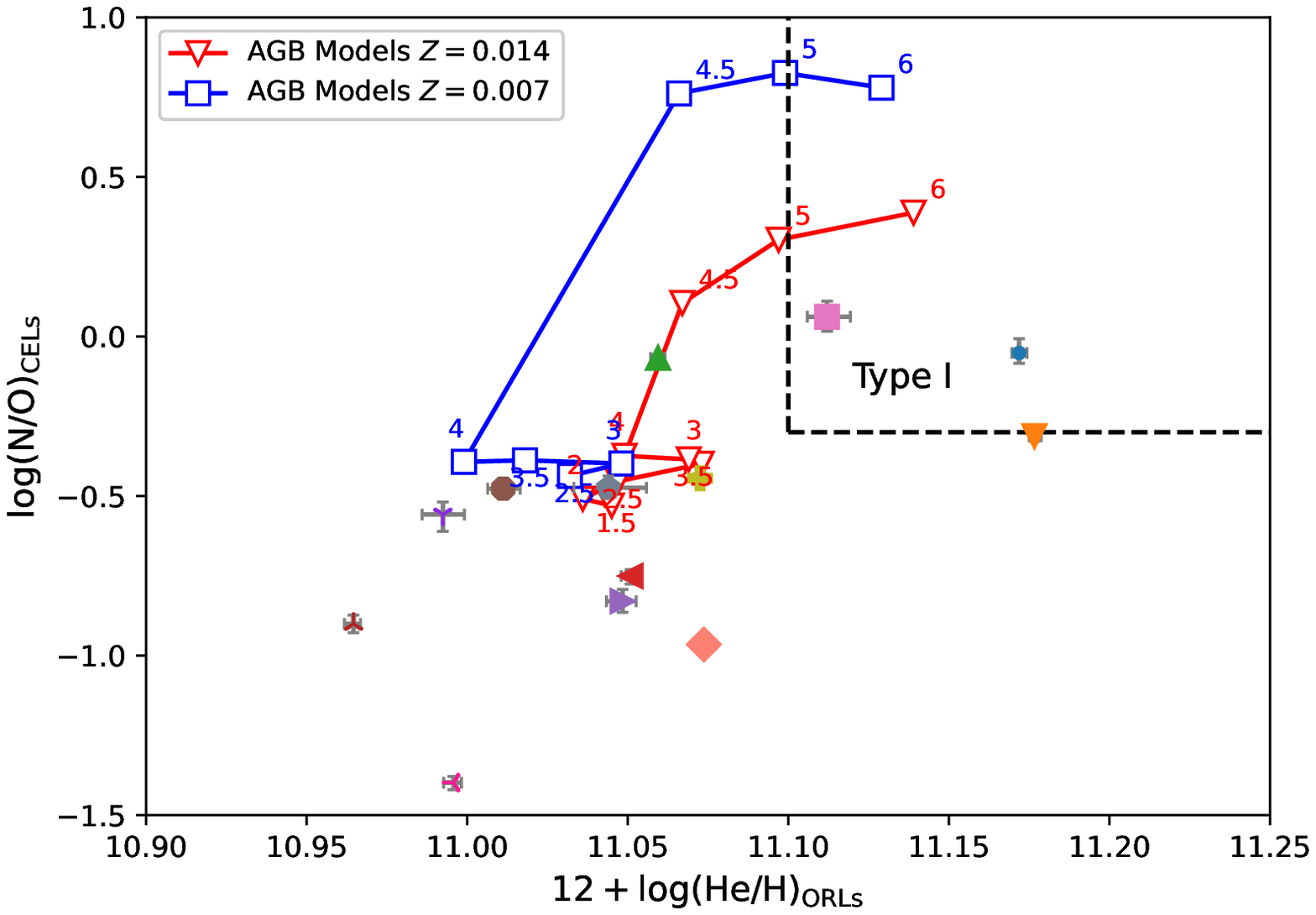}\\
\includegraphics[width=0.47\textwidth, trim = 0 0 15 5, clip, angle=0]{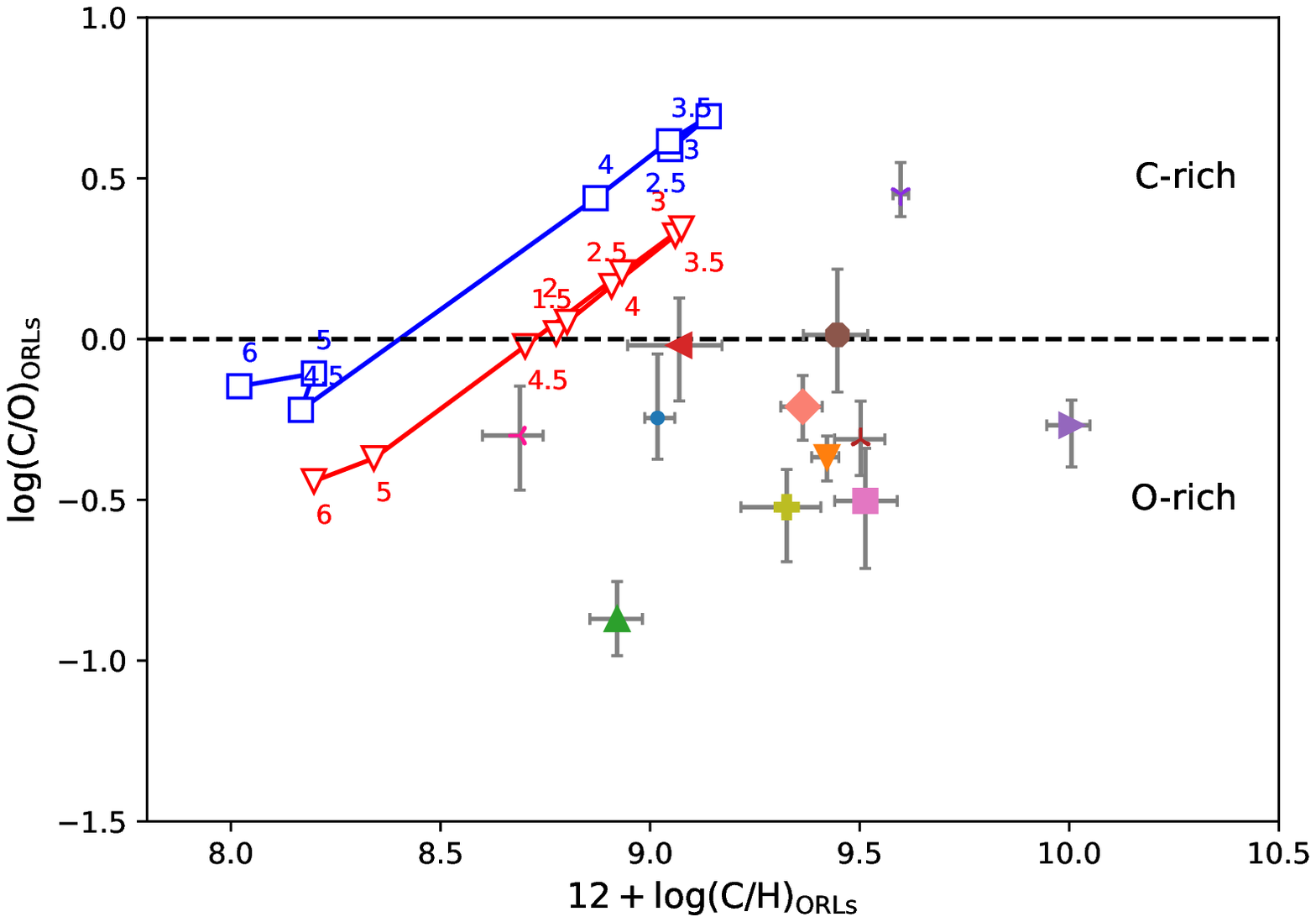}%
\caption{ 
\textit{Top Panel:}
$\log$(N/O)$_{\rm CELs}$ plotted against $12+\log$(N/H)$_{\rm CELs}$ derived from the PNe in our sample and
elemental yields predicted by the AGB models with initial masses 1.5--6${\rm M}_{\odot}$ at the solar metallicity ($Z=0.014$; red triangles) and 2.5--6${\rm M}_{\odot}$ at the sub-solar metallicity ($Z=0.007$; blue squares) from \citet{Karakas2016}.
The dashed line is a least-squares fit to our data, discussed in the text, with the 90\% confidence level shown by the shaded area.
\textit{Middle Panel:}
$\log$(N/O)$_{\rm CELs}$ versus $12+\log$(He/H)$_{\rm ORLs}$ for our sample PN and the AGB model yields with $Z=0.014$ and $0.007$.
The top-right dashed-line box shows the boundaries of Type I PNe as defined by \citet{Peimbert1983}.
\textit{Bottom Panel:}
$\log$(C/O)$_{\rm ORLs}$ versus $12+\log$(C/H)$_{\rm ORLs}$ for our PNe and the AGB model predictions ($Z=0.014$ and $0.007$).
The horizontal dashed line distinguishes between the PNe with C-rich and O-rich abundances derived from ORLs.
\label{wc:agb:model}%
}%
\end{center}
\end{figure}

\section{Comparison with AGB Models}
\label{wc:sec:agbmodels}

The elemental abundances of PNe are associated with the products of the nucleosynthesis and mixing processes that happened in previous evolutionary phases. The richest nucleosynthesis occurs during the AGB, where the third dredge up (TDU) mixes carbon and other helium burning products to the surface. Hot bottom burning (HBB) also occurs in intermediate-mass AGB stars with masses $\gtrsim 4.5 M_{\odot}$. Although we have a good qualitative picture of the evolution of low and intermediate-mass stars, the details of the mixing and nucleosynthesis during the AGB are uncertain \citep[e.g.][]{Busso1999,Herwig2005,Karakas2014}. Elemental abundances derived from PNe can be used as a tool to trace these uncertain mixing processes, and to gain insight into non-standard physics such as rotation \citep[e.g.][]{Charbonnel2010}.

For comparison in Figure~\ref{wc:agb:model}, we plot the logarithmic N/O abundance ratio from CELs against $12+\log$(N/H)$_{\rm CELs}$ (top) and  $12+\log$(He/H)$_{\rm ORLs}$ (middle), and the logarithmic C/O abundance ratio from ORLs against $12+\log$(C/H)$_{\rm ORLs}$ (bottom) for our PNe and the integrated yields for AGB models with progenitor masses between 1.5--6${\rm M}_{\odot}$ at solar metallicity $Z=0.014$ (red triangles) and 2.5--6${\rm M}_{\odot}$ at $Z=0.007$ (blue squares) from \citet{Karakas2016}. We consider the elemental abundances calculated using the conventional \textit{icf} formulas (KB94).  
The composition of the AGB model yields were weighted toward the tip of the AGB, which is when most of the mass is lost, so they are suitable for comparison to PNe. 
The initial composition of the $Z= 0.014$ models assumed the proto-solar composition from \citet{Asplund2009}.  
The initial composition in the $Z = 0.007$ models are the scaled-solar abundances from \citet{Anders1989}.  
The models with masses below 6${\rm M}_{\odot}$ were included, as more massive AGB models likely evolve too quickly to form PNe. Likewise, models below  1.5${\rm M}_{\odot}$ likely evolve too slowly during their post-AGB evolution \citep{Bloecker1995b}. Models with HBB, which produces Type I PNe \citep[N/O $> 0.8$;][]{Kingsburgh1994}, include masses $\ge$ 4.5${\rm M}_{\odot}$. 

We compare the predictions for the models with $Z = 0.014$ and $0.007$ to our elemental abundances (see KB94 in  Table~\ref{wc:abund:summary}). The oxygen abundance is used as a metallicity indicator.
According to the O/H elemental abundance derived from CELs, 9 objects have around solar metallicity (PB\,6, M\,3-30, IC\,1297, Pe\,1-1, M\,1-32, Hen\,2-142, Hen 2-113, M\,2-42, and NGC\,6629), 4 objects with slightly over solar metallicity (Hb\,4, M\,3-15, M\,1-25, and NGC\,6578), 
two PNe with super-solar metallicity (Th\,2-A, and Hen\,3-1333) and the remaining 3 PNe having half-solar to LMC metallicities
(K\,2-16, NGC 6567, and Sa 3-107). 

Comparing N/H, N/O and He/H with those derived from AGB models plotted in Figure~\ref{wc:agb:model}, 
M\,1-25, Hen\,2-142, Hen 2-113, Pe\,1-1, and M\,2-42
probably evolved from AGB stars with $Z=0.014$ and initial masses between 1 and 2\,M$_{\odot}$, 
whereas M\,3-30 and M\,3-15 from progenitor stars with $Z=0.014$ and initial mass between 2 and 3.5\,M$_{\odot}$.
Similarly, an AGB model with $Z=0.007$ and a progenitor mass of 2.5--4M$_{\odot}$ could likely produce the abundance patterns of K\,2-16. 
Fig.~\ref{wc:agb:model} (middle panel) shows the regions of Type I PNe (He/H\,$\geqslant$\,0.125 and $\log$\,N/O\,$\geqslant-0.3$) as classified by \citet{Peimbert1983}. The abundance patterns of PB\,6 and M\,1-32 (also possibly M\,3-30) suggest that they could be Type I PNe and likely evolved from AGB stars with initial masses of about 4--5M$_{\odot}$ in order to undergo the HBB phase \citep[see][]{Karakas2014a}. 
Moreover, the abundance ratio N/O $>0.5$ in Hb 4 could also correspond to HBB in an AGB star with a progenitor mass of $\sim 4.5$M$_{\odot}$
Similarly, the nebulae M\,3-15, M 1-25, and Pe\,1-1 with N/O $>0.5$ (DMS14) may be produced by AGB stars with $Z=0.014$ and initial masses of $\sim 4.5$\,M$_{\odot}$ if we consider the elemental abundances calculated by \textit{icf} formulas from \citet{Delgado-Inglada2014} (see DMS14 in Table~\ref{wc:abund:summary}). 
IC\,1297, Th\,2-A, and NGC\,6578 are likely associated with progenitor mass $\lesssim 1.5$\,M$_{\odot}$ and metallicity $Z \gtrsim 0.014$. Note that we do not have any calculated models for super-solar metallicity (Hen\,3-1333) and  very poor metallicity (Sa\,3-107).

Figure~\ref{wc:agb:model} (bottom panel) also shows the C/O abundance ratio as a function of the C/H abundance derived from ORLs.
We see that C/O predicted by ORLs have C/O\,$<1$, except for Pe\,1-1 and M\,2-42, implying that ORLs are mostly emitted from O-rich environments. 
However, the C/H abundances derived from ORLs in our PN sample are higher than the yields predicted by AGB stellar models, 
whereas their abundances from CELs are mostly consistent with AGB models with progenitor masses 1.5--5M$_{\odot}$.
From our results in \S\,\ref{wc:sec:adfs} and the C/H in Figure~\ref{wc:agb:model}, the ORLs could originate from cool, oxygen-rich clumpy structures 
having the chemical composition that cannot directly be associated with heavy elements produced by AGB stars. 

For our PN sample, a linear fit to the N/O abundance ratio  and the N/H abundance of the 17 PNe plotted in Figure~\ref{wc:agb:model} 
(top panel) yields 
$\log($N/O$) = (0.75 \pm 0.15) \times [12+\log($N/H$)]-(6.64\pm1.24) $
 with $r=0.79$ and $p=0.0002$, in agreement with $\log($N/O$) = 0.73 \times [12+\log($N/H$)]-6.50$ found by \citet{Garcia-Rojas2013}. It can be seen that the most of PNe with solar metallicity are close to this linear correlation (dashed in Figure~\ref{wc:agb:model}, top). 
However, NGC\,6567 and NGC\,6629 around \textit{wels} are not consistent with the AGB predictions, while 
they still follow the N/O vs. N/H correlation.   

AGB modeling parameters such as mass loss and convection are highly uncertain, and variations in either would alter the predicted elemental yields. For example, higher mass loss rates will lead to a shorter AGB lifetime and less TDU episodes, so a smaller C/O ratio is produced from a model with $\sim$3M$_{\odot}$ \citep{Marigo2002,Stancliffe2007,Karakas2010a}. More efficient convection would lead to hotter temperatures during HBB and a larger N/O ratio \citep[e.g.][]{Ventura2005,Ventura2013}. The neon abundance can be enhanced in lower mass AGB models (2.5 and 3$M_{\odot}$) via the partial mixing of protons and third dredge-up \citep{Karakas2003,Karakas2009}, and mildly (by $\sim0.3$ dex) through HBB and dredge-up in the metal-poor AGB star models ($Z=0.007$) with initial masses of 5 and 6$M_{\odot}$. Neutron captures mildly increase chlorine abundance in the metal-poor massive ($\gtrsim3M_{\odot}$) AGB models \citep{Karakas2009}. The sulfur abundance can be reduced via the depletion into dust \citep{Pottasch2006,Henry2012}. 

\section{Conclusions and Discussions}
\label{wc:sec:conclusion}

We have carried out detailed plasma diagnostics and abundance analyses for a sample of 18 Galactic PNe surrounding [WR]-type and \textit{wels} stars, using both CELs and ORLs. The electron density derived from CELs are closely
correlated with the intrinsic nebular H$\beta$ surface brightness, which agrees with the theoretical prediction 
of $S({\rm H}\beta) \propto \varepsilon\,r\,N_{\rm e}^{2}$ \citep{Odell1962}. The nebular H$\beta$ surface brightness is associated with the nebular evolution, and it drops as the nebula is expanding. Moreover, the electron temperature ratio $T_{\rm e}(\mbox{[O\,{\sc iii}]})$/$T_{\rm e}(\mbox{[N\,{\sc ii}]})$ derived from CELs is correlated with the excitation class (EC), which agrees with the previous results \citep{Kingsburgh1994,Wang2007}. Self-consistent plasma diagnostics of ORLs using atomic data of N${}^{2+}$ \citep{Fang2013a} and O${}^{2+}$ \citep{Storey2017} also suggest the presence of cool ($\lesssim$\,7000\,K), dense ($\sim$\,$10^4$--$10^5$\,cm$^{-3}$) small-scale structures in some PNe (see Table~\ref{wc:plasma:summary}).
However, we caution that some N\,{\sc ii} ORLs in low-excited PNe could be 
potentially contaminated by fluorescence \citep[see e.g.][]{Escalante2012}. 

Our abundance analyses indicate that there could be a dependence of the ORL/CEL ADFs of O$^{2+}$ upon the difference between temperatures derived from forbidden lines and those from He\,{\sc i} recombination lines (see Fig.\,\ref{wc:tedef_adf} top panels). 
Previously, the ADFs were found to be closely correlated with the difference between $T_{e}$([O\,{\sc iii}]) and $T_{\rm e}$(BJ) \citep{Liu2001,Liu2004b,Tsamis2004,Wesson2005,Wang2007}. 
We also found a weak correlation between the ADFs of O$^{2+}$ and N$^{2+}$ (Fig.\,\ref{wc:adf_n2:adf_o2}). 
The weak anti-correlations between the CEL--He\,{\sc i} temperature dichotomies and nebular surface brightness found here (see Fig.\,\ref{wc:tedef_adf} bottom panels) are also consistent with \citet{Tsamis2004}. However, we did not obtain any statistically significant correlations between the ADFs and surface brightness, which were found in the previous studies \citep{Liu2004b,Tsamis2004,Wang2007}. The correlations between ADF(O$^{2+}$) and CEL--He\,{\sc i} temperature dichotomies possibly imply that the observed O\,{\sc ii} ORLs could originate from cool ionized gases located in some oxygen-rich clumpy structures of high density within the diffuse warm nebula.

The correlations between ADF and temperature dichotomy has been predicted by the bi-abundance model proposed by \citet{Liu2000} that includes a fraction of cool metal-rich knots within the diffuse warm nebula. The feasibility of this model has been demonstrated using several photoionization models \citep{Ercolano2003b,Tsamis2005,Yuan2011,Danehkar2018a,Gomez-Llanos2020}. Nevertheless, the presence and origin of metal-rich inclusion are not well understood. The born-again scenario \citep{Iben1983,Iben1983b} is one possibility, whereby H-deficient material would have been ejected from the stellar surface during the (very-) late thermal pulse, such as hydrogen-poor ejecta in Abell\,30, Abell\,58 (V605 Aql), and Abell\,78 \citep[e.g.][]{Hazard1980,Jacoby1983,Pollacco1992,Borkowski1993,Guerrero1996,Fang2014}.
The detailed abundance analyses of the `born-again' PNe Abell\,30 \citep{Wesson2003} and Abell\,58 \citep[][]{Wesson2008}  
supported the idea that they contain some very cold hydrogen-deficient knots. 
In the both cases, the knots were found to be oxygen-rich and also neon-rich in contrast to the expectations of a very late thermal pulse. However, the infrared spectroscopic and imaging studies of the born-again PNe Abell\,30 and Abell\,78 by \citet{Toala2021} indicate that the C/O mass ratios of their hydrogen-poor ejecta are larger than 1 in agreement with the very late thermal pulse model, 
and they also contain hot carbon-rich dust grains spatially coincident with the hydrogen-poor ejecta.
Alternatively, \citet{Liu2003} suggested that H-deficient material could be introduced by the evaporation and destruction of planets by stars. However, \citet{Henney2010} argued that the destruction of solid bodies during the PN phase cannot produce enough gas-phase metallicity to explain the observed abundance discrepancies in PNe, though the sublimation of solid bodies during the final stages of the AGB phase might generate enough metal-rich material. The implications of these scenarios need more observations and detailed abundance analysis of a large sample of PNe. 

\citet{Nicholls2012,Nicholls2013} also proposed that a $\kappa$-distribution of electron energies could explain the abundance discrepancy and temperature dichotomy. In this scenario, the electrons in the gas have a non-thermal equilibrium energy distribution whose departure from the Maxwell-Boltzmann distribution is characterized by a $\kappa$ index. \citet{Dopita2013} concluded that $\kappa$-distributions with $\kappa\sim20$, or somewhat larger, are able to predict the abundance discrepancy and temperature dichotomy in H\,{\sc ii} Regions. However, \citet{Storey2013} found that Maxwell-Boltzmann distributions fit their data best but without statistically ruling out $\kappa$-distributions. Moreover, \citet{Mendoza2014} found that in some cases $\kappa$-distributed electrons can reconcile observations with theory but that this is in a minority of cases, and Maxwell-Boltzmann distributions are not firmly ruled out even then. Furthermore, \citet{Zhang2014} found that both the scenarios, bi-abundance models and $\kappa$-distributed electrons, are adequately consistent with observations of four PNe with very large ADFs, and concluded the spectra are emitted from cold and low-$\kappa$ plasmas rather than a single Maxwell-Boltzmann electron energy distribution. Nevertheless, \citet{Zhang2016} did not find any conclusive results for supporting $\kappa$-distributed electrons in PNs. Recently, $\kappa$-distributed recombination coefficients for hydrogen \citep{Storey2015a}, and  collision strengths for \foiii\ \citep{Storey2015b} became available. Upcoming $\kappa$-distributed recombination coefficients for \oii\ ORLs shall allow us to assess whether  $\kappa$-distributed electrons have a key role in the ADF of O${}^{2+}$ ion. It is unclear whether the metal-rich inclusion could introduce non-Maxwell-Boltzmann equilibrium electrons to the nebula. The relations between both the scenarios should be evaluated further. 

Moreover, abundance discrepancy factors could be related to binarity \citep[e.g.][]{Lau2011,Corradi2015,Bautista2018,Wesson2018}. 
\citet{Lau2011} envisaged that H-deficient material in V605 Aql might be ejected by either merger of a main-sequence
star and a massive white dwarf, or a classical nova. Moreover, \citet{Bautista2018} proposed that resonant temperature fluctuations produced in the nebula photoionized by short-period binary stars are responsible for temperature and abundance discrepancies. 
Deep spectroscopic studies of 3 PNe with post common-envelope binary stars showed very large abundance discrepancies:  
ADFs\,$\sim$\,120 in Abell\,46 and $\gtrsim300$ in its inner part, ADF\,$\sim$\,50 in Ou\,5, and ADF\,$\sim$\,10 in Abell\,63 \citep{Corradi2015}.
A statistical analysis of PNe by \citet{Wesson2018} also demonstrated that extreme abundance discrepancies could be linked to binary central stars and possible implications of a nova-like eruption shortly after the common-envelope phase.
\citet{Jones2016} obtained ADF(O$^{2+}$)\,$\approx18$ in the PN NGC 6778 around a short-period binary.
Moreover, the direct imaging of NGC 6778 by \citet{Garcia-Rojas2016} also revealed that 
weak O\,{\sc ii} $\lambda\lambda$4649+50 recombination emission originates from the central region of this PN where it is not spatially associated with
strong [O\,{\sc iii}] $\lambda$5007 emission. Recently, \citet{Jacoby2020} suggested the presence of 
a binary central star in the born-again PN Abell\,30 based on its light curves.

In conclusion, our detailed analyses of PNe around [WR]-type and \textit{wels} stars have allowed us to 
determine physical conditions and chemical abundances from forbidden and recombination lines. The correlations between the ADF(O${}^{2+}$) and CEL--He\,{\sc i} temperature dichotomy, as well as the physical properties derived from plasma diagnostics of heavy element ORLs, likely suggest that some dense metal-rich clumpy structures may be present in ionized gaseous nebulae. The nebular evolution does not seem to play a central role in the abundance discrepancy problem. Therefore, the ORL/CEL problem could be due to a hitherto unknown component or mechanism within PNe or/and complicated evolutionary history of CSPNe. 

\begin{acknowledgments}

This article is partially based on the dissertation by the author \citep[][]{Danehkar2014b}, supported by a Macquarie University Research Excellence Scholarship (MQRES) and a Sigma Xi Grants-in-Aid of Research (GIAR). 
The author would like to thank the referee for helpful comments and suggestions that greatly improved the paper, 
Quentin~Parker for supporting the 2010 ANU observations, 
Amanda Karakas for valuable comments on AGB models, 
Roger~Wesson for helpful discussions, David~Frew for helping with the observing proposal, the staff at the Siding Spring Observatory and Kyle~DePew for undertaking the WiFeS observations in 2010, Bruce~Balick for sharing the long-slit data of Hb 4 taken with the Palomar 5.1-m telescope in July 1993. 

\end{acknowledgments}



\software{IDL Astronomy User's Library \citep{Landsman1993}, IDL Coyote Library \citep{Fanning2011}, NumPy \citep{Harris2020}, SciPy \citep{Virtanen2020}, Matplotlib \citep{Hunter2007}.}
   
\facility{ATT (WiFeS).}   
      

\begin{appendix}

\section{Ionization Correction Factors}
\label{wc:sec:icf:2}


The total elemental He abundance relative to H is often obtained by simply taking the sum of He$^{+}$/H$^{+}$ and He$^{2+}$/H$^{+}$ ionic abundances.

Following \citet{Kingsburgh1994} and \citet{Wang2007}, the ORL carbon is derived, correcting for the unseen stages of ionization for the case if only C$^{2+}$ is measured using:
\begin{equation}
\left( \frac{{\rm C}}{{\rm H}} \right)_{\rm ORLs} =\left(\frac{{\rm C}^{2+}}{{\rm H}^{+}} \right) \left(\frac{{\rm O} }{{\rm O}^{2+}}\right)_{\rm CELs},
\label{wc:icf:c_orl}
\end{equation}
and if C$^{2+}$ and C$^{3+}$ are observed:
\begin{equation}
\left( \frac{{\rm C}}{{\rm H}} \right)_{\rm ORLs}=\left(\frac{{\rm C}^{2+}}{{\rm H}^{+}} + \frac{{\rm C}^{3+}}{{\rm H}^{+}}  \right)  \left(\frac{{\rm O}^{+} + {\rm O}^{2+} }{{\rm O}^{2+}}\right)_{\rm CELs}.
\label{wc:icf:c_orl_2}
\end{equation}
However, carbon ionic ratios derived from ORLs are generally not equal to those derived from CELs \citep{Tsamis2003a,Tsamis2004}, and oxygen ionic abundances derived from CELs may not be suitable choices for the ionization correction factor of the ORL carbon. 

For ORL nitrogen, when only N$^{2+}$ was measurable, the unobserved ionization stages are corrected for by assuming ${\rm N}/{\rm N}^{2+} = {\rm O}/{\rm O}^{2+}$, so
\begin{equation}
\left( \frac{{\rm N}}{{\rm H}} \right)_{\rm ORLs} =\left(\frac{{\rm N}^{2+}}{{\rm H}^{+}} \right) \left(\frac{{\rm O} }{{\rm O}^{2+}}\right)_{\rm CELs},
\label{wc:icf:n_orl1}
\end{equation}
while when both N$^{2+}$ and N$^{3+}$ were measured, the elemental abundance is derived assuming ${\rm N}/{\rm N}^{+} = {\rm O}/{\rm O}^{+}$, with N/H then given by
\begin{equation}
\left( \frac{{\rm N}}{{\rm H}} \right)_{\rm ORLs} =\left(\frac{{\rm N}^{2+}}{{\rm H}^{+}} + \frac{{\rm N}^{3+}}{{\rm H}^{+}}  \right) \bigg[ 1 -  \left(\frac{{\rm O}^{+} }{{\rm O}}\right)_{\rm CELs}\bigg]^{-1}.
\label{wc:icf:n_orl_2}
\end{equation}

The ionization correction factor for O is $({\rm He}/{\rm He}^{+})^{2/3}$ \citep{Kingsburgh1994}. However, only O$^{2+}$ is measured from ORLs. Following \citet{Wesson2005} and \citet{Wang2007}, we assume that O$^{+}$/O$^{2+}$ derived from CELs is applicable to ORLs, so instead the elemental abundance is derived using
\begin{equation}
\left( \frac{{\rm O}}{{\rm H}} \right)_{\rm ORLs} =\left(\frac{{\rm O}^{2+}}{{\rm H}^{+}}   \right) \left(\frac{{\rm He} }{{\rm He}^{+}}\right)^{2/3} \bigg[ 1 + \left(\frac{{\rm O}^{+} }{{\rm O}^{2+}}\right)_{\rm CELs}\bigg].
\label{wc:icf:o_orl}
\end{equation}

The CEL oxygen abundance is calculated from the O$^{+}$/H$^{+}$ and O$^{2+}$/H$^{+}$ ratios, correcting for the unseen O$^{3+}$/H$^{+}$ using,
\begin{equation}
\left( \frac{{\rm O}}{{\rm H}} \right)_{\rm CELs} =\left(\frac{{\rm O}^{+}}{{\rm H}^{+}} + \frac{{\rm O}^{2+}}{{\rm H}^{+}} \right)
\left(\frac{{\rm He} }{{\rm He}^{+}}\right)^{2/3}_{\rm ORLs},
\label{wc:icf:o_cel}
\end{equation}
where the He$^{+}$ and He abundances are derived from ORLs.

The CEL nitrogen is calculated from the N$^{+}$/H$^{+}$ ratio, correcting for the unseen N$^{2+}$/H$^{+}$ and N$^{3+}$/H$^{+}$ using,
\begin{equation}
\left( \frac{{\rm N}}{{\rm H}} \right)_{\rm CELs} =\left(\frac{{\rm N}^{+}}{{\rm H}^{+}}\right) \left(\frac{{\rm O}}{{\rm O}^{+}}\right).
\label{wc:icf:n_cel}
\end{equation}
To estimate the CEL/ORL neon, the unseen Ne$^{+}$/H$^{+}$ is corrected for, using
\begin{equation}
 \frac{{\rm Ne}}{{\rm H}}  =\left(\frac{{\rm Ne}^{2+}}{{\rm H}^{+}} \right)
\left(\frac{{\rm O}}{{\rm O}^{2+}}\right).
\end{equation}

For the case where both S$^{+}$  and S$^{2+}$ observed, the elemental S abundance is derived using
\begin{equation}
\left( \frac{{\rm S}}{{\rm H}} \right)_{\rm CELs} =  \left(\frac{{\rm S}^{+}}{{\rm H}^{+}} + \frac{{\rm S}^{2+}}{{\rm H}^{+}} \right) \left[1-\left(1-\frac{{\rm O}^{+}}{{\rm O}}\right)^{3}\right]^{-1/3}.
\label{wc:icf:s_cel}
\end{equation}
For the case where only S$^{+}$ observed, the elemental S abundance is derived as follows \citep{Kingsburgh1994}:
\begin{align}
\left( \frac{{\rm S}}{{\rm H}} \right)_{\rm CELs} = & \left(\frac{{\rm S}^{+}}{{\rm H}^{+}} \right)
\left[1+4.677\left(\frac{{\rm O}^{2+}}{{\rm O}^{+}}\right)^{0.433}\right] \notag \\
& \times \left[1-\left(1-\frac{{\rm O}^{+}}{{\rm O}}\right)^{3}\right]^{-1/3}.
\label{wc:icf:s_cel_2}
\end{align}
For the case where both Ar$^{2+}$ and Ar$^{3+}$ observed, the elemental Ar abundance is derived using the following equation, assuming Ar$^{+}$/Ar~=~N$^{+}$/N:
\begin{equation}
\left( \frac{{\rm Ar}}{{\rm H}} \right)_{\rm CELs} =\left(\frac{{\rm Ar}^{2+}}{{\rm H}^{+}} +\frac{{\rm Ar}^{3+}}{{\rm H}^{+}} + \frac{{\rm Ar}^{4+}}{{\rm H}^{+}} \right)
\left(1-\frac{{\rm N}^{+}}{{\rm N}}\right)^{-1}.
\label{wc:icf:ar_cel}
\end{equation}
For the case where only Ar$^{2+}$ observed, the elemental Ar abundance is derived as follows:
\begin{equation}
\left( \frac{{\rm Ar}}{{\rm H}} \right)_{\rm CELs} =1.87 \left(\frac{{\rm Ar}^{2+}}{{\rm H}^{+}} \right).
\label{wc:icf:ar_cel_2}
\end{equation}
For elemental Cl abundance, we use the following equation given by \citet{Liu2000} according to the ionization potential of Cl and S ion stages:
\begin{equation}
\left( \frac{{\rm Cl}}{{\rm H}} \right)_{\rm CELs} =\left(\frac{{\rm Cl}^{2+}}{{\rm H}^{+}} \right) 
\left(\frac{{\rm S}}{{\rm S}^{2+}}\right).
\label{wc:icf:cl_cel}
\end{equation}
Following \citet{Izotov1994}, the elemental Fe abundance is estimated when only Fe$^{2+}$ is observed as follows:
\begin{equation}
\left( \frac{{\rm Fe}}{{\rm H}} \right)_{\rm CELs} =\left(\frac{{\rm Fe}^{2+}}{{\rm H}^{+}} \right) \left[ 1.25 \left( \frac{{\rm O}}{{\rm O}^{+}} \right) \right],
\label{wc:icf:fe_cel_2}
\end{equation}
which is based on photoionization H\,{\sc ii} models \citep{Stasinska1990a}.

\section{Supplementary Data}
\label{appendix:a}

The following tables and figure sets are available for the electronic edition of this article:
\\
\textbf{Table~\ref{wc:tab:nebula:observations}.} Observed and dereddened line fluxes on a scale relative to H$\beta$, where H$\beta=100$. 
\\
\textbf{Table~\ref{wc:tab:diagnostic:cels}.} Plasma diagnostics based on CELs.
\\
\textbf{Table~\ref{wc:tab:diagnostic:hei}.} Plasma diagnostics based on \ionic{He}{i} lines.
\\
\textbf{Table~\ref{wc:tab:diagnostic:orls}.} Plasma diagnostics based on heavy element ORLs.
\\
\textbf{Table~\ref{wc:tab:abundances:cels}.} Ionic abundances derived from CELs.
\\
\textbf{Table~\ref{wc:tab:abundances:orls}.} Ionic abundances derived from ORLs.
\\
\textbf{Table~\ref{wc:tab:abundances:total}.} Mean ionic and total elemental abundances derived from ORLs and CELs.  
\\
\textbf{Fig. Set~\ref{wc:fig:dd:cels}.} $N_{\rm e}$--$T_{\rm e}$ diagnostic diagrams based on CELs.
\\
\textbf{Fig. Set~\ref{wc:fig:dd:orls}.} $T_{\rm e}$ diagnostic diagrams based on C\,{\sc ii} ORLs, and $N_{\rm e}$--$T_{\rm e}$ diagnostic diagrams based on O\,{\sc ii} and N\,{\sc ii} ORLs.

\smallskip

\noindent Tables~\ref{wc:tab:nebula:observations}, \ref{wc:tab:diagnostic:cels}--\ref{wc:tab:diagnostic:orls}, and \ref{wc:tab:abundances:cels}--\ref{wc:tab:abundances:total} are published in the machine-readable format. 

\end{appendix}

{ \small 
\begin{center}
\textbf{ORCID iDs}
\end{center}
\vspace{-5pt}

\noindent A.~Danehkar \orcidauthor{0000-0003-4552-5997} \url{https://orcid.org/0000-0003-4552-5997}

}


\newpage

\onecolumngrid

\begin{appendix}

\onecolumngrid

\setcounter{section}{1}

\vspace{20pt}

\section{Supplementary Data} 

\bigskip

\noindent The following tables and figure sets are available for the electronic edition of this article:
\medskip\newline
\textbf{Table~\ref{wc:tab:nebula:observations:2}.} Observed and dereddened line fluxes on a scale relative to H$\beta$, where H$\beta=100$. 
\medskip\newline
\textbf{Table~\ref{wc:tab:diagnostic:cels:2}.} Plasma diagnostics based on CELs.
\medskip\newline
\textbf{Table~\ref{wc:tab:diagnostic:hei:2}.} Plasma diagnostics based on \ionic{He}{i} lines.
\medskip\newline
\textbf{Table~\ref{wc:tab:diagnostic:orls:2}.} Plasma diagnostics based on heavy element ORLs.
\medskip\newline
\textbf{Table~\ref{wc:tab:abundances:cels:2}.}  Ionic abundances derived from CELs.
\medskip\newline
\textbf{Table~\ref{wc:tab:abundances:orls:2}.} Ionic abundances derived from ORLs.
\medskip\newline
\textbf{Table~\ref{wc:tab:abundances:total:2}.} Mean ionic and total elemental abundances derived from ORLs and CELs.
\medskip\newline
\textbf{Fig. Set~\ref{wc:fig:dd:cels:2}.} $N_{\rm e}$--$T_{\rm e}$ diagnostic diagrams based on CELs.
\medskip\newline
\textbf{Fig. Set~\ref{wc:fig:dd:orls:2}.} $T_{\rm e}$ diagnostic diagrams based on C\,{\sc ii} ORLs, and $N_{\rm e}$--$T_{\rm e}$ diagnostic diagrams based on O\,{\sc ii} and N\,{\sc ii} ORLs.
\medskip\newline
Tables~\ref{wc:tab:nebula:observations:2}, \ref{wc:tab:diagnostic:cels:2}--\ref{wc:tab:diagnostic:orls:2}, and \ref{wc:tab:abundances:cels:2}--\ref{wc:tab:abundances:total:2} are published in the machine-readable format.  

\end{appendix}



\newpage

\onecolumngrid

\renewcommand{\arraystretch}{0.68}


\newpage
\normalsize

\setcounter{table}{3}
\begin{center}
\begin{longtable}{lllccccccccc}
\caption{Observed and dereddened line fluxes on a scale relative to H$\beta$, where H$\beta=100$. Observed fluxes are denoted by $F(\lambda)$ and dereddened fluxes by $I(\lambda)$.
  The symbol `*' in the observed and dereddened fluxes indicates that the listed line is blended with the above listed line.
\label{wc:tab:nebula:observations:2}
}\\
\hline\hline
\noalign{\smallskip}
$\lambda_{\rm lab}$  & Ion  & $\lambda_{\rm obs}$ & $F(\lambda)$ & $\varepsilon_{F(\lambda)}$(\%) & $I(\lambda)$ & $\varepsilon_{I(\lambda)}$(\%) & Mult & Lower term & Upper term & g1 & g2 \\
\noalign{\smallskip}
\hline
\noalign{\smallskip}
\multicolumn{12}{c}{PB 6          (PNG278.8$+$04.9) }\\
\noalign{\smallskip}
\hline
\noalign{\smallskip}
   3726.03 & [O\,{\sc ii}]$^{\mathrm{a}}$   &    3726.03 &     50.700 & $\pm  0.0$ &     69.898 & $\pm  2.6$ & F1       & 2p3 4S*          & 2p3 2D*          & 4        & 4        \\
\noalign{\smallskip}
   3728.82 & [O\,{\sc ii}]$^{\mathrm{a}}$   &    3728.82 &          * &       &          * &       & F1       & 2p3 4S*          & 2p3 2D*          & 4        & 6        \\
\noalign{\smallskip}
   3868.75 & [Ne\,{\sc iii}]$^{\mathrm{a}}$ &    3868.75 &     84.000 & $\pm  0.0$ &    112.065 & $\pm  2.3$ & F1       & 2p4 3P           & 2p4 1D           & 5        & 5        \\
\noalign{\smallskip}
   3889.05 & H\,{\sc 8}$^{\mathrm{a}}$      &    3889.05 &     11.500 & $\pm  0.0$ &     15.266 & $_{ -1.9}^{+  2.3}$ & H8       & 2p+ 2P*          & 8d+ 2D           & 8        & *        \\
\noalign{\smallskip}
   3967.46 & [Ne\,{\sc iii}]$^{\mathrm{a}}$ &    3967.46 &     30.800 & $\pm  0.0$ &     40.079 & $_{ -1.8}^{+  2.1}$ & F1       & 2p4 3P           & 2p4 1D           & 3        & 5        \\
\hline
\end{longtable}

\small
\begin{tablenotes}
\item[1]\textbf{$^{\mathrm{a}}$} Fluxes adopted from the literature: 
PB 6 (K91), 
M 3-30 (P01,G07),
Hb 4 shell (P01,G07), 
Hb 4 N-knot (H97), 
Hb 4 S-knot (H97), 
IC 1297 (M02), 
Th 2-A (M02), 
Pe 1-1 (G12), 
M 1-32 (G12), 
M 3-15 (P01), 
M 1-25 (G07), 
Hen 2-142 (G07), 
Hen 3-1333 (D97), 
Hen 2-113 (D97), 
K 2-16 (P01), 
NGC 6578 (K03), 
M 2-42 (W07), 
NGC 6567 (K03), 
NGC 6629 (M02).  
References for line fluxes from the literature are as follows: D97, De Marco et al. (1997); G07, Girard et al. (2007); G12, Garc{\'{\i}}a-Rojas et al. (2012); H97, Hajian et al. (1997); K91, Kaler et al. (1991); K03, Kwitter et al. (2003); M02, Milingo et al. (2002); P01, Pe{\~n}a et al. (2001) and Pe{\~n}a et al. (1998); W07, Wang \& Liu (2007).
\end{tablenotes}
\end{center}

\newpage
\FloatBarrier
\normalsize

\setcounter{table}{5}
\begin{center}
\begin{longtable}{lllc}
\caption{Plasma diagnostics based on CELs.
\label{wc:tab:diagnostic:cels:2}
}\\
\hline\hline
\noalign{\smallskip}
Type& Ion  & Diagnostic & Value \\
\noalign{\smallskip}
\hline
\noalign{\smallskip}
\multicolumn{4}{c}{PB 6          (PNG278.8$+$04.9) }\\
\noalign{\smallskip}
\hline
\noalign{\smallskip}
$T_{\rm e}$                &[N~{\sc ii}]     &6548.10+6583.50     /5754.60              & $     11300_{      -260}^{+       200}$ \\
\noalign{\smallskip}
$T_{\rm e}$\,$_{\rm rc}$   &[N~{\sc ii}]     &6548.10+6583.50     /5754.60              & $     10270_{      -220}^{+       220}$ \\
\noalign{\smallskip}
$T_{\rm e}$                &[O~{\sc iii}]    &4958.91+5006.84     /4363.21              & $     14220_{       -90}^{+       110}$ \\
\noalign{\smallskip}
$T_{\rm e}$\,$_{\rm rc}$   &[O~{\sc iii}]    &4958.91+5006.84     /4363.21              & $     14040_{      -130}^{+        90}$ \\
\noalign{\smallskip}
$N_{\rm e}$                &[S~{\sc ii}]     &6730.82             /6716.44              & $      1800_{      -320}^{+       400}$ \\
\hline
\end{longtable}

\small
\begin{tablenotes}
\item[1]\textbf{Note:}  The label ``rc'' indicates that the auroral lines, $[$N\,{\sc ii}$]$ $\lambda5755$ and $[$O\,{\sc iii}$]$ $\lambda4363$, are corrected for recombination contribution.
\end{tablenotes}
\end{center}

\newpage
\FloatBarrier
\normalsize

\setcounter{table}{6}
\begin{center}
\begin{longtable}{lllc}
\caption{Plasma diagnostics based on \ionic{H}{i} lines.
\label{wc:tab:diagnostic:hei:2}
}\\
\hline\hline
\noalign{\smallskip}
Type& Ion  & Diagnostic & Value \\
\noalign{\smallskip}
\hline
\noalign{\smallskip}
\multicolumn{4}{c}{PB 6          (PNG278.8$+$04.9) }\\
\noalign{\smallskip}
\hline
\noalign{\smallskip}
$T_{\rm e}$                &He~{\sc i}       &5875.66             /4471.50              & $      4160_{      -620}^{+       540}$ \\
\noalign{\smallskip}
$T_{\rm e}$                &He~{\sc i}       &6678.16             /4471.50              & $      6990_{     -1190}^{+      1690}$ \\
\noalign{\smallskip}
$T_{\rm e}$                &He~{\sc i}       &Mean                                     & $      5580_{     -1020}^{+      1110}$ \\
\noalign{\smallskip}
\hline
\noalign{\smallskip}
\multicolumn{4}{c}{M 3-30        (PNG017.9$-$04.8) }\\
\noalign{\smallskip}
\hline
\noalign{\smallskip}
$T_{\rm e}$                &He~{\sc i}       &5875.66             /4471.50              & $      2310_{      -330}^{+       340}$ \\
\hline
\end{longtable}

\end{center}

\newpage
\FloatBarrier
\normalsize


\setcounter{table}{7}
\begin{center}
\begin{longtable}{lllc}
\caption{Plasma diagnostics based on heavy element ORLs.
\label{wc:tab:diagnostic:orls:2}
}\\
\hline\hline
\noalign{\smallskip}
Type& Ion  & Diagnostic & Value \\
\noalign{\smallskip}
\hline
\noalign{\smallskip}
\multicolumn{4}{c}{PB 6          (PNG278.8$+$04.9) }\\
\noalign{\smallskip}
\hline
\noalign{\smallskip}
$T_{\rm e}$                &N~{\sc ii}       &4613.87,4788.13,5676.02,5931.78                                                       & $      6300_{     -3100}^{+      9700}$ \\
\noalign{\smallskip}
$N_{\rm e}$                &N~{\sc ii}       &4613.87,4788.13,5676.02,5931.78                                                       & $     80600_{    -79900}^{+         0}$ \\
\noalign{\smallskip}
\hline
\noalign{\smallskip}
\multicolumn{4}{c}{M 3-30        (PNG017.9$-$04.8) }\\
\noalign{\smallskip}
\hline
\noalign{\smallskip}
$T_{\rm e}$                &N~{\sc ii}       &4601.48,5927.81                                                                       & $      6700_{     -1300}^{+      3100}$ \\
\noalign{\smallskip}
$N_{\rm e}$                &N~{\sc ii}       &4601.48,5927.81                                                                       & $     80600_{    -53200}^{+         0}$ \\
\hline
\end{longtable}

\begin{tablenotes}
\item[1]\textbf{Notes:} The electron temperatures and densities are determined using least squares minimization
on logarithmic $T_{\rm e}$--$N_{\rm e}$ grids. 
\end{tablenotes}
\end{center}

\newpage
\FloatBarrier
\normalsize

\setcounter{table}{10}
\begin{center}
\begin{longtable}{llcc}
\caption{Ionic abundances derived from CELs.
\label{wc:tab:abundances:cels:2}
}\\
\hline\hline
\noalign{\smallskip}
Ion  & Line & Weight & Abund. \\
\noalign{\smallskip}
\hline
\noalign{\smallskip}
\multicolumn{4}{c}{PB 6          (PNG278.8$+$04.9) }\\
\noalign{\smallskip}
\hline
\noalign{\smallskip}
N$^{+}$    &[N~{\sc ii}]     $\lambda$6548.10  &    1 & $ 4.225_{-0.164}^{+ 0.142} \times 10^{  -5}$ \\
\noalign{\smallskip}
N$^{+}$    &[N~{\sc ii}]     $\lambda$6583.50  &    3 & $ 4.426_{-0.188}^{+ 0.185} \times 10^{  -5}$ \\
\noalign{\smallskip}
O$^{0}$    &[O~{\sc i}]      $\lambda$6300.34  &    3 & $ 6.456_{-0.240}^{+ 0.203} \times 10^{  -6}$ \\
\noalign{\smallskip}
O$^{0}$    &[O~{\sc i}]      $\lambda$6363.78  &    1 & $ 6.947_{-0.273}^{+ 0.255} \times 10^{  -6}$ \\
\noalign{\smallskip}
O$^{+}$    &[O~{\sc ii}]     $\lambda$3726.03  &    1 & $ 4.928_{-0.142}^{+ 0.164} \times 10^{  -5}$ \\
\hline
\end{longtable}

\end{center}
\newpage
\FloatBarrier
\normalsize

\setcounter{table}{11}
\begin{center}
\begin{longtable}{llcc}
\caption{Ionic abundances derived from ORLs.
\label{wc:tab:abundances:orls:2}
}\\
\hline\hline
\noalign{\smallskip}
Ion  & Line & Weight & Abund. \\
\noalign{\smallskip}
\hline
\noalign{\smallskip}
\multicolumn{4}{c}{PB 6          (PNG278.8$+$04.9) }\\
\noalign{\smallskip}
\hline
\noalign{\smallskip}
He$^{+}$   &He~{\sc i}       $\lambda$4471.50  &    1 & $ 4.508_{-0.078}^{+ 0.065} \times 10^{  -2}$ \\
\noalign{\smallskip}
He$^{+}$   &He~{\sc i}       $\lambda$5875.66  &    3 & $ 4.670_{-0.064}^{+ 0.057} \times 10^{  -2}$ \\
\noalign{\smallskip}
He$^{+}$   &He~{\sc i}       $\lambda$6678.16  &    1 & $ 4.338_{-0.123}^{+ 0.096} \times 10^{  -2}$ \\
\noalign{\smallskip}
He$^{2+}$  &He~{\sc ii}      $\lambda$4685.68  &    1 & $ 1.028_{-0.004}^{+ 0.003} \times 10^{  -1}$ \\
\noalign{\smallskip}
C$^{2+}$   &C~{\sc ii}       $\lambda$6461.95  &    1 & $ 3.546_{-0.197}^{+ 0.221} \times 10^{  -4}$ \\
\hline
\end{longtable}

\end{center}

\newpage
\FloatBarrier
\normalsize

\setcounter{table}{12}
\begin{center}
\begin{longtable}{llcc}
\caption{Mean ionic and total elemental abundances relative to hydrogen derived from ORLs and CELs.
\label{wc:tab:abundances:total:2}
}\\
\hline\hline
\noalign{\smallskip}
Ion  & Type & Ref. & Abund. \\
\noalign{\smallskip}
\hline
\noalign{\smallskip}
\multicolumn{4}{c}{PB 6          (PNG278.8$+$04.9) }\\
\noalign{\smallskip}
\hline
\noalign{\smallskip}
He$^{+}$/H       & ORL     &          & $       4.571_{      -0.057}^{+       0.051} \times 10^{  -2}$ \\
\noalign{\smallskip}
He$^{2+}$/H      & ORL     &          & $       1.028_{      -0.004}^{+       0.003} \times 10^{  -1}$ \\
\noalign{\smallskip}
He/H             & ORL     &          & $       1.486_{      -0.008}^{+       0.008} \times 10^{  -1}$ \\
\noalign{\smallskip}
\noalign{\medskip}
C$^{2+}$/H       & ORL     &          & $       3.546_{      -0.197}^{+       0.221} \times 10^{  -4}$ \\
\noalign{\smallskip}
{\it icf}(C)     & ORL     & WL07     & $       2.938_{      -0.081}^{+       0.103}$ \\
\hline
\end{longtable}

\small
\begin{tablenotes}
\item[1]\textbf{Notes:} References for \textit{icf} formulas are as follows: 
Delgado-Inglada et al. (2014); ITL94 -- Izotov et al. (1994); KB94 -- Kingsburgh \&
Barlow (1994); L00 -- Liu et al. (2000); WL07 -- Wang \& Liu (2007).
For Sa\,3-107, the CEL O${}^{+}$ ionic abundance estimated by using the correlation 
${\rm S}^{2+}/{\rm S}^{+} = 4.892 \, ({\rm O}^{2+}/{\rm O}^{+})^{0.419}$ derived in Section 4.1.
\end{tablenotes}
\end{center}

\newpage
\FloatBarrier
\normalsize


\setcounter{figure}{1}

\figurenum{2}
\begin{figure*}
\begin{center}
\includegraphics[width=0.8\linewidth]{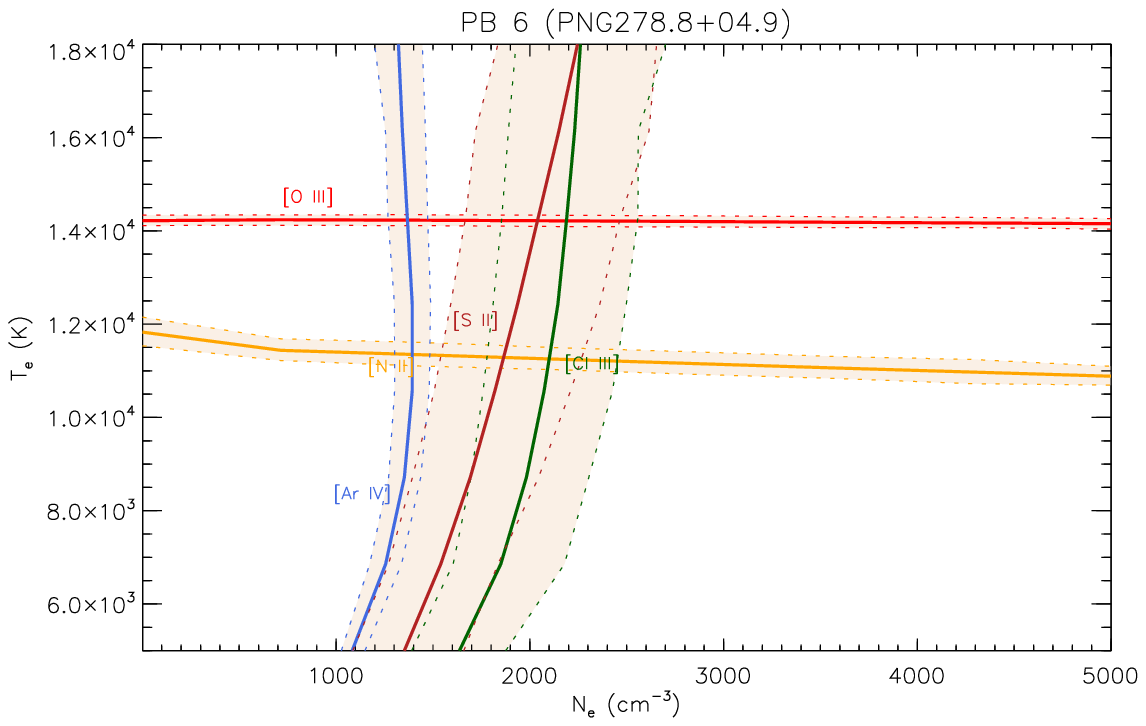}\\
\includegraphics[width=0.8\linewidth]{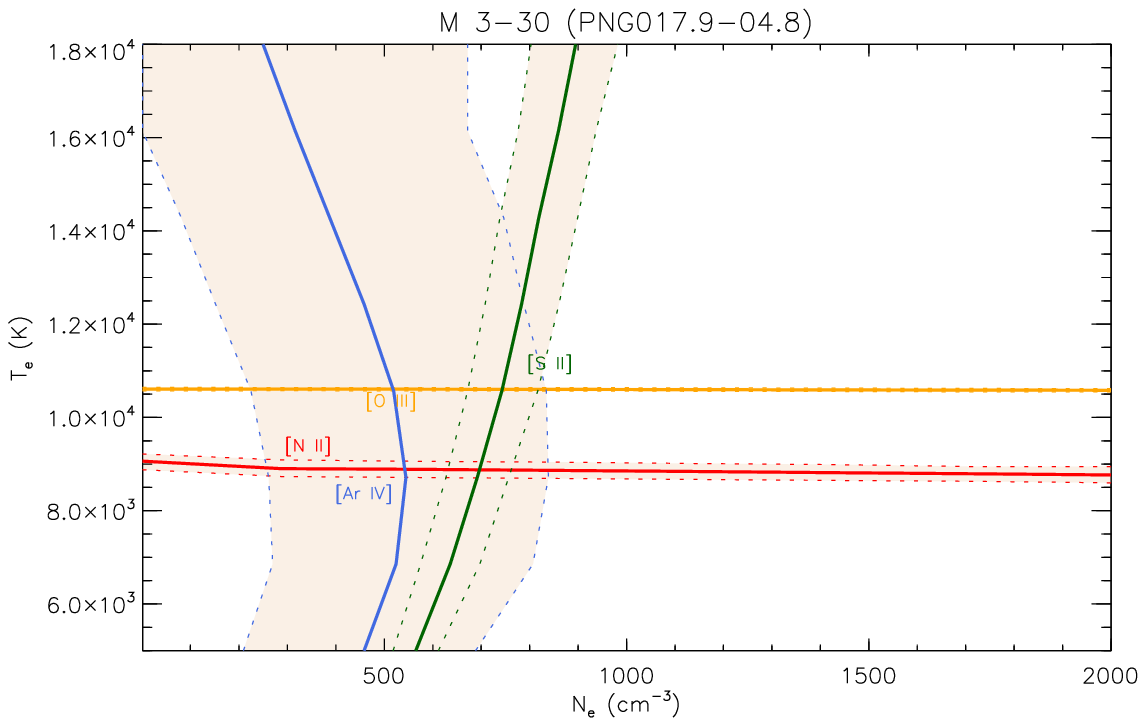}\\
\caption{$N_{\rm e}$--$T_{\rm e}$ diagnostic diagrams based on CELs. The upper and lower limits at the 90\% confidence level are plotted by the dotted lines.  
\label{wc:fig:dd:cels:2}
}
\end{center}
\end{figure*}

\figurenum{2}
\begin{figure*}
\begin{center}
\includegraphics[width=0.8\linewidth]{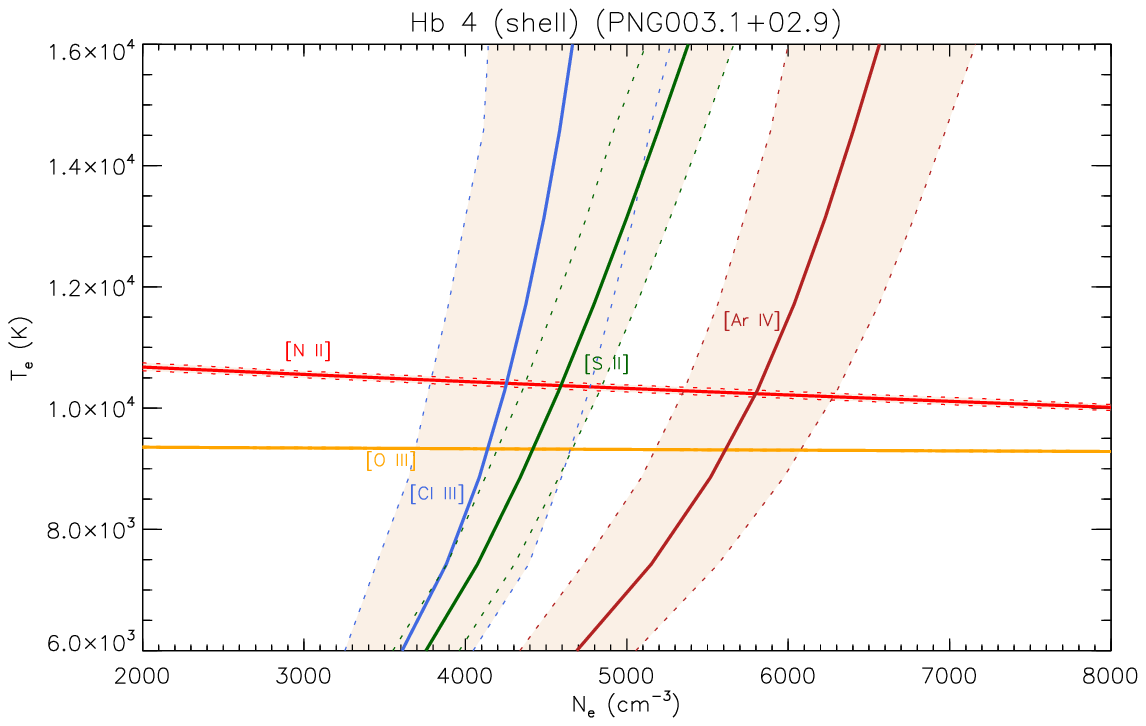}\\
\includegraphics[width=0.8\linewidth]{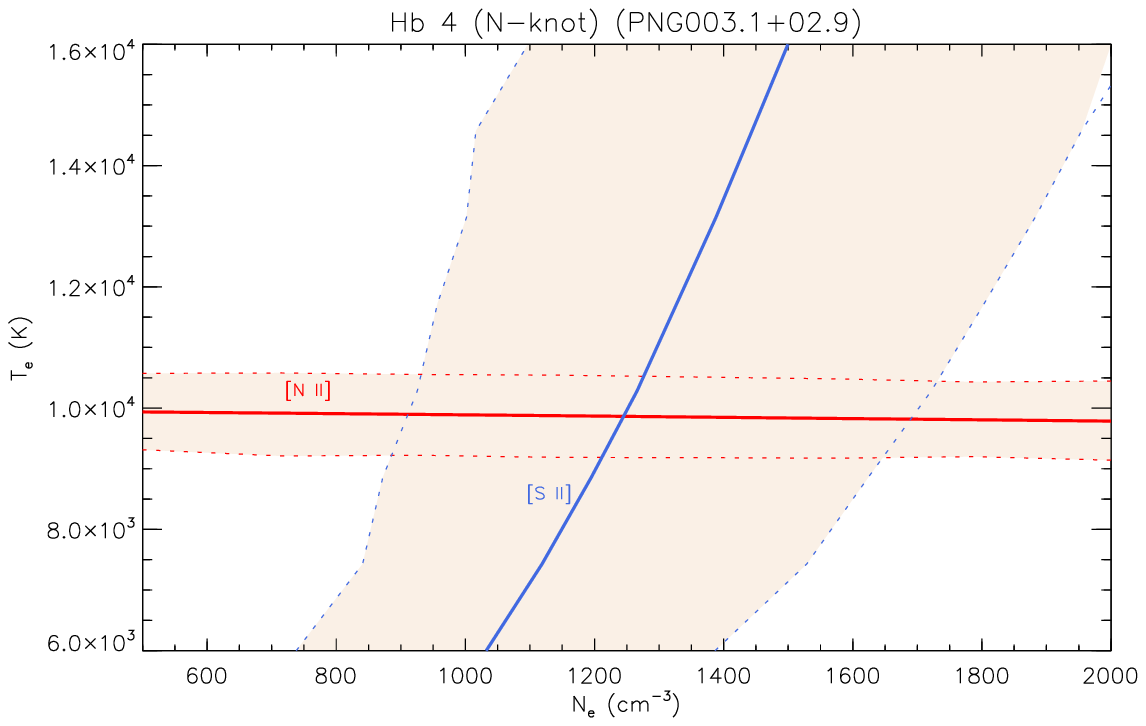}\\
\caption{\textit{-- continued}}
\end{center}
\end{figure*}

\figurenum{2}
\begin{figure*}
\begin{center}
\includegraphics[width=0.8\linewidth]{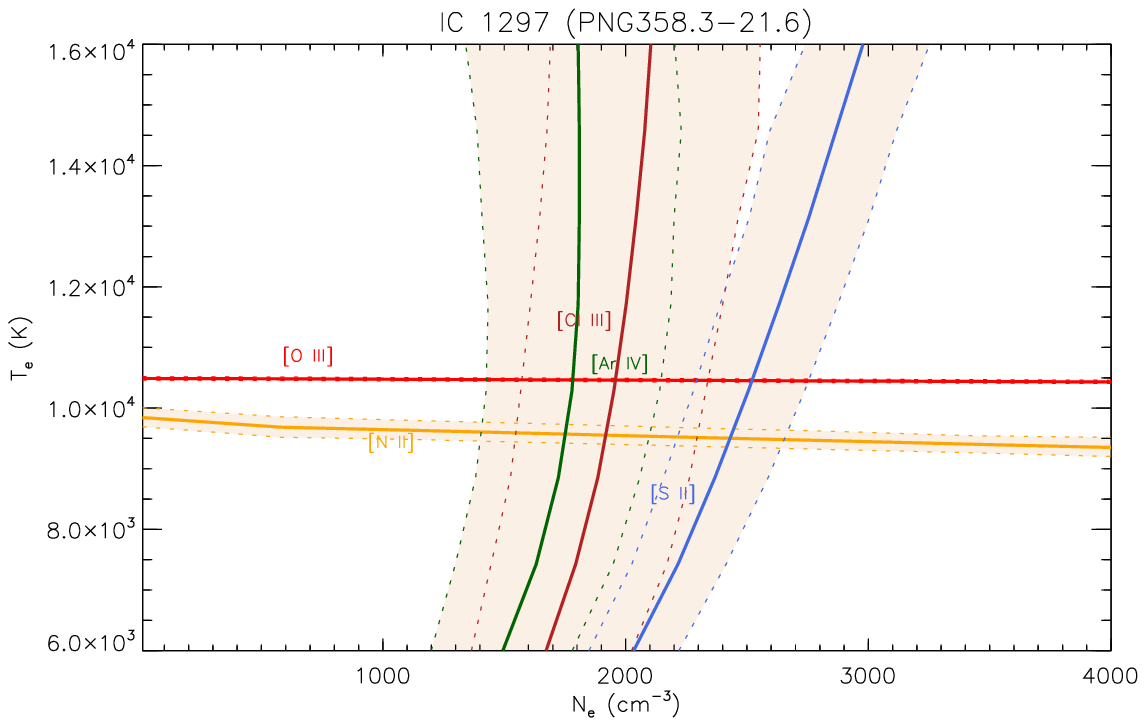}\\
\includegraphics[width=0.8\linewidth]{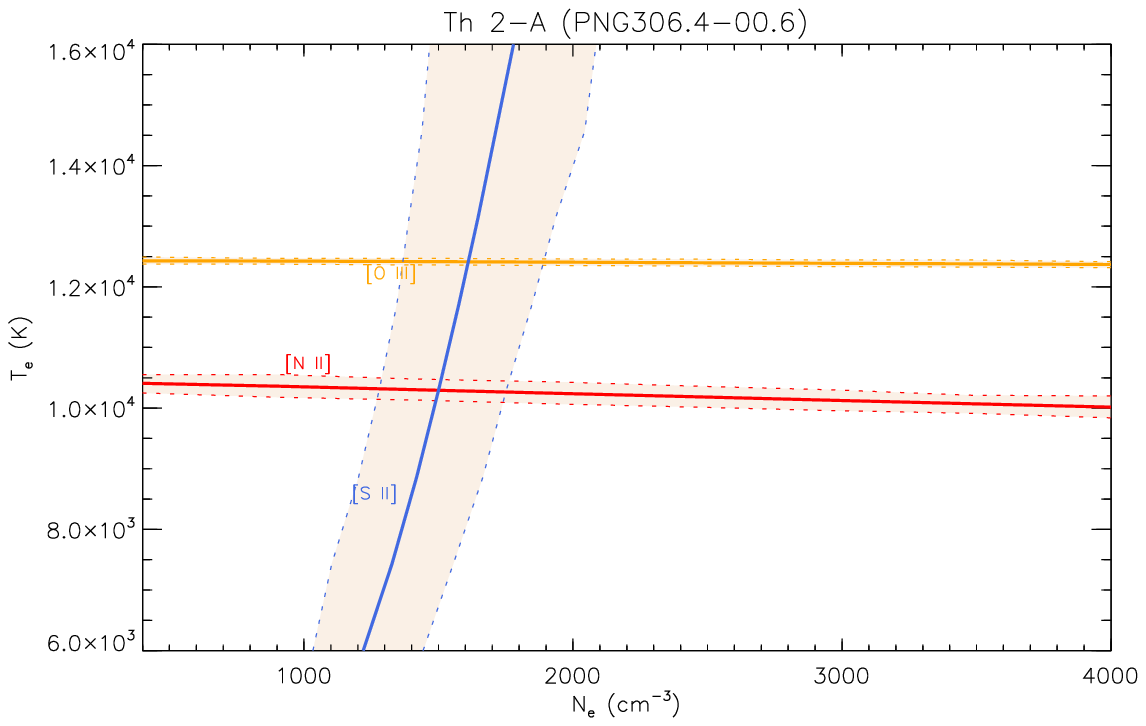}\\
\caption{\textit{-- continued}}
\end{center}
\end{figure*}

\figurenum{2}
\begin{figure*}
\begin{center}
\includegraphics[width=0.8\linewidth]{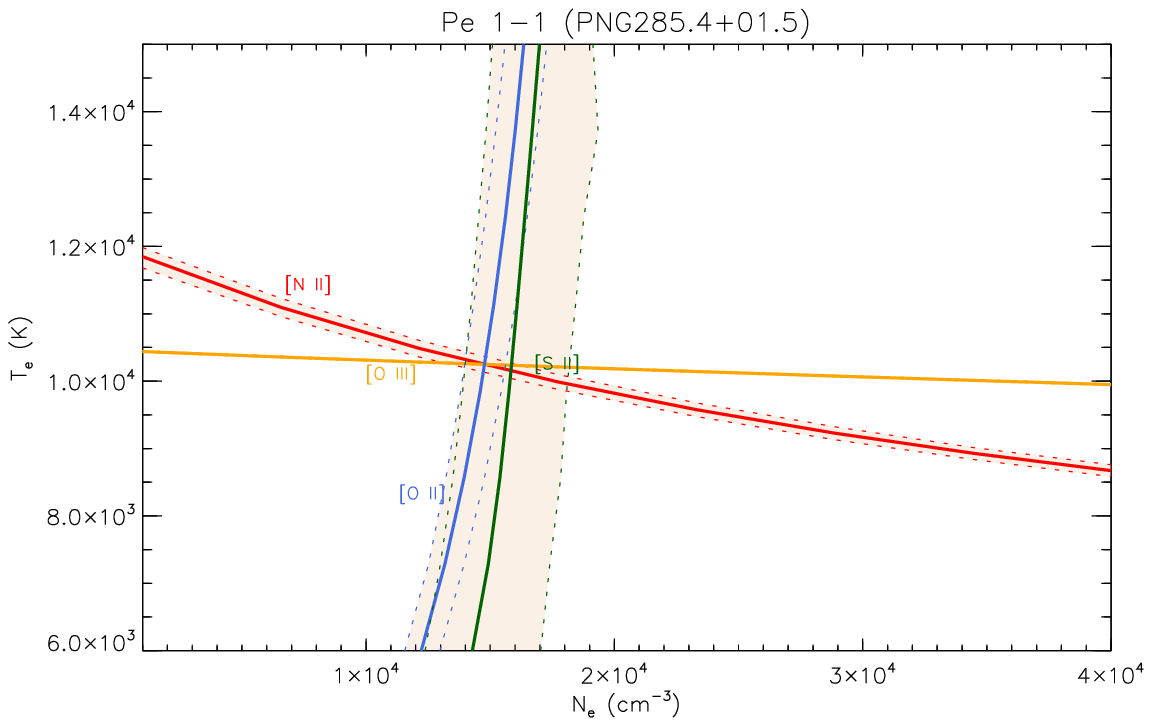}\\
\includegraphics[width=0.8\linewidth]{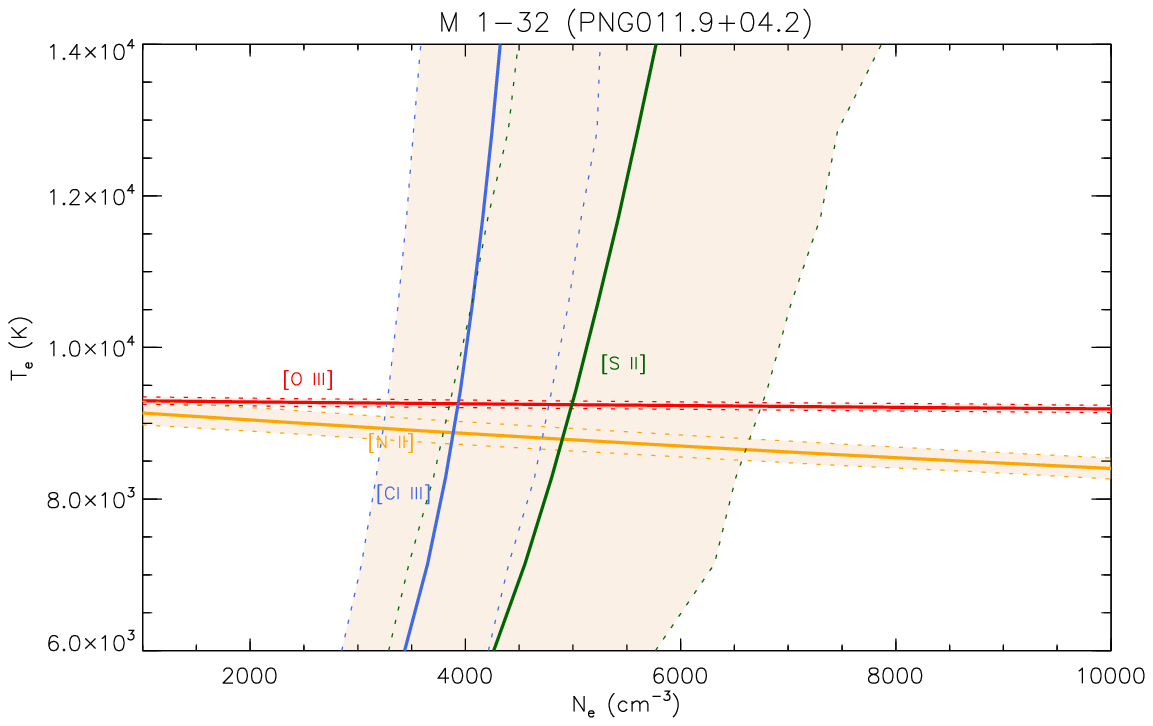}\\
\caption{\textit{-- continued}}
\end{center}
\end{figure*}

\figurenum{2}
\begin{figure*}
\begin{center}
\includegraphics[width=0.8\linewidth]{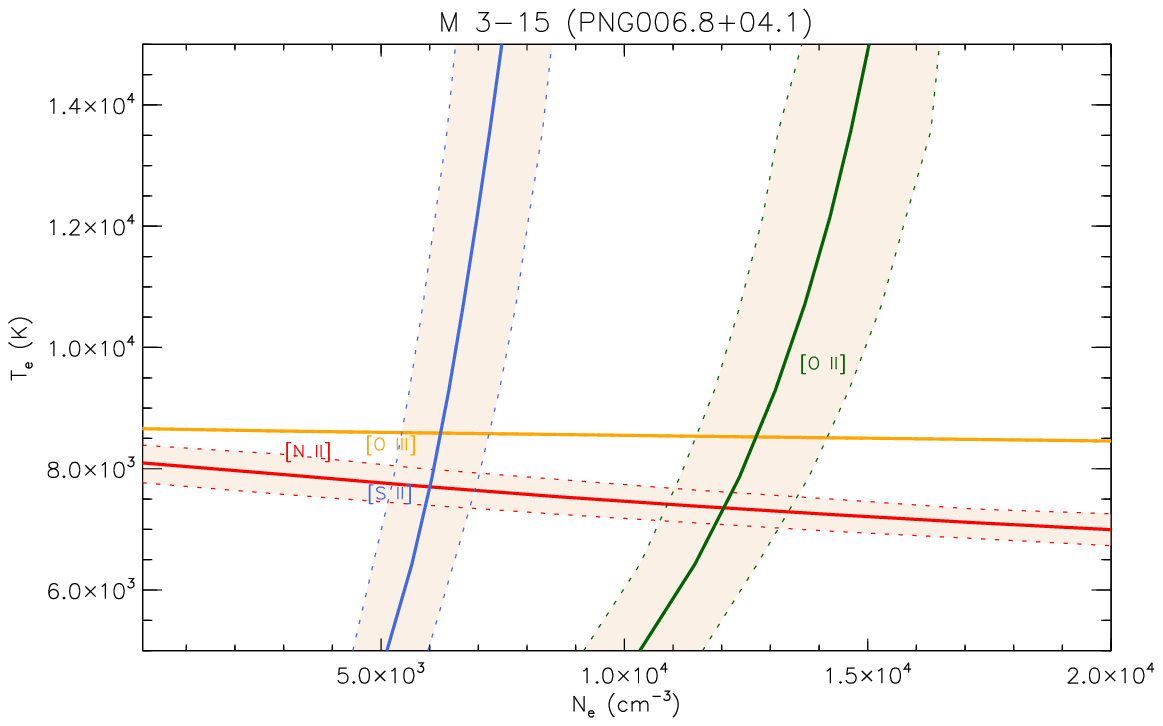}\\
\includegraphics[width=0.8\linewidth]{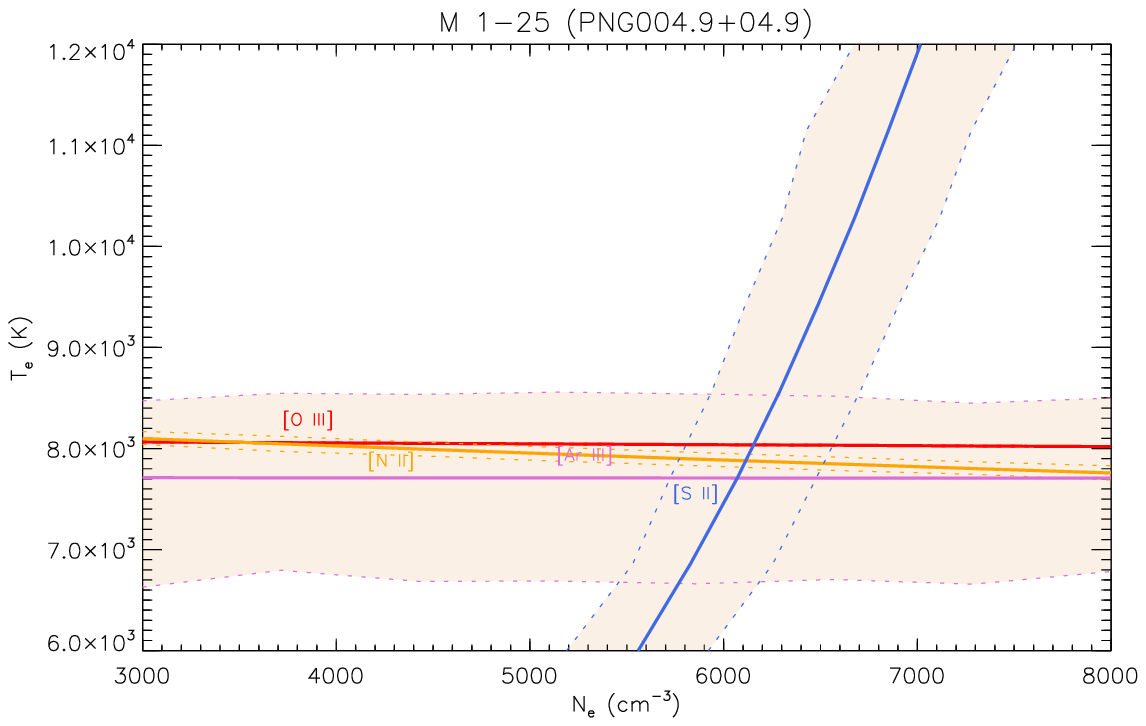}\\
\caption{\textit{-- continued}}
\end{center}
\end{figure*}

\figurenum{2}
\begin{figure*}
\begin{center}
\includegraphics[width=0.8\linewidth]{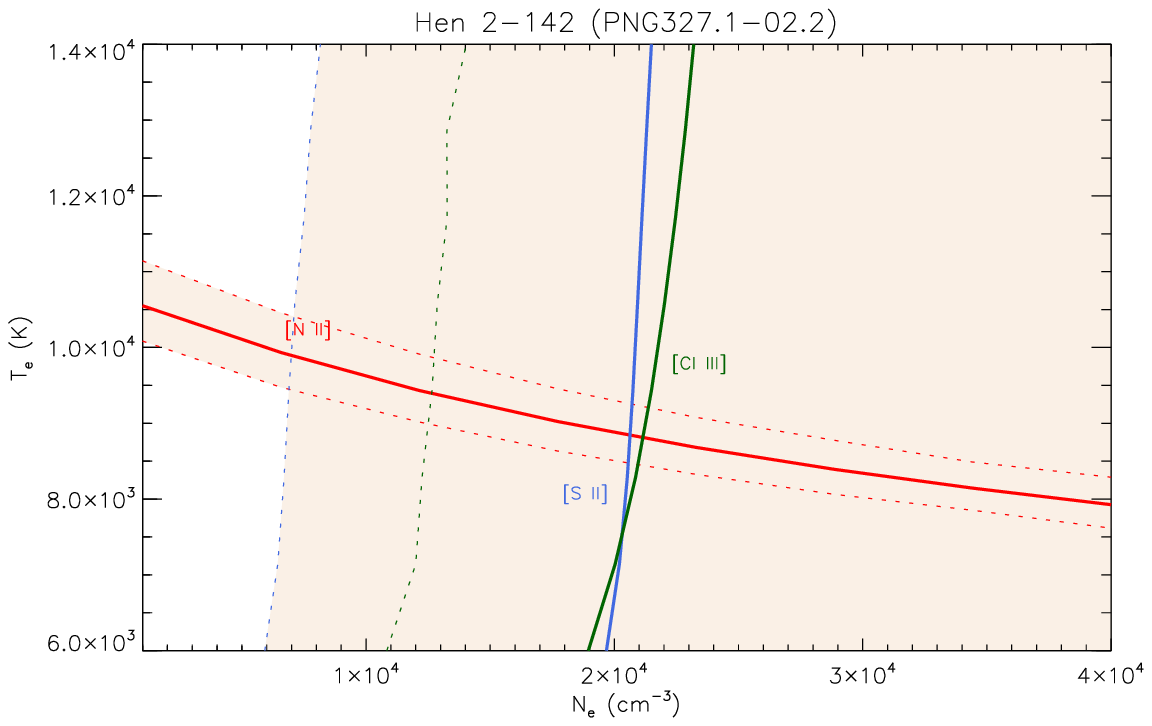}\\
\includegraphics[width=0.8\linewidth]{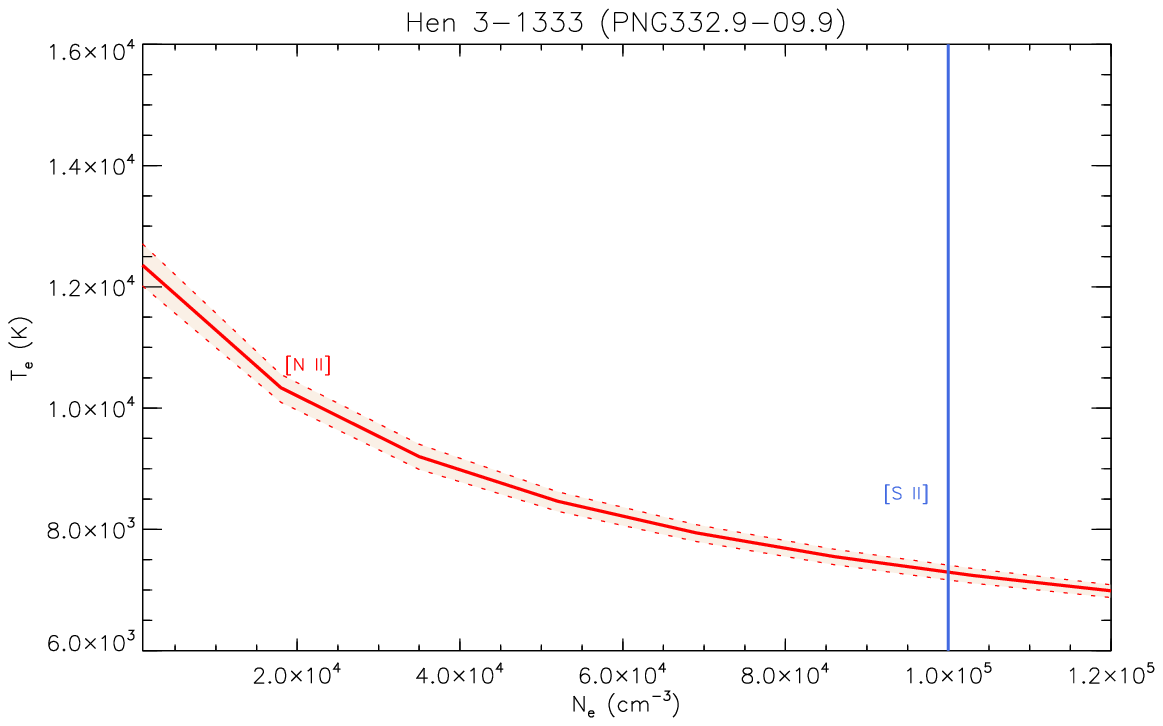}\\
\caption{\textit{-- continued}}
\end{center}
\end{figure*}

\figurenum{2}
\begin{figure*}
\begin{center}
\includegraphics[width=0.8\linewidth]{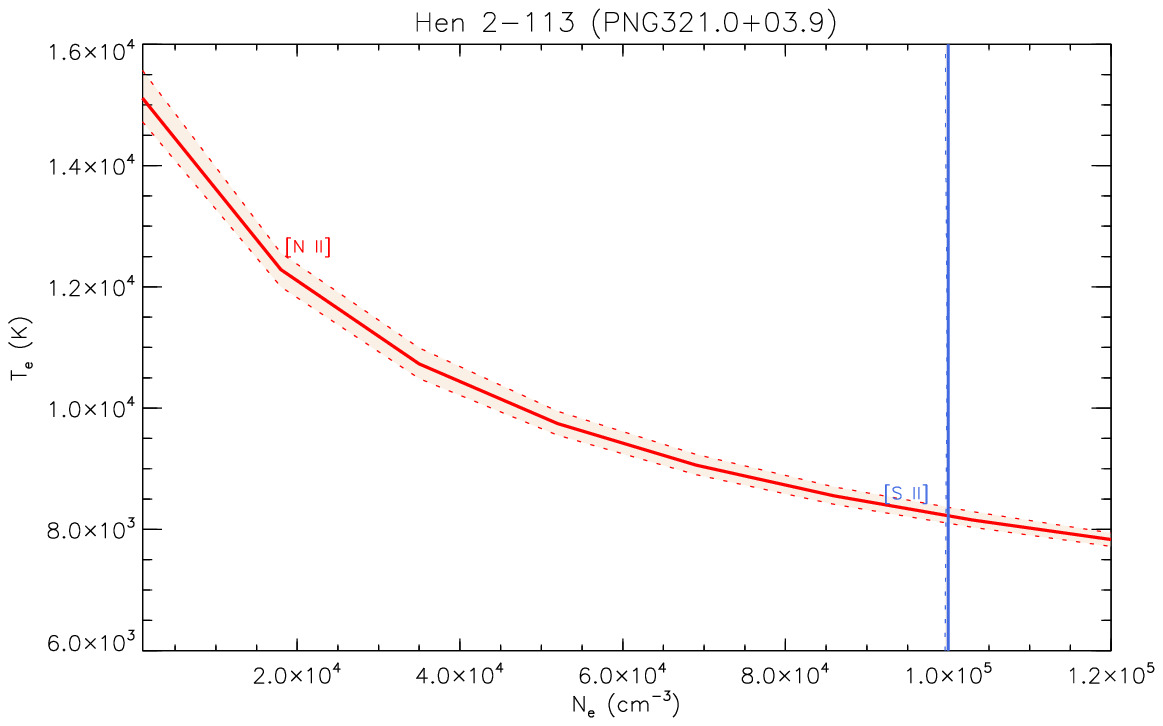}\\
\includegraphics[width=0.8\linewidth]{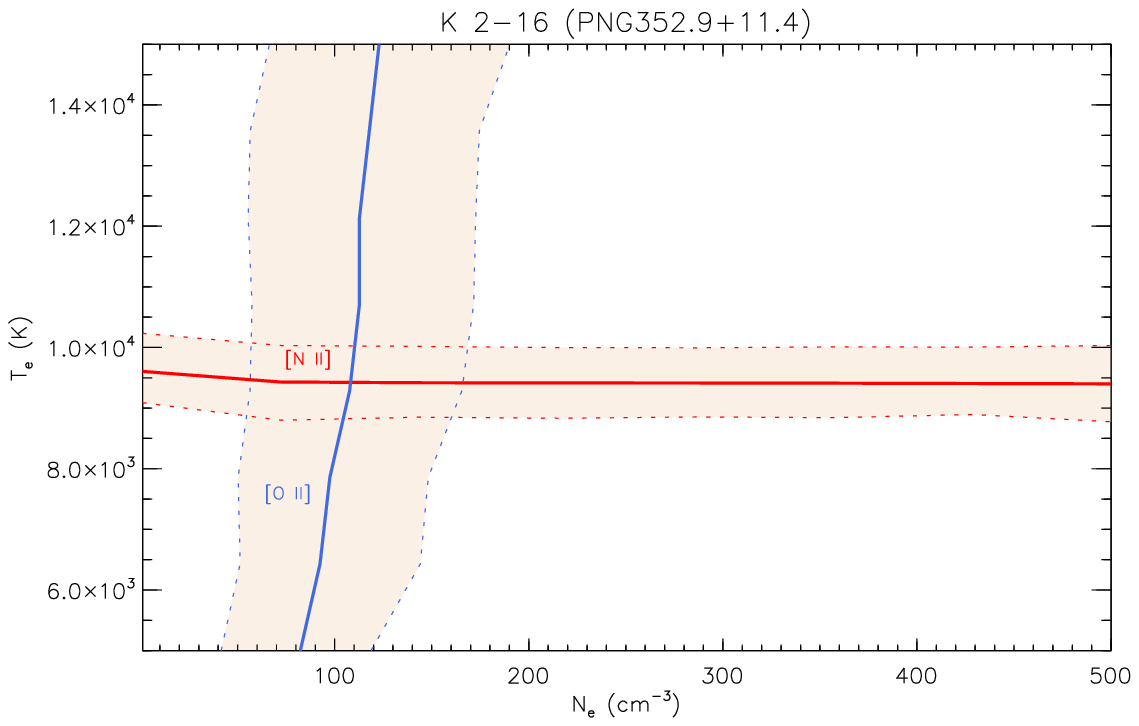}\\
\caption{\textit{-- continued}}
\end{center}
\end{figure*}

\figurenum{2}
\begin{figure*}
\begin{center}
\includegraphics[width=0.8\linewidth]{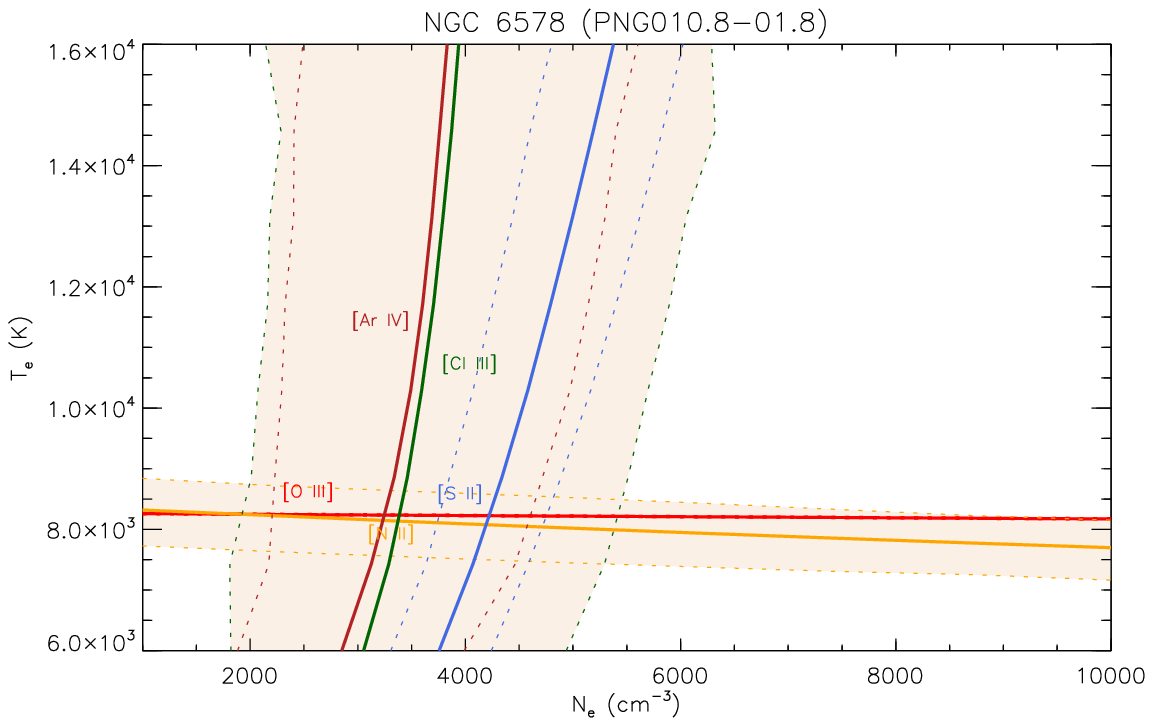}\\
\includegraphics[width=0.8\linewidth]{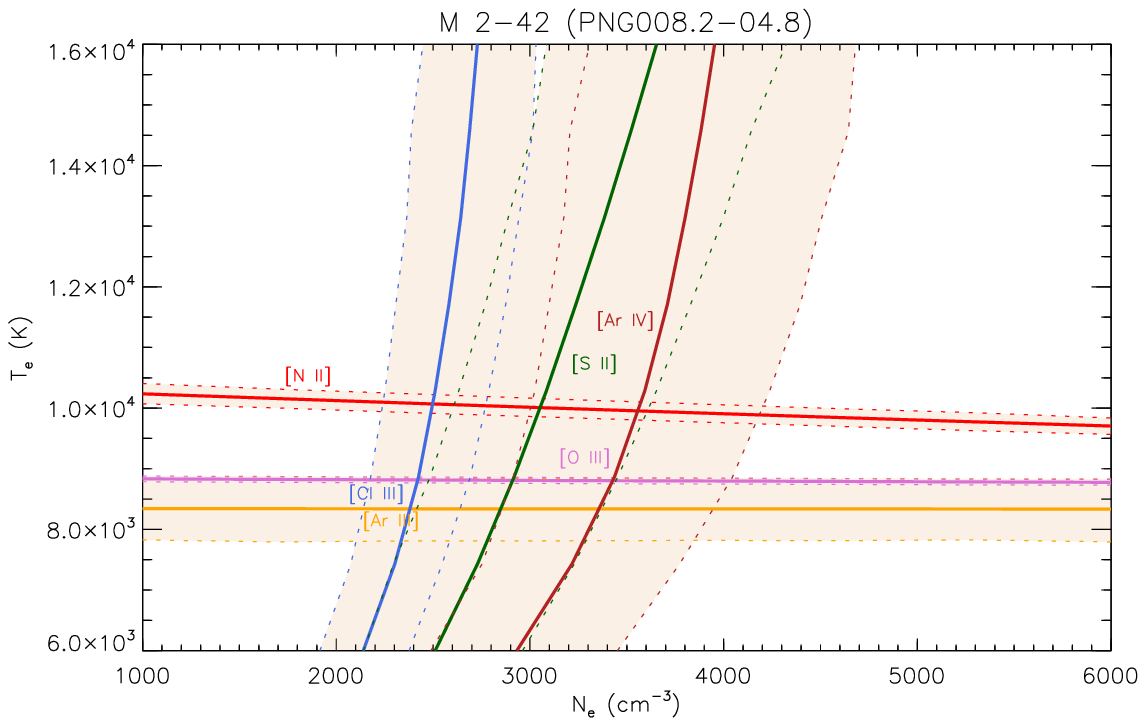}\\
\caption{\textit{-- continued}}
\end{center}
\end{figure*}

\figurenum{2}
\begin{figure*}
\begin{center}
\includegraphics[width=0.8\linewidth]{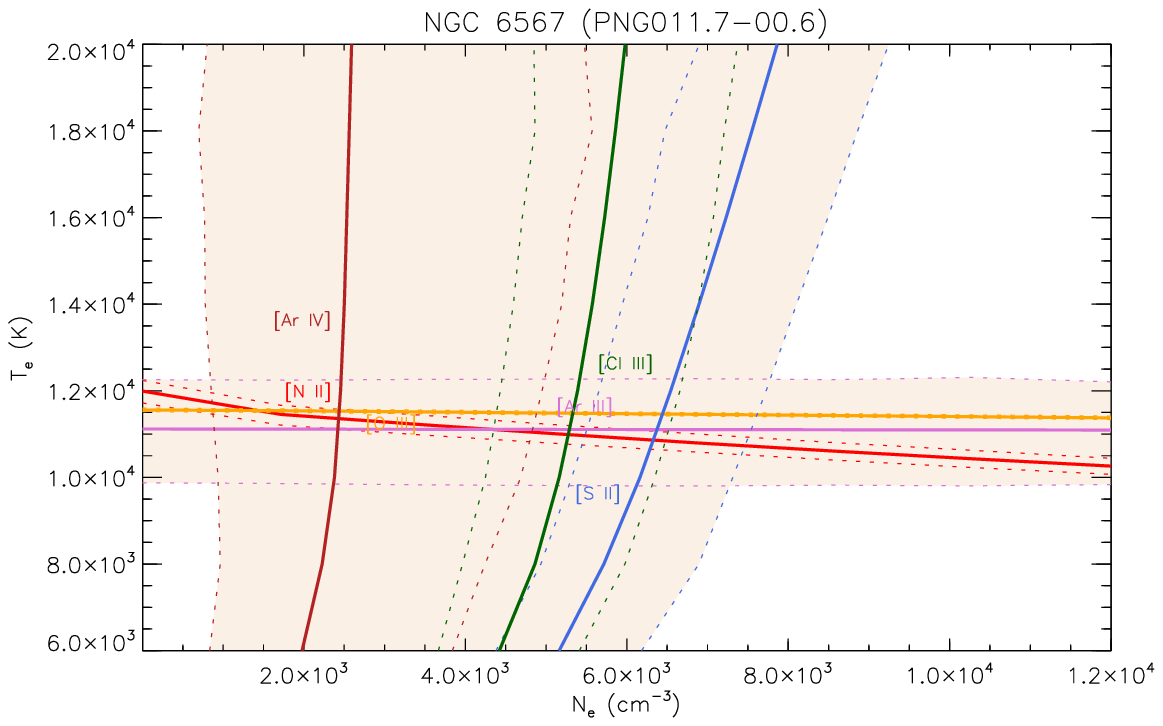}\\
\includegraphics[width=0.8\linewidth]{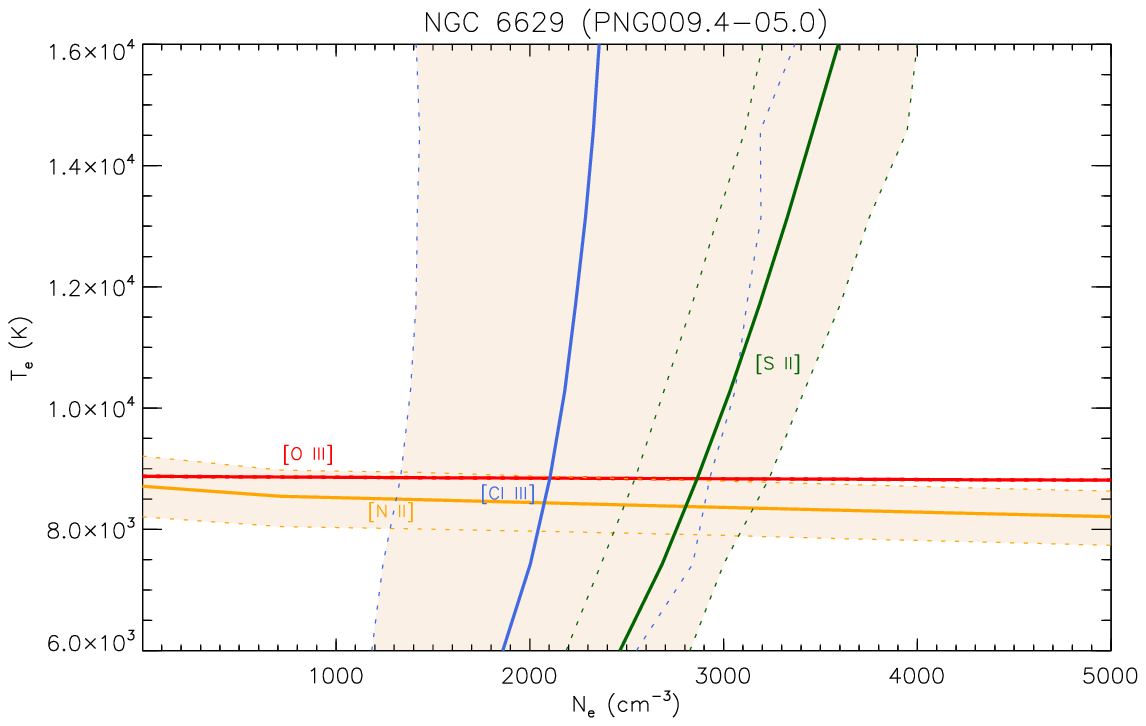}\\
\caption{\textit{-- continued}}
\end{center}
\end{figure*}

\figurenum{2}
\begin{figure*}
\begin{center}
\includegraphics[width=0.8\linewidth]{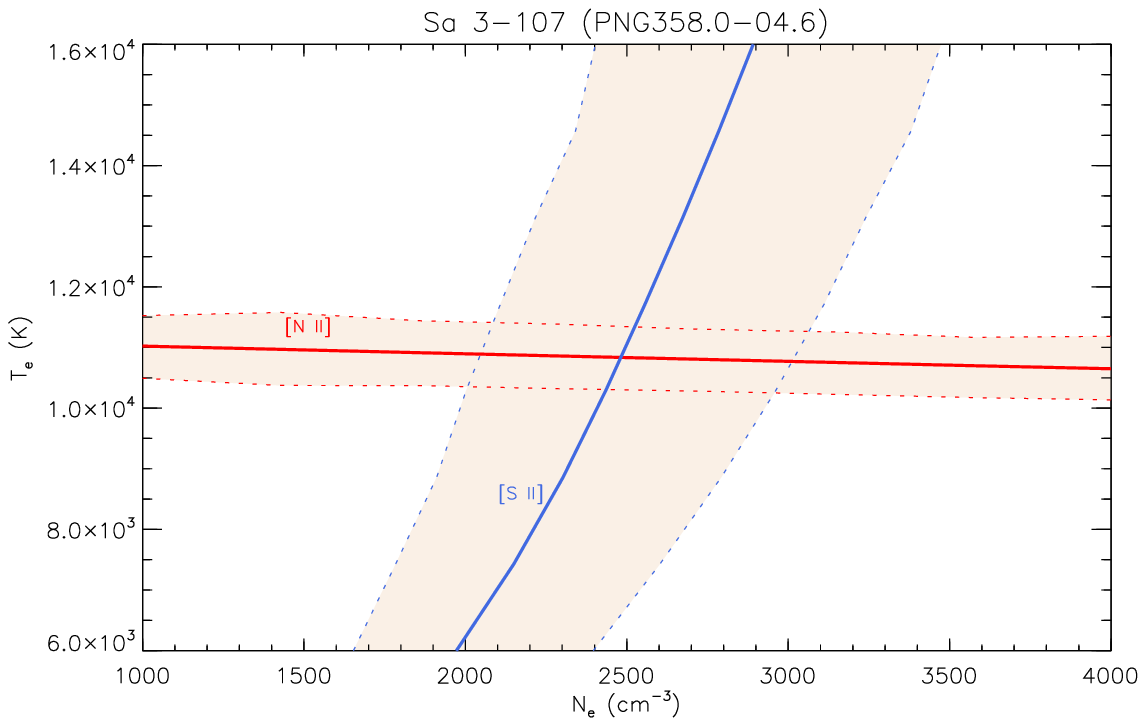}\\
\caption{\textit{-- continued}}
\end{center}
\end{figure*}


\setcounter{figure}{3}

\figurenum{4}
\begin{figure*}
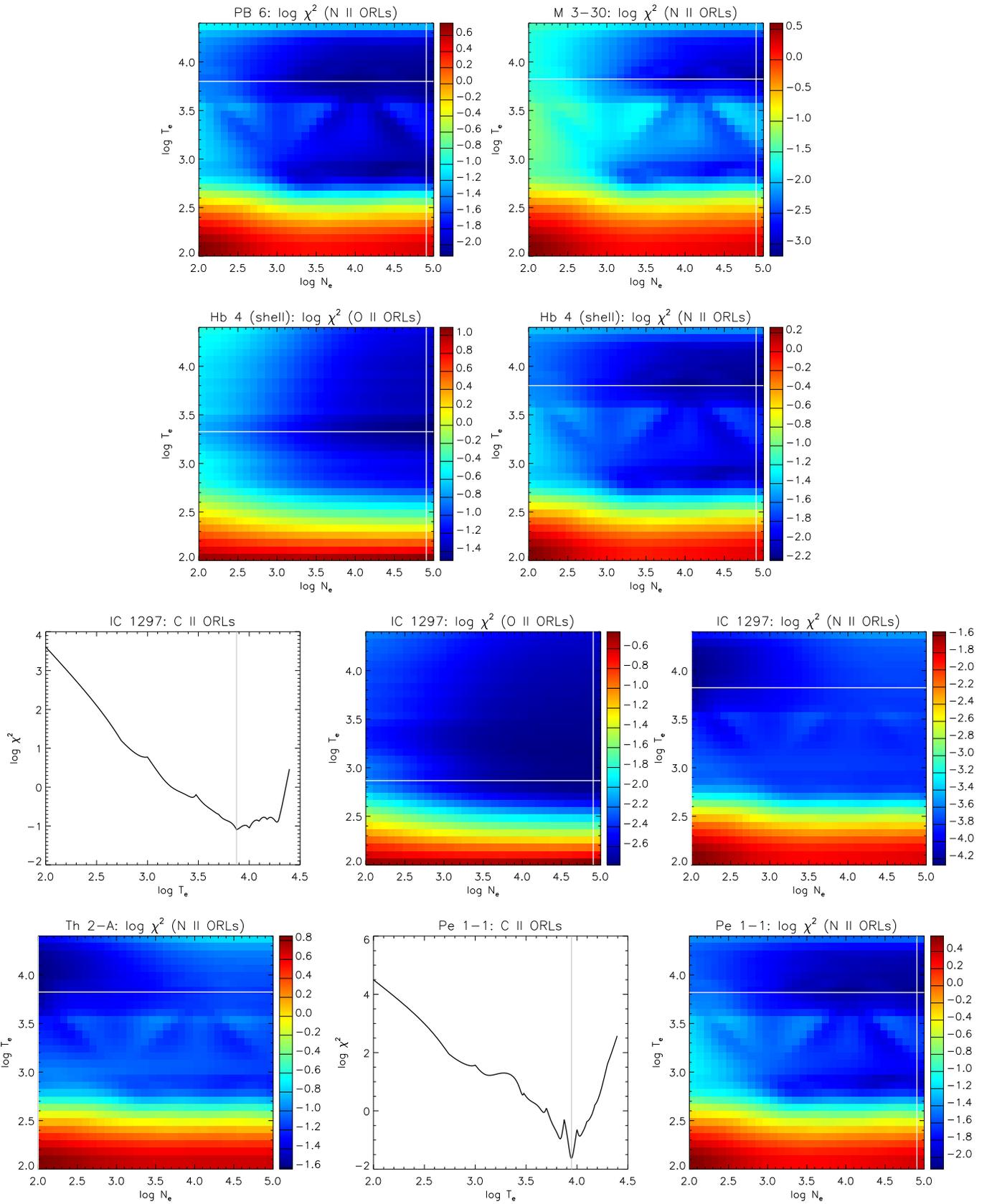

\begin{center}
\includegraphics[width=0.325\linewidth]{figure4/fig4_pb6_n_ii_final.eps}
\includegraphics[width=0.325\linewidth]{figure4/fig4_m330_n_ii_final.eps}\\
\includegraphics[width=0.325\linewidth]{figure4/fig4_hb4shell_o_ii_final.eps}
\includegraphics[width=0.325\linewidth]{figure4/fig4_hb4shell_n_ii_final.eps}\\
\includegraphics[width=0.325\linewidth]{figure4/fig4_ic1297_c_ii_final.eps}
\includegraphics[width=0.325\linewidth]{figure4/fig4_ic1297_o_ii_final.eps}%
\includegraphics[width=0.325\linewidth]{figure4/fig4_ic1297_n_ii_final.eps}\\
\includegraphics[width=0.325\linewidth]{figure4/fig4_th2a_n_ii_final.eps}
\includegraphics[width=0.325\linewidth]{figure4/fig4_pe11_c_ii_final.eps}%
\includegraphics[width=0.325\linewidth]{figure4/fig4_pe11_n_ii_final.eps}\\
\caption{$T_{\rm e}$ diagnostic diagrams based on C\,{\sc ii} ORLs, and $N_{\rm e}$--$T_{\rm e}$ diagnostic diagrams based on O\,{\sc ii} and N\,{\sc ii} ORLs. The best-fitting physical conditions at $\chi_{\rm min}^2$ are shown by the solid lines. 
\label{wc:fig:dd:orls:2}
}
\end{center}
\end{figure*}

\figurenum{4}
\begin{figure*}
\begin{center}
\includegraphics[width=0.325\linewidth]{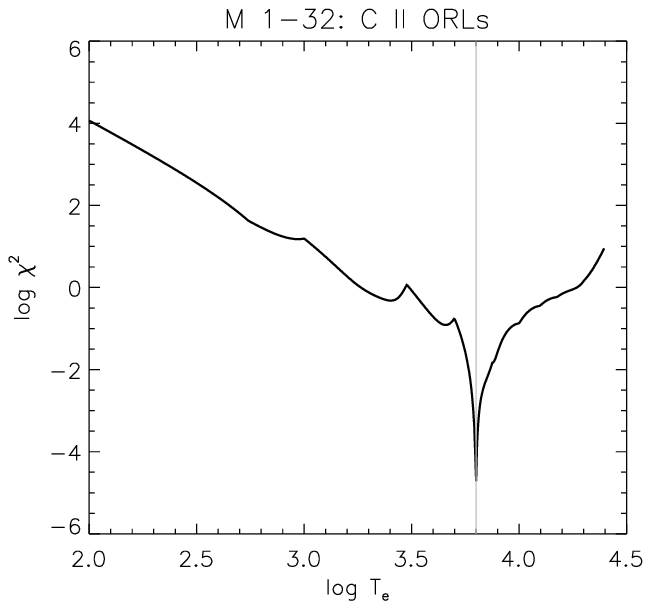}%
\includegraphics[width=0.325\linewidth]{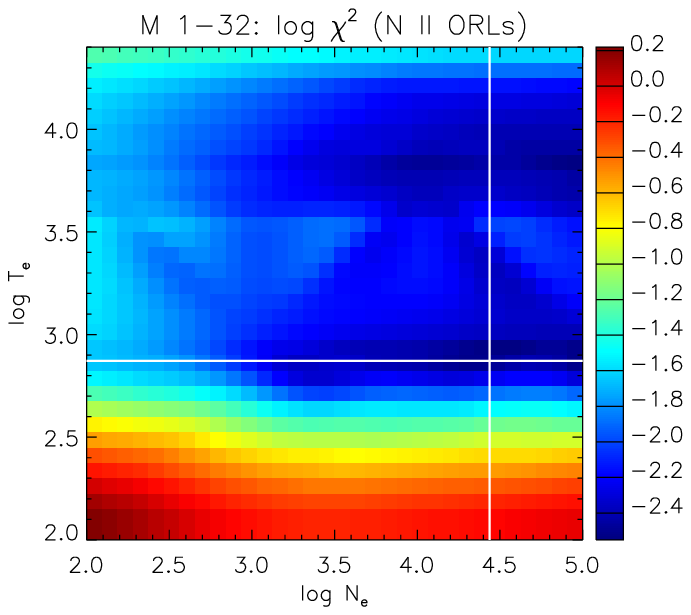}
\includegraphics[width=0.325\linewidth]{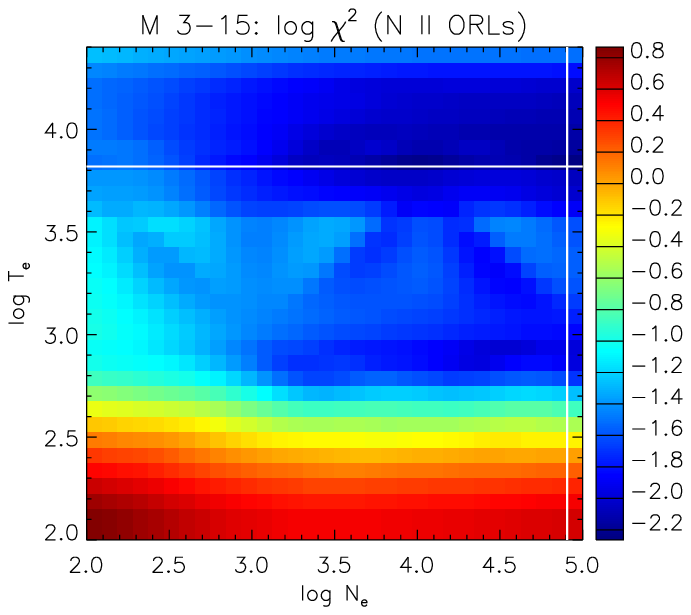}\\
\includegraphics[width=0.325\linewidth]{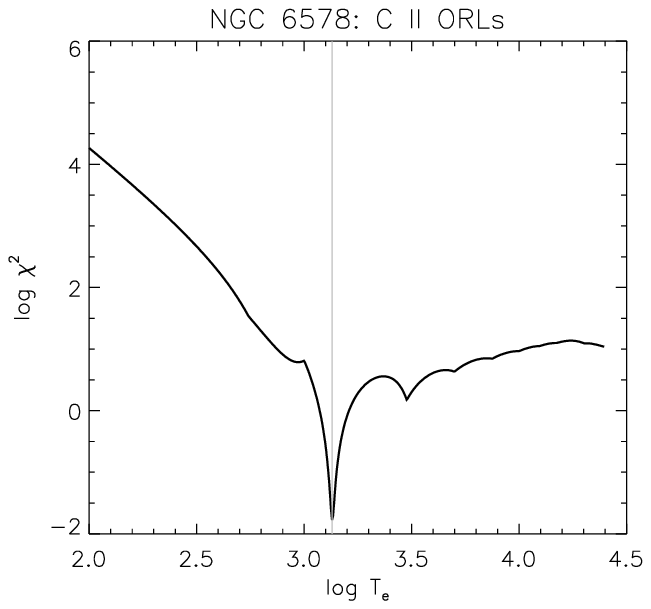}%
\includegraphics[width=0.325\linewidth]{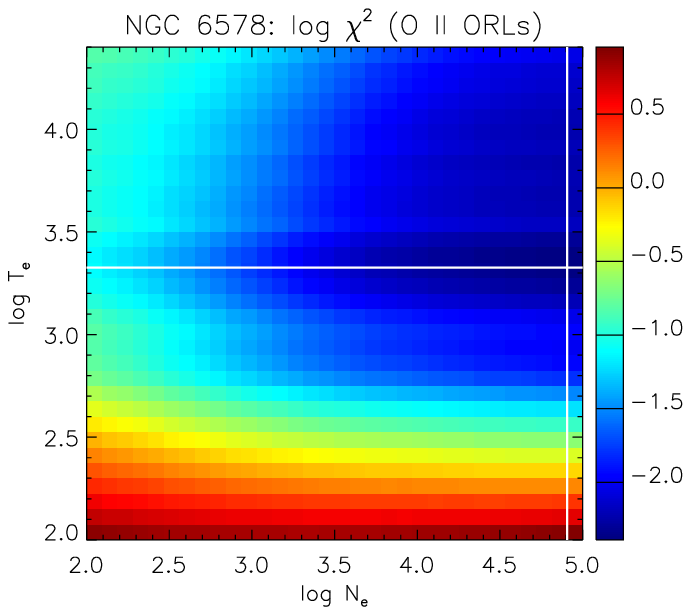}%
\includegraphics[width=0.325\linewidth]{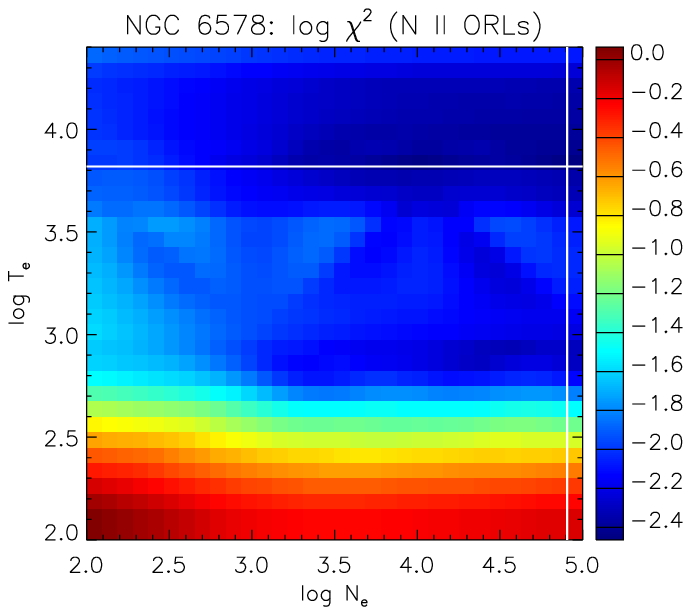}\\
\includegraphics[width=0.325\linewidth]{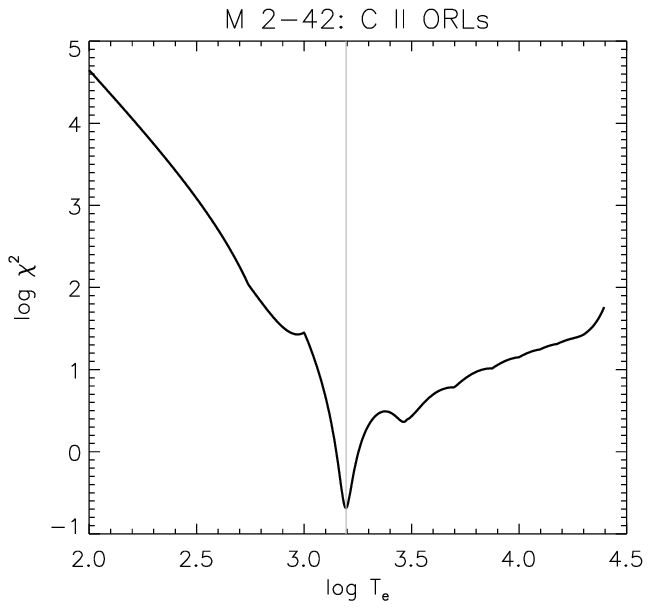}%
\includegraphics[width=0.325\linewidth]{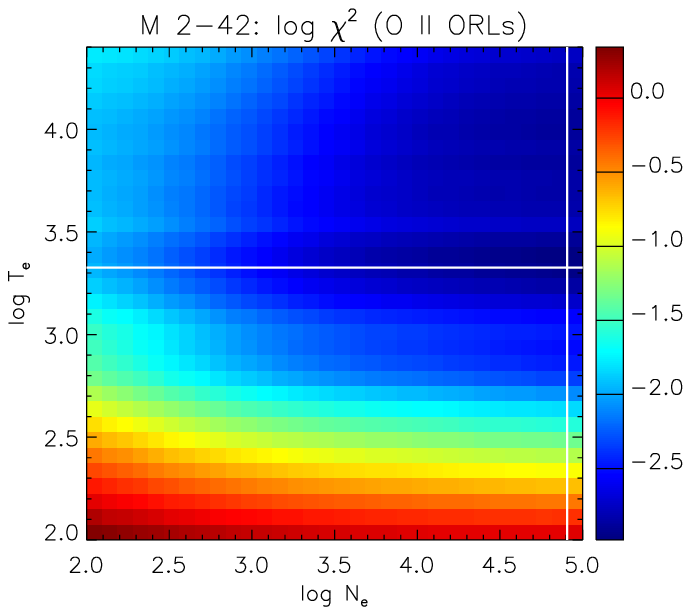}%
\includegraphics[width=0.325\linewidth]{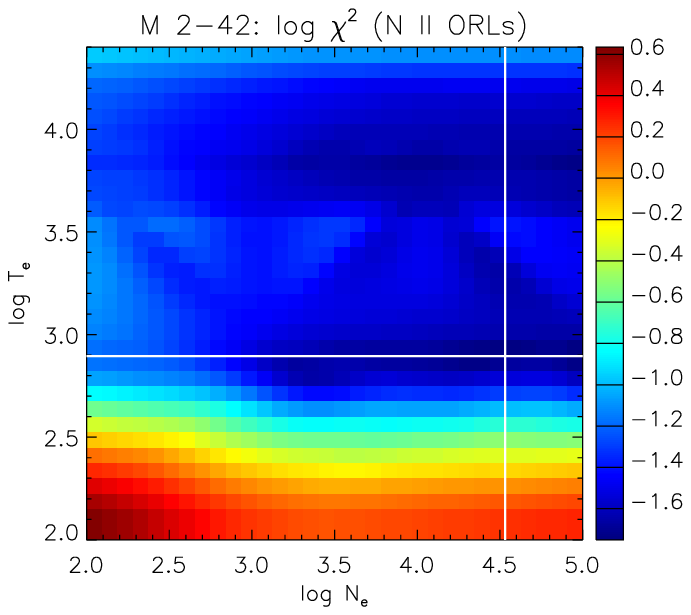}\\
\includegraphics[width=0.325\linewidth]{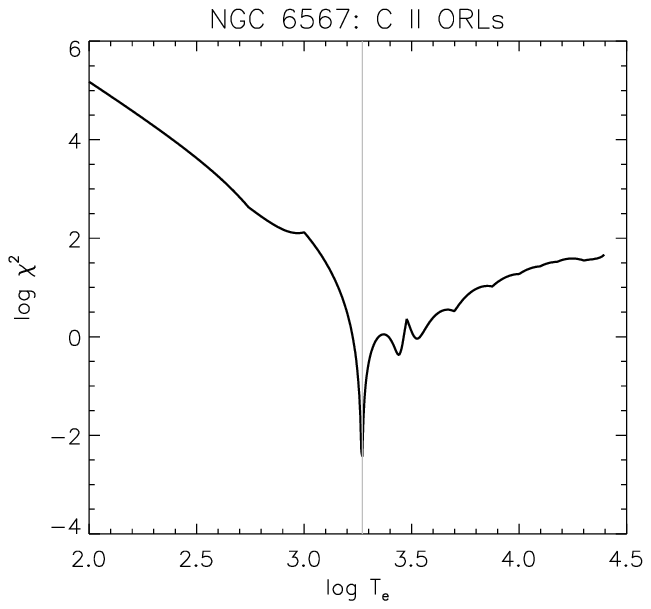}
\includegraphics[width=0.325\linewidth]{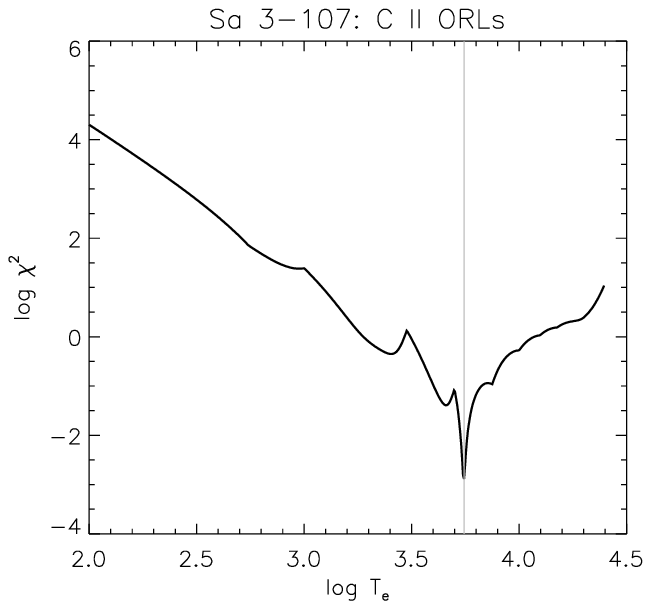}%
\caption{\textit{-- continued}}
\end{center}
\end{figure*}

\FloatBarrier





\begin{thebibliography}{211}
\expandafter\ifx\csname natexlab\endcsname\relax\def\natexlab#1{#1}\fi

\bibitem[{{Acker} {et~al.}(2002){Acker}, {Gesicki}, {Grosdidier}, \&
  {Durand}}]{Acker2002}
{Acker}, A., {Gesicki}, K., {Grosdidier}, Y., \& {Durand}, S. 2002,
  {\href{https://dx.doi.org/10.1051/0004-6361:20020009}{\color{magenta}\aap}},
  \href{https://ui.adsabs.harvard.edu/abs/2002A%26A...384..620A}{384, 620}

\bibitem[{{Acker} {et~al.}(1992){Acker}, {Marcout}, {Ochsenbein}, {Stenholm},
  {Tylenda}, \& {Schohn}}]{Acker1992}
{Acker}, A., {Marcout}, J., {Ochsenbein}, F., {et~al.} 1992, {The
  Strasbourg-ESO Catalogue of Galactic Planetary Nebulae. Parts I, II.}

\bibitem[{{Acker} \& {Neiner}(2003)}]{Acker2003}
{Acker}, A. \& {Neiner}, C. 2003,
  {\href{https://dx.doi.org/10.1051/0004-6361:20030391}{\color{magenta}\aap}},
  \href{https://ui.adsabs.harvard.edu/abs/2003A%26A...403..659A}{403, 659}

\bibitem[{{Acker} {et~al.}(1991){Acker}, {Raytchev}, {Koeppen}, \&
  {Stenholm}}]{Acker1991b}
{Acker}, A., {Raytchev}, B., {Koeppen}, J., \& {Stenholm}, B. 1991, \aaps,
  \href{https://ui.adsabs.harvard.edu/abs/1991A%26AS...89..237A}{89, 237}

\bibitem[{{Akras} \& {Gon{\c{c}}alves}(2016)}]{Akras2016}
{Akras}, S. \& {Gon{\c{c}}alves}, D.~R. 2016,
  {\href{https://dx.doi.org/10.1093/mnras/stv2139}{\color{magenta}\mnras}},
  \href{https://ui.adsabs.harvard.edu/abs/2016MNRAS.455..930A}{455, 930}

\bibitem[{{Akras} \& {L{\'o}pez}(2012)}]{Akras2012}
{Akras}, S. \& {L{\'o}pez}, J.~A. 2012,
  {\href{https://dx.doi.org/10.1111/j.1365-2966.2012.21578.x}{\color{magenta}\mnras}},
  \href{https://ui.adsabs.harvard.edu/abs/2012MNRAS.425.2197A}{425, 2197}

\bibitem[{{Akras} {et~al.}(2020){Akras}, {Monteiro}, {Aleman}, {Farias}, {May},
  \& {Pereira}}]{Akras2020}
{Akras}, S., {Monteiro}, H., {Aleman}, I., {et~al.} 2020,
  {\href{https://dx.doi.org/10.1093/mnras/staa383}{\color{magenta}\mnras}},
  \href{https://ui.adsabs.harvard.edu/abs/2020MNRAS.493.2238A}{493, 2238}

\bibitem[{{Ali} {et~al.}(2015){Ali}, {Amer}, {Dopita}, {Vogt}, \&
  {Basurah}}]{Ali2015}
{Ali}, A., {Amer}, M.~A., {Dopita}, M.~A., {Vogt}, F.~P.~A., \& {Basurah},
  H.~M. 2015,
  {\href{https://dx.doi.org/10.1051/0004-6361/201526223}{\color{magenta}\aap}},
  \href{https://ui.adsabs.harvard.edu/abs/2015A%26A...583A..83A}{583, A83}

\bibitem[{{Ali} {et~al.}(2016){Ali}, {Dopita}, {Basurah}, {Amer}, {Alsulami},
  \& {Alruhaili}}]{Ali2016}
{Ali}, A., {Dopita}, M.~A., {Basurah}, H.~M., {et~al.} 2016,
  {\href{https://dx.doi.org/10.1093/mnras/stw1744}{\color{magenta}\mnras}},
  \href{https://ui.adsabs.harvard.edu/abs/2016MNRAS.462.1393A}{462, 1393}

\bibitem[{{Aller} \& {Czyzak}(1983)}]{Aller1983}
{Aller}, L.~H. \& {Czyzak}, S.~J. 1983,
  {\href{https://dx.doi.org/10.1086/190846}{\color{magenta}\apjs}},
  \href{https://ui.adsabs.harvard.edu/abs/1983ApJS...51..211A}{51, 211}

\bibitem[{{Aller} \& {Menzel}(1945)}]{Aller1945}
{Aller}, L.~H. \& {Menzel}, D.~H. 1945,
  {\href{https://dx.doi.org/10.1086/144757}{\color{magenta}\apj}},
  \href{https://ui.adsabs.harvard.edu/abs/1945ApJ...102..239A}{102, 239}

\bibitem[{{Anders} \& {Grevesse}(1989)}]{Anders1989}
{Anders}, E. \& {Grevesse}, N. 1989,
  {\href{https://dx.doi.org/10.1016/0016-7037(89)90286-X}{\color{magenta}\gca}},
  \href{https://ui.adsabs.harvard.edu/abs/1989GeCoA..53..197A}{53, 197}

\bibitem[{{Asplund} {et~al.}(2009){Asplund}, {Grevesse}, {Sauval}, \&
  {Scott}}]{Asplund2009}
{Asplund}, M., {Grevesse}, N., {Sauval}, A.~J., \& {Scott}, P. 2009,
  {\href{https://dx.doi.org/10.1146/annurev.astro.46.060407.145222}{\color{magenta}\araa}},
  \href{https://ui.adsabs.harvard.edu/abs/2009ARA%26A..47..481A}{47, 481}

\bibitem[{{Basurah} {et~al.}(2016){Basurah}, {Ali}, {Dopita}, {Alsulami},
  {Amer}, \& {Alruhaili}}]{Basurah2016}
{Basurah}, H.~M., {Ali}, A., {Dopita}, M.~A., {et~al.} 2016,
  {\href{https://dx.doi.org/10.1093/mnras/stw468}{\color{magenta}\mnras}},
  \href{https://ui.adsabs.harvard.edu/abs/2016MNRAS.458.2694B}{458, 2694}

\bibitem[{{Bautista} \& {Ahmed}(2018)}]{Bautista2018}
{Bautista}, M.~A. \& {Ahmed}, E.~E. 2018,
  {\href{https://dx.doi.org/10.3847/1538-4357/aad95a}{\color{magenta}\apj}},
  \href{https://ui.adsabs.harvard.edu/abs/2018ApJ...866...43B}{866, 43}

\bibitem[{{Bell} {et~al.}(1998){Bell}, {Berrington}, \& {Thomas}}]{Bell1998}
{Bell}, K.~L., {Berrington}, K.~A., \& {Thomas}, M.~R.~J. 1998,
  {\href{https://dx.doi.org/10.1046/j.1365-8711.1998.01364.x}{\color{magenta}\mnras}},
  \href{https://ui.adsabs.harvard.edu/abs/1998MNRAS.293L..83B}{293, L83}

\bibitem[{{Benjamin} {et~al.}(1999){Benjamin}, {Skillman}, \&
  {Smits}}]{Benjamin1999}
{Benjamin}, R.~A., {Skillman}, E.~D., \& {Smits}, D.~P. 1999,
  {\href{https://dx.doi.org/10.1086/306923}{\color{magenta}\apj}},
  \href{https://ui.adsabs.harvard.edu/abs/1999ApJ...514..307B}{514, 307}

\bibitem[{{Bhatia} \& {Kastner}(1993)}]{Bhatia1993}
{Bhatia}, A.~K. \& {Kastner}, S.~O. 1993,
  {\href{https://dx.doi.org/10.1006/adnd.1993.1011}{\color{magenta}\adndt}},
  \href{https://ui.adsabs.harvard.edu/abs/1993ADNDT..54..133B}{54, 133}

\bibitem[{{Biemont} \& {Bromage}(1983)}]{Biemont1983}
{Biemont}, E. \& {Bromage}, G.~E. 1983, \mnras,
  \href{https://ui.adsabs.harvard.edu/abs/1983MNRAS.205.1085B}{205, 1085}

\bibitem[{{Bi{\'e}mont} \& {Hansen}(1986)}]{Biemont1986}
{Bi{\'e}mont}, E. \& {Hansen}, J.~E. 1986,
  {\href{https://dx.doi.org/10.1088/0031-8949/34/2/005}{\color{magenta}\physscr}},
  \href{https://ui.adsabs.harvard.edu/abs/1986PhyS...34..116B}{34, 116}

\bibitem[{{Bl\"{o}cker}(1995)}]{Bloecker1995b}
{Bl\"{o}cker}, T. 1995, \aap,
  \href{https://ui.adsabs.harvard.edu/abs/1995A%26A...299..755B}{299, 755}

\bibitem[{{Bl{\"o}cker}(2001)}]{Blocker2001}
{Bl{\"o}cker}, T. 2001, \apss,
  \href{https://ui.adsabs.harvard.edu/abs/2001Ap%26SS.275....1B}{275, 1}

\bibitem[{{Borkowski} {et~al.}(1993){Borkowski}, {Harrington}, {Tsvetanov}, \&
  {Clegg}}]{Borkowski1993}
{Borkowski}, K.~J., {Harrington}, J.~P., {Tsvetanov}, Z., \& {Clegg}, R. E.~S.
  1993, {\href{https://dx.doi.org/10.1086/187029}{\color{magenta}\apjl}},
  \href{https://ui.adsabs.harvard.edu/abs/1993ApJ...415L..47B}{415, L47}

\bibitem[{{Busso} {et~al.}(1999){Busso}, {Gallino}, \&
  {Wasserburg}}]{Busso1999}
{Busso}, M., {Gallino}, R., \& {Wasserburg}, G.~J. 1999,
  {\href{https://dx.doi.org/10.1146/annurev.astro.37.1.239}{\color{magenta}\araa}},
  \href{https://ui.adsabs.harvard.edu/abs/1999ARA%26A..37..239B}{37, 239}

\bibitem[{{Cahn} {et~al.}(1992){Cahn}, {Kaler}, \& {Stanghellini}}]{Cahn1992}
{Cahn}, J.~H., {Kaler}, J.~B., \& {Stanghellini}, L. 1992, \aaps,
  \href{https://ui.adsabs.harvard.edu/abs/1992A%26AS...94..399C}{94, 399}

\bibitem[{{Charbonnel} \& {Lagarde}(2010)}]{Charbonnel2010}
{Charbonnel}, C. \& {Lagarde}, N. 2010,
  {\href{https://dx.doi.org/10.1051/0004-6361/201014432}{\color{magenta}\aap}},
  \href{https://ui.adsabs.harvard.edu/abs/2010A%26A...522A..10C}{522, A10}

\bibitem[{{Condon} \& {Kaplan}(1998)}]{Condon1998}
{Condon}, J.~J. \& {Kaplan}, D.~L. 1998,
  {\href{https://dx.doi.org/10.1086/313128}{\color{magenta}\apjs}},
  \href{https://ui.adsabs.harvard.edu/abs/1998ApJS..117..361C}{117, 361}

\bibitem[{{Condon} {et~al.}(1999){Condon}, {Kaplan}, \& {Terzian}}]{Condon1999}
{Condon}, J.~J., {Kaplan}, D.~L., \& {Terzian}, Y. 1999,
  {\href{https://dx.doi.org/10.1086/313236}{\color{magenta}\apjs}},
  \href{https://ui.adsabs.harvard.edu/abs/1999ApJS..123..219C}{123, 219}

\bibitem[{{Corradi} {et~al.}(2015){Corradi}, {Garc{\'\i}a-Rojas}, {Jones}, \&
  {Rodr{\'\i}guez-Gil}}]{Corradi2015}
{Corradi}, R. L.~M., {Garc{\'\i}a-Rojas}, J., {Jones}, D., \&
  {Rodr{\'\i}guez-Gil}, P. 2015,
  {\href{https://dx.doi.org/10.1088/0004-637X/803/2/99}{\color{magenta}\apj}},
  \href{https://ui.adsabs.harvard.edu/abs/2015ApJ...803...99C}{803, 99}

\bibitem[{{Crowther} {et~al.}(1998){Crowther}, {De Marco}, \&
  {Barlow}}]{Crowther1998}
{Crowther}, P.~A., {De Marco}, O., \& {Barlow}, M.~J. 1998,
  {\href{https://dx.doi.org/10.1046/j.1365-8711.1998.01360.x}{\color{magenta}\mnras}},
  \href{https://ui.adsabs.harvard.edu/abs/1998MNRAS.296..367C}{296, 367}

\bibitem[{{Danehkar}(2014)}]{Danehkar2014b}
{Danehkar}, A. 2014,
  \href{https://ui.adsabs.harvard.edu/abs/2014PhDT........76D}{\href{https://ui.adsabs.harvard.edu/abs/2014PhDT........76D}{\color{magenta}{Evolution
  of Planetary Nebulae with WR-type Central Stars}}}, PhD thesis, Macquarie
  University

\bibitem[{{Danehkar}(2015)}]{Danehkar2015}
{Danehkar}, A. 2015,
  {\href{https://dx.doi.org/10.1088/0004-637X/815/1/35}{\color{magenta}\apj}},
  \href{https://ui.adsabs.harvard.edu/abs/2015ApJ...815...35D}{815, 35}

\bibitem[{{Danehkar}(2018{\natexlab{a}})}]{Danehkar2018a}
{Danehkar}, A. 2018{\natexlab{a}},
  {\href{https://dx.doi.org/10.1017/pasa.2018.1}{\color{magenta}\pasa}},
  \href{https://ui.adsabs.harvard.edu/abs/2018PASA...35....5D}{35, e005}

\bibitem[{{Danehkar}(2018{\natexlab{b}})}]{Danehkar2018b}
{Danehkar}, A. 2018{\natexlab{b}},
  {\href{https://dx.doi.org/10.21105/joss.00899}{\color{magenta}J. Open Source
  Softw.}}, \href{https://ui.adsabs.harvard.edu/abs/2018JOSS....3..899D}{3,
  899}

\bibitem[{{Danehkar}(2019)}]{Danehkar2019}
{Danehkar}, A. 2019,
  {\href{https://dx.doi.org/10.21105/joss.00898}{\color{magenta}J. Open Source
  Softw.}}, \href{https://ui.adsabs.harvard.edu/abs/2019JOSS....4..898D}{4,
  898}

\bibitem[{{Danehkar}(2021)}]{Danehkar2021}
{Danehkar}, A. 2021,
  \href{https://ui.adsabs.harvard.edu/abs/2021arXiv210703994D}{arXiv e-prints,
  arXiv:2107.03994}, \apjs, submitted

\bibitem[{{Danehkar} {et~al.}(2018){Danehkar}, {Karovska}, {Maksym}, \&
  {Montez}}]{Danehkar2018}
{Danehkar}, A., {Karovska}, M., {Maksym}, W.~P., \& {Montez}, Rodolfo, J. 2018,
  {\href{https://dx.doi.org/10.3847/1538-4357/aa9e8c}{\color{magenta}\apj}},
  \href{https://ui.adsabs.harvard.edu/abs/2018ApJ...852...87D}{852, 87}

\bibitem[{{Danehkar} \& {Parker}(2015)}]{Danehkar2015a}
{Danehkar}, A. \& {Parker}, Q.~A. 2015,
  {\href{https://dx.doi.org/10.1093/mnrasl/slv022}{\color{magenta}\mnras}},
  \href{https://ui.adsabs.harvard.edu/abs/2015MNRAS.449L..56D}{449, L56}

\bibitem[{{Danehkar} {et~al.}(2013){Danehkar}, {Parker}, \&
  {Ercolano}}]{Danehkar2013a}
{Danehkar}, A., {Parker}, Q.~A., \& {Ercolano}, B. 2013,
  {\href{https://dx.doi.org/10.1093/mnras/stt1116}{\color{magenta}\mnras}},
  \href{https://ui.adsabs.harvard.edu/abs/2013MNRAS.434.1513D}{434, 1513}

\bibitem[{{Danehkar} {et~al.}(2016){Danehkar}, {Parker}, \&
  {Steffen}}]{Danehkar2016}
{Danehkar}, A., {Parker}, Q.~A., \& {Steffen}, W. 2016,
  {\href{https://dx.doi.org/10.3847/0004-6256/151/2/38}{\color{magenta}\aj}},
  \href{https://ui.adsabs.harvard.edu/abs/2016AJ....151...38D}{151, 38}

\bibitem[{{Danehkar} {et~al.}(2014){Danehkar}, {Todt}, {Ercolano}, \&
  {Kniazev}}]{Danehkar2014a}
{Danehkar}, A., {Todt}, H., {Ercolano}, B., \& {Kniazev}, A.~Y. 2014,
  {\href{https://dx.doi.org/10.1093/mnras/stu203}{\color{magenta}\mnras}},
  \href{https://ui.adsabs.harvard.edu/abs/2014MNRAS.439.3605D}{439, 3605}

\bibitem[{{Davey} {et~al.}(2000){Davey}, {Storey}, \& {Kisielius}}]{Davey2000}
{Davey}, A.~R., {Storey}, P.~J., \& {Kisielius}, R. 2000,
  {\href{https://dx.doi.org/10.1051/aas:2000139}{\color{magenta}\aaps}},
  \href{https://ui.adsabs.harvard.edu/abs/2000A%26AS..142...85D}{142, 85}

\bibitem[{{Daw} {et~al.}(2000){Daw}, {Parkinson}, {Smith}, \&
  {Calamai}}]{Daw2000}
{Daw}, A., {Parkinson}, W.~H., {Smith}, P.~L., \& {Calamai}, A.~G. 2000,
  {\href{https://dx.doi.org/10.1086/312605}{\color{magenta}\apjl}},
  \href{https://ui.adsabs.harvard.edu/abs/2000ApJ...533L.179D}{533, L179}

\bibitem[{{De Marco} {et~al.}(1997){De Marco}, {Barlow}, \&
  {Storey}}]{DeMarco1997}
{De Marco}, O., {Barlow}, M.~J., \& {Storey}, P.~J. 1997, \mnras,
  \href{https://ui.adsabs.harvard.edu/abs/1997MNRAS.292...86D}{292, 86}

\bibitem[{{De Marco} \& {Crowther}(1998)}]{DeMarco1998a}
{De Marco}, O. \& {Crowther}, P.~A. 1998,
  {\href{https://dx.doi.org/10.1046/j.1365-8711.1998.01379.x}{\color{magenta}\mnras}},
  \href{https://ui.adsabs.harvard.edu/abs/1998MNRAS.296..419D}{296, 419}

\bibitem[{{Delgado-Inglada} {et~al.}(2014){Delgado-Inglada}, {Morisset}, \&
  {Stasi{\'n}ska}}]{Delgado-Inglada2014}
{Delgado-Inglada}, G., {Morisset}, C., \& {Stasi{\'n}ska}, G. 2014,
  {\href{https://dx.doi.org/10.1093/mnras/stu341}{\color{magenta}\mnras}},
  \href{https://ui.adsabs.harvard.edu/abs/2014MNRAS.440..536D}{440, 536}

\bibitem[{{Depew} {et~al.}(2011){Depew}, {Parker}, {Miszalski}, {De Marco},
  {Frew}, {Acker}, {Kovacevic}, \& {Sharp}}]{Depew2011}
{Depew}, K., {Parker}, Q.~A., {Miszalski}, B., {et~al.} 2011,
  {\href{https://dx.doi.org/10.1111/j.1365-2966.2011.18337.x}{\color{magenta}\mnras}},
  \href{https://ui.adsabs.harvard.edu/abs/2011MNRAS.414.2812D}{414, 2812}

\bibitem[{{Dere} {et~al.}(2019){Dere}, {Del Zanna}, {Young}, {Landi}, \&
  {Sutherland}}]{Dere2019}
{Dere}, K.~P., {Del Zanna}, G., {Young}, P.~R., {Landi}, E., \& {Sutherland},
  R.~S. 2019,
  {\href{https://dx.doi.org/10.3847/1538-4365/ab05cf}{\color{magenta}\apjs}},
  \href{https://ui.adsabs.harvard.edu/abs/2019ApJS..241...22D}{241, 22}

\bibitem[{{Derlopa} {et~al.}(2019){Derlopa}, {Akras}, {Boumis}, \&
  {Steffen}}]{Derlopa2019}
{Derlopa}, S., {Akras}, S., {Boumis}, P., \& {Steffen}, W. 2019,
  {\href{https://dx.doi.org/10.1093/mnras/stz193}{\color{magenta}\mnras}},
  \href{https://ui.adsabs.harvard.edu/abs/2019MNRAS.484.3746D}{484, 3746}

\bibitem[{{Dopita} {et~al.}(2007){Dopita}, {Hart}, {McGregor}, {Oates},
  {Bloxham}, \& {Jones}}]{Dopita2007}
{Dopita}, M., {Hart}, J., {McGregor}, P., {et~al.} 2007,
  {\href{https://dx.doi.org/10.1007/s10509-007-9510-z}{\color{magenta}\apss}},
  \href{https://ui.adsabs.harvard.edu/abs/2007Ap%26SS.310..255D}{310, 255}

\bibitem[{{Dopita} {et~al.}(2010){Dopita}, {Rhee}, {Farage}, {McGregor},
  {Bloxham}, {Green}, {Roberts}, {Neilson}, {Wilson}, {Young}, {Firth},
  {Busarello}, \& {Merluzzi}}]{Dopita2010}
{Dopita}, M., {Rhee}, J., {Farage}, C., {et~al.} 2010,
  {\href{https://dx.doi.org/10.1007/s10509-010-0335-9}{\color{magenta}\apss}},
  \href{https://ui.adsabs.harvard.edu/abs/2010Ap%26SS.327..245D}{327, 245}

\bibitem[{{Dopita} {et~al.}(2017){Dopita}, {Ali}, {Sutherland}, {Nicholls}, \&
  {Amer}}]{Dopita2017}
{Dopita}, M.~A., {Ali}, A., {Sutherland}, R.~S., {Nicholls}, D.~C., \& {Amer},
  M.~A. 2017,
  {\href{https://dx.doi.org/10.1093/mnras/stx1166}{\color{magenta}\mnras}},
  \href{https://ui.adsabs.harvard.edu/abs/2017MNRAS.470..839D}{470, 839}

\bibitem[{{Dopita} \& {Meatheringham}(1990)}]{Dopita1990}
{Dopita}, M.~A. \& {Meatheringham}, S.~J. 1990,
  {\href{https://dx.doi.org/10.1086/168900}{\color{magenta}\apj}},
  \href{https://ui.adsabs.harvard.edu/abs/1990ApJ...357..140D}{357, 140}

\bibitem[{{Dopita} \& {Meatheringham}(1991)}]{Dopita1991}
{Dopita}, M.~A. \& {Meatheringham}, S.~J. 1991,
  {\href{https://dx.doi.org/10.1086/170377}{\color{magenta}\apj}},
  \href{https://ui.adsabs.harvard.edu/abs/1991ApJ...377..480D}{377, 480}

\bibitem[{{Dopita} {et~al.}(2013){Dopita}, {Sutherland}, {Nicholls}, {Kewley},
  \& {Vogt}}]{Dopita2013}
{Dopita}, M.~A., {Sutherland}, R.~S., {Nicholls}, D.~C., {Kewley}, L.~J., \&
  {Vogt}, F.~P.~A. 2013,
  {\href{https://dx.doi.org/10.1088/0067-0049/208/1/10}{\color{magenta}\apjs}},
  \href{https://ui.adsabs.harvard.edu/abs/2013ApJS..208...10D}{208, 10}

\bibitem[{{Ercolano} {et~al.}(2003){Ercolano}, {Barlow}, {Storey}, {Liu},
  {Rauch}, \& {Werner}}]{Ercolano2003b}
{Ercolano}, B., {Barlow}, M.~J., {Storey}, P.~J., {et~al.} 2003,
  {\href{https://dx.doi.org/10.1046/j.1365-8711.2003.06892.x}{\color{magenta}\mnras}},
  \href{https://ui.adsabs.harvard.edu/abs/2003MNRAS.344.1145E}{344, 1145}

\bibitem[{{Ercolano} {et~al.}(2008){Ercolano}, {Young}, {Drake}, \&
  {Raymond}}]{Ercolano2008}
{Ercolano}, B., {Young}, P.~R., {Drake}, J.~J., \& {Raymond}, J.~C. 2008,
  {\href{https://dx.doi.org/10.1086/524378}{\color{magenta}\apjs}},
  \href{https://ui.adsabs.harvard.edu/abs/2008ApJS..175..534E}{175, 534}

\bibitem[{{Escalante} \& {Morisset}(2005)}]{Escalante2005}
{Escalante}, V. \& {Morisset}, C. 2005,
  {\href{https://dx.doi.org/10.1111/j.1365-2966.2005.09217.x}{\color{magenta}\mnras}},
  \href{https://ui.adsabs.harvard.edu/abs/2005MNRAS.361..813E}{361, 813}

\bibitem[{{Escalante} {et~al.}(2012){Escalante}, {Morisset}, \&
  {Georgiev}}]{Escalante2012}
{Escalante}, V., {Morisset}, C., \& {Georgiev}, L. 2012,
  {\href{https://dx.doi.org/10.1111/j.1365-2966.2012.21862.x}{\color{magenta}\mnras}},
  \href{https://ui.adsabs.harvard.edu/abs/2012MNRAS.426.2318E}{426, 2318}

\bibitem[{{Fang} {et~al.}(2014){Fang}, {Guerrero}, {Marquez-Lugo}, {Toal{\'a}},
  {Arthur}, {Chu}, {Blair}, {Gruendl}, {Hamann}, {Oskinova}, \&
  {Todt}}]{Fang2014}
{Fang}, X., {Guerrero}, M.~A., {Marquez-Lugo}, R.~A., {et~al.} 2014,
  {\href{https://dx.doi.org/10.1088/0004-637X/797/2/100}{\color{magenta}\apj}},
  \href{https://ui.adsabs.harvard.edu/abs/2014ApJ...797..100F}{797, 100}

\bibitem[{{Fang} {et~al.}(2011){Fang}, {Storey}, \& {Liu}}]{Fang2011}
{Fang}, X., {Storey}, P.~J., \& {Liu}, X.-W. 2011,
  {\href{https://dx.doi.org/10.1051/0004-6361/201116511}{\color{magenta}\aap}},
  \href{https://ui.adsabs.harvard.edu/abs/2011A%26A...530A..18F}{530, A18}

\bibitem[{{Fang} {et~al.}(2013){Fang}, {Storey}, \& {Liu}}]{Fang2013a}
{Fang}, X., {Storey}, P.~J., \& {Liu}, X.-W. 2013,
  {\href{https://dx.doi.org/10.1051/0004-6361/201116511e}{\color{magenta}\aap}},
  \href{https://ui.adsabs.harvard.edu/abs/2013A%26A...550C...2F}{550, C2}

\bibitem[{{Fanning}(2011)}]{Fanning2011}
{Fanning}, D.~W. 2011, {Coyote's Guide to Traditional IDL Graphics} (Coyote
  Book Publishing)

\bibitem[{{Ferland}(1992)}]{Ferland1992}
{Ferland}, G.~J. 1992,
  {\href{https://dx.doi.org/10.1086/186349}{\color{magenta}\apjl}},
  \href{https://ui.adsabs.harvard.edu/abs/1992ApJ...389L..63F}{389, L63}

\bibitem[{{Frew} {et~al.}(2013){Frew}, {Boji{\v c}i{\'c}}, \&
  {Parker}}]{Frew2013a}
{Frew}, D.~J., {Boji{\v c}i{\'c}}, I.~S., \& {Parker}, Q.~A. 2013,
  {\href{https://dx.doi.org/10.1093/mnras/sts393}{\color{magenta}\mnras}},
  \href{https://ui.adsabs.harvard.edu/abs/2013MNRAS.431....2F}{431, 2}

\bibitem[{{Frew} {et~al.}(2016){Frew}, {Parker}, \&
  {Boji{\v{c}}i{\'c}}}]{Frew2016}
{Frew}, D.~J., {Parker}, Q.~A., \& {Boji{\v{c}}i{\'c}}, I.~S. 2016,
  {\href{https://dx.doi.org/10.1093/mnras/stv1516}{\color{magenta}\mnras}},
  \href{https://ui.adsabs.harvard.edu/abs/2016MNRAS.455.1459F}{455, 1459}

\bibitem[{{Froese Fischer} \& {Tachiev}(2004)}]{FroeseFischer2004}
{Froese Fischer}, C. \& {Tachiev}, G. 2004,
  {\href{https://dx.doi.org/10.1016/j.adt.2004.02.001}{\color{magenta}\adndt}},
  \href{https://ui.adsabs.harvard.edu/abs/2004ADNDT..87....1F}{87, 1}

\bibitem[{{Froese Fischer} {et~al.}(2006){Froese Fischer}, {Tachiev}, \&
  {Irimia}}]{FroeseFischer2006}
{Froese Fischer}, C., {Tachiev}, G., \& {Irimia}, A. 2006,
  {\href{https://dx.doi.org/10.1016/j.adt.2006.03.001}{\color{magenta}\adndt}},
  \href{https://ui.adsabs.harvard.edu/abs/2006ADNDT..92..607F}{92, 607}

\bibitem[{{Galavis} {et~al.}(1995){Galavis}, {Mendoza}, \&
  {Zeippen}}]{Galavis1995}
{Galavis}, M.~E., {Mendoza}, C., \& {Zeippen}, C.~J. 1995, \aaps,
  \href{https://ui.adsabs.harvard.edu/abs/1995A%26AS..111..347G}{111, 347}

\bibitem[{{Garc{\'\i}a-Rojas} {et~al.}(2016){Garc{\'\i}a-Rojas}, {Corradi},
  {Monteiro}, {Jones}, {Rodr{\'\i}guez-Gil}, \&
  {Cabrera-Lavers}}]{Garcia-Rojas2016}
{Garc{\'\i}a-Rojas}, J., {Corradi}, R. L.~M., {Monteiro}, H., {et~al.} 2016,
  {\href{https://dx.doi.org/10.3847/2041-8205/824/2/L27}{\color{magenta}\apjl}},
  \href{https://ui.adsabs.harvard.edu/abs/2016ApJ...824L..27G}{824, L27}

\bibitem[{{Garc{\'{\i}}a-Rojas} {et~al.}(2013){Garc{\'{\i}}a-Rojas},
  {Pe{\~n}a}, {Morisset}, {Delgado-Inglada}, {Mesa-Delgado}, \&
  {Ruiz}}]{Garcia-Rojas2013}
{Garc{\'{\i}}a-Rojas}, J., {Pe{\~n}a}, M., {Morisset}, C., {et~al.} 2013,
  {\href{https://dx.doi.org/10.1051/0004-6361/201322354}{\color{magenta}\aap}},
  \href{https://ui.adsabs.harvard.edu/abs/2013A%26A...558A.122G}{558, A122}

\bibitem[{{Garc{\'{\i}}a-Rojas} {et~al.}(2012){Garc{\'{\i}}a-Rojas},
  {Pe{\~n}a}, {Morisset}, {Mesa-Delgado}, \& {Ruiz}}]{Garcia-Rojas2012}
{Garc{\'{\i}}a-Rojas}, J., {Pe{\~n}a}, M., {Morisset}, C., {Mesa-Delgado}, A.,
  \& {Ruiz}, M.~T. 2012,
  {\href{https://dx.doi.org/10.1051/0004-6361/201118217}{\color{magenta}\aap}},
  \href{https://ui.adsabs.harvard.edu/abs/2012A%26A...538A..54G}{538, A54}

\bibitem[{{Garc{\'{\i}}a-Rojas} {et~al.}(2009){Garc{\'{\i}}a-Rojas},
  {Pe{\~n}a}, \& {Peimbert}}]{Garcia-Rojas2009}
{Garc{\'{\i}}a-Rojas}, J., {Pe{\~n}a}, M., \& {Peimbert}, A. 2009,
  {\href{https://dx.doi.org/10.1051/0004-6361:200811185}{\color{magenta}\aap}},
  \href{https://ui.adsabs.harvard.edu/abs/2009A%26A...496..139G}{496, 139}

\bibitem[{{Garnett}(1992)}]{Garnett1992}
{Garnett}, D.~R. 1992,
  {\href{https://dx.doi.org/10.1086/116146}{\color{magenta}\aj}},
  \href{https://ui.adsabs.harvard.edu/abs/1992AJ....103.1330G}{103, 1330}

\bibitem[{{Girard} {et~al.}(2007){Girard}, {K{\"o}ppen}, \&
  {Acker}}]{Girard2007}
{Girard}, P., {K{\"o}ppen}, J., \& {Acker}, A. 2007,
  {\href{https://dx.doi.org/10.1051/0004-6361:20052807}{\color{magenta}\aap}},
  \href{https://ui.adsabs.harvard.edu/abs/2007A%26A...463..265G}{463, 265}

\bibitem[{{G{\'o}mez-Llanos} \& {Morisset}(2020)}]{Gomez-Llanos2020}
{G{\'o}mez-Llanos}, V. \& {Morisset}, C. 2020,
  {\href{https://dx.doi.org/10.1093/mnras/staa2157}{\color{magenta}\mnras}},
  \href{https://ui.adsabs.harvard.edu/abs/2020MNRAS.497.3363G}{497, 3363}

\bibitem[{{Goodman} \& {Weare}(2010)}]{Goodman2010}
{Goodman}, J. \& {Weare}, J. 2010, Comm. App. Math. \& Comp. Sci., 5, 5

\bibitem[{{Gorny} {et~al.}(1997){Gorny}, {Stasi{\'n}ska}, \&
  {Tylenda}}]{Gorny1997}
{Gorny}, S.~K., {Stasi{\'n}ska}, G., \& {Tylenda}, R. 1997, \aap,
  \href{https://ui.adsabs.harvard.edu/abs/1997A%26A...318..256G}{318, 256}

\bibitem[{{Guerrero} \& {Manchado}(1996)}]{Guerrero1996}
{Guerrero}, M.~A. \& {Manchado}, A. 1996,
  {\href{https://dx.doi.org/10.1086/178101}{\color{magenta}\apj}},
  \href{https://ui.adsabs.harvard.edu/abs/1996ApJ...472..711G}{472, 711}

\bibitem[{{Hajian} {et~al.}(1997){Hajian}, {Balick}, {Terzian}, \&
  {Perinotto}}]{Hajian1997}
{Hajian}, A.~R., {Balick}, B., {Terzian}, Y., \& {Perinotto}, M. 1997,
  {\href{https://dx.doi.org/10.1086/304598}{\color{magenta}\apj}},
  \href{https://ui.adsabs.harvard.edu/abs/1997ApJ...487..304H}{487, 304}

\bibitem[{{Hambly} {et~al.}(2001){Hambly}, {MacGillivray}, {Read}, {Tritton},
  {Thomson}, {Kelly}, {Morgan}, {Smith}, {Driver}, {Williamson}, {Parker},
  {Hawkins}, {Williams}, \& {Lawrence}}]{Hambly2001}
{Hambly}, N.~C., {MacGillivray}, H.~T., {Read}, M.~A., {et~al.} 2001,
  {\href{https://dx.doi.org/10.1111/j.1365-8711.2001.04660.x}{\color{magenta}\mnras}},
  \href{https://ui.adsabs.harvard.edu/abs/2001MNRAS.326.1279H}{326, 1279}

\bibitem[{{Harris} {et~al.}(2020){Harris}, {Millman}, {van der Walt},
  {Gommers}, {Virtanen}, {Cournapeau}, {Wieser}, {Taylor}, {Berg}, {Smith},
  {Kern}, {Picus}, {Hoyer}, {van Kerkwijk}, {Brett}, {Haldane}, {del R{\'\i}o},
  {Wiebe}, {Peterson}, {G{\'e}rard-Marchant}, {Sheppard}, {Reddy}, {Weckesser},
  {Abbasi}, {Gohlke}, \& {Oliphant}}]{Harris2020}
{Harris}, C.~R., {Millman}, K.~J., {van der Walt}, S.~J., {et~al.} 2020,
  {\href{https://dx.doi.org/10.1038/s41586-020-2649-2}{\color{magenta}\nat}},
  \href{https://ui.adsabs.harvard.edu/abs/2020Natur.585..357H}{585, 357}

\bibitem[{{Hazard} {et~al.}(1980){Hazard}, {Terlevich}, {Morton}, {Sargent}, \&
  {Ferland}}]{Hazard1980}
{Hazard}, C., {Terlevich}, R., {Morton}, D.~C., {Sargent}, W.~L.~W., \&
  {Ferland}, G. 1980,
  {\href{https://dx.doi.org/10.1038/285463a0}{\color{magenta}\nat}},
  \href{https://ui.adsabs.harvard.edu/abs/1980Natur.285..463H}{285, 463}

\bibitem[{{Henney} \& {Stasi{\'n}ska}(2010)}]{Henney2010}
{Henney}, W.~J. \& {Stasi{\'n}ska}, G. 2010,
  {\href{https://dx.doi.org/10.1088/0004-637X/711/2/881}{\color{magenta}\apj}},
  \href{https://ui.adsabs.harvard.edu/abs/2010ApJ...711..881H}{711, 881}

\bibitem[{{Henry} {et~al.}(2004){Henry}, {Kwitter}, \& {Balick}}]{Henry2004}
{Henry}, R.~B.~C., {Kwitter}, K.~B., \& {Balick}, B. 2004,
  {\href{https://dx.doi.org/10.1086/382242}{\color{magenta}\aj}},
  \href{https://ui.adsabs.harvard.edu/abs/2004AJ....127.2284H}{127, 2284}

\bibitem[{{Henry} {et~al.}(2012){Henry}, {Speck}, {Karakas}, {Ferland}, \&
  {Maguire}}]{Henry2012}
{Henry}, R.~B.~C., {Speck}, A., {Karakas}, A.~I., {Ferland}, G.~J., \&
  {Maguire}, M. 2012,
  {\href{https://dx.doi.org/10.1088/0004-637X/749/1/61}{\color{magenta}\apj}},
  \href{https://ui.adsabs.harvard.edu/abs/2012ApJ...749...61H}{749, 61}

\bibitem[{{Herwig}(2001)}]{Herwig2001}
{Herwig}, F. 2001, \apss,
  \href{https://ui.adsabs.harvard.edu/abs/2001Ap%26SS.275...15H}{275, 15}

\bibitem[{{Herwig}(2005)}]{Herwig2005}
{Herwig}, F. 2005,
  {\href{https://dx.doi.org/10.1146/annurev.astro.43.072103.150600}{\color{magenta}\araa}},
  \href{https://ui.adsabs.harvard.edu/abs/2005ARA%26A..43..435H}{43, 435}

\bibitem[{{Howarth}(1983)}]{Howarth1983}
{Howarth}, I.~D. 1983, \mnras,
  \href{https://ui.adsabs.harvard.edu/abs/1983MNRAS.203..301H}{203, 301}

\bibitem[{{Hudson} {et~al.}(2012){Hudson}, {Ramsbottom}, \&
  {Scott}}]{Hudson2012}
{Hudson}, C.~E., {Ramsbottom}, C.~A., \& {Scott}, M.~P. 2012,
  {\href{https://dx.doi.org/10.1088/0004-637X/750/1/65}{\color{magenta}\apj}},
  \href{https://ui.adsabs.harvard.edu/abs/2012ApJ...750...65H}{750, 65}

\bibitem[{{Hunter}(2007)}]{Hunter2007}
{Hunter}, J.~D. 2007,
  {\href{https://dx.doi.org/10.1109/MCSE.2007.55}{\color{magenta}Comput. Sci.
  Eng.}}, \href{https://ui.adsabs.harvard.edu/abs/2007CSE.....9...90H}{9, 90}

\bibitem[{{Iben} {et~al.}(1983){Iben}, {Kaler}, {Truran}, \&
  {Renzini}}]{Iben1983b}
{Iben}, Jr., I., {Kaler}, J.~B., {Truran}, J.~W., \& {Renzini}, A. 1983,
  {\href{https://dx.doi.org/10.1086/160631}{\color{magenta}\apj}},
  \href{https://ui.adsabs.harvard.edu/abs/1983ApJ...264..605I}{264, 605}

\bibitem[{{Iben} \& {Renzini}(1983)}]{Iben1983}
{Iben}, Jr., I. \& {Renzini}, A. 1983,
  {\href{https://dx.doi.org/10.1146/annurev.aa.21.090183.001415}{\color{magenta}\araa}},
  \href{https://ui.adsabs.harvard.edu/abs/1983ARA%26A..21..271I}{21, 271}

\bibitem[{{Izotov} {et~al.}(1994){Izotov}, {Thuan}, \&
  {Lipovetsky}}]{Izotov1994}
{Izotov}, Y.~I., {Thuan}, T.~X., \& {Lipovetsky}, V.~A. 1994,
  {\href{https://dx.doi.org/10.1086/174843}{\color{magenta}\apj}},
  \href{https://ui.adsabs.harvard.edu/abs/1994ApJ...435..647I}{435, 647}

\bibitem[{{Jacoby} \& {Ford}(1983)}]{Jacoby1983}
{Jacoby}, G.~H. \& {Ford}, H.~C. 1983,
  {\href{https://dx.doi.org/10.1086/160779}{\color{magenta}\apj}},
  \href{https://ui.adsabs.harvard.edu/abs/1983ApJ...266..298J}{266, 298}

\bibitem[{{Jacoby} {et~al.}(2020){Jacoby}, {Hillwig}, \& {Jones}}]{Jacoby2020}
{Jacoby}, G.~H., {Hillwig}, T.~C., \& {Jones}, D. 2020,
  {\href{https://dx.doi.org/10.1093/mnrasl/slaa138}{\color{magenta}\mnras}},
  \href{https://ui.adsabs.harvard.edu/abs/2020MNRAS.498L.114J}{498, L114}

\bibitem[{{Jones} {et~al.}(2016){Jones}, {Wesson}, {Garc{\'\i}a-Rojas},
  {Corradi}, \& {Boffin}}]{Jones2016}
{Jones}, D., {Wesson}, R., {Garc{\'\i}a-Rojas}, J., {Corradi}, R.~L.~M., \&
  {Boffin}, H.~M.~J. 2016,
  {\href{https://dx.doi.org/10.1093/mnras/stv2519}{\color{magenta}\mnras}},
  \href{https://ui.adsabs.harvard.edu/abs/2016MNRAS.455.3263J}{455, 3263}

\bibitem[{{Kaler} {et~al.}(1991){Kaler}, {Shaw}, {Feibelman}, \&
  {Imhoff}}]{Kaler1991}
{Kaler}, J.~B., {Shaw}, R.~A., {Feibelman}, W.~A., \& {Imhoff}, C.~L. 1991,
  {\href{https://dx.doi.org/10.1086/132796}{\color{magenta}\pasp}},
  \href{https://ui.adsabs.harvard.edu/abs/1991PASP..103...67K}{103, 67}

\bibitem[{{Karakas}(2010)}]{Karakas2010a}
{Karakas}, A.~I. 2010,
  {\href{https://dx.doi.org/10.1111/j.1365-2966.2009.16198.x}{\color{magenta}\mnras}},
  \href{https://ui.adsabs.harvard.edu/abs/2010MNRAS.403.1413K}{403, 1413}

\bibitem[{{Karakas}(2014)}]{Karakas2014a}
{Karakas}, A.~I. 2014,
  {\href{https://dx.doi.org/10.1093/mnras/stu1727}{\color{magenta}\mnras}},
  \href{https://ui.adsabs.harvard.edu/abs/2014MNRAS.445..347K}{445, 347}

\bibitem[{{Karakas} \& {Lattanzio}(2003)}]{Karakas2003}
{Karakas}, A.~I. \& {Lattanzio}, J.~C. 2003,
  {\href{https://dx.doi.org/10.1071/AS03059}{\color{magenta}\pasa}},
  \href{https://ui.adsabs.harvard.edu/abs/2003PASA...20..393K}{20, 393}

\bibitem[{{Karakas} \& {Lattanzio}(2014)}]{Karakas2014}
{Karakas}, A.~I. \& {Lattanzio}, J.~C. 2014,
  {\href{https://dx.doi.org/10.1017/pasa.2014.21}{\color{magenta}\pasa}},
  \href{https://ui.adsabs.harvard.edu/abs/2014PASA...31...30K}{31, 30}

\bibitem[{{Karakas} \& {Lugaro}(2016)}]{Karakas2016}
{Karakas}, A.~I. \& {Lugaro}, M. 2016,
  {\href{https://dx.doi.org/10.3847/0004-637X/825/1/26}{\color{magenta}\apj}},
  \href{https://ui.adsabs.harvard.edu/abs/2016ApJ...825...26K}{825, 26}

\bibitem[{{Karakas} {et~al.}(2009){Karakas}, {van Raai}, {Lugaro}, {Sterling},
  \& {Dinerstein}}]{Karakas2009}
{Karakas}, A.~I., {van Raai}, M.~A., {Lugaro}, M., {Sterling}, N.~C., \&
  {Dinerstein}, H.~L. 2009,
  {\href{https://dx.doi.org/10.1088/0004-637X/690/2/1130}{\color{magenta}\apj}},
  \href{https://ui.adsabs.harvard.edu/abs/2009ApJ...690.1130K}{690, 1130}

\bibitem[{{Kingsburgh} \& {Barlow}(1994)}]{Kingsburgh1994}
{Kingsburgh}, R.~L. \& {Barlow}, M.~J. 1994, \mnras,
  \href{https://ui.adsabs.harvard.edu/abs/1994MNRAS.271..257K}{271, 257}

\bibitem[{{Kisielius} {et~al.}(2009){Kisielius}, {Storey}, {Ferland}, \&
  {Keenan}}]{Kisielius2009}
{Kisielius}, R., {Storey}, P.~J., {Ferland}, G.~J., \& {Keenan}, F.~P. 2009,
  {\href{https://dx.doi.org/10.1111/j.1365-2966.2009.14989.x}{\color{magenta}\mnras}},
  \href{https://ui.adsabs.harvard.edu/abs/2009MNRAS.397..903K}{397, 903}

\bibitem[{{Koesterke}(2001)}]{Koesterke2001}
{Koesterke}, L. 2001, \apss,
  \href{https://ui.adsabs.harvard.edu/abs/2001Ap%26SS.275...41K}{275, 41}

\bibitem[{{Koesterke} \& {Hamann}(1997)}]{Koesterke1997}
{Koesterke}, L. \& {Hamann}, W.-R. 1997, \aap,
  \href{https://ui.adsabs.harvard.edu/abs/1997A%26A...320...91K}{320, 91}

\bibitem[{{Kwitter} \& {Henry}(2001)}]{Kwitter2001}
{Kwitter}, K.~B. \& {Henry}, R.~B.~C. 2001,
  {\href{https://dx.doi.org/10.1086/322505}{\color{magenta}\apj}},
  \href{https://ui.adsabs.harvard.edu/abs/2001ApJ...562..804K}{562, 804}

\bibitem[{{Kwitter} {et~al.}(2003){Kwitter}, {Henry}, \&
  {Milingo}}]{Kwitter2003}
{Kwitter}, K.~B., {Henry}, R.~B.~C., \& {Milingo}, J.~B. 2003,
  {\href{https://dx.doi.org/10.1086/345108}{\color{magenta}\pasp}},
  \href{https://ui.adsabs.harvard.edu/abs/2003PASP..115...80K}{115, 80}

\bibitem[{{Landi} \& {Bhatia}(2005)}]{Landi2005}
{Landi}, E. \& {Bhatia}, A.~K. 2005,
  {\href{https://dx.doi.org/10.1016/j.adt.2005.02.003}{\color{magenta}\adndt}},
  \href{https://ui.adsabs.harvard.edu/abs/2005ADNDT..89..195L}{89, 195}

\bibitem[{{Landi} {et~al.}(2012){Landi}, {Del Zanna}, {Young}, {Dere}, \&
  {Mason}}]{Landi2012}
{Landi}, E., {Del Zanna}, G., {Young}, P.~R., {Dere}, K.~P., \& {Mason}, H.~E.
  2012,
  {\href{https://dx.doi.org/10.1088/0004-637X/744/2/99}{\color{magenta}\apj}},
  \href{https://ui.adsabs.harvard.edu/abs/2012ApJ...744...99L}{744, 99}

\bibitem[{{Landman} {et~al.}(1982){Landman}, {Roussel-Dupre}, \&
  {Tanigawa}}]{Landman1982}
{Landman}, D.~A., {Roussel-Dupre}, R., \& {Tanigawa}, G. 1982,
  {\href{https://dx.doi.org/10.1086/160383}{\color{magenta}\apj}},
  \href{https://ui.adsabs.harvard.edu/abs/1982ApJ...261..732L}{261, 732}

\bibitem[{{Landsman}(1993)}]{Landsman1993}
{Landsman}, W.~B. 1993, in ASP Conf. Ser., Vol.~52, Astronomical Data Analysis
  Software and Systems II, ed. R.~J. {Hanisch}, R.~J.~V. {Brissenden}, \&
  J.~{Barnes} (San Francisco, CA: ASP),
  \href{https://ui.adsabs.harvard.edu/abs/1993ASPC...52..246L}{246}

\bibitem[{{Lau} {et~al.}(2011){Lau}, {De Marco}, \& {Liu}}]{Lau2011}
{Lau}, H. H.~B., {De Marco}, O., \& {Liu}, X.~W. 2011,
  {\href{https://dx.doi.org/10.1111/j.1365-2966.2010.17568.x}{\color{magenta}\mnras}},
  \href{https://ui.adsabs.harvard.edu/abs/2011MNRAS.410.1870L}{410, 1870}

\bibitem[{{Lennon} \& {Burke}(1994)}]{Lennon1994}
{Lennon}, D.~J. \& {Burke}, V.~M. 1994, \aaps,
  \href{https://ui.adsabs.harvard.edu/abs/1994A%26AS..103..273L}{103, 273}

\bibitem[{{Lenz} \& {Ayres}(1992)}]{Lenz1992}
{Lenz}, D.~D. \& {Ayres}, T.~R. 1992,
  {\href{https://dx.doi.org/10.1086/133096}{\color{magenta}\pasp}},
  \href{https://ui.adsabs.harvard.edu/abs/1992PASP..104.1104L}{104, 1104}

\bibitem[{{Leuenhagen} \& {Hamann}(1998)}]{Leuenhagen1998}
{Leuenhagen}, U. \& {Hamann}, W.-R. 1998, \aap,
  \href{https://ui.adsabs.harvard.edu/abs/1998A%26A...330..265L}{330, 265}

\bibitem[{{Leuenhagen} {et~al.}(1996){Leuenhagen}, {Hamann}, \&
  {Jeffery}}]{Leuenhagen1996}
{Leuenhagen}, U., {Hamann}, W.-R., \& {Jeffery}, C.~S. 1996, \aap,
  \href{https://ui.adsabs.harvard.edu/abs/1996A%26A...312..167L}{312, 167}

\bibitem[{{Liu}(2003)}]{Liu2003}
{Liu}, X.-W. 2003, in IAU Symposium, Vol. 209, Planetary Nebulae: Their
  Evolution and Role in the Universe, ed. S.~{Kwok}, M.~{Dopita}, \&
  R.~{Sutherland},
  \href{https://ui.adsabs.harvard.edu/abs/2003IAUS..209..339L}{339}

\bibitem[{{Liu} {et~al.}(2001){Liu}, {Luo}, {Barlow}, {Danziger}, \&
  {Storey}}]{Liu2001}
{Liu}, X.-W., {Luo}, S.-G., {Barlow}, M.~J., {Danziger}, I.~J., \& {Storey},
  P.~J. 2001,
  {\href{https://dx.doi.org/10.1046/j.1365-8711.2001.04676.x}{\color{magenta}\mnras}},
  \href{https://ui.adsabs.harvard.edu/abs/2001MNRAS.327..141L}{327, 141}

\bibitem[{{Liu} {et~al.}(2000){Liu}, {Storey}, {Barlow}, {Danziger}, {Cohen},
  \& {Bryce}}]{Liu2000}
{Liu}, X.-W., {Storey}, P.~J., {Barlow}, M.~J., {et~al.} 2000,
  {\href{https://dx.doi.org/10.1046/j.1365-8711.2000.03167.x}{\color{magenta}\mnras}},
  \href{https://ui.adsabs.harvard.edu/abs/2000MNRAS.312..585L}{312, 585}

\bibitem[{{Liu} {et~al.}(2004){Liu}, {Liu}, {Barlow}, \& {Luo}}]{Liu2004b}
{Liu}, Y., {Liu}, X.-W., {Barlow}, M.~J., \& {Luo}, S.-G. 2004,
  {\href{https://dx.doi.org/10.1111/j.1365-2966.2004.08156.x}{\color{magenta}\mnras}},
  \href{https://ui.adsabs.harvard.edu/abs/2004MNRAS.353.1251L}{353, 1251}

\bibitem[{{Luo} {et~al.}(2001){Luo}, {Liu}, \& {Barlow}}]{Luo2001}
{Luo}, S.-G., {Liu}, X.-W., \& {Barlow}, M.~J. 2001,
  {\href{https://dx.doi.org/10.1046/j.1365-8711.2001.04631.x}{\color{magenta}\mnras}},
  \href{https://ui.adsabs.harvard.edu/abs/2001MNRAS.326.1049L}{326, 1049}

\bibitem[{{Marigo}(2002)}]{Marigo2002}
{Marigo}, P. 2002,
  {\href{https://dx.doi.org/10.1051/0004-6361:20020304}{\color{magenta}\aap}},
  \href{https://ui.adsabs.harvard.edu/abs/2002A%26A...387..507M}{387, 507}

\bibitem[{{McLaughlin} \& {Bell}(2000)}]{McLaughlin2000}
{McLaughlin}, B.~M. \& {Bell}, K.~L. 2000,
  {\href{https://dx.doi.org/10.1088/0953-4075/33/4/301}{\color{magenta}J. Phys.
  B}}, \href{https://ui.adsabs.harvard.edu/abs/2000JPhB...33..597M}{33, 597}

\bibitem[{{McNabb} {et~al.}(2013){McNabb}, {Fang}, {Liu}, {Bastin}, \&
  {Storey}}]{McNabb2013}
{McNabb}, I.~A., {Fang}, X., {Liu}, X.-W., {Bastin}, R.~J., \& {Storey}, P.~J.
  2013,
  {\href{https://dx.doi.org/10.1093/mnras/sts283}{\color{magenta}\mnras}},
  \href{https://ui.adsabs.harvard.edu/abs/2013MNRAS.428.3443M}{428, 3443}

\bibitem[{{Mendoza} \& {Bautista}(2014)}]{Mendoza2014}
{Mendoza}, C. \& {Bautista}, M.~A. 2014,
  {\href{https://dx.doi.org/10.1088/0004-637X/785/2/91}{\color{magenta}\apj}},
  \href{https://ui.adsabs.harvard.edu/abs/2014ApJ...785...91M}{785, 91}

\bibitem[{{Mendoza} \& {Zeippen}(1982)}]{Mendoza1982}
{Mendoza}, C. \& {Zeippen}, C.~J. 1982, \mnras,
  \href{https://ui.adsabs.harvard.edu/abs/1982MNRAS.198..127M}{198, 127}

\bibitem[{{Merkelis} {et~al.}(1999){Merkelis}, {Martinson}, {Kisielius}, \&
  {Vilkas}}]{Merkelis1999}
{Merkelis}, G., {Martinson}, I., {Kisielius}, R., \& {Vilkas}, M.~J. 1999,
  {\href{https://dx.doi.org/10.1238/Physica.Regular.059a00122}{\color{magenta}\physscr}},
  \href{https://ui.adsabs.harvard.edu/abs/1999PhyS...59..122M}{59, 122}

\bibitem[{{Milingo} {et~al.}(2002){Milingo}, {Kwitter}, {Henry}, \&
  {Cohen}}]{Milingo2002a}
{Milingo}, J.~B., {Kwitter}, K.~B., {Henry}, R.~B.~C., \& {Cohen}, R.~E. 2002,
  {\href{https://dx.doi.org/10.1086/324291}{\color{magenta}\apjs}},
  \href{https://ui.adsabs.harvard.edu/abs/2002ApJS..138..279M}{138, 279}

\bibitem[{{Milne} \& {Aller}(1975)}]{Milne1975}
{Milne}, D.~K. \& {Aller}, L.~H. 1975, \aap,
  \href{https://ui.adsabs.harvard.edu/abs/1975A%26A....38..183M}{38, 183}

\bibitem[{{Nicholls} {et~al.}(2012){Nicholls}, {Dopita}, \&
  {Sutherland}}]{Nicholls2012}
{Nicholls}, D.~C., {Dopita}, M.~A., \& {Sutherland}, R.~S. 2012,
  {\href{https://dx.doi.org/10.1088/0004-637X/752/2/148}{\color{magenta}\apj}},
  \href{https://ui.adsabs.harvard.edu/abs/2012ApJ...752..148N}{752, 148}

\bibitem[{{Nicholls} {et~al.}(2013){Nicholls}, {Dopita}, {Sutherland},
  {Kewley}, \& {Palay}}]{Nicholls2013}
{Nicholls}, D.~C., {Dopita}, M.~A., {Sutherland}, R.~S., {Kewley}, L.~J., \&
  {Palay}, E. 2013,
  {\href{https://dx.doi.org/10.1088/0067-0049/207/2/21}{\color{magenta}\apjs}},
  \href{https://ui.adsabs.harvard.edu/abs/2013ApJS..207...21N}{207, 21}

\bibitem[{{O'Dell}(1962)}]{Odell1962}
{O'Dell}, C.~R. 1962,
  {\href{https://dx.doi.org/10.1086/147277}{\color{magenta}\apj}},
  \href{https://ui.adsabs.harvard.edu/abs/1962ApJ...135..371O}{135, 371}

\bibitem[{{Parker} {et~al.}(2005){Parker}, {Phillipps}, {Pierce}, \&
  et~al.}]{Parker2005}
{Parker}, Q.~A., {Phillipps}, S., {Pierce}, M., \& et~al. 2005,
  {\href{https://dx.doi.org/10.1111/j.1365-2966.2005.09350.x}{\color{magenta}\mnras}},
  \href{https://ui.adsabs.harvard.edu/abs/2005MNRAS.362..689P}{362, 689}

\bibitem[{{Pe{\~n}a} {et~al.}(1998){Pe{\~n}a}, {Stasi{\'n}ska}, {Esteban},
  {Koesterke}, {Medina}, \& {Kingsburgh}}]{Pena1998}
{Pe{\~n}a}, M., {Stasi{\'n}ska}, G., {Esteban}, C., {et~al.} 1998, \aap,
  \href{https://ui.adsabs.harvard.edu/abs/1998A%26A...337..866P}{337, 866}

\bibitem[{{Pe{\~n}a} {et~al.}(2001){Pe{\~n}a}, {Stasi{\'n}ska}, \&
  {Medina}}]{Pena2001}
{Pe{\~n}a}, M., {Stasi{\'n}ska}, G., \& {Medina}, S. 2001,
  {\href{https://dx.doi.org/10.1051/0004-6361:20000497}{\color{magenta}\aap}},
  \href{https://ui.adsabs.harvard.edu/abs/2001A%26A...367..983P}{367, 983}

\bibitem[{{Peimbert}(1967)}]{Peimbert1967}
{Peimbert}, M. 1967,
  {\href{https://dx.doi.org/10.1086/149385}{\color{magenta}\apj}},
  \href{https://ui.adsabs.harvard.edu/abs/1967ApJ...150..825P}{150, 825}

\bibitem[{{Peimbert}(1971)}]{Peimbert1971}
{Peimbert}, M. 1971, Boletin de los Observatorios Tonantzintla y Tacubaya,
  \href{https://ui.adsabs.harvard.edu/abs/1971BOTT....6...29P}{6, 29}

\bibitem[{{Peimbert} \& {Torres-Peimbert}(1983)}]{Peimbert1983}
{Peimbert}, M. \& {Torres-Peimbert}, S. 1983, in IAU Symposium, Vol. 103,
  Planetary Nebulae, ed. D.~R. {Flower},
  \href{https://ui.adsabs.harvard.edu/abs/1983IAUS..103..233P}{233--241}

\bibitem[{{P{\'e}quignot} {et~al.}(1991){P{\'e}quignot}, {Petitjean}, \&
  {Boisson}}]{Pequignot1991}
{P{\'e}quignot}, D., {Petitjean}, P., \& {Boisson}, C. 1991, \aap,
  \href{https://ui.adsabs.harvard.edu/abs/1991A%26A...251..680P}{251, 680}

\bibitem[{{P{\'e}quignot} {et~al.}(2000){P{\'e}quignot}, {Walsh}, {Zijlstra},
  \& {Dudziak}}]{Pequignot2000}
{P{\'e}quignot}, D., {Walsh}, J.~R., {Zijlstra}, A.~A., \& {Dudziak}, G. 2000,
  \aap, \href{https://ui.adsabs.harvard.edu/abs/2000A%26A...361L...1P}{361, L1}

\bibitem[{{Pollacco} {et~al.}(1992){Pollacco}, {Lawson}, {Clegg}, \&
  {Hill}}]{Pollacco1992}
{Pollacco}, D.~L., {Lawson}, W.~A., {Clegg}, R.~E.~S., \& {Hill}, P.~W. 1992,
  {\href{https://dx.doi.org/10.1093/mnras/257.1.33P}{\color{magenta}\mnras}},
  \href{https://ui.adsabs.harvard.edu/abs/1992MNRAS.257P..33P}{257, 33P}

\bibitem[{{Porter} {et~al.}(2013){Porter}, {Ferland}, {Storey}, \&
  {Detisch}}]{Porter2013}
{Porter}, R.~L., {Ferland}, G.~J., {Storey}, P.~J., \& {Detisch}, M.~J. 2013,
  {\href{https://dx.doi.org/10.1093/mnrasl/slt049}{\color{magenta}\mnras}},
  \href{https://ui.adsabs.harvard.edu/abs/2013MNRAS.433L..89P}{433, L89}

\bibitem[{{Pottasch} \& {Bernard-Salas}(2006)}]{Pottasch2006}
{Pottasch}, S.~R. \& {Bernard-Salas}, J. 2006,
  {\href{https://dx.doi.org/10.1051/0004-6361:20065548}{\color{magenta}\aap}},
  \href{https://ui.adsabs.harvard.edu/abs/2006A%26A...457..189P}{457, 189}

\bibitem[{{Pottasch} {et~al.}(2011){Pottasch}, {Surendiranath}, \&
  {Bernard-Salas}}]{Pottasch2011}
{Pottasch}, S.~R., {Surendiranath}, R., \& {Bernard-Salas}, J. 2011,
  {\href{https://dx.doi.org/10.1051/0004-6361/201116669}{\color{magenta}\aap}},
  \href{https://ui.adsabs.harvard.edu/abs/2011A%26A...531A..23P}{531, A23}

\bibitem[{{Preite-Martinez} {et~al.}(1989){Preite-Martinez}, {Acker},
  {Koeppen}, \& {Stenholm}}]{Preite-Martinez1989}
{Preite-Martinez}, A., {Acker}, A., {Koeppen}, J., \& {Stenholm}, B. 1989,
  \aaps, \href{https://ui.adsabs.harvard.edu/abs/1989A%26AS...81..309P}{81,
  309}

\bibitem[{{Purton} {et~al.}(1982){Purton}, {Feldman}, {Marsh}, {Allen}, \&
  {Wright}}]{Purton1982}
{Purton}, C.~R., {Feldman}, P.~A., {Marsh}, K.~A., {Allen}, D.~A., \& {Wright},
  A.~E. 1982, \mnras,
  \href{https://ui.adsabs.harvard.edu/abs/1982MNRAS.198..321P}{198, 321}

\bibitem[{{Ramsbottom} \& {Bell}(1997)}]{Ramsbottom1997a}
{Ramsbottom}, C.~A. \& {Bell}, K.~L. 1997,
  {\href{https://dx.doi.org/10.1006/adnd.1997.0741}{\color{magenta}\adndt}},
  \href{https://ui.adsabs.harvard.edu/abs/1997ADNDT..66...65R}{66, 65}

\bibitem[{{Ramsbottom} {et~al.}(1997){Ramsbottom}, {Bell}, \&
  {Keenan}}]{Ramsbottom1997}
{Ramsbottom}, C.~A., {Bell}, K.~L., \& {Keenan}, F.~P. 1997, \mnras,
  \href{https://ui.adsabs.harvard.edu/abs/1997MNRAS.284..754R}{284, 754}

\bibitem[{{Ramsbottom} {et~al.}(1998){Ramsbottom}, {Bell}, \&
  {Keenan}}]{Ramsbottom1998}
{Ramsbottom}, C.~A., {Bell}, K.~L., \& {Keenan}, F.~P. 1998,
  {\href{https://dx.doi.org/10.1046/j.1365-8711.1998.01054.x}{\color{magenta}\mnras}},
  \href{https://ui.adsabs.harvard.edu/abs/1998MNRAS.293..233R}{293, 233}

\bibitem[{{Ramsbottom} {et~al.}(2001){Ramsbottom}, {Bell}, \&
  {Keenan}}]{Ramsbottom2001}
{Ramsbottom}, C.~A., {Bell}, K.~L., \& {Keenan}, F.~P. 2001,
  {\href{https://dx.doi.org/10.1006/adnd.2000.0846}{\color{magenta}\adndt}},
  \href{https://ui.adsabs.harvard.edu/abs/2001ADNDT..77...57R}{77, 57}

\bibitem[{{Ramsbottom} {et~al.}(1996){Ramsbottom}, {Bell}, \&
  {Stafford}}]{Ramsbottom1996}
{Ramsbottom}, C.~A., {Bell}, K.~L., \& {Stafford}, R.~P. 1996,
  {\href{https://dx.doi.org/10.1006/adnd.1996.0009}{\color{magenta}\adndt}},
  \href{https://ui.adsabs.harvard.edu/abs/1996ADNDT..63...57R}{63, 57}

\bibitem[{{Rodr{\'\i}guez}(2020)}]{Rodriguez2020}
{Rodr{\'\i}guez}, M. 2020,
  {\href{https://dx.doi.org/10.1093/mnras/staa1286}{\color{magenta}\mnras}},
  \href{https://ui.adsabs.harvard.edu/abs/2020MNRAS.495.1016R}{495, 1016}

\bibitem[{{Sawey} \& {Berrington}(1993)}]{Sawey1993}
{Sawey}, P.~M.~J. \& {Berrington}, K.~A. 1993,
  {\href{https://dx.doi.org/10.1006/adnd.1993.1017}{\color{magenta}\adndt}},
  \href{https://ui.adsabs.harvard.edu/abs/1993ADNDT..55...81S}{55, 81}

\bibitem[{{Seaton}(1979)}]{Seaton1979a}
{Seaton}, M.~J. 1979, \mnras,
  \href{https://ui.adsabs.harvard.edu/abs/1979MNRAS.187..785S}{187, 785}

\bibitem[{{Shaw} \& {Kaler}(1989)}]{Shaw1989}
{Shaw}, R.~A. \& {Kaler}, J.~B. 1989,
  {\href{https://dx.doi.org/10.1086/191320}{\color{magenta}\apjs}},
  \href{https://ui.adsabs.harvard.edu/abs/1989ApJS...69..495S}{69, 495}

\bibitem[{{Smits}(1996)}]{Smits1996}
{Smits}, D.~P. 1996, \mnras,
  \href{https://ui.adsabs.harvard.edu/abs/1996MNRAS.278..683S}{278, 683}

\bibitem[{{Stancliffe} \& {Jeffery}(2007)}]{Stancliffe2007}
{Stancliffe}, R.~J. \& {Jeffery}, C.~S. 2007,
  {\href{https://dx.doi.org/10.1111/j.1365-2966.2006.11363.x}{\color{magenta}\mnras}},
  \href{https://ui.adsabs.harvard.edu/abs/2007MNRAS.375.1280S}{375, 1280}

\bibitem[{{Stanghellini} \& {Haywood}(2010)}]{Stanghellini2010}
{Stanghellini}, L. \& {Haywood}, M. 2010,
  {\href{https://dx.doi.org/10.1088/0004-637X/714/2/1096}{\color{magenta}\apj}},
  \href{https://ui.adsabs.harvard.edu/abs/2010ApJ...714.1096S}{714, 1096}

\bibitem[{{Stanghellini} {et~al.}(2003){Stanghellini}, {Shaw}, {Balick},
  {Mutchler}, {Blades}, \& {Villaver}}]{Stanghellini2003}
{Stanghellini}, L., {Shaw}, R.~A., {Balick}, B., {et~al.} 2003,
  {\href{https://dx.doi.org/10.1086/378042}{\color{magenta}\apj}},
  \href{https://ui.adsabs.harvard.edu/abs/2003ApJ...596..997S}{596, 997}

\bibitem[{{Stanghellini} {et~al.}(2002){Stanghellini}, {Shaw}, {Mutchler},
  {Palen}, {Balick}, \& {Blades}}]{Stanghellini2002}
{Stanghellini}, L., {Shaw}, R.~A., {Mutchler}, M., {et~al.} 2002,
  {\href{https://dx.doi.org/10.1086/341146}{\color{magenta}\apj}},
  \href{https://ui.adsabs.harvard.edu/abs/2002ApJ...575..178S}{575, 178}

\bibitem[{{Stanghellini} {et~al.}(2008){Stanghellini}, {Shaw}, \&
  {Villaver}}]{Stanghellini2008}
{Stanghellini}, L., {Shaw}, R.~A., \& {Villaver}, E. 2008,
  {\href{https://dx.doi.org/10.1086/592395}{\color{magenta}\apj}},
  \href{https://ui.adsabs.harvard.edu/abs/2008ApJ...689..194S}{689, 194}

\bibitem[{{Stasi{\'n}ska}(1990)}]{Stasinska1990a}
{Stasi{\'n}ska}, G. 1990, \aaps,
  \href{https://ui.adsabs.harvard.edu/abs/1990A%26AS...83..501S}{83, 501}

\bibitem[{{Stasi{\'n}ska}(2005)}]{Stasinska2005}
{Stasi{\'n}ska}, G. 2005,
  {\href{https://dx.doi.org/10.1051/0004-6361:20042216}{\color{magenta}\aap}},
  \href{https://ui.adsabs.harvard.edu/abs/2005A%26A...434..507S}{434, 507}

\bibitem[{{Stasi{\'n}ska} {et~al.}(2013){Stasi{\'n}ska}, {Pe{\~n}a},
  {Bresolin}, \& {Tsamis}}]{Stasinska2013}
{Stasi{\'n}ska}, G., {Pe{\~n}a}, M., {Bresolin}, F., \& {Tsamis}, Y.~G. 2013,
  {\href{https://dx.doi.org/10.1051/0004-6361/201220345}{\color{magenta}\aap}},
  \href{https://ui.adsabs.harvard.edu/abs/2013A%26A...552A..12S}{552, A12}

\bibitem[{{Stasi{\'n}ska} {et~al.}(1998){Stasi{\'n}ska}, {Richer}, \&
  {McCall}}]{Stasinska1998}
{Stasi{\'n}ska}, G., {Richer}, M.~G., \& {McCall}, M.~L. 1998, \aap,
  \href{https://ui.adsabs.harvard.edu/abs/1998A%26A...336..667S}{336, 667}

\bibitem[{{Storey} \& {Hummer}(1995)}]{Storey1995}
{Storey}, P.~J. \& {Hummer}, D.~G. 1995, \mnras,
  \href{https://ui.adsabs.harvard.edu/abs/1995MNRAS.272...41S}{272, 41}

\bibitem[{{Storey} \& {Sochi}(2013)}]{Storey2013}
{Storey}, P.~J. \& {Sochi}, T. 2013,
  {\href{https://dx.doi.org/10.1093/mnras/sts660}{\color{magenta}\mnras}},
  \href{https://ui.adsabs.harvard.edu/abs/2013MNRAS.430..599S}{430, 599}

\bibitem[{{Storey} \& {Sochi}(2015{\natexlab{a}})}]{Storey2015b}
{Storey}, P.~J. \& {Sochi}, T. 2015{\natexlab{a}},
  {\href{https://dx.doi.org/10.1093/mnras/stv484}{\color{magenta}\mnras}},
  \href{https://ui.adsabs.harvard.edu/abs/2015MNRAS.449.2974S}{449, 2974}

\bibitem[{{Storey} \& {Sochi}(2015{\natexlab{b}})}]{Storey2015a}
{Storey}, P.~J. \& {Sochi}, T. 2015{\natexlab{b}},
  {\href{https://dx.doi.org/10.1093/mnras/stu2243}{\color{magenta}\mnras}},
  \href{https://ui.adsabs.harvard.edu/abs/2015MNRAS.446.1864S}{446, 1864}

\bibitem[{{Storey} {et~al.}(2017){Storey}, {Sochi}, \& {Bastin}}]{Storey2017}
{Storey}, P.~J., {Sochi}, T., \& {Bastin}, R. 2017,
  {\href{https://dx.doi.org/10.1093/mnras/stx1189}{\color{magenta}\mnras}},
  \href{https://ui.adsabs.harvard.edu/abs/2017MNRAS.470..379S}{470, 379}

\bibitem[{{Storey} \& {Zeippen}(2000)}]{Storey2000}
{Storey}, P.~J. \& {Zeippen}, C.~J. 2000,
  {\href{https://dx.doi.org/10.1046/j.1365-8711.2000.03184.x}{\color{magenta}\mnras}},
  \href{https://ui.adsabs.harvard.edu/abs/2000MNRAS.312..813S}{312, 813}

\bibitem[{{Straniero} {et~al.}(1997){Straniero}, {Chieffi}, {Limongi}, {Busso},
  {Gallino}, \& {Arlandini}}]{Straniero1997}
{Straniero}, O., {Chieffi}, A., {Limongi}, M., {et~al.} 1997,
  {\href{https://dx.doi.org/10.1086/303794}{\color{magenta}\apj}},
  \href{https://ui.adsabs.harvard.edu/abs/1997ApJ...478..332S}{478, 332}

\bibitem[{{Tachiev} \& {Froese Fischer}(2001)}]{Tachiev2001}
{Tachiev}, G. \& {Froese Fischer}, C. 2001,
  {\href{https://dx.doi.org/10.1139/p01-059}{\color{magenta}\cajpj}},
  \href{https://ui.adsabs.harvard.edu/abs/2001CaJPh..79..955T}{79, 955}

\bibitem[{{Tachiev} \& {Froese Fischer}(2002)}]{Tachiev2002}
{Tachiev}, G.~I. \& {Froese Fischer}, C. 2002,
  {\href{https://dx.doi.org/10.1051/0004-6361:20011816}{\color{magenta}\aap}},
  \href{https://ui.adsabs.harvard.edu/abs/2002A&A...385..716T}{385, 716}

\bibitem[{{Tayal}(1997)}]{Tayal1997}
{Tayal}, S.~S. 1997,
  {\href{https://dx.doi.org/10.1006/adnd.1997.0753}{\color{magenta}\adndt}},
  \href{https://ui.adsabs.harvard.edu/abs/1997ADNDT..67..331T}{67, 331}

\bibitem[{{Tayal}(2006)}]{Tayal2006}
{Tayal}, S.~S. 2006,
  {\href{https://dx.doi.org/10.1086/499337}{\color{magenta}\apjs}},
  \href{https://ui.adsabs.harvard.edu/abs/2006ApJS..163..207T}{163, 207}

\bibitem[{{Tayal}(2011)}]{Tayal2011}
{Tayal}, S.~S. 2011,
  {\href{https://dx.doi.org/10.1088/0067-0049/195/2/12}{\color{magenta}\apjs}},
  \href{https://ui.adsabs.harvard.edu/abs/2011ApJS..195...12T}{195, 12}

\bibitem[{{Toal{\'a}} {et~al.}(2021){Toal{\'a}}, {Jim{\'e}nez-Hern{\'a}ndez},
  {Rodr{\'\i}guez-Gonz{\'a}lez}, {Estrada-Dorado}, {Guerrero},
  {G{\'o}mez-Gonz{\'a}lez}, {Ramos-Larios}, {Garc{\'\i}a-Hern{\'a}ndez}, \&
  {Todt}}]{Toala2021}
{Toal{\'a}}, J.~A., {Jim{\'e}nez-Hern{\'a}ndez}, P.,
  {Rodr{\'\i}guez-Gonz{\'a}lez}, J.~B., {et~al.} 2021,
  {\href{https://dx.doi.org/10.1093/mnras/stab593}{\color{magenta}\mnras}},
  \href{https://ui.adsabs.harvard.edu/abs/2021MNRAS.503.1543T}{503, 1543}

\bibitem[{{Tsamis} {et~al.}(2003{\natexlab{a}}){Tsamis}, {Barlow}, {Liu},
  {Danziger}, \& {Storey}}]{Tsamis2003a}
{Tsamis}, Y.~G., {Barlow}, M.~J., {Liu}, X.-W., {Danziger}, I.~J., \& {Storey},
  P.~J. 2003{\natexlab{a}},
  {\href{https://dx.doi.org/10.1046/j.1365-8711.2003.06972.x}{\color{magenta}\mnras}},
  \href{https://ui.adsabs.harvard.edu/abs/2003MNRAS.345..186T}{345, 186}

\bibitem[{{Tsamis} {et~al.}(2003{\natexlab{b}}){Tsamis}, {Barlow}, {Liu},
  {Danziger}, \& {Storey}}]{Tsamis2003}
{Tsamis}, Y.~G., {Barlow}, M.~J., {Liu}, X.-W., {Danziger}, I.~J., \& {Storey},
  P.~J. 2003{\natexlab{b}},
  {\href{https://dx.doi.org/10.1046/j.1365-8711.2003.06081.x}{\color{magenta}\mnras}},
  \href{https://ui.adsabs.harvard.edu/abs/2003MNRAS.338..687T}{338, 687}

\bibitem[{{Tsamis} {et~al.}(2004){Tsamis}, {Barlow}, {Liu}, {Storey}, \&
  {Danziger}}]{Tsamis2004}
{Tsamis}, Y.~G., {Barlow}, M.~J., {Liu}, X.-W., {Storey}, P.~J., \& {Danziger},
  I.~J. 2004,
  {\href{https://dx.doi.org/10.1111/j.1365-2966.2004.08140.x}{\color{magenta}\mnras}},
  \href{https://ui.adsabs.harvard.edu/abs/2004MNRAS.353..953T}{353, 953}

\bibitem[{{Tsamis} \& {P{\'e}quignot}(2005)}]{Tsamis2005}
{Tsamis}, Y.~G. \& {P{\'e}quignot}, D. 2005,
  {\href{https://dx.doi.org/10.1111/j.1365-2966.2005.09595.x}{\color{magenta}\mnras}},
  \href{https://ui.adsabs.harvard.edu/abs/2005MNRAS.364..687T}{364, 687}

\bibitem[{{Tsamis} {et~al.}(2008){Tsamis}, {Walsh}, {P{\'e}quignot}, {Barlow},
  {Danziger}, \& {Liu}}]{Tsamis2008}
{Tsamis}, Y.~G., {Walsh}, J.~R., {P{\'e}quignot}, D., {et~al.} 2008,
  {\href{https://dx.doi.org/10.1111/j.1365-2966.2008.13051.x}{\color{magenta}\mnras}},
  \href{https://ui.adsabs.harvard.edu/abs/2008MNRAS.386...22T}{386, 22}

\bibitem[{{Tylenda} {et~al.}(1993){Tylenda}, {Acker}, \&
  {Stenholm}}]{Tylenda1993}
{Tylenda}, R., {Acker}, A., \& {Stenholm}, B. 1993, \aaps,
  \href{https://ui.adsabs.harvard.edu/abs/1993A%26AS..102..595T}{102, 595}

\bibitem[{{Tylenda} {et~al.}(2003){Tylenda}, {Si{\'o}dmiak}, {G{\'o}rny},
  {Corradi}, \& {Schwarz}}]{Tylenda2003}
{Tylenda}, R., {Si{\'o}dmiak}, N., {G{\'o}rny}, S.~K., {Corradi}, R.~L.~M., \&
  {Schwarz}, H.~E. 2003,
  {\href{https://dx.doi.org/10.1051/0004-6361:20030645}{\color{magenta}\aap}},
  \href{https://ui.adsabs.harvard.edu/abs/2003A%26A...405..627T}{405, 627}

\bibitem[{{van der Hucht}(2001)}]{vanderHucht2001}
{van der Hucht}, K.~A. 2001,
  {\href{https://dx.doi.org/10.1016/S1387-6473(00)00112-3}{\color{magenta}\nar}},
  \href{https://ui.adsabs.harvard.edu/abs/2001NewAR..45..135V}{45, 135}

\bibitem[{{van der Hucht} {et~al.}(1981){van der Hucht}, {Conti}, {Lundstrom},
  \& {Stenholm}}]{vanderHucht1981}
{van der Hucht}, K.~A., {Conti}, P.~S., {Lundstrom}, I., \& {Stenholm}, B.
  1981, {\href{https://dx.doi.org/10.1007/BF00173260}{\color{magenta}\ssr}},
  \href{https://ui.adsabs.harvard.edu/abs/1981SSRv...28..227V}{28, 227}

\bibitem[{{Ventura} \& {D'Antona}(2005)}]{Ventura2005}
{Ventura}, P. \& {D'Antona}, F. 2005,
  {\href{https://dx.doi.org/10.1051/0004-6361:20041917}{\color{magenta}\aap}},
  \href{https://ui.adsabs.harvard.edu/abs/2005A%26A...431..279V}{431, 279}

\bibitem[{{Ventura} {et~al.}(2013){Ventura}, {Di Criscienzo}, {Carini}, \&
  {D'Antona}}]{Ventura2013}
{Ventura}, P., {Di Criscienzo}, M., {Carini}, R., \& {D'Antona}, F. 2013,
  {\href{https://dx.doi.org/10.1093/mnras/stt444}{\color{magenta}\mnras}},
  \href{https://ui.adsabs.harvard.edu/abs/2013MNRAS.431.3642V}{431, 3642}

\bibitem[{{Viegas} \& {Clegg}(1994)}]{Viegas1994}
{Viegas}, S.~M. \& {Clegg}, R.~E.~S. 1994, \mnras,
  \href{https://ui.adsabs.harvard.edu/abs/1994MNRAS.271..993V}{271, 993}

\bibitem[{{Virtanen} {et~al.}(2020){Virtanen}, {Gommers}, {Oliphant},
  {Haberland}, {Reddy}, {Cournapeau}, {Burovski}, {Peterson}, {Weckesser},
  {Bright}, {van der Walt}, {Brett}, {Wilson}, {Millman}, {Mayorov}, {Nelson},
  {Jones}, {Kern}, {Larson}, {Carey}, {Polat}, {Feng}, {Moore}, {VanderPlas},
  {Laxalde}, {Perktold}, {Cimrman}, {Henriksen}, {Quintero}, {Harris},
  {Archibald}, {Ribeiro}, {Pedregosa}, {van Mulbregt}, \& {SciPy 1. 0
  Contributors}}]{Virtanen2020}
{Virtanen}, P., {Gommers}, R., {Oliphant}, T.~E., {et~al.} 2020,
  {\href{https://dx.doi.org/10.1038/s41592-019-0686-2}{\color{magenta}Nature
  Methods}}, \href{https://ui.adsabs.harvard.edu/abs/2020NatMe..17..261V}{17,
  261}

\bibitem[{{Wang} \& {Liu}(2007)}]{Wang2007}
{Wang}, W. \& {Liu}, X.-W. 2007,
  {\href{https://dx.doi.org/10.1111/j.1365-2966.2007.12198.x}{\color{magenta}\mnras}},
  \href{https://ui.adsabs.harvard.edu/abs/2007MNRAS.381..669W}{381, 669}

\bibitem[{{Weidmann} {et~al.}(2008){Weidmann}, {Gamen}, {D{\'{\i}}az}, \&
  {Niemela}}]{Weidmann2008}
{Weidmann}, W.~A., {Gamen}, R., {D{\'{\i}}az}, R.~J., \& {Niemela}, V.~S. 2008,
  {\href{https://dx.doi.org/10.1051/0004-6361:200809989}{\color{magenta}\aap}},
  \href{https://ui.adsabs.harvard.edu/abs/2008A%26A...488..245W}{488, 245}

\bibitem[{{Werner} \& {Herwig}(2006)}]{Werner2006}
{Werner}, K. \& {Herwig}, F. 2006,
  {\href{https://dx.doi.org/10.1086/500443}{\color{magenta}\pasp}},
  \href{https://ui.adsabs.harvard.edu/abs/2006PASP..118..183W}{118, 183}

\bibitem[{{Wesson}(2016)}]{Wesson2016}
{Wesson}, R. 2016,
  {\href{https://dx.doi.org/10.1093/mnras/stv2946}{\color{magenta}\mnras}},
  \href{https://ui.adsabs.harvard.edu/abs/2016MNRAS.456.3774W}{456, 3774}

\bibitem[{{Wesson} {et~al.}(2008){Wesson}, {Barlow}, {Liu}, {Storey},
  {Ercolano}, \& {De Marco}}]{Wesson2008}
{Wesson}, R., {Barlow}, M.~J., {Liu}, X.-W., {et~al.} 2008,
  {\href{https://dx.doi.org/10.1111/j.1365-2966.2007.12683.x}{\color{magenta}\mnras}},
  \href{https://ui.adsabs.harvard.edu/abs/2008MNRAS.383.1639W}{383, 1639}

\bibitem[{{Wesson} {et~al.}(2018){Wesson}, {Jones}, {Garc{\'\i}a-Rojas},
  {Boffin}, \& {Corradi}}]{Wesson2018}
{Wesson}, R., {Jones}, D., {Garc{\'\i}a-Rojas}, J., {Boffin}, H.~M.~J., \&
  {Corradi}, R.~L.~M. 2018,
  {\href{https://dx.doi.org/10.1093/mnras/sty1871}{\color{magenta}\mnras}},
  \href{https://ui.adsabs.harvard.edu/abs/2018MNRAS.480.4589W}{480, 4589}

\bibitem[{{Wesson} \& {Liu}(2004)}]{Wesson2004}
{Wesson}, R. \& {Liu}, X.-W. 2004,
  {\href{https://dx.doi.org/10.1111/j.1365-2966.2004.07856.x}{\color{magenta}\mnras}},
  \href{https://ui.adsabs.harvard.edu/abs/2004MNRAS.351.1026W}{351, 1026}

\bibitem[{{Wesson} {et~al.}(2003){Wesson}, {Liu}, \& {Barlow}}]{Wesson2003}
{Wesson}, R., {Liu}, X.-W., \& {Barlow}, M.~J. 2003,
  {\href{https://dx.doi.org/10.1046/j.1365-8711.2003.06289.x}{\color{magenta}\mnras}},
  \href{https://ui.adsabs.harvard.edu/abs/2003MNRAS.340..253W}{340, 253}

\bibitem[{{Wesson} {et~al.}(2005){Wesson}, {Liu}, \& {Barlow}}]{Wesson2005}
{Wesson}, R., {Liu}, X.-W., \& {Barlow}, M.~J. 2005,
  {\href{https://dx.doi.org/10.1111/j.1365-2966.2005.09325.x}{\color{magenta}\mnras}},
  \href{https://ui.adsabs.harvard.edu/abs/2005MNRAS.362..424W}{362, 424}

\bibitem[{{Wyse}(1942)}]{Wyse1942}
{Wyse}, A.~B. 1942,
  {\href{https://dx.doi.org/10.1086/144409}{\color{magenta}\apj}},
  \href{https://ui.adsabs.harvard.edu/abs/1942ApJ....95..356W}{95, 356}

\bibitem[{{Yuan} {et~al.}(2011){Yuan}, {Liu}, {P{\'e}quignot}, {Rubin},
  {Ercolano}, \& {Zhang}}]{Yuan2011}
{Yuan}, H.-B., {Liu}, X.-W., {P{\'e}quignot}, D., {et~al.} 2011,
  {\href{https://dx.doi.org/10.1111/j.1365-2966.2010.17732.x}{\color{magenta}\mnras}},
  \href{https://ui.adsabs.harvard.edu/abs/2011MNRAS.411.1035Y}{411, 1035}

\bibitem[{{Zatsarinny} \& {Tayal}(2003)}]{Zatsarinny2003}
{Zatsarinny}, O. \& {Tayal}, S.~S. 2003,
  {\href{https://dx.doi.org/10.1086/377354}{\color{magenta}\apjs}},
  \href{https://ui.adsabs.harvard.edu/abs/2003ApJS..148..575Z}{148, 575}

\bibitem[{{Zeippen}(1982)}]{Zeippen1982}
{Zeippen}, C.~J. 1982,
  {\href{https://dx.doi.org/10.1093/mnras/198.1.111}{\color{magenta}\mnras}},
  \href{https://ui.adsabs.harvard.edu/abs/1982MNRAS.198..111Z}{198, 111}

\bibitem[{{Zhang} {et~al.}(2005){Zhang}, {Liu}, {Liu}, \& {Rubin}}]{Zhang2005a}
{Zhang}, Y., {Liu}, X.-W., {Liu}, Y., \& {Rubin}, R.~H. 2005,
  {\href{https://dx.doi.org/10.1111/j.1365-2966.2005.08810.x}{\color{magenta}\mnras}},
  \href{https://ui.adsabs.harvard.edu/abs/2005MNRAS.358..457Z}{358, 457}

\bibitem[{{Zhang} {et~al.}(2004){Zhang}, {Liu}, {Wesson}, {Storey}, {Liu}, \&
  {Danziger}}]{Zhang2004}
{Zhang}, Y., {Liu}, X.-W., {Wesson}, R., {et~al.} 2004,
  {\href{https://dx.doi.org/10.1111/j.1365-2966.2004.07838.x}{\color{magenta}\mnras}},
  \href{https://ui.adsabs.harvard.edu/abs/2004MNRAS.351..935Z}{351, 935}

\bibitem[{{Zhang} {et~al.}(2014){Zhang}, {Liu}, \& {Zhang}}]{Zhang2014}
{Zhang}, Y., {Liu}, X.-W., \& {Zhang}, B. 2014,
  {\href{https://dx.doi.org/10.1088/0004-637X/780/1/93}{\color{magenta}\apj}},
  \href{https://ui.adsabs.harvard.edu/abs/2014ApJ...780...93Z}{780, 93}

\bibitem[{{Zhang} {et~al.}(2016){Zhang}, {Zhang}, \& {Liu}}]{Zhang2016}
{Zhang}, Y., {Zhang}, B., \& {Liu}, X.-W. 2016,
  {\href{https://dx.doi.org/10.3847/0004-637X/817/1/68}{\color{magenta}\apj}},
  \href{https://ui.adsabs.harvard.edu/abs/2016ApJ...817...68Z}{817, 68}

\end{thebibliography}

\end{document}